%% file: main.tex
\documentclass[preprint2]{aastex631}
\usepackage{CJK}
\usepackage{textgreek}
\usepackage{hyperref}

\DeclareTextSymbolDefault{\dh}{T1}

\usepackage{pifont}
\newcommand{\checktikz}{\ding{51}}%

\shortauthors{Ca\~nas et al.}
\shorttitle{Low-mass KOI companions characterized with APOGEE-N}

\graphicspath{{./}{figures/}}

\newcommand{\UA}{Steward Observatory, The University of Arizona, 933 N.\ Cherry Avenue, Tucson, AZ 85721, USA}
\newcommand{\PSUAA}{Department of Astronomy \& Astrophysics, The Pennsylvania State University, 525 Davey Laboratory, University Park, PA 16802, USA}
\newcommand{\PSUCEHW}{Center for Exoplanets and Habitable Worlds, The Pennsylvania State University, 525 Davey Laboratory, University Park, PA 16802, USA}
\newcommand{\Princeton}{Department of Astrophysical Sciences, Princeton University, 4 Ivy Lane, Princeton, NJ 08540, USA}
\newcommand{\RUSSELL}{Henry Norris Russell Fellow}
\newcommand{\UVA}{Department of Astronomy, University of Virginia, Charlottesville, VA 22904, USA}
\newcommand{\Vanderbilt}{Department of Physics \& Astronomy, Vanderbilt University, Nashville, TN 37235, USA}
\newcommand{\GSFC}{NASA Goddard Space Flight Center, 8800 Greenbelt Road, Greenbelt, MD 20771, USA}

\newcommand{\figsetcapnum}{}
\newcommand{\figsetcaptitle}{}
\renewcommand{\figsetstart}{{\bf Fig. Set}}
\renewcommand{\figsetend}{}
\renewcommand{\figsetnum}[1]{{\bf #1.}}
\renewcommand{\figsettitle}[1]{ {\bf #1}}
\renewcommand{\figsetplot}[1]{\plotone{#1}}
\renewcommand{\figsetgrpnote}[1]{\caption{#1}}
\renewcommand{\figsetgrpnum}[1]{ \renewcommand{\figsetcapnum}{#1}}
\renewcommand{\figsetgrptitle}[1]{ \renewcommand{\figsetcaptitle}{#1} }
\renewcommand{\figsetgrpstart}{\begin{figure*}[!h]\renewcommand{\figurename}{Fig.}\renewcommand{\thefigure}{Set \figsetcapnum~$-$ \figsetcaptitle}
}
\renewcommand{\figsetgrpend}{\end{figure*}}
\begin{document}
\begin{CJK*}{UTF8}{gbsn}

\title{Characterization of low-mass companions to \textit{Kepler} objects of interest observed with APOGEE-N}
\correspondingauthor{Caleb I. Ca\~nas}
\email{c.canas@nasa.gov}

\author[0000-0003-4835-0619]{Caleb I. Ca\~nas}
\altaffiliation{NASA Postdoctoral Program Fellow}
\affiliation{\GSFC}
\affiliation{\PSUAA}
\affiliation{\PSUCEHW}

\author[0000-0003-4384-7220]{Chad F.\ Bender}
\affil{\UA}

\author[0000-0001-9596-7983]{Suvrath Mahadevan}
\affil{\PSUAA}
\affil{\PSUCEHW}
\affil{ETH Zurich, Institute for Particle Physics \& Astrophysics, Zurich, Switzerland}

\author[0000-0002-3601-133X]{Dmitry Bizyaev}
\affiliation{Apache Point Observatory and New Mexico State University, P.O. Box 59, Sunspot, NM, 88349-0059, USA}
\affiliation{Sternberg Astronomical Institute, Moscow State University, Moscow}

\author[0000-0002-3657-0705]{Nathan De Lee}
\affiliation{Department of Physics, Geology, and Engineering Technology, Northern Kentucky University, Highland Heights, KY 41099, USA}
\affiliation{\Vanderbilt}

\author[0000-0003-0556-027X]{Scott W. Fleming}
\affiliation{Space Telescope Science Institute, 3700 San Martin Dr., Baltimore, MD 21218, USA}

\author[0000-0002-1664-3102]{Fred Hearty}
\affil{\PSUAA}
\affil{\PSUCEHW}

\author[0000-0003-2025-3147]{Steven R. Majewski}
\affiliation{\UVA}

\author[0000-0003-4752-4365]{Christian Nitschelm}
\affiliation{Centro de Astronom{\'i}a (CITEVA), Universidad de Antofagasta, Avenida Angamos 601, Antofagasta 1270300, Chile}

\author[0000-0001-7240-7449]{Donald P. Schneider}
\affil{\PSUAA}
\affil{\PSUCEHW}

\author[0000-0001-7351-6540]{Javier Serna}
\affiliation{Instituto de Astronom\'{i}a, Universidad Aut\'{o}noma de M\'{e}xico, Ensenada, B.C, M\'{e}xico}

\author[0000-0002-3481-9052]{Keivan G. Stassun}
\affiliation{\Vanderbilt}

\author[0000-0001-7409-5688]{Gu\dh mundur Stef\'ansson}
\altaffiliation{\RUSSELL}
\affiliation{\Princeton}

\author[0000-0003-1479-3059]{Guy S. Stringfellow}
\affiliation{Center for Astrophysics and Space Astronomy, Department of Astrophysical and Planetary Sciences, University of Colorado, Boulder, CO 80309, USA}

\author{John C. Wilson}
\affiliation{\UVA}

\begin{abstract}

We report the characterization of 28 low-mass ($0.02\mathrm{~M_\odot}\le\mathrm{~M_{2}}\le0.25\mathrm{~M_\odot}$) companions to \textit{Kepler} objects of interest (KOIs), eight of which were previously designated confirmed planets. These objects were detected as transiting companions to Sun-like stars (G and F dwarfs) by the \textit{Kepler} mission and are confirmed as single-lined spectroscopic binaries in the current work using the northern multiplexed Apache Point Observatory Galactic Evolution Experiment near-infrared spectrograph (APOGEE-N) as part of the third and fourth Sloan Digital Sky Surveys. We have observed hundreds of KOIs using APOGEE-N and collected a total of 43,175 spectra with a median of 19 visits and a median baseline of $\sim1.9$ years per target. We jointly model the \textit{Kepler} photometry and APOGEE-N radial velocities to derive fundamental parameters for this subset of 28 transiting companions. The radii for most of these low-mass companions are over-inflated (by $\sim10\%$) when compared to theoretical models. Tidally locked M dwarfs on short period orbits show the largest amount of inflation, but inflation is also evident for companions that are well separated from the host star. We demonstrate that APOGEE-N data provides reliable radial velocities when compared to precise high-resolution spectrographs that enable detailed characterization of individual systems and the inference of orbital elements for faint ($H>12$) KOIs. The data from the entire APOGEE-KOI program is public and presents an opportunity to characterize an extensive subset of the binary population observed by \textit{Kepler}. 
\end{abstract}

\section{Introduction} \label{sec:intro}
Precise physical parameters, including the mass and radius, are important to understand the formation and evolution low-mass M dwarfs and brown dwarfs. M dwarfs are primary targets for recent spectroscopic and photometric surveys, such as Mearth \citep{Irwin2015}, NGTS \citep{Wheatley2018}, SPECULOOS \citep{Delrez2018}, CARMENES \citep{Reiners2018}, and the Habitable-zone Planet Finder \citep[HPF;][]{Mahadevan2012,Mahadevan2014}, because their smaller radii and masses relative to Sun-like stars yield deeper transits and larger RV amplitudes, which facilitates the detection of exoplanets in these systems. M dwarfs are also lucrative targets for the search for habitable planets because their habitable zones are closer-in than Sun-like stars \citep[e.g.,][]{Kopparapu2013,Kopparapu2017,Wandel2018}. 

The characterization of any exoplanetary system is fundamentally limited by how well the stellar parameters (including the mass, radius, and age) can be constrained. Accurate stellar radii and masses are required to derive accurate planet radii and masses, which are necessary to understand the population of exoplanets. Some examples which require precise planetary parameters include refining the mass-radius relationship for exoplanets \citep[e.g.,][]{Chen2017,Wolfgang2016,Kanodia2019,Ulmer-Moll2019}, inferring atmospheric properties \citep{Batalha2019}, or recovering the planetary interior structure \citep[e.g.,][]{Dorn2015,Otegi2020}. 

Brown dwarfs are objects with masses spanning $\sim13\mathrm{~M_J}<M<\sim80\mathrm{~M_J}$, where the lower and upper mass limits are the deuterium-burning and hydrogen-burning mass limits, respectively \citep[e.g.,][]{Chabrier2014,Baraffe2015}. The ``brown dwarf desert'' is the low occurrence rate ($\lesssim1\%$) of brown dwarfs as companions to Sun-like stars on close orbits ($<5$ au) and it has been extensively studied via radial velocity surveys \citep[e.g.,][]{Grether2006,Troup2016,Grieves2017,Triaud2017,Kiefer2019,Kiefer2021} and transit surveys \citep{Sahlmann2011,Csizmadia2016,Santerne2016}. This feature may be correlated with the transition in the formation processes of gas giants and stars but the limited number of brown dwarfs with precisely determined ($>3\sigma$) properties has prevented a detailed statistical analysis of the population \citep[e.g.,][]{Ma2014,Maldonado2017}. Additional brown dwarf systems with precise masses and radii to increase the population of transiting brown dwarfs will be useful to examine if formation mechanisms have a role in the observed desert. 

The physical parameters for low-mass stars and brown dwarfs are often derived using evolutionary models \citep[e.g.,][]{Baraffe2015} and the uncertainties in these models or relationships will extend to the derived planetary parameters. The theoretical mass-radius relationships for M dwarfs are known to be insufficient to accurately derive the parameters of M dwarfs for exoplanetary studies \citep[e.g.,][]{Lopez-Morales2007,Parsons2018} and have motivated empirical mass-radius relationships \citep[e.g.,][]{Stassun2012,Mann2015,Mann2019} for low-mass stars. Theoretical models for brown dwarfs have similarly shown scatter in the mass-radius relationship \citep[e.g.,][]{Burrows2011,Marley2021}. A larger sample of very low-mass M dwarfs and brown dwarfs with accurate fundamental parameters, including mass and radius, are required to reliably calibrate and improve the predictions from existing theoretical relationships. Direct measurements of these parameters are beneficial for our theoretical understanding of these stars and substellar companions and can constrain the set of plausible stellar evolutionary models \citep[e.g.,][]{Torres2010,Stevens2018}.

In this paper, we present stellar parameters for a subset of objects from a spectroscopic survey of \textit{Kepler} objects of interest (KOIs) outlined by \cite{Fleming2015} and conducted as part of the Sloan Digital Sky survey IV \citep[][]{Blanton2017}. Our sample includes brown dwarfs and fully convective M dwarf companions to F and G host stars. We present a summary of the spectroscopic survey, derive stellar and orbital parameters for the systems, and compare the fundamental parameters to the predictions of evolutionary models. The paper is structured as follows: Section \ref{sec:targets} describes the spectroscopic survey and the subset of targets presented in this work while Section \ref{sec:observations} provides a description of all observations. Section \ref{sec:modelfit} provides a discussion of the models and fit to photometry and RVs. In Section \ref{sec:modelcomp} we compare the derived properties to theoretical models for brown dwarfs and low-mass stars while in Section \ref{sec:discussion} we provide further discussion of the nature of these transiting brown dwarfs and M dwarfs. We end with a summary of our key results in Section \ref{sec:summary}.

\section{The APOGEE-KOI program} \label{sec:targets}
The \textit{Kepler} mission \citep{Borucki2010,Koch2010} was launched in 2009 to examine the frequency of Earth-sized exoplanets within the habitable zone of their host stars \citep[][]{Bryson2021}. It has revolutionized our understanding of exoplanets and stars by providing nearly continuous observations of \(\sim 200,000\) stars. The photometric observations of most stars were conducted in long cadence mode (\(\sim 29.4\) minutes) with exquisite precision \citep[\(\sim 80\) ppm for 6 hour timescales of Sun-like stars; see][]{Christiansen2012} that are well suited for planetary and asteroseismic analysis. In its final data release \citep[DR25;][]{Thompson2018}, \textit{Kepler} identified $>8000$ \textit{Kepler} objects of interest (KOIs), or periodic transit-like events that were most likely astrophysical in nature. 

To address the difficulty of removing false positive signals from genuine planetary candidates, candidate vetting and validation are used to identify the signals with the highest probability of being genuine planets. Vetting of KOIs can be performed without additional data by relying on high-quality space-based photometry \citep[e.g., the existence of a deep secondary eclipse, centroid offsets, or ellipsoidal variations,][]{Bryson2013,Mullally2016,Thompson2018} or with additional data from programs designed to characterize KOIs, including spectroscopic observations \citep[e.g.,][]{Ehrenreich2011,Matson2017,Petigura2017}, photometric deblending \citep[e.g.,][]{Torres2011,Kirk2016}, or adaptive-optic observations \citep[][]{Law2014,Ziegler2017,Ziegler2018a}. Candidate validation evaluates a transiting signal for common signs of false positive scenarios (e.g., background eclipsing binaries or hierarchical eclipsing binaries) using constraints from photometry or external datasets including RVs and high-contrast imaging. Algorithms designed to statistically validate KOIs include \texttt{BLENDER} \citep{Torres2011}, \texttt{VESPA} \citep{Morton2012}, and \texttt{PASTIS} \citep{Diaz2014a,Santerne2015}. 

Only \texttt{VESPA} has been applied to the entire KOI sample \citep{Morton2016} but not all validated signals are genuine planets \citep[e.g.,][]{Canas2018}. The mass of the transiting companion is one of the most important dynamical probes to determine if a KOI is a genuine planet or an astrophysical false positive \citep[a stellar binary, e.g.,][]{Fleming2015,Parviainen2019}. A traditional RV spectrograph observes one object at a time and limits surveys to either extensively observe a small subset of KOIs \citep[e.g.,][]{Santerne2016} or to sparsely observe ($\le 2$ epochs) a much larger population \citep[e.g.,][]{Petigura2017}. We designed the APOGEE-KOI program \citep[][]{Fleming2015} to address the bottleneck in spectroscopic follow-up of the KOI program. It was a pilot program under the third Sloan Digital Sky Survey \citep[SDSS-III;][]{Eisenstein2011} and an ancillary program as part of the fourth Sloan Digital Sky Survey \citep[SDSS-IV;][]{Blanton2017}. 

This program used the northern Apache Point Observatory (APO) Galactic Evolution Experiment (APOGEE) spectrograph \citep[APOGEE-N;][]{Wilson2012,Wilson2019} located on the Sloan 2.5-meter telescope \citep{Gunn2006} at APO. APOGEE-N was designed for the original APOGEE survey to conduct a detailed chemical and kinematic study of the galactic stellar population by observing \(\sim146,000\) red giant stars \citep{Majewski2017,Zasowski2017}. APOGEE-N is a multi-object, fiber-fed, near-infrared spectrograph capable of observing up to three hundred objects simultaneously at high-resolution (\(R\sim 22,500\)) in the H-band (\(1.514-1.696\) \textmu{}m). It achieves a radial velocity precision of \(\lesssim0.5~\mathrm{km~s^{-1}}\) for most stars with $H<15$ and an RV precision of $\sim100\mathrm{~m~s^{-1}}$ for stars with $H<11$ \citep[see][]{Joensson2020}. APOGEE-N achieves an average signal-to-noise ratio per pixel of \(100\) on a \(H = 11\) star in a single visit with one hour of total integration time. 

As discussed in \cite{Majewski2017}, the design of the APOGEE-N spectrograph and the bandpass enable various ancillary science projects within the survey. The field-of-view (FOV) for APOGEE-N has a radius of \(1.49^{\circ}\) which coincidentally provides an opportunity to use the multiplexing capabilities of APOGEE-N and simultaneously observe multiple KOIs from one \textit{Kepler} CCD module. A \textit{Kepler} CCD module is approximately the same size as the APOGEE-N FOV. The multiplexing also makes APOGEE-N more efficient \citep[$\sim3$ times more efficient;][]{Fleming2015} than high-precision single-target spectrographs at achieving an RV precision of $\gtrsim100\mathrm{~m~s^{-1}}$. The primary science goals of the APOGEE-KOI program are presented in \cite{Fleming2015} and include (i) dynamical vetting and refinement of KOIs, (ii) searching for any dependencies on disposition with stellar or candidate parameters, and (iii) improving our understanding of binarity rates and the effects of stellar multiplicity on the planet host population. 

\begin{figure*}[!ht]
\epsscale{1.15}
\plotone{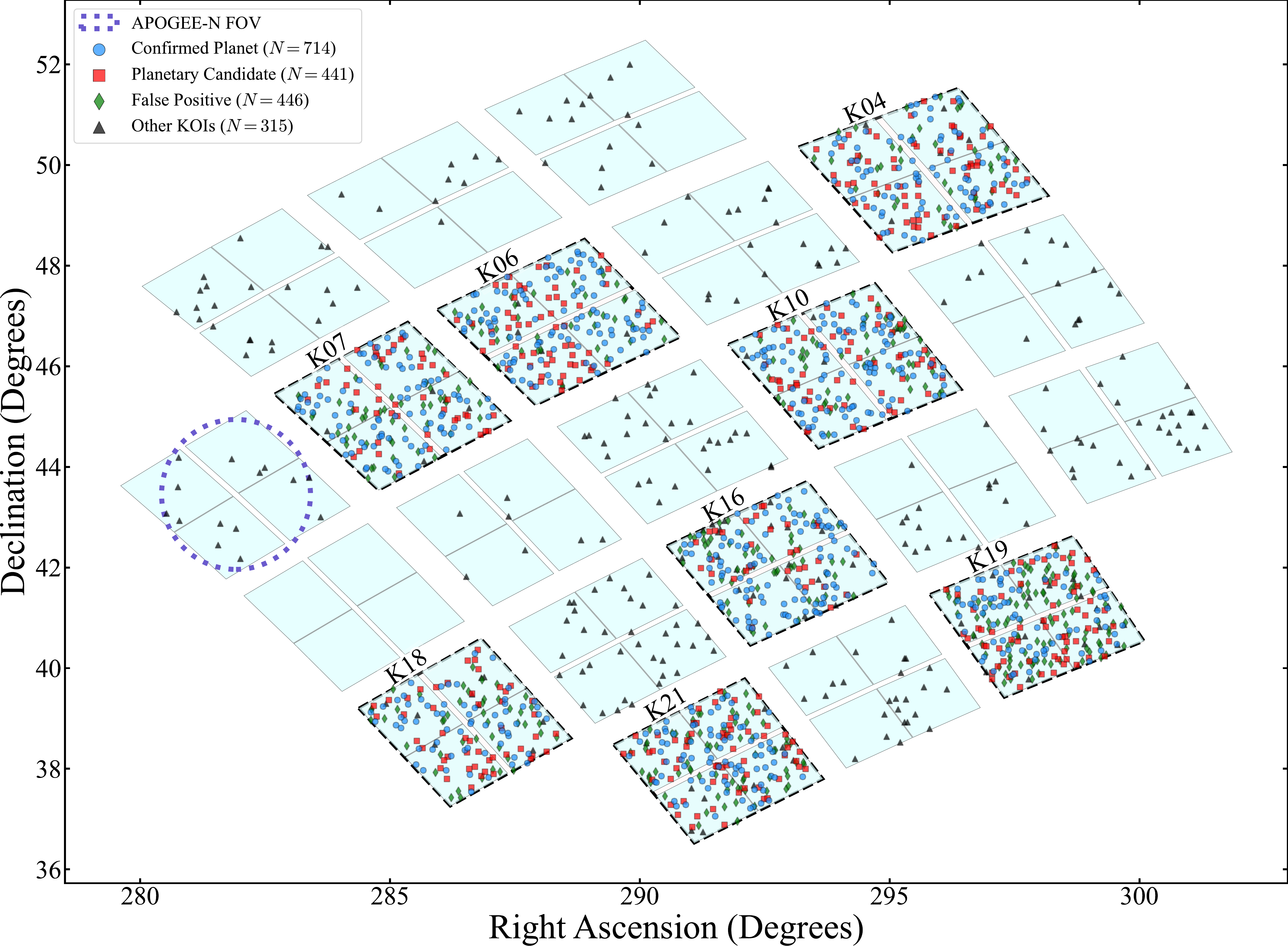}
\caption{A plot of the KOIs listed in Table \ref{tab:koilist} and observed by our program overlaid on the \textit{Kepler} footprint from \cite{Mullally2016a}. The APOGEE-KOI field is listed above the CCD modules from the \textit{Kepler} field observed as part of our program between $2013-2020$. The $1.49^\circ$ APOGEE-N FOV is plotted as a dotted circle. The marker reflects the DR25 disposition from the NASA Exoplanet Archive. The APOGEE-KOI program observed 714 confirmed planets (blue circles), 441 planetary candidates (red squares), and 446 false positives (green diamonds) for a median of 19 visits. For reference, KOIs not observed as part of this program but observed multiple times with APOGEE-N are shown as black triangles.}
\label{fig:koi}
\end{figure*}

The APOGEE-KOI program began in 2013 under SDSS-III and completed observations in 2020 as part of SDSS-IV. It targeted $\sim1600$ KOIs for a median of 19 epochs per target and a median baseline of 683 days (1.87 years). The \textit{Kepler} field is only accessible for observations over a limited range of local sidereal time (LST) for ground-based observations. The bulk of the observations with the APOGEE-N spectrograph for our program occurred between LSTs of $18-20$ hours where over-subscription with other APOGEE programs ultimately set the temporal baseline for the various APOGEE-KOI fields. 

The targets for the APOGEE-KOI program include confirmed planets, planetary candidates, and false positive systems. Figure \ref{fig:koi} displays the observed \textit{Kepler} footprint\footnote{\url{https://keplergo.github.io/KeplerScienceWebsite/the-kepler-space-telescope.html\#field-of-view}} along with the targets observed with APOGEE-N. The KOIs in these fields were chosen based on various characteristics. Selection of a KOI for the survey required (i) a disposition from the NASA Exoplanet Archive\footnote{The KOI catalogs used to identify targets were dated July 2013 for K16, K10, and K21, March 2016 for K04, February 2017 for K06 and K07, and April 2018 for K18 and K19.} of either planetary candidate, confirmed planet, or false positive, (ii) an H-band magnitude $<14$, (iii) a position within the field of view of the instrument (\(1.5^{\circ}\) from the center of each module), (iv) a position greater than 100\arcsec{} from the center of the plug plates, and (v) a separation from other targets larger than $\sim72$\arcsec. These requirements were placed to ensure targets (i) would have a signal-to-noise ratio (SNR) that could ensure single-visit precision $<0.5\mathrm{~km~s^{-1}}$, (ii) lie within the plate boundary, and (iii) that fibers were separated enough to physically allow for drilling the position on aluminum plug plates \citep[see][]{Owen1994}. Fiber collision errors (when the targets were separated by less than $\sim72$\arcsec) were avoided by prioritizing targets using \textit{Kepler} disposition and magnitude. Confirmed planets had the highest priority, followed by planetary candidates, and then false positive systems with a preference for cooler host stars. If any two conflicting targets had the same disposition, we preferentially selected the KOI with a brighter \textit{Kepler} magnitude. In each field we were able to accommodate all confirmed and planetary candidate KOIs, barring any fiber collisions.  This selection criteria provides $>200$ KOIs per plate with a small number of fibers ($10-30$) allocated to stars observed by \textit{Kepler} but not known to have transiting companions (no disposition in the respective KOI list), other APOGEE programs, or for telluric and sky calibration. We note that the selection criterion slightly changed as the APOGEE-KOI program progressed because the \textit{Kepler} team refined its KOI pipeline. 

CCD Module K16 was observed as part of SDSS-III and a description of the APOGEE-KOI program with preliminary results was reported by \cite{Fleming2015}. The observations for CCD module K16 began in September 2013 and were completed in June 2014 while observations for CCD modules K10 and K21 began in October 2014 and were completed in April 2017. Observations for CCD modules K04, K06, and K07 began in May 2017 and ended early 2019. An expansion of APOGEE-N programs \citep[see][]{Beaton2021} to fill an excess of bright time allowed for the inclusion of two additional fields, K18 and K19, which were observed between 2019 and 2020. The targets in our program can be identified in the DR17 \citep{Abdurrouf2022} \texttt{allStar} catalog\footnote{\url{https://data.sdss.org/datamodel/files/APOGEE\_ASPCAP/APRED\_VERS/ASPCAP\_VERS/allStar.html}} with the target flags \texttt{APOGEE\_RV\_MONITOR\_KEPLER} (for module K16) or \texttt{APOGEE2\_KOI} for all other CCD modules. For ease of reference, Table \ref{tab:koilist} provides a list of the stars, including non-KOIs, observed as part of the APOGEE-KOI program and various identifiers (2MASS, KIC, TIC, Gaia DR3). 

\begin{deluxetable*}{ccccccccccc}
{\tabletypesize{\scriptsize }
\rotate
\tablecaption{Stars observed by the APOGEE-KOI program. \label{tab:koilist}}
\tablehead{
\colhead{APOGEE ID} &
\colhead{KIC ID} &
\colhead{KOI}  &
\colhead{\textit{Kepler} Name$^a$}  &
\colhead{TIC}  &
\colhead{Gaia DR3}  &
\colhead{Visits} &
\colhead{Temporal Baseline (days)} &
\colhead{S/N$^b$} &
\colhead{Disposition$^c$} &
\colhead{Module$^d$}
}
\startdata
2M19494889+4100395 &   5812701 &    12 &    Kepler-448  &  169461816 &  2073860662955260416 &  18 &   768 &  109 &       CONFIRMED &  K19\\
2M19075308+4652061 &   9941662 &    13 &     Kepler-13  &  158324245 &  2130632159136095104 &  25 &  2664 &  155 &       CONFIRMED &  K06\\
2M18523991+4524110 &   9071386 &    23 &                &  164458426 &  2107001760170220032 &  18 &   683 &   68 &  FALSE POSITIVE &  K07\\
2M19170279+4748558 &  10593759 &    25 &                &  158984684 &  2127940314153495680 &  18 &   661 &   41 &  FALSE POSITIVE &  K06\\
2M19083956+3922369 &   4247791 &    28 &                &  121121622 &  2100409638208481536 &  15 &  1623 &  123 &  FALSE POSITIVE &  K18\\
2M19475229+4055363 &   5725087 &    33 &                &  168813467 &  2076775949311326592 &  20 &  1506 &  342 &  FALSE POSITIVE &  K19\\
2M19253263+4159249 &   6521045 &    41 &    Kepler-100  &  159654016 &  2101733244046205568 &  28 &   452 &  104 &       CONFIRMED &  K16\\
2M18523616+4508233 &   8866102 &    42 &  Kepler-410 A  &  164458714 &  2106904148451706752 &  23 &  2503 &  109 &       CONFIRMED &  K07\\
2M19285977+4609535 &   9527334 &    49 &    Kepler-461  &   63206513 &  2126801563705846528 &  20 &  1861 &   33 &       CONFIRMED &  K10\\
2M19254039+3840204 &   3544595 &    69 &     Kepler-93  &  137151335 &  2052747119115620352 &  21 &  1928 &  287 &       CONFIRMED &  K21\\
\multicolumn{8}{c}{$\vdots$}  \\
2M19554332+3959497 &   4866028 &       &                &  171500649 &  2073514248077061632 &   7 &  1116 &   95 &                 &  K19\\
2M19560588+4127518 &   6152072 &       &                &  171512615 &                      &   7 &  1116 &  112 &                 &  K19\\
2M19563075+4149418 &   6471048 &       &                &  171968838 &  2075370945254291200 &   7 &  2244 &   84 &                 &  K19\\
2M19563429+4014532 &   5130305 &       &                &  171878493 &  2073541387961862272 &   7 &  1116 &  120 &                 &  K19\\
2M19565344+4202296 &   6637066 &       &                &  268493357 &  2075380565981543680 &  29 &  2690 &  257 &                 &  K19\\
2M19570991+4022505 &   5219533 &       &                &  171977166 &  2073549432450598656 &  20 &  1506 &  208 &                 &  K19\\
2M19572477+4038084 &   5479821 &       &                &  171975736 &  2075057721882739456 &   7 &  1116 &  122 &                 &  K19\\
2M19580129+4139138 &   6315593 &       &                &  172378715 &  2075172345974035840 &  22 &  1506 &  179 &                 &  K19\\
2M19582041+4012465 &   5132589 &       &                &  172426403 &  2074282596228937856 &   7 &  1116 &  138 &                 &  K19\\
2M19582996+4037459 &   5481306 &       &                &  172423983 &  2075048376034233600 &  19 &   768 &    8 &                 &  K19\\
2M19583347+4028296 &   5309121 &       &                &  172424885 &  2075037621435984000 &   7 &  1116 &  132 &                 &  K19\\
\enddata
\tablenotetext{a}{The \textit{Kepler} name from the NASA Exoplanet Archive\footnote{\url{https://exoplanetarchive.ipac.caltech.edu/docs/API_keplernames_columns.html}}.}
\tablenotetext{b}{The median S/N per pixel of all visits from the \texttt{allVisit} catalog\footnote{\url{https://data.sdss.org/datamodel/files/APOGEE\_ASPCAP/APRED\_VERS/ASPCAP\_VERS/allVisit.html}}.}
\tablenotetext{c}{NASA Exoplanet Archive disposition of the first transiting signal from the supplemental DR25 table \footnote{\url{https://exoplanetarchive.ipac.caltech.edu/docs/PurposeOfKOITable.html\#q1-q17_sup_dr25}}.}
\tablenotetext{d}{The \textit{Kepler} CCD module as identified in Figure \ref{fig:koi}.}
}
\tablecomments{This table is published in its entirety in the machine-readable format and contains 2279 entries. A portion is shown here for guidance regarding its form and content.}
\end{deluxetable*}

\section{Observations} \label{sec:observations}
In this manuscript, we present an analysis of a subset of the APOGEE-KOI program: the 28 KOIs listed in Table \ref{tab:koiphot}. These KOIs were selected for further analysis based on a fit of the Keplerian radial velocity (RV) curve to the derived APOGEE-N RVs (see Section \ref{sec:apogeervs}) using \texttt{radvel} \citep{Fulton2018}. In this fit, the ephemeris was fixed to the value contained in the DR25 KOI catalog. Using the best-fitting orbital parameters and the DR25 stellar table provided by \cite{Mathur2017}, we estimated the $M_{2}\sin i$ for the observed KOI sample. The targets selected for analysis were KOIs found to have low-mass ($M_2<0.25~\mathrm{M_\odot}$) transiting companions that had (i) no indication of secondary light in the APOGEE-N spectra (e.g., no additional peaks in the cross-correlation function) and (ii) deep occultations identified by the \textit{Kepler} team. We selected this sample to ensure the system had a negligible flux ratio and could be modeled as a single-lined spectroscopic binary.

\subsection{Photometry with \textit{Kepler}}
\textit{Kepler} observed our targets for the entirety of the original mission in long-cadence mode (30 min cadence) with data from 2009 May 13 through 2013 May 11. Some targets were observed in short-cadence (2 min cadence) mode. A summary of the photometry used in this work is included in Table \ref{tab:koiphot}. The \textit{Kepler} team used a fully automated vetting pipeline \citep[see][]{Coughlin2016,Mullally2016,Twicken2016} designed to maximize the reliability of the final catalog. 

For the analysis in this manuscript, we use the entire pre-search data-conditioned \citep[PDCSAP;][]{Stumpe2012,Smith2012} light curves available at the Mikulski Archive for Space Telescopes (MAST). We use the PDCSAP light curves from all available quarters of \textit{Kepler} and exclude observations with non-zero data quality flags. These flags indicate poor-quality data due to conditions such as spacecraft events or cosmic ray hits and are described in the \textit{Kepler} Archive Manual \citep[see Table 2-3 in][]{Thompson2016}. We do not perform additional processing or apply outlier rejection beyond the data quality flags. The detrending of the raw photometry is described in detail in Appendix \ref{app:a}. General information for each target is listed in Table \ref{tab:koigeneral}.

\subsection{Adaptive optics imaging with Robo-AO}
23 KOIs (see Table \ref{tab:koigeneral}) were observed as part of the Robo-AO \textit{Kepler} planetary candidate survey \citep[][]{Law2014,Baranec2016,Ziegler2017,Ziegler2018a}.\footnote{Data is publicly available on \url{http://roboaokepler.com/}} These observations were performed using the Robo-AO laser adaptive optics system \citep{Baranec2013,Baranec2014} on the 2.1-m telescope at Kitt Peak National Observatory \citep{Jensen-Clem2018} with a 1.85-m circular aperture mask on the primary mirror. The typical seeing at the Kitt Peak Observatory is between $0.8-1.6\arcsec$, with a median seeing of $1.3\arcsec$ \citep{Jensen-Clem2018} while the typical diffraction limited FWHM resolution of the Robo-AO system is $0.15\arcsec$ \citep{Ziegler2018a}. Robo-AO observed most of these KOIs using a long-pass filter with a hard cut off at 600 nm that was designed to approximate the \textit{Kepler} bandpass at redder wavelengths and suppress blue wavelengths to minimize the impact on adaptive optics performance. KOI-1356 was observed using the Sloan $i\prime$ filter. The adaptive optics data set acquired by Robo-AO was reduced using the Robo-AO pipeline \citep{Law2014} which (i) performs PSF subtraction, (ii) perform an automated search for companions, and (iii) calculates constraints of the nearby star sensitivity with a $5\sigma$ contrast curve. \cite{Ziegler2018a} presents a detailed description of the data reduction pipeline. 

\subsection{Doppler spectroscopy with APOGEE-N} \label{sec:apogeervs}
For this work, we use the publicly available DR17 data\footnote{See \url{https://www.sdss.org/dr17/irspec/spectro_data/} to access the data for individual systems.}. The APOGEE data pipeline \citep{Nidever2015} performs sky subtraction, telluric and barycentric correction, and wavelength and flux calibration for each observation and has been shown to achieve a typical RV precision of $\sim100~\mathrm{m~s^{-1}}$ for most stars with $H<11$ \citep{Joensson2020}. In this work, we do not use archival DR17 RVs but instead derive RVs using the processed DR17 spectra (the \texttt{apVisit} files\footnote{See \url{https://data.sdss.org/datamodel/files/APOGEE_REDUX/APRED_VERS/visit/TELESCOPE/FIELD/PLATE_ID/MJD5/apVisit.html} for more information.}). While the APOGEE data pipeline provides radial velocity measurements, we performed additional post-processing on the spectrum to remove residual sky emission lines prior to analysis and derive RVs following the procedure described in \cite{Canas2019}. Briefly, we identified the best-fit synthetic spectrum by cross-correlating the highest S/N spectra using synthetic spectra generated from MARCs models \citep{Gustafsson2008} that were specifically generated for the APOGEE-N survey \citep[see][]{Meszaros2012,Zamora2015,Holtzman2018}. The best-matching synthetic spectrum with the largest cross-correlation value was used to derive the reported radial velocities. The values for the best-fitting spectra are reported in Table \ref{tab:koimarcsmods}. The uncertainties for each observation were calculated by following the maximum-likelihood approach presented by \cite{Zucker2003}. The median RV precision of our sample with APOGEE-N is 262 $\mathrm{m~s^{-1}}$. The derived RVs, the \(1\sigma\) uncertainties, and the S/N per pixel are presented in Table \ref{tab:koirvs}. 

\setcounter{table}{3}
\begin{deluxetable*}{lllcccc}
\tablecaption{MARCS template parameters used to derive APOGEE-N RVs. \label{tab:koimarcsmods}}
\tablehead{
\colhead{APOGEE ID} &
\colhead{KIC ID} &
\colhead{KOI ID}  &
\colhead{$T_e$} &
\colhead{$\log g_\star$} &
\colhead{[Fe/H]} &
\colhead{$v\sin i_\star^a$}
\\
\colhead{} & 
\colhead{} & 
\colhead{} &
\colhead{(K)} &
\colhead{(dex)} &
\colhead{(dex)} &
\colhead{($\mathrm{km~s~^{-1}}$)} 
}
\startdata
2M18523991+4524110 &   9071386 &    23 &  6250 & 4.5 & 0.25 & 17 \\
2M19395458+3840421 &   3558981 &    52 &  5250 & 4.5 & 0.25 & 18 \\
2M19480226+5022203 &  11974540 &   129 &  6250 & 4.5 & -0.25 & 23 \\
2M19492647+4025473 &   5297298 &   130 &  6000 & 4.5 & 0.0 & 0 \\
2M19424111+4035566 &   5376836 &   182 &  5750 & 4.5 & 0.0 & 22 \\
2M19485138+4139505 &   6305192 &   219 &  5750 & 4.5 & 0.0 & 8 \\
2M19223275+3842276 &   3642741 &   242 &  5750 & 4.5 & 0.25 & 6 \\
2M19073111+3922421 &   4247092 &   403 &  6000 & 4.5 & 0.0 & 11 \\
2M19331345+4136229 &   6289650 &   415 &  5750 & 4.5 & 0.0 & 5 \\
2M19043647+4519572 &   9008220 &   466 &  5750 & 4.5 & 0.0 & 0 \\
2M19214782+3951172 &   4742414 &   631 &  5500 & 4.5 & 0.25 & 0 \\
2M19371604+5004488 &  11818800 &   777 &  5250 & 4.0 & -0.5 & 13 \\
2M19473316+4123459 &   6061119 &   846 &  5750 & 4.5 & 0.0 & 8\\
2M19270249+4156386 &   6522242 &   855 &  5000 & 4.5 & -0.5 & 0 \\
2M19001520+4410043 &   8218274 &  1064 &  6500 & 4.5 & -0.25 & 46 \\
2M18535277+4503088 &   8801343 &  1247 &  6000 & 4.5 & -0.25 & 37 \\
2M19160484+4807113 &  10790387 &  1288 &  6200 & 5.0 & 0.25 & 0 \\
2M19320489+4230318 &   7037540 &  1347 &  6000 & 4.5 & -0.25 & 7 \\
2M19282877+4255540 &   7363829 &  1356 &  5750 & 4.5 & 0.25 & 5 \\
2M19460177+4927262 &  11517719 &  1416 &  5750 & 4.5 & -0.5 & 36 \\
2M19191325+4629301 &   9705459 &  1448 &  5250 & 4.0 & 0.0 & 21 \\
2M19344052+4622453 &   9653622 &  2513 &  5750 & 4.5 & -0.25 & 9 \\
2M19254244+4209507 &   6690171 &  3320 &  5000 & 4.0 & 0.0 & 6\\
2M19273337+3921423 &   4263529 &  3358 &  5250 & 4.5 & 0.0 & 7 \\
2M19520793+3952594 &   4773392 &  4367 &  6250 & 5.0 & 0.0 & 5 \\
2M19543478+4217089 &   6805414 &  5329 &  6000 & 4.5 & 0.0 & 11 \\
2M19480000+4117241 &   5979863 &  6018 &  5500 & 4.5 & -0.25 & 1 \\
2M19352118+4207199 &   6698670 &  6760 &  5500 & 4.5 & 0.0 & 7 \\
\enddata
\tablenotetext{}{The step sizes for $T_e$, $\log g_\star$, and [Fe/H] are from the APOGEE-N MARCS library grid ($\Delta T_e=250$ K, $\Delta \log g_\star=0.5$ dex, $\Delta\mathrm{[Fe/H]}=0.25$ dex) while search adopts $\Delta v\sin i_\star=1\mathrm{~km~s^{-1}}$. We note these are nominal values that are the best fit to the highest S/N observation and not to the co-added template used by ASPCAP to derive calibrated spectroscopic parameters.}
\tablenotetext{a}{We caution that values of $v\sin i_\star\lesssim8\mathrm{~km~s^{-1}}$ are below the APOGEE-N resolution limit \citep{Gilhool2018}.}
\end{deluxetable*}

\begin{deluxetable*}{lrcccc}
{\tabletypesize{\scriptsize }
\tablecaption{RVs of the KOI systems listed in Table \ref{tab:koiphot}. \label{tab:koirvs}}
\tablehead{
\colhead{$\mathrm{BJD_{TDB}}$}  &
\colhead{RV} &
\colhead{$\sigma$} & 
\colhead{S/N$^a$} &
\colhead{Instrument}\\
& 
\colhead{$(\mathrm{m~s^{-1}})$} & 
\colhead{$(\mathrm{m~s^{-1}})$} & 
&
\colhead{} & 
}
\startdata
\hline\multicolumn{5}{l}{\hspace{-0.2cm} KOI-23:} \\ 
~~2457879.87238 &   $-58634$ &   159 &   74 &  APOGEE-N \\
~~2457908.78365 &   $-71305$ &   199 &   53 &  APOGEE-N \\
~~2457918.77059 &   $-70905$ &   148 &   75 &  APOGEE-N \\
~~2457919.78831 &   $-50636$ &   173 &   60 &  APOGEE-N \\
~~2457920.79793 &   $-37443$ &   133 &  105 &  APOGEE-N \\
~~2457938.76514 &   $-46787$ &   181 &   51 &  APOGEE-N \\
~~2457940.71654 &   $-54188$ &   173 &   59 &  APOGEE-N \\
~~2457941.71519 &   $-72086$ &   163 &   78 &  APOGEE-N \\
~~2458007.74254 &   $-72976$ &   142 &   73 &  APOGEE-N \\
~~2458188.01380 &   $-38493$ &   179 &   48 &  APOGEE-N \\
~~2458209.98654 &   $-67836$ &   214 &   39 &  APOGEE-N \\
~~2458234.98148 &   $-37938$ &   157 &   62 &  APOGEE-N \\
~~2458237.92302 &   $-70895$ &   143 &   78 &  APOGEE-N \\
~~2458238.91768 &   $-50766$ &   157 &   74 &  APOGEE-N \\
~~2458261.83229 &   $-64566$ &   200 &   48 &  APOGEE-N \\
~~2458290.77684 &   $-45956$ &   149 &   74 &  APOGEE-N \\
~~2458385.59155 &   $-38125$ &   144 &   84 &  APOGEE-N \\
~~2458562.98452 &   $-46216$ &   134 &   74 &  APOGEE-N \\ \hline
\multicolumn{5}{c}{$\vdots$}  \\ \hline
\multicolumn{5}{l}{\hspace{-0.2cm} KOI-6760:} \\                                       
~~2456557.73352 &   $-41353$ &   120 &   72 &  APOGEE-N \\
~~2456559.72345 &   $-23038$ &   112 &   77 &  APOGEE-N \\
~~2456560.72118 &   $-19307$ &   116 &   77 &  APOGEE-N \\
~~2456584.63236 &   $-38816$ &   110 &   76 &  APOGEE-N \\
~~2456585.63086 &   $-46714$ &   118 &   74 &  APOGEE-N \\
~~2456757.89303 &   $-40635$ &   124 &   63 &  APOGEE-N \\
~~2456758.90238 &   $-47634$ &   512 &   13 &  APOGEE-N \\
~~2456760.90580 &   $-52692$ &   142 &   54 &  APOGEE-N \\
~~2456761.87290 &   $-50127$ &   385 &   16 &  APOGEE-N \\
~~2456762.86869 &   $-44404$ &   129 &   59 &  APOGEE-N \\
~~2456763.88121 &   $-35845$ &   134 &   52 &  APOGEE-N \\
~~2456783.83569 &   $-48448$ &   156 &   36 &  APOGEE-N \\
~~2456784.82203 &   $-42122$ &   143 &   44 &  APOGEE-N \\
~~2456785.82551 &   $-32454$ &   132 &   59 &  APOGEE-N \\
~~2456786.79852 &   $-23422$ &   139 &   52 &  APOGEE-N \\
~~2456787.80942 &   $-19312$ &   122 &   59 &  APOGEE-N \\
~~2456788.84316 &   $-25087$ &   183 &   25 &  APOGEE-N \\
~~2456812.74627 &   $-46955$ &   135 &   52 &  APOGEE-N \\
~~2456814.75554 &   $-52803$ &   128 &   54 &  APOGEE-N \\
~~2456815.78559 &   $-50712$ &   118 &   67 &  APOGEE-N \\
~~2456816.76635 &   $-45836$ &   212 &   26 &  APOGEE-N \\
~~2456817.76205 &   $-37487$ &   128 &   56 &  APOGEE-N \\
~~2456818.76465 &   $-27841$ &   123 &   59 &  APOGEE-N \\
~~2456819.76229 &   $-20577$ &   107 &   56 &  APOGEE-N \\
~~2456820.75608 &   $-21332$ &   108 &   67 &  APOGEE-N \\
\enddata
\tablenotetext{a}{All S/N estimates are per pixel. The APOGEE-N, SOPHIE, and HPF S/N are the median values per 1D extracted pixel at 1600nm (``green'' chip), 555nm (order index 26), and 1070nm (order index 18) with resolution elements of \(\sim2\) pixels, \(\sim5\) pixels, and \(\sim3\) pixels, respectively.}
}
\tablecomments{This table is published in its entirety in the machine-readable format. A portion is shown here for guidance regarding its form and content.}
\end{deluxetable*}

\subsection{High-resolution Doppler spectroscopy with SOPHIE}
KOI-129, KOI-219, KOI-415, KOI-466, KOI-855, and KOI-1288 were observed with the SOPHIE spectrograph as part of observations of the \textit{Kepler} field \citep{Ehrenreich2011}. SOPHIE is a cross-dispersed, environmentally stabilized echelle spectrograph covering the wavelength region of $3872-6943$ \AA{} that is located on the 1.93m telescope at the Observatoire de Haute Provence \citep{Bouchy2009,Perruchot2008}. The observations for these targets were acquired between 2013 June and 2018 September and were obtained using an exposure time of 1800s in high-efficiency mode, which provides a resolution of $R\sim40,000$. The high-efficiency mode collects 2.5 times more light than high-resolution mode. 

Briefly, the SOPHIE pipeline \citep{Bouchy2009} performs bias subtraction, optimal extraction using the Horne algorithm \citep{Horne1986}, cosmic-ray rejection, spectral flat-field correction, and wavelength-calibration. The wavelength-calibrated spectra are cross-correlated with a grid of numerical binary masks for various spectral types (F0, G2, K0, K5, M4) consisting of 1 and 0 value-zones, where the non-zero regions correspond to the theoretical positions and widths of the stellar absorption lines at zero velocity \citep[see further discussion in][]{Baranne1996,Pepe2002}. The reported RV is the minimum of the cross-correlation function and is determined by fitting a Gaussian function. The corresponding uncertainties are derived semi-empirically using the cross-correlation function and account for photon noise, uncertainties in the wavelength calibration, and systematic instrumental error. The RVs for KOI-415 were obtained from \cite{Moutou2013} while all other RVs were retrieved using the SOPHIE archive\footnote{\url{http://atlas.obs-hp.fr/sophie/}}. The median RV precision of our sample with SOPHIE is 31 $\mathrm{m~s^{-1}}$. The derived RVs, the \(1\sigma\) uncertainties, and the S/N per pixel at 555nm (order index 26) are presented in Table \ref{tab:koirvs}.

\subsection{High-resolution Doppler spectroscopy with HPF}
KOI-631 was observed with the HPF spectrograph between 2019 March 3 and 2019 July 17. HPF is a high-resolution ($R\sim55,000$), fiber-fed \citep{Kanodia2018a}, temperature controlled \citep{Stefansson2016},  near-infrared (\(\lambda\sim8080-12780\)\ \AA) spectrograph located on the 10m Hobby-Eberly Telescope \citep[HET;][]{Ramsey1998,Hill2021} at McDonald Observatory in Texas. Observations are executed in a queue by the HET resident astronomers \citep{Shetrone2007}.

We used the \texttt{HxRGproc} tool\footnote{\url{https://github.com/indiajoe/HxRGproc}} \citep{Ninan2018} to process the data and perform bias noise removal, nonlinearity correction, cosmic-ray correction, and slope/flux and variance image calculation. The one-dimensional spectra were extracted following the procedures in \cite{Ninan2018}, \cite{Kaplan2019}, and \cite{Metcalf2019}. The wavelength solution and drift correction were extrapolated using laser frequency comb (LFC) frames obtained from routine calibrations \citep[see Appendix A in][]{Stefansson2020}. The HPF RVs and the uncertainties were derived analogously to the APOGEE-N RVs albeit using the synthetic spectra from the PHOENIX-generated library by \cite{Husser2013}. We performed the (i) telluric correction using the \texttt{TERRASPEC} code \citep[see][]{Bender2012,Lockwood2014} and (ii) barycentric correction using \texttt{barycorrpy} \citep{Kanodia2018}, a Python implementation of the algorithms from \cite{wright2014}. The median RV precision of our sample with SOPHIE is 31 $\mathrm{m~s^{-1}}$. The derived RVs, the \(1\sigma\) uncertainties, and the S/N per pixel are presented in Table \ref{tab:koirvs}. 

\section{SED, photometric, and RV modeling} \label{sec:modelfit}
We employ the {\tt EXOFASTv2} analysis package \citep{Eastman2019} to jointly model the photometry, RVs, and the spectral energy density. The SED model uses the precomputed bolometric corrections\footnote{\url{http://waps.cfa.harvard.edu/MIST/model_grids.html\#bolometric}} from the MIST model grids \citep{Dotter2016,Choi2016}. \texttt{EXOFASTv2} models the RVs with a standard Keplerian curve in which the eccentricity and argument of periastron are modeled as $\sqrt{e}\sin\omega_\star$ and $\sqrt{e}\cos\omega_\star$. The light curves are modeled following the formalism from \cite{Mandel2002} with a quadratic limb-darkening law. For the long-cadence \textit{Kepler} photometry, we use the \texttt{longcadence} option to supersample the light curve model to the average of 10 samples over a duration of 29.425 minutes centered on the input time. The photometric model for targets with an occultation includes an additional component in the form of a thermal emission parameter and assumes uniform limb darkening for the occultation. \texttt{EXOFASTv2} accounts for light travel time in the models but ignores any relativistic effects. We use the default convergence criteria described in \cite{Eastman2019} for all systems.

The SED fit for each target used Gaussian priors on the (i) broadband photometry listed in Table \ref{tab:koibroadband}, (ii) host star spectroscopic parameters from the APOGEE Stellar Parameter and Chemical Abundances Pipeline \citep[ASPCAP;][]{GarciaPerez2016}, (iii) maximum visual extinction from estimates of Galactic dust extinction \citep{Green2018}, and (iv) distance estimates from \cite{Bailer-Jones2018}. The spectroscopic parameters from ASPCAP are derived using the combined visit APOGEE-N spectrum, empirically calibrated, and determined to be reliable \citep[see][]{Wilson2018,Joensson2020,Wilson2022}. The uncertainties for all calibrated spectroscopic parameters (\texttt{TEFF\_ERR}, \texttt{LOGG\_ERR}, and \texttt{FE\_H\_ERR} in the DR17 \texttt{allStar} file) are underestimated for our targets because the systematic errors (e.g., due to imperfect LSF-matching, synthetic model atmospheres, synthetic spectra, etc.) may be larger than the statistical errors \citep[see discussions in][]{GarciaPerez2016,Holtzman2015,Holtzman2018}. We employ the equations in Section 5.4 of \cite{Joensson2020} to estimate more reliable uncertainties for these spectroscopic parameters. 

We note that our models ignore the impact of ellipsoidal variations, reflected light, and star spots. These effects are presumed to be removed in the reduction of \textit{Kepler} photometry, which relies on a Gaussian process model to remove spot-induced photometric variability and does not incorporate a star spot model nor account for contamination from persistent spots or faculae that may be present in all transits \citep[e.g.,][]{Irwin2011a,McCullough2014,Rackham2017,Rackham2018}. The effect is most significant for active M dwarfs but for Sun-like stars it can produce contamination of tens of ppm \citep{Rackham2019} and for fairly active stars with evolving photometric variability (such as KOI-1416), the rotation signal can produce significant transit depth modulations \citep[e.g.,][]{Croll2015a,Pan2020}. The SED fit treats these systems as single-star systems even in the WISE bands and although these are single-lined binaries in the APOGEE-N $H$-band spectra, some contamination may occur in the infrared. The SED fit also relies on theoretical model grids for the SED analysis which provide another source of systematic uncertainty \citep[e.g.,][]{Serenelli2021,Dieterich2021} and, together with the simple treatment of photometric variability, may be a source of systematic uncertainties that are comparable to the quoted statistical uncertainties. 

\begin{figure*}[!ht]
    \figurenum{2a}
    \epsscale{1.15}
    \plotone{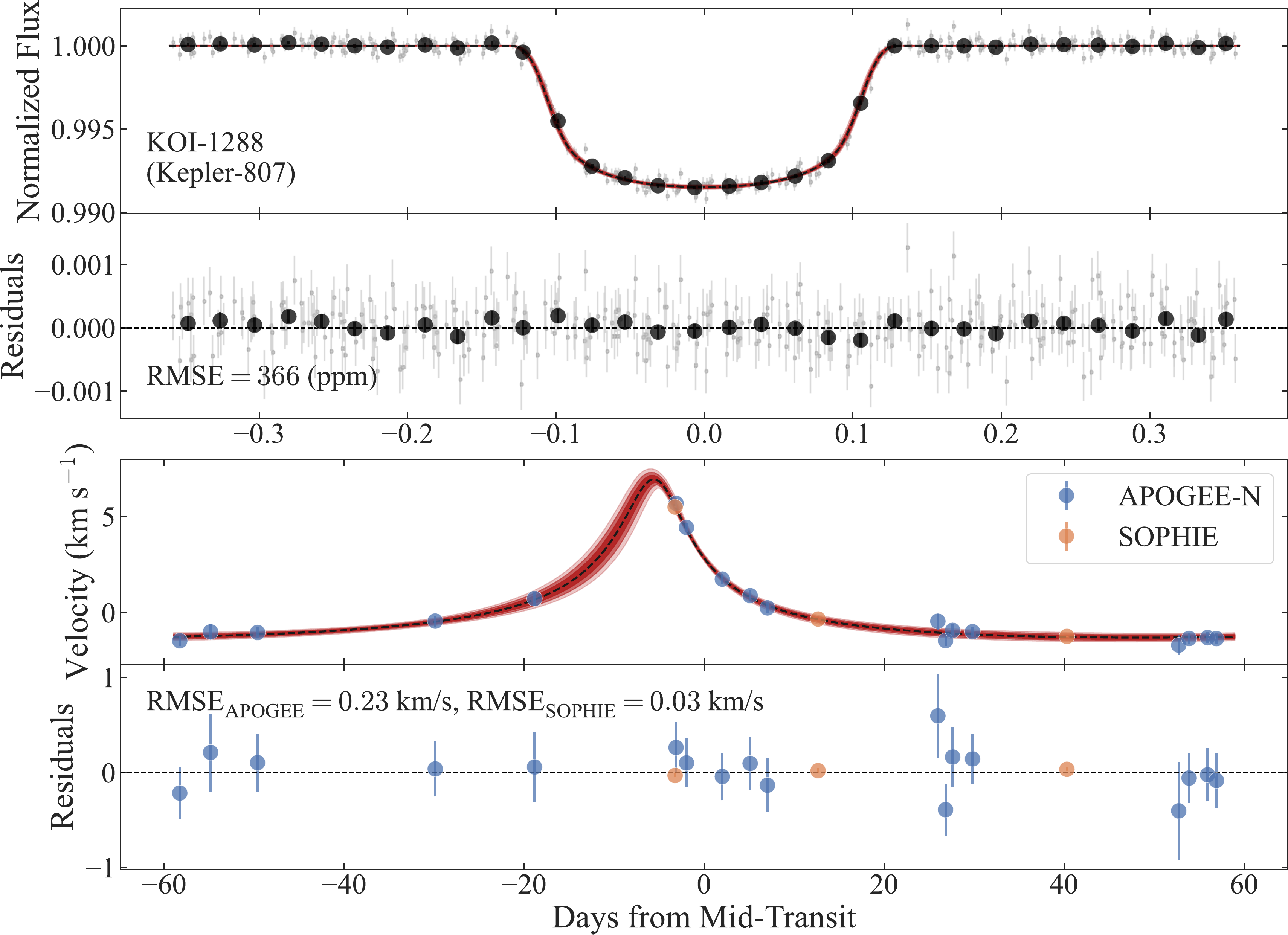}
    \caption{\textbf{Top}. The \textit{Kepler} photometry for KOI-1288 after phase-folding to the derived ephemeris. The large circles represent 30 min bins of the raw photometry. \textbf{Bottom}. The RVs after removing instrumental offsets and phase-folding the data to the derived ephemeris. In each panel, the $1\sigma$ (darkest), $2\sigma$, and $3\sigma$ (brightest) extent of the models are shown for reference. The complete figure set for the KOIs (28 images) is available in the online journal. 
\label{fig:keplerfit1}}
\end{figure*}

\begin{figure*}[!ht]
    \figurenum{2b}
    \epsscale{1.15}
    \plotone{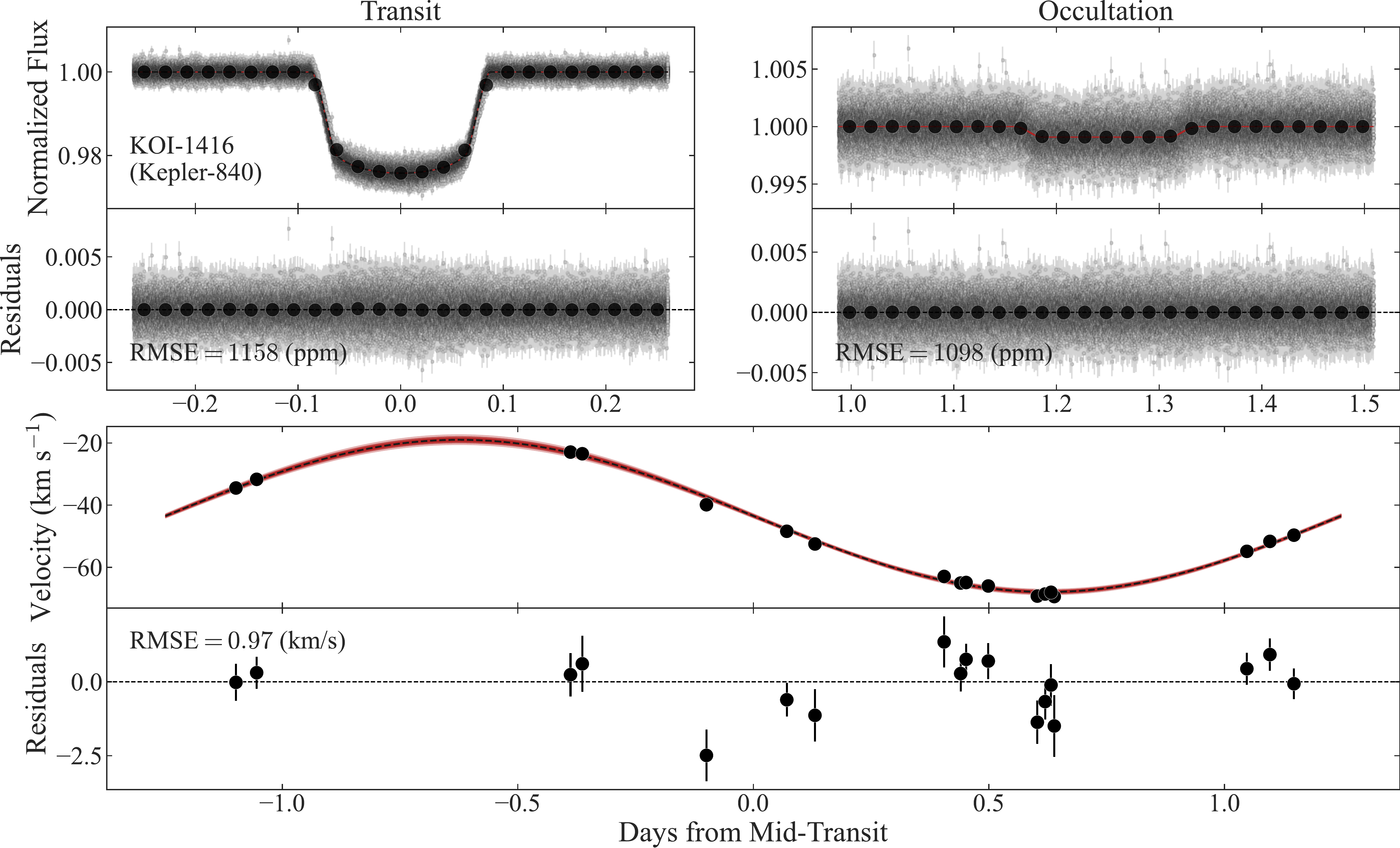}
    \caption{Figure 2 (cont.). \textbf{Top}. The \textit{Kepler} photometry for KOI-1416 centered on the transit and occultation after phase-folding to the derived ephemeris. The large circles represent 30 min bins of the raw data. \textbf{Bottom}. The APOGEE-N RVs after phase-folding the data to the derived ephemeris. In each panel, the $1\sigma$ (darkest), $2\sigma$, and $3\sigma$ (brightest) extent of the models are shown for reference. The complete figure set for the KOIs (28 images) is available in the online journal. \label{fig:keplerfit2}}
\end{figure*}

The derived stellar parameters are listed in Table \ref{tab:koistellar}, the model system parameters are listed in Table \ref{tab:koiorbital}\footnote{Our fits use the $\sqrt{e}\sin\omega_\star$ and $\sqrt{e}\cos\omega_\star$ parameterization; $e$ and $\omega_\star$ are not direct model parameters but derived values}, and the derived physical parameters are listed in Table \ref{tab:koiphysical}. Two examples of the model fit for the brown dwarf KOI-1288.01 and the M dwarf KOI-1416.01 are shown in Figures \ref{fig:keplerfit1} and \ref{fig:keplerfit2}, respectively.

\section{Comparison to theoretical models}\label{sec:modelcomp}
\begin{figure*}[!ht]
    \figurenum{3}
    \epsscale{1.15}
    \plotone{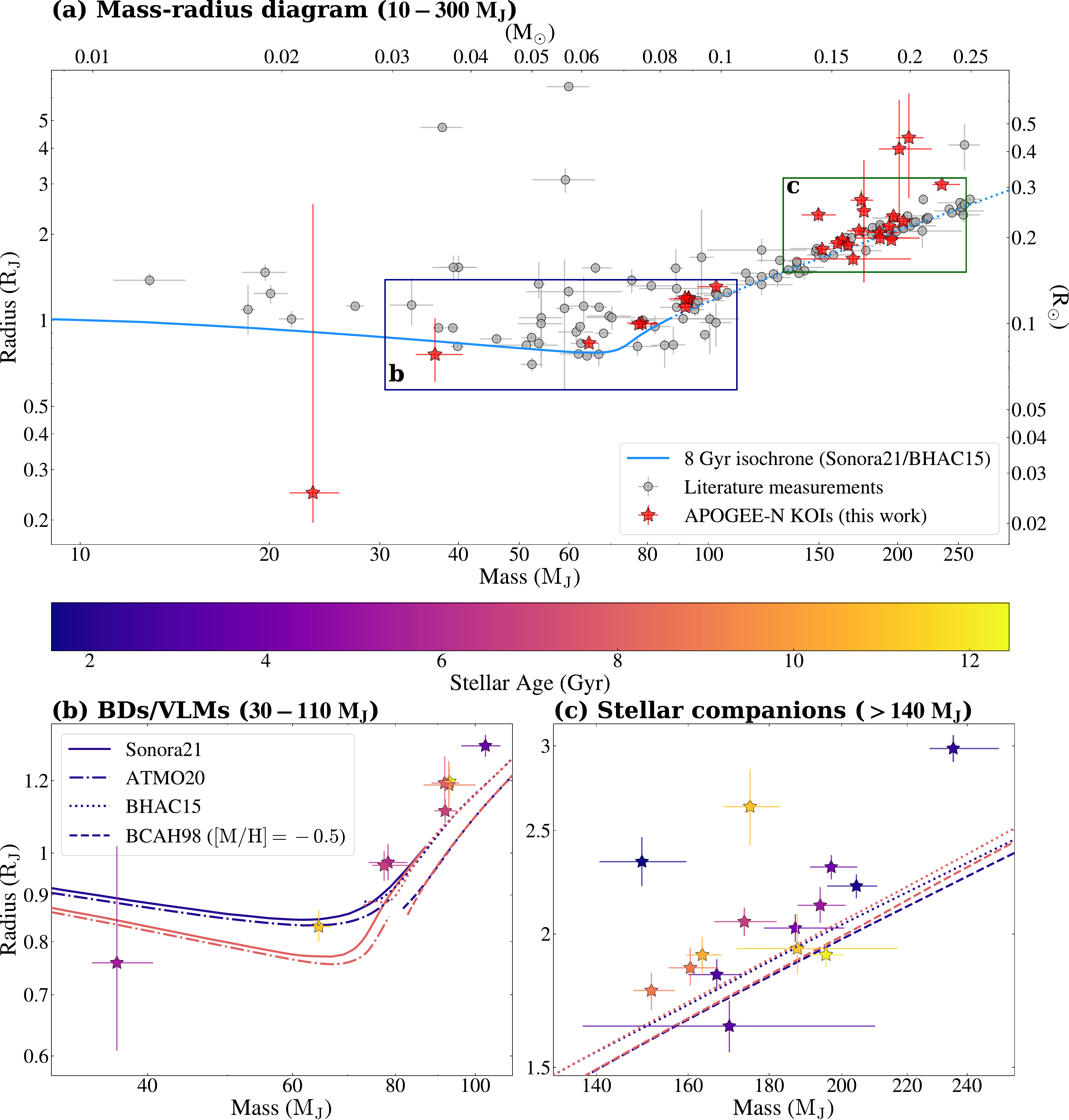}
    \caption{\textbf{(a)}. The mass-radius diagram spanning the regime of substellar companions and convective M dwarfs. The KOIs characterized in this work are shown as red stars while the published objects are grey circles. A nominal isochrone of 8 Gyrs (the median SED age for the 28 KOIs) is plotted for reference. 24 objects are non-grazing and have larger radii than theoretical models predict. The regions shown in panels (b) and (c) are marked with rectangles. \textbf{(b)}. The mass-radius diagram for the nine non-grazing KOI companions spanning $30\mathrm{~M_J}<M_{2}<110\mathrm{~M_J}$. \textbf{(c)} The mass-radius diagram for the 15 non-grazing KOI companions spanning $140\mathrm{~M_J}<M_{2}<250\mathrm{~M_J}$. The objects in panels (b) and (c) are colored based on the age estimated from the SED fit. Evolutionary tracks for 2 and 8 Gyrs from models by \cite{Marley2021} (Sonora21), \cite{Phillips2020} (ATMO20), \cite{Baraffe1998} (BCAH98), and \cite{Baraffe2015} (BHAC15) are plotted for reference and use the same color scale as the stars.  \label{fig:massrad}}
\end{figure*}

The derived masses ($M_2$) and radii ($R_2$) of the 28 KOI companions are displayed in Figure \ref{fig:massrad} along with the published set of masses and radii \citep[compiled from][]{Parsons2018,Chaturvedi2018,Carmichael2020,Grieves2021,Acton2021,vanRoestel2021,Canas2022a} for objects spanning $10-300~\mathrm{M_J}$ ($\sim0.10-0.29~\mathrm{M_\odot}$). We compare the measured masses and radii to the values predicted using evolution tracks from \cite{Marley2021} (Sonora21), \cite{Phillips2020} (ATMO20), \cite{Baraffe1998} (BCAH98) and \cite{Baraffe2015} (BHAC15). All evolutionary models are calculated for solar metallicity ($\mathrm{[M/H]}=0$) except the BCAH98 models which are calculated for metal-poor stars ($\mathrm{[M/H]}=-0.5$). 

There are four systems (KOI-403, KOI-2513, KOI-3358, KOI-4367) where the radius is poorly constrained ($\sigma_{R_2}>1~\mathrm{R_J}$) due to the grazing nature of the transits. Of the 24 remaining systems, all but four are larger than evolutionary tracks of comparable age would suggest. Figures \hyperref[fig:massrad]{\ref*{fig:massrad}(b)} and \hyperref[fig:massrad]{\ref*{fig:massrad}(c)} display the 24 non-grazing substellar and very-low mass ($10-110~\mathrm{M_J}$) and M dwarf ($>140~\mathrm{M_J}$) companions, respectively, colored to the median age derived from the SED fit. 

\begin{figure*}[!ht]
    \figurenum{4}
    \epsscale{1.15}
    \plotone{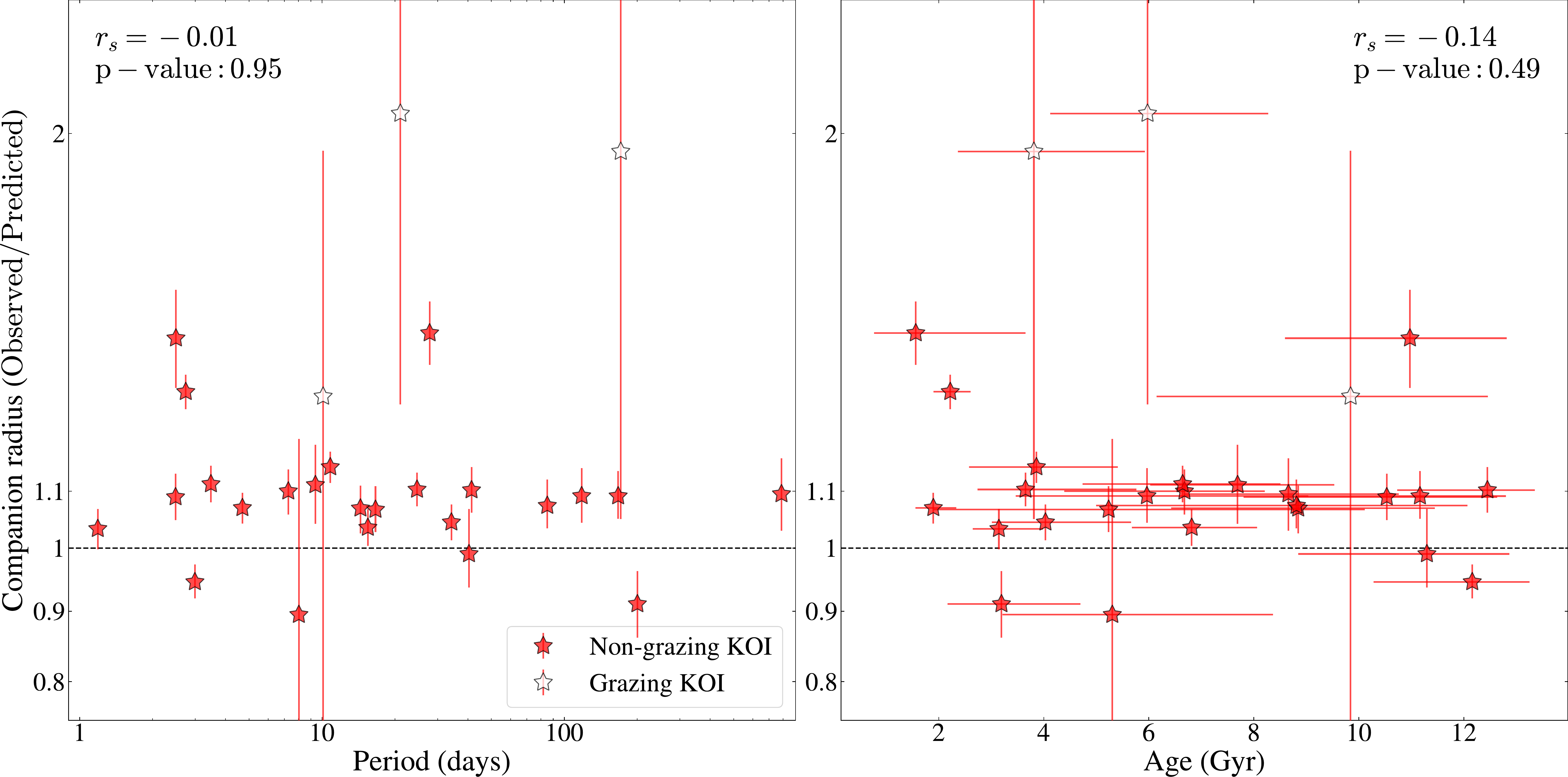}
    \caption{The ratio between the measured radius and the predicted radius from the evolutionary models of \cite{Baraffe2015} and \cite{Marley2021} as a function of the period (left panel) and stellar age (right panel). In each panel, non-grazing KOIs are plotted as red stars while grazing KOIs are shown as white stars. For clarity, KOI-2513 is not shown on this plot because of the large uncertainty on radius. Each system is compared to the prediction from evolutionary models at the median age determined from the SED fit. The Spearman correlation ($r_s$) and associated p-value are listed for each panel. On average, the observed radii are $9.1\%$ larger than the predicted radii. \label{fig:perrad}}
\end{figure*}

For the sample of APOGEE-N KOIs, the disagreement between predicted and observed radii is within $2\sigma$ of the measured radius. Regardless of the choice of evolutionary model (Sonora21, ATMO20, BCAH98, BHAC15), almost all non-grazing KOIs systems are discrepant by $9.1\%$ when compared to a track of comparable age. The M dwarf companions to KOI-846, KOI-1416, and KOI-1247 are inflated by more than $\sim10\%$ compared to the other M dwarfs in Figure \hyperref[fig:massrad]{\ref*{fig:massrad}(c)} and have $R_{2}>2.3~\mathrm{R_J}$. KOI-1247 is tidally locked and the light curve reveals ellipsoidal variations while KOI-1416 is almost tidally locked, such that tidal interactions with the respective host stars may be the source for inflation. Our photometric model ignores the effects of ellipsoidal variability such that the reported radii may be biased and underestimate the formal uncertainties. The \textit{Kepler} photometry for KOI-846 reveals a short rotation period (see Appendix \ref{app:rotation}) $<3$ days, which is smaller than the average rotation period of $\gtrsim4$ days for a G0/F8 dwarf seen in the \textit{Kepler} sample \citep[e.g.,][]{Nielsen2013,McQuillan2014}. The rapid rotation period is in agreement with the young age derived from the SED, suggesting that the companion may be a young, rapidly rotating, active M dwarf which are thought to be inflated due to magnetic activity \citep[][]{Morales2008,Jackson2018}. The amount of inflation for all other non-grazing systems is smaller than KOI-1247, KOI-1416, and KOI-846 but persists regardless of orbital period (see Figure \ref{fig:perrad}) such that tidal interactions cannot account for the inflation seen in the sample. Two systems, KOI-52 and KOI-5329, are $\sim7\%$ smaller than predicted but do not have any indication (e.g., a high RUWE, RV trend, or companions detected with Robo-AO) that an unresolved stellar companion could exist and be a source of dilution. The median offset of the APOGEE-N KOIs is $0.13~\mathrm{R_J}$ ($9.1\%$) and does not correlate with orbital period or stellar age (see Figure \ref{fig:perrad}). 

The inflation regardless of age or tidal interactions is in agreement with the observed discrepancy in measurements from detached eclipsing M dwarf binaries. \cite{Parsons2018} used a sample of detached eclipsing M dwarf binaries to demonstrate that the mass-radius relationship for M dwarfs contained a lot of ``scatter'' relative to theoretical predictions from evolutionary models by \cite{Baraffe2003} and \cite{Baraffe2015}. This discrepancy between theoretical models and measured values was first observed with high-precision data for low-mass eclipsing binaries \citep[e.g.,][]{Torres2002,Ribas2003,Lopez-Morales2005,Morales2009} where the measured radii were seen to exceed the predictions at fixed masses from evolutionary models by $5-10\%$. Additional mass and radius measurements of eclipsing M dwarf binaries \citep[e.g.,][]{Kraus2011,Birkby2012} and fully convective M dwarfs with measured rotation periods \citep[][]{Kesseli2018} similarly reported a radius inflation of $10-15\%$ relative to theoretical models that could not be ascribed to binarity or age. 

For the subset of short-period binary systems, tidal interactions with the host star are thought to be the source of the inconsistency with stellar models. Tidal interactions with a close stellar companion may enhance stellar activity due to magnetic locking, disk disruption, tidal effects, or angular momentum exchange \citep[][]{Lopez-Morales2007,Chabrier2007a,Morgan2012,Stassun2012}, which could increase star-spot coverage and/or increase the magnetic field strength. Strong magnetic fields could inhibit convective heat transport \citep[e.g.,][]{Chabrier2007a,Strassmeier2009,Feiden2013,Feiden2016}, provide magnetic pressure support \citep[e.g.,][]{MacDonald2017}, or result in dark magnetic spots \citep[e.g.,][]{Spruit1992,Chabrier2007a,Morales2010,Somers2015} which could serve to increase the stellar radius. Most of the KOIs analyzed in this work are well-separated ($a/R_1>10$) from their host stars (see Figure \ref{fig:perrad}) and are not tidally locked (see Appendix \ref{app:rotation}), such that enhanced magnetic activity induced due to binarity cannot explain the inflation. 

The stellar companions ($M_2\gtrsim80 \mathrm{M_J}$) to the KOIs in this work are mid- to late-M dwarfs and observations of such low-mass stars \citep[e.g.,][]{West2015,Newton2017} have shown this population can be magnetically active. In this low-mass regime, the inflation has also been shown to correlate with metrics of stellar and magnetic activity \citep[e.g.,][]{Stassun2012}. The scatter on the mass-radius diagram for these 28 KOI companions is comparable to previous work on M dwarf eclipsing binaries \citep[e.g.,][]{Parsons2018,Chaturvedi2018} and may be endemic to low-mass stars due to their intrinsic stellar activity. 
 
\section{Discussion}\label{sec:discussion}
\subsection{Detailing the five brown dwarfs observed}
The sample of 28 KOI systems includes 5 brown dwarf companions: KOI-2513, KOI-219, KOI-415, KOI-242, and KOI-1288. Two of these, KOI-415 \citep[][]{Moutou2013} and KOI-242 \citep[][]{Canas2018}, were previously published but are included in this work to provide updated system parameters. The other three objects are newly discovered substellar companions orbiting early G or late F dwarfs. These objects are described in greater detail below. 

\begin{itemize}
    \item \textbf{KOI-2513} is a faint ($V=14.99$, $H=13.45$), metal-poor ($\mathrm{[Fe/H]}=-0.37\pm0.02$) G dwarf with a grazing companion on a 19-day orbit detected by \textit{Kepler}. While the grazing nature precludes a precise measurement of the radius, it does allow us to constrain the mass. The Keplerian fit to the APOGEE-N RVs indicate the transiting companion is a low-mass brown dwarf with $M_{2}=23_{-2}^{+3}\mathrm{~M_J}$ on a slightly eccentric orbit ($e=0.38\pm0.04$). No bright ($\Delta\mathrm{mag}<3$) companions are detected within 4\arcsec{} using Robo-AO. The distance is poorly constrained by Gaia and the RUWE is similarly large. Future releases from Gaia will improve these measurements to determine if there may be an unresolved, faint ($\Delta\mathrm{mag}>3$) companion nearby \citep[e.g.,][]{Belokurov2020}.
    \item \textbf{KOI-219 (Kepler-494)} is a faint ($V=14.39$, $H=12.54$), metal-rich ($\mathrm{[Fe/H]}=0.016\pm0.02$) G dwarf with a transiting companion on an 8-day orbit detected by \textit{Kepler}. This object was statistically validated by \cite{Morton2016} and given the designation Kepler-494 b. The Keplerian fit to the APOGEE-N RVs indicate the transiting companion is an intermediate-mass brown dwarf with $M_{2}=37_{-2}^{+4}\mathrm{~M_J}$ on a slightly eccentric orbit ($e=0.26\pm0.02$). No bright ($\Delta\mathrm{mag}<4$) companions are detected within 4\arcsec{} using Robo-AO. Like KOI-2513, the distance is poorly constrained by Gaia and the RUWE is similarly large and future releases from Gaia will improve these measurements.
    \item \textbf{KOI-415} is a faint ($V=14.34$, $H=12.67$), metal-poor ($\mathrm{[Fe/H]}=-0.24\pm0.02$), slightly evolved G dwarf with a transiting companion on a 167-day orbit detected by \textit{Kepler}. This object was initially published as a brown dwarf by \cite{Moutou2013} using SOPHIE data. We update the orbit and physical parameters using a joint fit with \textit{Kepler}, APOGEE-N, and SOPHIE. KOI-415.01 has a mass of $M_{2}=65_{-1}^{+2}\mathrm{~M_J}$, a radius of $R_{2}=0.83^{+0.04}_{-0.03}\mathrm{~R_J}$, and is on a very eccentric orbit ($e=0.701\pm0.002$). No bright ($\Delta\mathrm{mag}<4$) companions are detected within 4\arcsec{} using Robo-AO.     
    \item \textbf{KOI-1288 (Kepler-807)} is a faint ($V=15.30$, $H=13.77$), solar metallicity ($\mathrm{[Fe/H]}=0.04\pm0.02$) late F dwarf with a transiting companion on a 118-day orbit detected by \textit{Kepler}. This object was statistically validated by \cite{Morton2016} and given the designation Kepler-807 b. The Keplerian fit to the APOGEE-N RVs indicate the transiting companion is near the hydrogen-burning mass limit and is either a high-mass brown dwarf or a very low-mass star with $M_{2}=78\pm4\mathrm{~M_J}$ on a very eccentric orbit ($e=0.69\pm0.01$). No bright ($\Delta\mathrm{mag}<3$) companions are detected within 4\arcsec{} using Robo-AO. 
    \item \textbf{KOI-242} is a faint ($V=15.01$, $H=13.14$), metal-rich ($\mathrm{[Fe/H]}=0.26\pm0.01$), slightly evolved G dwarf with a transiting companion on a 7.25-day orbit detected by \textit{Kepler}. This object was initially published as an object near the hydrogen-burning mass limit by \cite{Canas2018} using APOGEE-N data (from DR14). We update the orbit and physical parameters using a joint fit with \textit{Kepler} and the complete set of APOGEE-N RVs. KOI-242.01 has a mass of $M_{2}=77_{-3}^{+5}\mathrm{~M_J}$, a radius of $R_{2}=0.97 \pm 0.04\mathrm{~R_J}$, and is on a circular orbit ($e=0.007_{-0.005}^{+0.008}$). No bright ($\Delta\mathrm{mag}<3$) companions are detected within 4\arcsec{} using Robo-AO.
\end{itemize}

None of the aforementioned objects show secondary light in the APOGEE-N spectra. The host stars of these objects are evolving off the main sequence with KOI-242 being the only object formally on the subgiant branch. KOI-242.01 is the only brown dwarf in our sample found to have a circular orbit and this may be due to the subgiant nature of its host star ($a/R_\star<10$).

\subsection{Tidal evolution in low-mass binaries}

\begin{figure*}[!ht]
    \figurenum{5}
    \epsscale{1.15}
    \plotone{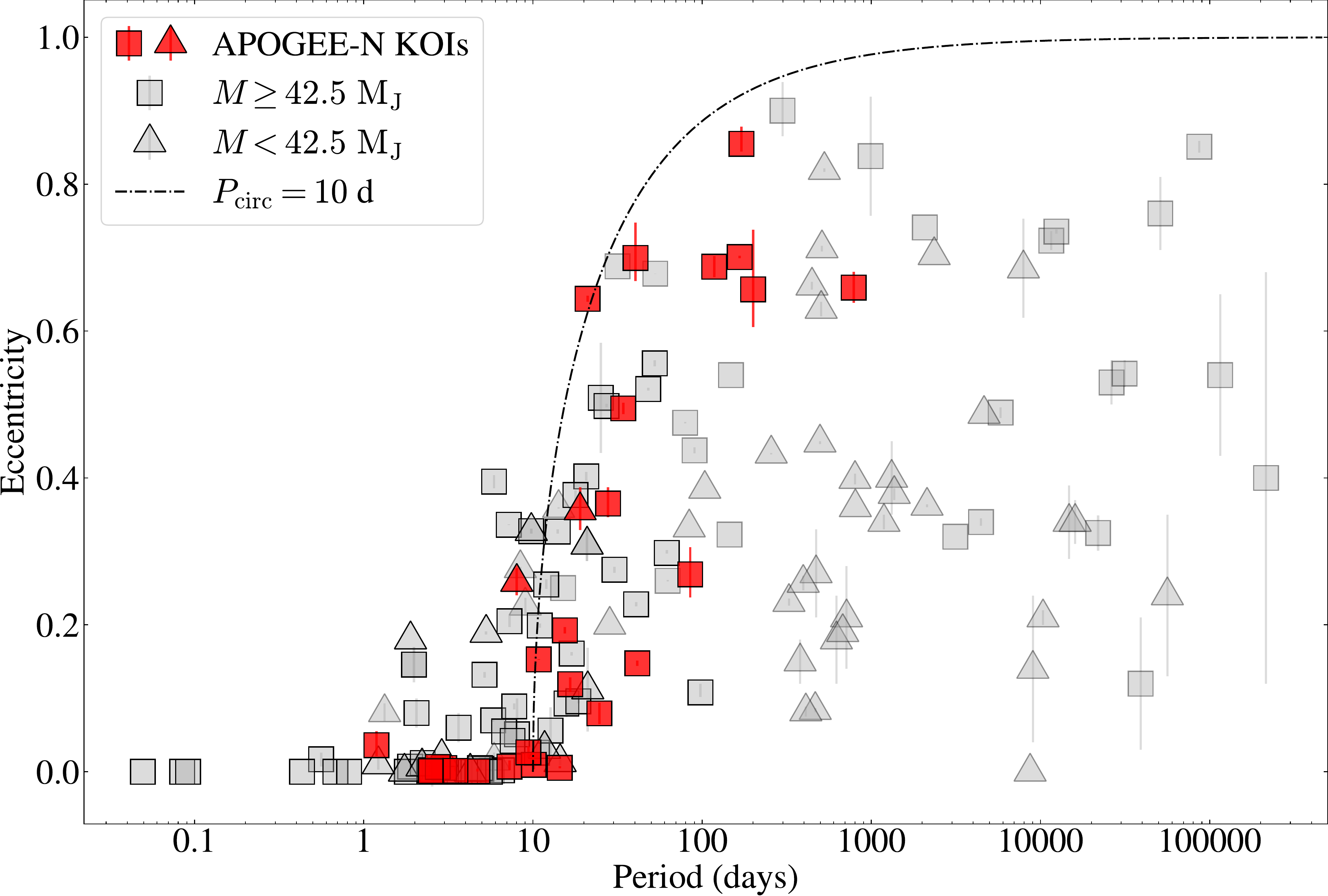}
    \caption{The eccentricity as a function of the period for published brown dwarf-hosting systems. Grey triangles denote the brown dwarfs with $M_{2}\sin i < 42.5\mathrm{~M_J}$ while grey squares are larger than this mass. The dashed line indicates the maximum eccentricity for systems unaffected by tides when adopting a circularization period \citep[see][]{Halbwachs2005} of 10 days. The KOIs are denoted as red squares or triangles, depending on the mass. \label{fig:perecc}}
\end{figure*}

A comparison of the KOI sample on the period-eccentricity diagram (see Figure \ref{fig:perecc}) with published substellar and M dwarf companions shows that the overall population is consistent with the circularization period of $\sim10$ days for nearby field binaries orbiting Sun-like stars \citep[e.g.,][]{Duquennoy1991,Meibom2005,Raghavan2010}. We compare the sample of KOIs with the maximum eccentricity a system unaffected by tides may have \citep{Halbwachs2005} assuming a circularization period of 10 days. With the exception of the intermediate-mass brown dwarf KOI-219.01, the KOI sample shows the expected circularization for short-period systems caused by tidal forces. The M dwarf companions ($M_{2}>140\mathrm{~M_J}$) to the KOIs in our sample adhere to the trend that short-period binary systems of comparable age are primarily in circular orbits.

An analysis of the brown dwarf population by \citep{Ma2014} showed different eccentricity distributions for low-mass and high-mass brown dwarfs and postulated the different eccentricity distributions may be a result of different formation mechanisms for two populations. The low-mass regime of brown dwarfs $M<42.5\mathrm{M_J}$ may form similar to planets and follow the eccentricity distribution of gas giant planets \citep[circularization periods $<5$ days;][]{Halbwachs2005,Pont2011,Bonomo2017} while a high-mass regime of brown dwarfs with $M>42.5\mathrm{M_J}$ may form like stellar binaries and adhere to the eccentricity distribution of field binary stars. The short period and high eccentricity of KOI-219 may simply be an imprint of migration and scattering after formation \citep[see][]{Whitworth2007,Chabrier2014,Whitworth2018,Forgan2018}. 

\subsection{Low-mass stellar companions masquerading as planets}
As part of this survey, we prioritized known planets \citep{Fleming2015} to complement other methods of statistical validation \citep[e.g.,][]{Torres2011,Morton2016}. Seven of the KOI companions (see Table \ref{tab:koiphot}) were previously listed as genuine planet candidates. These systems were statistically validated as part of a false-positive analysis of thousands of KOIs by \cite{Morton2016} using an algorithm \cite{Morton2012} designed to assess the probability a KOI is a genuine planet by simulating and determining the likelihood of a range of astrophysical false-positive scenarios, including background eclipsing binaries (BEBs), eclipsing binaries (EBs), and hierarchical eclipsing binaries (HEBs). The analysis by \cite{Morton2016} was performed prior to a Gaia data release such that the stellar parameters may have systematically underestimated the radius which is particularly relevant for the \textit{Kepler} field as it is known to contain a significant fraction of mildly evolved stars \citep{Bastien2014,Bastien2016}. 

Even with revised stellar parameters from Gaia \citep[e.g.,][]{Berger2020}, it is difficult to ascertain whether these systems are inflated Jupiters or genuine substellar and stellar companions. APOGEE-N data serves to eliminate the EB and HEB false positive scenarios while also placing limits on the presence of additional stellar companions. Data from the APOGEE-KOI program has already been used to vet a planetary system and confirm a unique example of a hot Jupiter with an interior super-Earth \citep[e.g.,][]{Canas2019} while in this work it serves to confirm the false positive nature of eight systems classified as confirmed planets. 

\subsection{Gaia metrics of the APOGEE-N KOIs}
The re-normalized unit weight error (RUWE) statistic from Gaia EDR3 has been shown to correlate with the existence of an unresolved stellar companion in recent studies of bright ($G<12$) stellar binaries \citep[e.g.,][]{Belokurov2020,Penoyre2020,Stassun2021}. The RUWE, or the square root of the reduced $\chi^2$ statistic that has been corrected for calibration errors, may be sensitive to the photocentric motions of unresolved objects \citep[see][]{Lindegren2021}. Stars with massive companions on orbital periods much shorter than the baseline of Gaia (34 months or 1035 days for EDR3) may exhibit deviations from a single-star astrometric solution that appears as noise as the primary star orbits the center of mass \citep[e.g.,][]{Kervella2019,Kiefer2019}. A threshold of \(\mathrm{RUWE}\gtrsim1.4\) has been suggested as an indicator of binarity \citep{Stassun2021,Kervella2022}.

Most of the KOIs analyzed (24 systems) have a RUWE below the threshold of 1.4, suggesting these systems do not appear discrepant from a single-star astrometric solution. KOI-219, KOI-2513, KOI-1247, and KOI-1416 are the four KOI systems with $\mathrm{RUWE}>1.4$. These systems have no indication of secondary light in the APOGEE-N spectra and no detection of bright companions from Robo-AO, such that the origin of the large RUWE may either be a spurious measurement due to the faintness of the objects or a system with a close (separations $<0.5\arcsec{}$), faint ($\Delta\mathrm{mag}>3$) on-sky stellar companion \citep[see examples in][]{Ziegler2020,Ziegler2021} that is beyond detection using existing data. 

In addition to bound companions, an excess RUWE may be caused by deformities between the model point-spread function and the observed image as a result of an unbound, unresolved companion \citep[e.g.,][]{Wood2021}, scattered light due to the presence of a circumstellar disk \citep[e.g.,][]{Fitton2022}, or stars with significant photometric or color variability \cite[e.g., RR Lyrae, long-period variable sources, young stars;][]{Belokurov2020,Lindegren2018a}. Both KOI-1416 and KOI-1247 have large radii and RUWE excess while also showing significant photometric variability in the \textit{Kepler} photometry. If the RUWE were due to an unresolved stellar companion, the transits would be diluted and the radius anomaly would not be as large when compared to the other KOI systems \citep[e.g.,][]{Ciardi2015,Bouma2018}. It may be that for systems with significant photometric variability, a RUWE excess does not necessitate an unresolved companion.

All of our targets are faint ($12<G<16$) and are located at a median distance of 1060 pc such that the astrometric amplitudes caused by the transiting companions are too small to be detected with the precision afforded by EDR3. Predictions for the detection of low-mass companions from the full and extended Gaia mission by \cite{Holl2022} suggest that Gaia will eventually have the precision to detect the presence of substellar companions down to $G\sim17$. The RVs from the APOGEE-KOI program will be useful to determine the orbit of long-period stellar companions that appear as trends in our program. 

\subsection{The synergy of APOGEE with other missions}
A subset of the 28 KOIs observed with APOGEE-N also have observations from other high-resolution spectrographs: six KOIs were observed with SOPHIE and one system with HPF. In either case, the joint fit with APOGEE-N demonstrates that, while APOGEE-N does not have the precision afforded by these instruments, the APOGEE-N RVs are reliable even for faint targets (median of $H=13$ for the sample observed by SOPHIE) and enable the determination of accurate orbital elements. For KOI-415 and KOI-855, the two systems with the largest number of observations with a high-precision instrument, the orbital elements derived from APOGEE-N and SOPHIE separately are within their respective $1\sigma$ uncertainties. While the precision afforded by SOPHIE and HPF RVs can enable tighter constraints on the derived orbital elements, obtaining as many visits for thousands of KOIs with these or similar high-precision spectrographs is unfeasible. The precision, cadence, and baseline afforded by \textit{Kepler} data and the multiplexing capabilities and high-efficiency of APOGEE-N present an opportunity to reliably analyze individual KOI systems and infer precise orbital elements solely from a joint fit with APOGEE-N RVs.

The objects presented in this work represent a small subset of the science enabled by the APOGEE-KOI program. We have shown that observations from the APOGEE-KOI program complement sparse observations of individual systems from high-precision spectroscopic surveys of the \textit{Kepler} field \citep[e.g.,][]{Ehrenreich2011}. Work by \cite{Wilson2022} has analyzed the sample as a population to leverage the abundances derived with ASPCAP and place constraints on the trends evident in the observed \textit{Kepler} planetary population. The APOGEE-KOI program also targeted a large fraction of KOIs designated as false positive in DR25, where the transit depth or the presence of ellipsoidal variations or a deep occultation reveal the stellar nature of the transiting companion. With the number of APOGEE-N RVs, it should be possible to study the population of single-lined and double-lined eclipsing binaries to revisit tidal evolution in the population of Sun-like stars, similar to the work by \cite{Triaud2017}. 

The subset of double-lined eclipsing binaries observed by APOGEE-N also enables model-independent mass and radii measurements for a large number of systems spanning all spectral types of MKGF dwarfs. In this work, we have ignored the effect of out-of-transit variability in the form of beaming, ellipsoidal variations, and reflection \citep[BEER,][]{Faigler2011,Engel2020} for short-period binaries, but a joint BEER analysis with RVs can facilitate very precise mass measurements of even non-transiting companions \citep[e.g.,][]{Tal-Or2015}. 

Beyond the \textit{Kepler} field, APOGEE-N has targeted various K2 \citep{Howell2014} campaigns and the Transiting Exoplanet Survey Satellite \citep[TESS;][]{Ricker2015} continuous-viewing zones. For many transiting candidates from these missions there may exist a few observations of K2OIs or TOIs where serendipitous APOGEE-N observations can effectively vet a candidate signal. In the context of TESS, the ongoing search for transits \cite[e.g.,][]{Jenkins2016,Huang2020,Huang2020b} will benefit from the coverage of APOGEE. RVs from APOGEE-N have already been used in tandem with data from other spectrographs to confirm a transiting hot Jupiter \citep{Canas2019a} and a transiting substellar companion \citep{Canas2022a}. An extensive list of ancillary science projects beyond planets can be found in \cite{Beaton2021}.

\section{Summary}\label{sec:summary}
In this manuscript we present an analysis of a subset of 28 KOIs observed by the APOGEE-KOI program and characterize these low-mass transiting companions using \textit{Kepler} photometry and RVs from APOGEE-N. We leverage the precision afforded by \textit{Kepler} and APOGEE-N RVs to derive masses and radii for these companions and compare this population to existing observations of brown dwarfs and M dwarfs. We find our sample is slightly inflated ($9.1\%$) relative to the radius predicted from stellar evolutionary models evaluated at the age determined from an SED fit, and this inflation does not correlate with period or stellar age. This inflation is consistent with previous extensive studies of brown dwarfs and eclipsing M dwarf binaries and we hypothesize it may be a result of stellar activity (e.g., strong magnetic fields or dark star spots) in the very low-mass companion. The sample also adheres to the circularization trends seen in field eclipsing binaries. The systems observed in this work represent a small subset of the APOGEE-KOI program and we highlight the utility of APOGEE RVs to complement data from high-precision RV surveys, validation of systems in transiting surveys, and eventually any detections of substellar and low-mass companions from the full Gaia mission. 

\section*{Acknowledgments}
We thank the anonymous referee for detailed and thoughtful feedback, which has improved the quality of this manuscript. CIC acknowledges support by NASA Headquarters under the NASA Earth and Space Science Fellowship Program through grant 80NSSC18K1114, the Alfred P. Sloan Foundation's Minority Ph.D. Program through grant G-2016-20166039, the Pennsylvania State University's Bunton-Waller program, and an appointment to the NASA Postdoctoral Program at the Goddard Space Flight Center, administered by USRA through a contract with NASA. 
The Center for Exoplanets and Habitable Worlds is supported by the Pennsylvania State University and the Eberly College of Science.
The computations for this research were performed on the Pennsylvania State University's Institute for Computational and Data Sciences' Roar supercomputer, including the CyberLAMP cluster supported by NSF grant MRI-1626251. This content is solely the responsibility of the authors and does not necessarily represent the views of the Institute for Computational and Data Sciences.

The Pennsylvania State University campuses are located on the original homelands of the Erie, Haudenosaunee (Seneca, Cayuga, Onondaga, Oneida, Mohawk, and Tuscarora), Lenape (Delaware Nation, Delaware Tribe, Stockbridge-Munsee), Shawnee (Absentee, Eastern, and Oklahoma), Susquehannock, and Wahzhazhe (Osage) Nations.  As a land grant institution, we acknowledge and honor the traditional caretakers of these lands and strive to understand and model their responsible stewardship. We also acknowledge the longer history of these lands and our place in that history.

We acknowledge support from NSF grants AST 1006676, AST 1126413, AST 1310875, AST 1310885, AST 2009554, AST 2009889, AST 2108512, AST 2108801 and the NASA Astrobiology Institute (NNA09DA76A) in our pursuit of precision RVs in the near-infrared. We acknowledge support from the Heising-Simons Foundation via grant 2017-0494.

Some of these results are based on observations obtained with HPF on the HET. The HET is a joint project of the University of Texas at Austin, the Pennsylvania State University, Ludwig-Maximilians-Universit\"at M\"unchen, and Georg-August Universit\"at Gottingen. The HET is named in honor of its principal benefactors, William P. Hobby and Robert E. Eberly. The HET collaboration acknowledges the support and resources from the Texas Advanced Computing Center. We are grateful to the HET Resident Astronomers and Telescope Operators for their valuable assistance in gathering our HPF data.
We would like to acknowledge that the HET is built on Indigenous land. Moreover, we would like to acknowledge and pay our respects to the Carrizo \& Comecrudo, Coahuiltecan, Caddo, Tonkawa, Comanche, Lipan Apache, Alabama-Coushatta, Kickapoo, Tigua Pueblo, and all the American Indian and Indigenous Peoples and communities who have been or have become a part of these lands and territories in Texas, here on Turtle Island.

Some of the data presented in this paper were obtained from MAST at STScI. The specific observations analyzed can be accessed via \dataset[DOI: 10.17909/nrwp-q285]{https://doi.org/10.17909/nrwp-q285}. Support for MAST for non-HST data is provided by the NASA Office of Space Science via grant NNX09AF08G and by other grants and contracts.
This work includes data collected by the \textit{Kepler} mission, which are publicly available from MAST. Funding for the \textit{Kepler} mission is provided by the NASA Science Mission directorate. 
This research made use of the (i) NASA Exoplanet Archive, which is operated by Caltech, under contract with NASA under the Exoplanet Exploration Program, (ii) SIMBAD database, operated at CDS, Strasbourg, France, (iii) NASA's Astrophysics Data System Bibliographic Services, (iv) data from 2MASS, a joint project of the University of Massachusetts and IPAC at Caltech, funded by NASA and the NSF, and (v) AAVSO Photometric All-Sky Survey (APASS), funded by the Robert Martin Ayers Sciences Fund and NSF AST-1412587. 
This work used data retrieved from the SOPHIE archive at Observatoire de Haute-Provence, available at \url{http://atlas.obs-hp.fr/sophie}.

Funding for the SDSS-IV has been provided by the Alfred P. Sloan Foundation, the U.S. Department of Energy Office of Science, and the Participating Institutions. SDSS-IV acknowledges support and resources from the Center for High Performance Computing at the University of Utah. The SDSS website is www.sdss.org. SDSS-IV is managed by the Astrophysical Research Consortium for the Participating Institutions of the SDSS Collaboration including the Brazilian Participation Group, the Carnegie Institution for Science, Carnegie Mellon University, Center for Astrophysics | Harvard \& Smithsonian, the Chilean Participation Group, the French Participation Group, Instituto de Astrof\'isica de Canarias, The Johns Hopkins University, Kavli Institute for the Physics and Mathematics of the Universe (IPMU) / University of Tokyo, the Korean Participation Group, Lawrence Berkeley National Laboratory, Leibniz Institut f\"ur Astrophysik Potsdam (AIP), Max-Planck-Institut f\"ur Astronomie (MPIA Heidelberg), Max-Planck-Institut f\"ur Astrophysik (MPA Garching), Max-Planck-Institut f\"ur Extraterrestrische Physik (MPE), National Astronomical Observatories of China, New Mexico State University, New York University, University of Notre Dame, Observat\'ario Nacional / MCTI, The Ohio State University, Pennsylvania State University, Shanghai Astronomical Observatory, United Kingdom Participation Group, Universidad Nacional Aut\'onoma de M\'exico, University of Arizona, University of Colorado Boulder, University of Oxford, University of Portsmouth, University of Utah, University of Virginia, University of Washington, University of Wisconsin, Vanderbilt University, and Yale University.

The Pan-STARRS1 Surveys (PS1) and the PS1 public science archive have been made possible through contributions by the Institute for Astronomy, the University of Hawaii, the Pan-STARRS Project Office, the Max-Planck Society and its participating institutes, the Max Planck Institute for Astronomy, Heidelberg and the Max Planck Institute for Extraterrestrial Physics, Garching, The Johns Hopkins University, Durham University, the University of Edinburgh, the Queen's University Belfast, the Harvard-Smithsonian Center for Astrophysics, the Las Cumbres Observatory Global Telescope Network Incorporated, the National Central University of Taiwan, the Space Telescope Science Institute, the National Aeronautics and Space Administration under Grant No. NNX08AR22G issued through the Planetary Science Division of the NASA Science Mission Directorate, the National Science Foundation Grant No. AST-1238877, the University of Maryland, Eotvos Lorand University (ELTE), the Los Alamos National Laboratory, and the Gordon and Betty Moore Foundation.

This work has made use of data from the European Space Agency (ESA) mission Gaia (\url{https://www.cosmos.esa.int/gaia}), processed by the Gaia Data Processing and Analysis Consortium (DPAC, \url{https://www.cosmos.esa.int/web/gaia/dpac/consortium}). Funding for the DPAC has been provided by national institutions, in particular the institutions participating in the Gaia Multilateral Agreement.

\facilities{Exoplanet Archive, Gaia, HET (HPF), \textit{Kepler}, KPNO:2.1m (Robo-AO), Sloan (APOGEE-N), MAST, OHP:1.93m (SOPHIE)} 
\software{
\texttt{astroquery} \citep{Ginsburg2019},
\texttt{astropy} \citep{AstropyCollaboration2018},
\texttt{ARC2} \citep{Aigrain2017},
\texttt{barycorrpy} \citep{Kanodia2018},
\texttt{dustmaps} \citep{Green2018a},
\texttt{dynesty} \citep{Speagle2020},
\texttt{EXOFASTv2} \citep{Eastman2019},
\texttt{HxRGproc} \citep{Ninan2018},
\texttt{K2fov} \citep{Mullally2016a},
\texttt{juliet} \citep{Espinoza2019},
\texttt{matplotlib} \citep{hunter2007},
\texttt{numpy} \citep{vanderwalt2011},
\texttt{pandas} \citep{McKinney2010},
\texttt{scipy} \citep{Virtanen2020},
\texttt{terraspec} \citep{Bender2012}
}

\setcounter{table}{1}
\begin{deluxetable*}{cccccc}
{\tabletypesize{\footnotesize }
\tablecaption{\textit{Kepler} photometry used for the 28 KOIs. \label{tab:koiphot}}
\tablehead{
\colhead{APOGEE ID} &
\colhead{KIC ID} &
\colhead{KOI}  &
\colhead{Designation$^a$}  &
\colhead{\textit{Kepler} Short Cadence} &
\colhead{\textit{Kepler} Long Cadence}
}
\startdata
2M18523991+4524110 &   9071386 &    23 &  \nodata             &  {\color{blue} \checktikz} & \nodata                                                           \\
2M19395458+3840421 &   3558981 &    52 &  \nodata             & \nodata                                                         &  {\color{orange} \checktikz} \\
2M19480226+5022203 &  11974540 &   129 &  Kepler-470 b &  {\color{blue} \checktikz} &  {\color{orange} \checktikz} \\
2M19492647+4025473 &   5297298 &   130 &  \nodata             &  {\color{blue} \checktikz} &  {\color{orange} \checktikz} \\
2M19424111+4035566 &   5376836 &   182 &  \nodata             &  {\color{blue} \checktikz} & \nodata                                                           \\
2M19485138+4139505 &   6305192 &   219 &  Kepler-494 b & \nodata                                                         &  {\color{orange} \checktikz} \\
2M19223275+3842276 &   3642741 &   242 &  \nodata             & \nodata                                                         &  {\color{orange} \checktikz} \\
2M19073111+3922421 &   4247092 &   403 &  \nodata             & \nodata                                                         &  {\color{orange} \checktikz} \\
2M19331345+4136229 &   6289650 &   415 &  \nodata             & \nodata                                                         &  {\color{orange} \checktikz} \\
2M19043647+4519572 &   9008220 &   466 &  \nodata             & \nodata                                                         &  {\color{orange} \checktikz} \\
2M19214782+3951172 &   4742414 &   631 &  Kepler-628 b &  {\color{blue} \checktikz} &  {\color{orange} \checktikz} \\
2M19371604+5004488 &  11818800 &   777 &  \nodata             & \nodata                                                         &  {\color{orange} \checktikz} \\
2M19473316+4123459 &   6061119 &   846 &  Kepler-699 b & \nodata                                                         &  {\color{orange} \checktikz} \\
2M19270249+4156386 &   6522242 &   855 &  Kepler-706 b & \nodata                                                         &  {\color{orange} \checktikz} \\
2M19001520+4410043 &   8218274 &  1064 &  \nodata             & \nodata                                                         &  {\color{orange} \checktikz} \\
2M18535277+4503088 &   8801343 &  1247 &  \nodata             &  {\color{blue} \checktikz} & \nodata                                                           \\
2M19160484+4807113 &  10790387 &  1288 &  Kepler-807 b & \nodata                                                         &  {\color{orange} \checktikz} \\
2M19320489+4230318 &   7037540 &  1347 &  \nodata             & \nodata                                                         &  {\color{orange} \checktikz} \\
2M19282877+4255540 &   7363829 &  1356 &  \nodata             &  {\color{blue} \checktikz} &  {\color{orange} \checktikz} \\
2M19460177+4927262 &  11517719 &  1416 &  Kepler-840 b &  {\color{blue} \checktikz} & \nodata                                                           \\
2M19191325+4629301 &   9705459 &  1448 &  \nodata             &  {\color{blue} \checktikz} & \nodata                                                           \\
2M19344052+4622453 &   9653622 &  2513 &  \nodata             & \nodata                                                         &  {\color{orange} \checktikz} \\
2M19254244+4209507 &   6690171 &  3320 &  \nodata             & \nodata                                                         &  {\color{orange} \checktikz} \\
2M19273337+3921423 &   4263529 &  3358 &  \nodata             & \nodata                                                         &  {\color{orange} \checktikz} \\
2M19520793+3952594 &   4773392 &  4367 &  \nodata             & \nodata                                                         &  {\color{orange} \checktikz} \\
2M19543478+4217089 &   6805414 &  5329 &  \nodata             & \nodata                                                         &  {\color{orange} \checktikz} \\
2M19480000+4117241 &   5979863 &  6018 &  \nodata             & \nodata                                                         &  {\color{orange} \checktikz} \\
2M19352118+4207199 &   6698670 &  6760 &  \nodata             &  {\color{blue} \checktikz} & \nodata                                                           \\
\enddata
\tablenotetext{a}{The \textit{Kepler} designation for targets on the NASA Exoplanet Archive with a disposition of ``confirmed planet'' as of 2022 June 5.}
}
\end{deluxetable*}

\setcounter{table}{2}
\begin{splitdeluxetable*}{ccccccBcccccc}
    {\tabletypesize{\tiny}
    \tablecaption{General information of the KOI host stars. \label{tab:koigeneral}}
    \tablehead{
    \colhead{APOGEE ID}  &
    \colhead{KIC ID}  &
    \colhead{KOI}  &
    \colhead{Gaia} & 
    \colhead{RA} & 
    \colhead{Dec} & 
    \colhead{Proper motion (RA)} &
    \colhead{Proper motion (Dec)} &
    \colhead{Distance$^a$} & 
    \colhead{RUWE} & 
    \colhead{Max $A_V$$^b$} &
    \colhead{Robo-AO Detection$^c$} 
    \\
    & 
    &
    (DR25) &
    (DR3) &
    (HH:MM:SS) &
    (DD:MM:SS) &
    \colhead{($\mathrm{mas~yr^{-1}}$)} & 
    \colhead{($\mathrm{mas~yr^{-1}}$)} & 
    \colhead{(pc)} & 
    (DR3) &
    (Green) &
    }
    \startdata
    2M18523991+4524110 &   9071386 &    23 &  2107001760170220032 &  18:52:39.91 &  45:24:10.96 &    $1.79 \pm 0.01$ &    $4.26 \pm 0.01$ &        $761_{-6}^{+5}$ &   0.95 &  0.19 &  \nodata                      \\ 
    2M19395458+3840421 &   3558981 &    52 &  2052143143639171072 &  19:39:54.58 &  38:40:41.87 &    $1.44 \pm 0.01$ &  $-13.50 \pm 0.02$ &            $474 \pm 3$ &   0.92 &  0.19 &  \nodata                        \\ 
    2M19480226+5022203 &  11974540 &   129 &  2135313669189075968 &  19:48:02.26 &  50:22:20.21 &   $-1.91 \pm 0.01$ &   $-9.88 \pm 0.01$ &            $929 \pm 7$ &   1.03 &  0.25 &  5.87 mag at 2.1 \arcsec{}, L14 \\
    2M19492647+4025473 &   5297298 &   130 &  2073737036615242240 &  19:49:26.46 &  40:25:47.09 &   $-8.98 \pm 0.01$ &  $-13.99 \pm 0.01$ &            $940 \pm 9$ &   0.97 &  0.52 &                      None, Z18 \\
    2M19424111+4035566 &   5376836 &   182 &  2076405620054037120 &  19:42:41.11 &  40:35:56.81 &    $2.99 \pm 0.02$ &    $6.68 \pm 0.02$ &     $1050_{-10}^{+20}$ &   1.09 &  0.39 &                      None, Z18 \\
    2M19485138+4139505 &   6305192 &   219 &  2076942628402548608 &  19:48:51.36 &  41:39:50.42 &     $-3.8 \pm 0.3$ &     $-6.4 \pm 0.4$ &  $1800_{-500}^{+1000}$ &  20.74 &  0.58 &                      None, L14 \\
    2M19223275+3842276 &   3642741 &   242 &  2052853703021795456 &  19:22:32.75 &  38:42:27.52 &   $-3.54 \pm 0.02$ &   $-3.36 \pm 0.02$ &     $1550_{-30}^{+40}$ &   0.95 &  0.38 &                      None, Z17 \\
    2M19073111+3922421 &   4247092 &   403 &  2100400949491004800 &  19:07:31.11 &  39:22:41.98 &   $-2.91 \pm 0.02$ &  $-11.06 \pm 0.02$ &          $1070 \pm 10$ &   0.94 &  0.66 &                      None, L14 \\
    2M19331345+4136229 &   6289650 &   415 &  2077596288060821120 &  19:33:13.45 &  41:36:22.69 &    $6.89 \pm 0.01$ &  $-16.77 \pm 0.01$ &       $920_{-10}^{+9}$ &   0.98 &  0.30 &                      None, L14 \\
    2M19043647+4519572 &   9008220 &   466 &  2106436649851315712 &  19:04:36.47 &  45:19:57.25 &   $-5.72 \pm 0.02$ &   $-3.24 \pm 0.02$ &     $1640_{-40}^{+50}$ &   0.93 &  0.19 &                      None, Z17 \\
    2M19214782+3951172 &   4742414 &   631 &  2101084708287433984 &  19:21:47.83 &  39:51:17.28 &    $6.68 \pm 0.01$ &    $1.33 \pm 0.01$ &        $870_{-8}^{+9}$ &   0.92 &  0.19 &                      None, Z18 \\
    2M19371604+5004488 &  11818800 &   777 &  2135105075517958144 &  19:37:16.03 &  50:04:48.65 &   $-4.47 \pm 0.03$ &   $-9.07 \pm 0.03$ &     $2140_{-80}^{+90}$ &   0.94 &  0.25 &                      None, Z17 \\
    2M19473316+4123459 &   6061119 &   846 &  2076838896353601152 &  19:47:33.15 &  41:23:45.92 &   $-1.27 \pm 0.03$ &   $-2.59 \pm 0.03$ &     $1900_{-80}^{+90}$ &   0.99 &  0.55 &                      None, B16 \\
    2M19270249+4156386 &   6522242 &   855 &  2101740257736205696 &  19:27:02.46 &  41:56:38.68 &  $-16.93 \pm 0.03$ &   $-0.45 \pm 0.03$ &      $840_{-10}^{+20}$ &   1.11 &  0.22 &                      None, Z17 \\
    2M19001520+4410043 &   8218274 &  1064 &  2105945644892917248 &  19:00:15.21 &  44:10:04.32 &    $7.65 \pm 0.01$ &   $-3.85 \pm 0.01$ &       $1012_{-8}^{+9}$ &   1.02 &  0.20 &                      None, Z18 \\
    2M18535277+4503088 &   8801343 &  1247 &  2106981148624921344 &  18:53:52.78 &  45:03:08.89 &    $7.06 \pm 0.03$ &    $5.44 \pm 0.03$ &        $624_{-8}^{+7}$ &   1.71 &  0.16 &                      None, Z18 \\
    2M19160484+4807113 &  10790387 &  1288 &  2130955518633883008 &  19:16:04.83 &  48:07:11.21 &   $-1.89 \pm 0.02$ &   $-5.65 \pm 0.03$ &          $1480 \pm 50$ &   1.02 &  0.22 &                      None, L14 \\
    2M19320489+4230318 &   7037540 &  1347 &  2077796914573293568 &  19:32:04.88 &  42:30:31.89 &   $-6.30 \pm 0.02$ &    $5.56 \pm 0.02$ &           $950 \pm 10$ &   1.04 &  0.22 & \nodata                         \\ 
    2M19282877+4255540 &   7363829 &  1356 &  2125814958179612544 &  19:28:28.77 &  42:55:53.87 &    $1.80 \pm 0.02$ &   $-6.08 \pm 0.03$ &    $1890_{-70}^{+100}$ &   0.94 &  0.27 &                      None, Z17 \\
    2M19460177+4927262 &  11517719 &  1416 &  2134884520357421824 &  19:46:01.76 &  49:27:26.02 &    $0.31 \pm 0.06$ &  $-13.27 \pm 0.07$ &         $1200 \pm 100$ &   4.13 &  0.36 &                      None, Z18 \\
    2M19191325+4629301 &   9705459 &  1448 &  2127712474727909504 &  19:19:13.24 &  46:29:30.14 &   $-0.80 \pm 0.03$ &   $-0.44 \pm 0.03$ &     $1210_{-30}^{+40}$ &   1.08 &  0.25 &                      None, Z18 \\
    2M19344052+4622453 &   9653622 &  2513 &  2128167225867000576 &  19:34:40.53 &  46:22:45.23 &      $2.5 \pm 0.2$ &     $-6.1 \pm 0.2$ &  $2100_{-500}^{+1000}$ &   9.97 &  0.44 &                      None, Z17 \\
    2M19254244+4209507 &   6690171 &  3320 &  2101758056080602112 &  19:25:42.45 &  42:09:50.59 &    $2.08 \pm 0.03$ &   $-7.27 \pm 0.03$ &          $1030 \pm 30$ &   0.93 &  0.27 &                      None, Z17 \\
    2M19273337+3921423 &   4263529 &  3358 &  2053170843410198656 &  19:27:33.41 &  39:21:42.24 &   $19.83 \pm 0.01$ &  $-10.31 \pm 0.02$ &            $508 \pm 3$ &   0.95 &  0.16 &                      None, Z17 \\
    2M19520793+3952594 &   4773392 &  4367 &  2073471985601809152 &  19:52:07.93 &  39:52:59.39 &   $-2.31 \pm 0.01$ &   $-1.45 \pm 0.01$ &          $1120 \pm 20$ &   0.97 &  0.36 &                      None, B16 \\
    2M19543478+4217089 &   6805414 &  5329 &  2075448082868939136 &  19:54:34.78 &  42:17:08.89 &   $-2.81 \pm 0.02$ &   $-5.80 \pm 0.03$ &    $2130_{-90}^{+100}$ &   1.03 &  0.63 &                      None, Z17 \\
    2M19480000+4117241 &   5979863 &  6018 &  2076790449121125120 &  19:47:59.99 &  41:17:24.01 &   $-5.45 \pm 0.03$ &   $-4.89 \pm 0.03$ &     $1090_{-20}^{+30}$ &   0.92 &  0.52 & \nodata                         \\ 
    2M19352118+4207199 &   6698670 &  6760 &  2077714932236541440 &  19:35:21.18 &  42:07:19.74 &    $2.22 \pm 0.02$ &  $-11.26 \pm 0.02$ &            $532 \pm 4$ &   1.09 &  0.22 & \nodata                         \\ 
    \enddata
    \tablerefs{B16 \citep{Baranec2016}, DR25 \citep{Thompson2018}, DR3 \citep[][]{GaiaCollaboration2022a}, Green \citep{Green2019}, L14 \citep{Law2014}, Z17 \citep{Ziegler2017}, Z18 \citep{Ziegler2018a}.}
    \tablenotetext{a}{The geometric distance from \cite{Bailer-Jones2021}.}
    \tablenotetext{b}{Maximum visual extinction determined from \cite{Green2019}.}
    \tablenotetext{c}{Empty rows are objects not observed as part of the Robo-AO \textit{Kepler} survey. KOI-1356 was observed in a Sloan $i\prime$ filter, all other observations were performed in the LP600 filter.}
    }
\end{splitdeluxetable*}   

\setcounter{table}{5}
\begin{splitdeluxetable*}{cccccccccccBcccccccccccc}
{\tabletypesize{\tiny}
\tablecaption{Broadband photometry of the KOI host stars. \label{tab:koibroadband}}
\tablehead{
\colhead{APOGEE ID}  &
\colhead{KIC ID} & 
\colhead{KOI} & 
\colhead{$U$} & 
\colhead{$B$} & 
\colhead{$V$} &
\colhead{$u^\prime$} & 
\colhead{$g^\prime$} & 
\colhead{$r^\prime$} & 
\colhead{$i^\prime$} & 
\colhead{$z^\prime$} & 
\colhead{$g_{PS}$} & 
\colhead{$r_{PS}$} & 
\colhead{$i_{PS}$} & 
\colhead{$z_{PS}$} & 
\colhead{$y_{PS}$} & 
\colhead{$J$} & 
\colhead{$H$} & 
\colhead{$K$} &
\colhead{W1} &
\colhead{W2} &
\colhead{W3} \\
\colhead{} &
\colhead{} &
\colhead{} &
\colhead{EHK} & 
\colhead{EHK} & 
\colhead{EHK} & 
\colhead{SDSS} & 
\colhead{APASS} & 
\colhead{APASS} & 
\colhead{APASS} & 
\colhead{SDSS} & 
\colhead{PS1} & 
\colhead{PS1} & 
\colhead{PS1} & 
\colhead{PS1} & 
\colhead{PS1} & 
\colhead{2MASS} & 
\colhead{2MASS} & 
\colhead{2MASS} &
\colhead{WISE} &
\colhead{WISE} &
\colhead{WISE}
}
\startdata
2M18523991+4524110 &   9071386 &    23 &  $12.89 \pm 0.02$ &  $12.88 \pm 0.03$ &  $12.42 \pm 0.02$ &                     &                     &                     &                     &                     &                     &                     &                     &                     &    $12.19 \pm 0.01$ &  $11.34 \pm 0.02$ &  $11.15 \pm 0.02$ &  $11.07 \pm 0.02$ &  $11.07 \pm 0.02$ &  $11.10 \pm 0.02$ &    $11.1 \pm 0.1$ \\ 
2M19395458+3840421 &   3558981 &    52 &  $15.38 \pm 0.02$ &  $14.92 \pm 0.03$ &  $14.08 \pm 0.02$ &                     &    $14.19 \pm 0.02$ &    $13.57 \pm 0.05$ &    $13.23 \pm 0.03$ &                     &    $14.27 \pm 0.07$ &    $13.70 \pm 0.06$ &                     &  $13.396 \pm 0.009$ &    $13.31 \pm 0.02$ &  $12.44 \pm 0.02$ &  $11.97 \pm 0.02$ &  $11.92 \pm 0.02$ &  $11.74 \pm 0.02$ &  $11.81 \pm 0.02$ &    $11.6 \pm 0.2$ \\ 
2M19480226+5022203 &  11974540 &   129 &  $13.75 \pm 0.02$ &  $13.79 \pm 0.02$ &  $13.30 \pm 0.02$ &                     &                     &                     &      $13.0 \pm 0.1$ &                     &                     &                     &                     &    $13.10 \pm 0.02$ &  $13.089 \pm 0.005$ &  $12.27 \pm 0.03$ &  $12.04 \pm 0.03$ &  $12.03 \pm 0.03$ &  $11.90 \pm 0.02$ &  $11.93 \pm 0.02$ &    $12.0 \pm 0.2$ \\ 
2M19492647+4025473 &   5297298 &   130 &                   &  $14.08 \pm 0.03$ &  $13.46 \pm 0.04$ &                     &                     &                     &                     &                     &    $13.69 \pm 0.02$ &    $13.31 \pm 0.02$ &                     &    $13.13 \pm 0.01$ &    $13.08 \pm 0.01$ &  $12.18 \pm 0.02$ &  $11.94 \pm 0.02$ &  $11.87 \pm 0.02$ &  $11.87 \pm 0.02$ &  $11.90 \pm 0.02$ &                   \\ 
2M19424111+4035566 &   5376836 &   182 &  $15.10 \pm 0.02$ &  $14.93 \pm 0.02$ &  $14.27 \pm 0.02$ &                     &    $14.44 \pm 0.04$ &  $14.019 \pm 0.006$ &    $13.88 \pm 0.03$ &                     &    $14.48 \pm 0.01$ &    $14.01 \pm 0.02$ &  $13.861 \pm 0.007$ &  $13.809 \pm 0.002$ &  $13.770 \pm 0.005$ &  $12.90 \pm 0.03$ &  $12.57 \pm 0.02$ &  $12.49 \pm 0.02$ &  $12.42 \pm 0.02$ &  $12.48 \pm 0.02$ &    $12.4 \pm 0.3$ \\ 
2M19485138+4139505 &   6305192 &   219 &  $15.31 \pm 0.02$ &  $15.12 \pm 0.02$ &  $14.39 \pm 0.02$ &                     &    $14.75 \pm 0.04$ &    $14.07 \pm 0.01$ &    $13.97 \pm 0.05$ &                     &  $14.611 \pm 0.005$ &  $14.105 \pm 0.008$ &    $13.95 \pm 0.05$ &  $13.838 \pm 0.008$ &    $13.76 \pm 0.01$ &  $12.90 \pm 0.02$ &  $12.54 \pm 0.03$ &  $12.47 \pm 0.02$ &  $12.42 \pm 0.02$ &  $12.44 \pm 0.02$ &                   \\ 
2M19223275+3842276 &   3642741 &   242 &  $16.06 \pm 0.02$ &  $15.75 \pm 0.02$ &  $15.01 \pm 0.02$ &  $16.908 \pm 0.008$ &  $15.288 \pm 0.003$ &  $14.668 \pm 0.003$ &  $14.454 \pm 0.003$ &  $14.357 \pm 0.004$ &  $15.239 \pm 0.006$ &  $14.681 \pm 0.005$ &    $14.50 \pm 0.01$ &  $14.407 \pm 0.006$ &  $14.359 \pm 0.008$ &  $13.46 \pm 0.03$ &  $13.14 \pm 0.03$ &  $13.05 \pm 0.03$ &  $12.96 \pm 0.02$ &  $12.98 \pm 0.03$ &    $12.4 \pm 0.4$ \\ 
2M19073111+3922421 &   4247092 &   403 &  $15.35 \pm 0.02$ &  $15.20 \pm 0.03$ &  $14.50 \pm 0.02$ &                     &    $14.70 \pm 0.01$ &    $14.00 \pm 0.02$ &    $13.87 \pm 0.06$ &                     &  $14.611 \pm 0.004$ &  $14.123 \pm 0.002$ &  $13.961 \pm 0.004$ &  $13.900 \pm 0.006$ &  $13.839 \pm 0.008$ &  $12.97 \pm 0.02$ &  $12.68 \pm 0.03$ &  $12.60 \pm 0.02$ &  $12.45 \pm 0.02$ &  $12.51 \pm 0.02$ &    $12.7 \pm 0.5$ \\ 
2M19331345+4136229 &   6289650 &   415 &  $15.04 \pm 0.02$ &  $14.93 \pm 0.02$ &  $14.34 \pm 0.02$ &                     &                     &                     &                     &                     &  $14.505 \pm 0.006$ &  $14.106 \pm 0.006$ &  $13.973 \pm 0.009$ &  $13.928 \pm 0.004$ &  $13.882 \pm 0.009$ &  $13.05 \pm 0.02$ &  $12.67 \pm 0.02$ &  $12.66 \pm 0.03$ &  $12.57 \pm 0.02$ &  $12.60 \pm 0.02$ &    $12.4 \pm 0.3$ \\ 
2M19043647+4519572 &   9008220 &   466 &  $15.57 \pm 0.02$ &  $15.48 \pm 0.03$ &  $14.87 \pm 0.02$ &                     &                     &                     &                     &                     &    $15.01 \pm 0.01$ &  $14.628 \pm 0.008$ &  $14.503 \pm 0.008$ &  $14.477 \pm 0.004$ &  $14.451 \pm 0.008$ &  $13.62 \pm 0.03$ &  $13.34 \pm 0.03$ &  $13.21 \pm 0.03$ &  $13.20 \pm 0.02$ &  $13.21 \pm 0.03$ &    $12.4 \pm 0.3$ \\ 
2M19214782+3951172 &   4742414 &   631 &  $14.08 \pm 0.02$ &  $13.95 \pm 0.02$ &  $13.30 \pm 0.02$ &                     &    $13.55 \pm 0.01$ &    $13.03 \pm 0.04$ &    $12.86 \pm 0.03$ &                     &  $13.425 \pm 0.003$ &                     &                     &                     &    $12.85 \pm 0.02$ &  $11.98 \pm 0.02$ &  $11.66 \pm 0.02$ &  $11.60 \pm 0.02$ &  $11.55 \pm 0.02$ &  $11.61 \pm 0.02$ &    $11.8 \pm 0.2$ \\ 
2M19371604+5004488 &  11818800 &   777 &  $16.80 \pm 0.02$ &  $16.58 \pm 0.03$ &  $15.81 \pm 0.02$ &                     &    $15.90 \pm 0.06$ &    $15.37 \pm 0.07$ &      $15.2 \pm 0.1$ &                     &  $15.990 \pm 0.009$ &    $15.42 \pm 0.02$ &    $15.19 \pm 0.01$ &    $15.08 \pm 0.01$ &    $14.99 \pm 0.03$ &  $14.03 \pm 0.03$ &  $13.63 \pm 0.03$ &  $13.49 \pm 0.04$ &  $13.46 \pm 0.02$ &  $13.50 \pm 0.03$ &                   \\ 
2M19473316+4123459 &   6061119 &   846 &  $16.50 \pm 0.03$ &  $16.42 \pm 0.03$ &  $15.78 \pm 0.02$ &                     &                     &                     &                     &                     &  $15.958 \pm 0.009$ &    $15.47 \pm 0.01$ &  $15.252 \pm 0.005$ &    $15.17 \pm 0.01$ &    $15.09 \pm 0.01$ &  $14.24 \pm 0.03$ &  $13.74 \pm 0.03$ &  $13.67 \pm 0.04$ &  $13.70 \pm 0.03$ &  $13.81 \pm 0.04$ &                   \\ 
2M19270249+4156386 &   6522242 &   855 &  $16.46 \pm 0.03$ &  $16.18 \pm 0.03$ &  $15.47 \pm 0.02$ &                     &    $15.76 \pm 0.02$ &    $15.19 \pm 0.04$ &    $14.95 \pm 0.05$ &                     &    $15.68 \pm 0.01$ &    $15.15 \pm 0.01$ &    $14.93 \pm 0.01$ &    $14.82 \pm 0.01$ &    $14.76 \pm 0.01$ &  $13.81 \pm 0.03$ &  $13.41 \pm 0.03$ &  $13.31 \pm 0.04$ &  $13.28 \pm 0.02$ &  $13.27 \pm 0.03$ &                   \\ 
2M19001520+4410043 &   8218274 &  1064 &  $13.59 \pm 0.02$ &  $13.69 \pm 0.03$ &  $13.23 \pm 0.02$ &                     &    $13.43 \pm 0.03$ &    $13.21 \pm 0.05$ &    $13.07 \pm 0.06$ &                     &  $13.356 \pm 0.002$ &                     &                     &    $13.12 \pm 0.02$ &  $13.133 \pm 0.005$ &  $12.35 \pm 0.02$ &  $12.14 \pm 0.02$ &  $12.09 \pm 0.03$ &  $12.08 \pm 0.02$ &  $12.10 \pm 0.02$ &    $12.2 \pm 0.3$ \\ 
2M18535277+4503088 &   8801343 &  1247 &  $12.25 \pm 0.02$ &  $12.30 \pm 0.02$ &  $11.82 \pm 0.02$ &                     &      $11.9 \pm 0.1$ &    $11.70 \pm 0.03$ &    $11.57 \pm 0.04$ &                     &                     &                     &                     &                     &                     &  $10.80 \pm 0.02$ &  $10.57 \pm 0.02$ &  $10.54 \pm 0.01$ &  $10.49 \pm 0.02$ &  $10.52 \pm 0.02$ &  $10.39 \pm 0.05$ \\ 
2M19160484+4807113 &  10790387 &  1288 &  $15.92 \pm 0.03$ &  $15.88 \pm 0.03$ &  $15.30 \pm 0.02$ &                     &    $15.53 \pm 0.03$ &    $15.14 \pm 0.09$ &      $15.0 \pm 0.2$ &                     &  $15.385 \pm 0.008$ &  $15.053 \pm 0.005$ &  $14.951 \pm 0.004$ &  $14.928 \pm 0.007$ &    $14.91 \pm 0.01$ &  $14.08 \pm 0.03$ &  $13.77 \pm 0.04$ &  $13.69 \pm 0.05$ &  $13.70 \pm 0.02$ &  $13.72 \pm 0.03$ &                   \\ 
2M19320489+4230318 &   7037540 &  1347 &  $15.43 \pm 0.02$ &  $15.40 \pm 0.03$ &  $14.75 \pm 0.02$ &                     &  $14.753 \pm 0.008$ &    $14.48 \pm 0.06$ &    $14.27 \pm 0.06$ &                     &    $14.87 \pm 0.01$ &  $14.460 \pm 0.008$ &    $14.32 \pm 0.01$ &    $14.28 \pm 0.02$ &  $14.232 \pm 0.009$ &  $13.36 \pm 0.03$ &  $13.06 \pm 0.03$ &  $12.93 \pm 0.03$ &  $12.95 \pm 0.02$ &  $12.98 \pm 0.03$ &                   \\ 
2M19282877+4255540 &   7363829 &  1356 &  $16.48 \pm 0.02$ &  $16.17 \pm 0.03$ &  $15.44 \pm 0.02$ &                     &    $15.58 \pm 0.04$ &    $15.26 \pm 0.03$ &  $15.009 \pm 0.008$ &                     &  $15.630 \pm 0.007$ &    $15.14 \pm 0.01$ &  $14.976 \pm 0.004$ &    $14.93 \pm 0.01$ &  $14.879 \pm 0.008$ &  $13.96 \pm 0.03$ &  $13.66 \pm 0.03$ &  $13.61 \pm 0.05$ &  $13.56 \pm 0.02$ &  $13.67 \pm 0.03$ &                   \\ 
2M19460177+4927262 &  11517719 &  1416 &  $14.94 \pm 0.02$ &  $14.88 \pm 0.03$ &  $14.36 \pm 0.02$ &                     &      $14.6 \pm 0.1$ &    $14.15 \pm 0.07$ &      $14.0 \pm 0.2$ &                     &    $14.50 \pm 0.01$ &    $14.11 \pm 0.02$ &    $13.95 \pm 0.01$ &  $13.909 \pm 0.004$ &    $13.86 \pm 0.01$ &  $13.00 \pm 0.02$ &  $12.68 \pm 0.02$ &  $12.60 \pm 0.03$ &  $12.56 \pm 0.02$ &  $12.58 \pm 0.02$ &    $12.2 \pm 0.2$ \\ 
2M19191325+4629301 &   9705459 &  1448 &  $16.74 \pm 0.03$ &  $16.43 \pm 0.03$ &  $15.70 \pm 0.02$ &                     &                     &                     &                     &                     &    $15.84 \pm 0.02$ &    $15.33 \pm 0.02$ &  $15.182 \pm 0.004$ &  $15.131 \pm 0.008$ &    $15.09 \pm 0.05$ &  $14.21 \pm 0.03$ &  $13.86 \pm 0.03$ &  $13.72 \pm 0.04$ &  $13.67 \pm 0.03$ &  $13.71 \pm 0.03$ &                   \\ 
2M19344052+4622453 &   9653622 &  2513 &                   &  $15.65 \pm 0.03$ &  $14.99 \pm 0.03$ &                     &    $15.25 \pm 0.02$ &    $14.79 \pm 0.02$ &    $14.75 \pm 0.06$ &                     &    $15.15 \pm 0.01$ &  $14.796 \pm 0.008$ &    $14.65 \pm 0.02$ &  $14.631 \pm 0.007$ &    $14.60 \pm 0.02$ &  $13.76 \pm 0.03$ &  $13.45 \pm 0.03$ &  $13.37 \pm 0.04$ &  $13.35 \pm 0.02$ &  $13.38 \pm 0.03$ &                   \\ 
2M19254244+4209507 &   6690171 &  3320 &  $18.16 \pm 0.06$ &  $17.19 \pm 0.03$ &  $16.29 \pm 0.02$ &                     &    $16.78 \pm 0.04$ &                     &                     &                     &  $16.613 \pm 0.003$ &    $15.88 \pm 0.01$ &  $15.639 \pm 0.009$ &  $15.527 \pm 0.009$ &    $15.44 \pm 0.01$ &  $14.44 \pm 0.03$ &  $13.97 \pm 0.03$ &  $13.84 \pm 0.04$ &  $13.90 \pm 0.03$ &  $14.04 \pm 0.04$ &                   \\ 
2M19273337+3921423 &   4263529 &  3358 &  $15.40 \pm 0.02$ &  $15.11 \pm 0.03$ &  $14.36 \pm 0.02$ &                     &    $14.45 \pm 0.05$ &    $13.87 \pm 0.08$ &    $13.70 \pm 0.08$ &                     &  $14.470 \pm 0.003$ &    $13.98 \pm 0.04$ &    $13.80 \pm 0.03$ &  $13.722 \pm 0.003$ &  $13.666 \pm 0.004$ &  $12.73 \pm 0.03$ &  $12.32 \pm 0.03$ &  $12.30 \pm 0.02$ &  $12.25 \pm 0.02$ &  $12.29 \pm 0.02$ &    $12.4 \pm 0.3$ \\ 
2M19520793+3952594 &   4773392 &  4367 &  $14.76 \pm 0.02$ &  $14.75 \pm 0.02$ &  $14.20 \pm 0.01$ &                     &                     &                     &                     &                     &  $14.288 \pm 0.009$ &  $13.993 \pm 0.002$ &    $13.86 \pm 0.02$ &  $13.838 \pm 0.009$ &  $13.811 \pm 0.007$ &  $12.98 \pm 0.02$ &  $12.73 \pm 0.02$ &  $12.69 \pm 0.02$ &  $12.64 \pm 0.03$ &  $12.69 \pm 0.03$ &    $10.8 \pm 0.2$ \\ 
2M19543478+4217089 &   6805414 &  5329 &  $16.44 \pm 0.02$ &  $16.26 \pm 0.03$ &  $15.59 \pm 0.02$ &                     &    $15.30 \pm 0.06$ &    $15.30 \pm 0.04$ &      $15.2 \pm 0.2$ &                     &  $15.790 \pm 0.006$ &  $15.341 \pm 0.009$ &  $15.150 \pm 0.005$ &  $15.090 \pm 0.007$ &    $15.02 \pm 0.01$ &  $14.13 \pm 0.03$ &  $13.83 \pm 0.04$ &  $13.77 \pm 0.04$ &  $13.67 \pm 0.02$ &  $13.71 \pm 0.03$ &                   \\ 
2M19480000+4117241 &   5979863 &  6018 &  $16.41 \pm 0.02$ &  $16.24 \pm 0.02$ &  $15.54 \pm 0.01$ &                     &    $15.76 \pm 0.03$ &    $15.27 \pm 0.04$ &    $15.21 \pm 0.02$ &                     &  $15.653 \pm 0.005$ &    $15.18 \pm 0.01$ &  $14.997 \pm 0.009$ &    $14.95 \pm 0.01$ &    $14.90 \pm 0.01$ &  $13.98 \pm 0.02$ &  $13.60 \pm 0.03$ &  $13.58 \pm 0.04$ &  $13.60 \pm 0.03$ &  $13.72 \pm 0.03$ &    $12.3 \pm 0.3$ \\ 
2M19352118+4207199 &   6698670 &  6760 &  $13.73 \pm 0.02$ &  $13.70 \pm 0.02$ &  $13.13 \pm 0.02$ &                     &    $13.37 \pm 0.05$ &    $12.94 \pm 0.05$ &    $12.78 \pm 0.06$ &                     &                     &                     &                     &                     &  $12.787 \pm 0.004$ &  $11.93 \pm 0.02$ &  $11.65 \pm 0.02$ &  $11.61 \pm 0.02$ &  $11.56 \pm 0.02$ &  $11.59 \pm 0.02$ &    $11.5 \pm 0.1$ \\ 
\enddata
\tablerefs{KIC \citep{Stassun2019}, EHK \citep[][]{Everett2012}, SDSS \citep{Alam2015}, APASS \citep{Henden2018}, PS1 \citep[][]{Magnier2020}, 2MASS \citep{Cutri2003}, WISE \citep{Wright2010}}
}
\end{splitdeluxetable*}

\begin{deluxetable*}{ccccccccccc}
{
\tabletypesize{\footnotesize}
\rotate
\tablecaption{Stellar parameters for the KOI host stars. \label{tab:koistellar}}
\tablehead{
\colhead{APOGEE ID}  &
\colhead{KIC ID} &
\colhead{KOI} & 
\colhead{$T_{e}$\(^a\)} &
\colhead{[Fe/H]\(^a\)} &
\colhead{$\log g_{\star}$} &
\colhead{$M_{\star}$} &
\colhead{$R_{\star}$} &
\colhead{$\rho_{\star}$} &
\colhead{Age} &
\colhead{$A_V$}\\
\colhead{} & 
\colhead{} & 
\colhead{} & 
\colhead{($\mathrm{K}$)} & 
\colhead{(dex)} & 
\colhead{(dex)} & 
\colhead{($\mathrm{M_{\odot}}$)} & 
\colhead{($\mathrm{R_{\odot}}$)} & 
\colhead{($\mathrm{g~cm^{-3}}$)} & 
\colhead{($\mathrm{Gyr}$)} & 
\colhead{}
}
\startdata
2M18523991+4524110 &   9071386 &    23 &   $6240 \pm 80$ &   $0.14 \pm 0.01$ &  $3.99_{-0.04}^{+0.03}$ &  $1.46_{-0.09}^{+0.07}$ &         $2.01 \pm 0.05$ &         $0.25 \pm 0.02$ &   $2.6_{-0.5}^{+0.9}$ &  $0.10_{-0.06}^{+0.05}$ \\
2M19395458+3840421 &   3558981 &    52 &   $5090 \pm 90$ &   $0.08 \pm 0.01$ &         $4.41 \pm 0.02$ &  $0.89_{-0.02}^{+0.03}$ &         $0.97 \pm 0.03$ &           $1.4 \pm 0.1$ &        $12_{-2}^{+1}$ &  $0.08_{-0.06}^{+0.07}$ \\
2M19480226+5022203 &  11974540 &   129 &  $6300 \pm 100$ &  $-0.15 \pm 0.02$ &  $4.08_{-0.05}^{+0.04}$ &  $1.25_{-0.12}^{+0.07}$ &         $1.69 \pm 0.05$ &         $0.36 \pm 0.04$ &   $3.7_{-0.9}^{+2.1}$ &  $0.20_{-0.06}^{+0.04}$ \\
2M19492647+4025473 &   5297298 &   130 &  $6100 \pm 100$ &  $-0.17 \pm 0.02$ &  $3.99_{-0.04}^{+0.05}$ &  $1.18_{-0.08}^{+0.13}$ &  $1.82_{-0.05}^{+0.06}$ &  $0.28_{-0.03}^{+0.04}$ &             $5 \pm 2$ &   $0.33_{-0.1}^{+0.09}$ \\
2M19424111+4035566 &   5376836 &   182 &  $5800 \pm 100$ &   $0.05 \pm 0.01$ &         $4.09 \pm 0.04$ &  $1.06_{-0.06}^{+0.07}$ &         $1.55 \pm 0.05$ &  $0.41_{-0.04}^{+0.05}$ &             $8 \pm 2$ &         $0.29 \pm 0.05$ \\
2M19485138+4139505 &   6305192 &   219 &  $6000 \pm 100$ &   $0.16 \pm 0.02$ &           $4.2 \pm 0.2$ &     $1.1_{-0.1}^{+0.2}$ &     $1.4_{-0.3}^{+0.5}$ &     $0.5_{-0.3}^{+0.4}$ &         $5_{-2}^{+3}$ &  $0.50_{-0.06}^{+0.05}$ \\
2M19223275+3842276 &   3642741 &   242 &  $5800 \pm 100$ &   $0.26 \pm 0.01$ &  $4.01_{-0.04}^{+0.05}$ &  $1.17_{-0.06}^{+0.13}$ &  $1.78_{-0.07}^{+0.06}$ &  $0.30_{-0.03}^{+0.04}$ &         $7_{-2}^{+1}$ &  $0.36_{-0.03}^{+0.02}$ \\
2M19073111+3922421 &   4247092 &   403 &  $6100 \pm 100$ &  $-0.01 \pm 0.02$ &         $4.13 \pm 0.04$ &  $1.09_{-0.07}^{+0.09}$ &  $1.50_{-0.04}^{+0.05}$ &         $0.46 \pm 0.05$ &             $7 \pm 2$ &           $0.4 \pm 0.1$ \\
2M19331345+4136229 &   6289650 &   415 &  $5700 \pm 100$ &  $-0.24 \pm 0.02$ &  $4.15_{-0.03}^{+0.04}$ &  $0.93_{-0.03}^{+0.05}$ &  $1.34_{-0.04}^{+0.05}$ &  $0.54_{-0.06}^{+0.07}$ &            $11 \pm 2$ &         $0.14 \pm 0.09$ \\
2M19043647+4519572 &   9008220 &   466 &  $5800 \pm 100$ &  $-0.01 \pm 0.02$ &  $4.02_{-0.04}^{+0.05}$ &  $1.08_{-0.06}^{+0.07}$ &         $1.69 \pm 0.07$ &  $0.32_{-0.04}^{+0.05}$ &             $8 \pm 2$ &  $0.12_{-0.06}^{+0.05}$ \\
2M19214782+3951172 &   4742414 &   631 &  $5800 \pm 100$ &  $-0.02 \pm 0.01$ &         $3.93 \pm 0.03$ &  $1.12_{-0.05}^{+0.06}$ &         $1.91 \pm 0.06$ &         $0.23 \pm 0.02$ &             $7 \pm 1$ &  $0.07_{-0.05}^{+0.07}$ \\
2M19371604+5004488 &  11818800 &   777 &   $5000 \pm 90$ &  $-0.55 \pm 0.02$ &         $3.67 \pm 0.04$ &  $0.87_{-0.02}^{+0.04}$ &           $2.3 \pm 0.1$ &         $0.11 \pm 0.02$ &        $12_{-2}^{+1}$ &  $0.14_{-0.08}^{+0.07}$ \\
2M19473316+4123459 &   6061119 &   846 &  $6000 \pm 100$ &   $0.15 \pm 0.02$ &         $4.16 \pm 0.05$ &  $1.15_{-0.08}^{+0.09}$ &         $1.48 \pm 0.08$ &  $0.50_{-0.07}^{+0.09}$ &         $6_{-2}^{+3}$ &  $0.50_{-0.06}^{+0.03}$ \\
2M19270249+4156386 &   6522242 &   855 &   $5010 \pm 90$ &  $-0.57 \pm 0.02$ &         $4.42 \pm 0.03$ &         $0.78 \pm 0.02$ &         $0.90 \pm 0.03$ &     $1.5_{-0.1}^{+0.2}$ &  $12.4_{-1.7}^{+0.9}$ &  $0.10_{-0.06}^{+0.07}$ \\
2M19001520+4410043 &   8218274 &  1064 &  $6000 \pm 300$ &    $-1.5 \pm 0.1$ &         $4.14 \pm 0.04$ &         $1.38 \pm 0.07$ &         $1.65 \pm 0.06$ &  $0.44_{-0.05}^{+0.06}$ &   $2.2_{-0.7}^{+0.9}$ &  $0.08_{-0.05}^{+0.07}$ \\
2M18535277+4503088 &   8801343 &  1247 &   $6040 \pm 90$ &  $-0.12 \pm 0.01$ &  $3.85_{-0.03}^{+0.05}$ &  $1.29_{-0.06}^{+0.14}$ &  $2.24_{-0.06}^{+0.07}$ &         $0.16 \pm 0.02$ &   $3.8_{-1.2}^{+0.7}$ &  $0.06_{-0.04}^{+0.06}$ \\
2M19160484+4807113 &  10790387 &  1288 &  $6100 \pm 100$ &   $0.04 \pm 0.02$ &         $4.31 \pm 0.04$ &         $1.05 \pm 0.06$ &         $1.19 \pm 0.05$ &           $0.9 \pm 0.1$ &         $6_{-2}^{+3}$ &  $0.13_{-0.07}^{+0.06}$ \\
2M19320489+4230318 &   7037540 &  1347 &  $5800 \pm 100$ &  $-0.34 \pm 0.02$ &         $4.27 \pm 0.04$ &  $0.89_{-0.03}^{+0.04}$ &  $1.15_{-0.05}^{+0.04}$ &           $0.8 \pm 0.1$ &        $11_{-3}^{+2}$ &  $0.13_{-0.07}^{+0.06}$ \\
2M19282877+4255540 &   7363829 &  1356 &  $5600 \pm 100$ &   $0.17 \pm 0.01$ &  $4.03_{-0.05}^{+0.06}$ &  $1.08_{-0.06}^{+0.07}$ &           $1.7 \pm 0.1$ &  $0.33_{-0.05}^{+0.07}$ &             $9 \pm 2$ &  $0.18_{-0.07}^{+0.06}$ \\
2M19460177+4927262 &  11517719 &  1416 &  $5500 \pm 100$ &  $-0.38 \pm 0.02$ &  $3.88_{-0.06}^{+0.07}$ &  $0.94_{-0.05}^{+0.06}$ &     $1.9_{-0.1}^{+0.2}$ &  $0.21_{-0.04}^{+0.05}$ &            $10 \pm 2$ &  $0.13_{-0.08}^{+0.09}$ \\
2M19191325+4629301 &   9705459 &  1448 &   $5300 \pm 90$ &   $0.09 \pm 0.01$ &         $4.37 \pm 0.03$ &  $0.92_{-0.03}^{+0.04}$ &         $1.04 \pm 0.04$ &           $1.2 \pm 0.1$ &        $11_{-3}^{+2}$ &  $0.11_{-0.06}^{+0.08}$ \\
2M19344052+4622453 &   9653622 &  2513 &  $5900 \pm 100$ &  $-0.37 \pm 0.02$ &           $4.3 \pm 0.2$ &   $0.91_{-0.06}^{+0.10}$ &     $1.2_{-0.2}^{+0.4}$ &     $0.8_{-0.5}^{+0.7}$ &             $8 \pm 3$ &  $0.13_{-0.08}^{+0.09}$ \\
2M19254244+4209507 &   6690171 &  3320 &   $5030 \pm 90$ &   $0.29 \pm 0.01$ &         $4.52 \pm 0.03$ &         $0.86 \pm 0.03$ &         $0.85 \pm 0.03$ &           $2.0 \pm 0.2$ &        $10_{-4}^{+3}$ &  $0.17_{-0.09}^{+0.07}$ \\
2M19273337+3921423 &   4263529 &  3358 &  $5300 \pm 100$ &  $-0.16 \pm 0.01$ &         $4.48 \pm 0.02$ &  $0.83_{-0.02}^{+0.03}$ &         $0.87 \pm 0.02$ &           $1.8 \pm 0.1$ &        $11_{-3}^{+2}$ &         $0.08 \pm 0.05$ \\
2M19520793+3952594 &   4773392 &  4367 &  $6300 \pm 100$ &  $-0.02 \pm 0.02$ &         $4.18 \pm 0.05$ &         $1.17 \pm 0.08$ &  $1.45_{-0.05}^{+0.06}$ &  $0.54_{-0.07}^{+0.08}$ &             $4 \pm 2$ &  $0.27_{-0.09}^{+0.06}$ \\
2M19543478+4217089 &   6805414 &  5329 &  $6200 \pm 100$ &   $0.26 \pm 0.02$ &         $4.12 \pm 0.05$ &  $1.34_{-0.08}^{+0.07}$ &         $1.67 \pm 0.09$ &  $0.41_{-0.06}^{+0.07}$ &   $3.1_{-0.9}^{+1.2}$ &  $0.55_{-0.08}^{+0.05}$ \\
2M19480000+4117241 &   5979863 &  6018 &  $5700 \pm 100$ &   $0.04 \pm 0.01$ &         $4.43 \pm 0.04$ &  $0.93_{-0.04}^{+0.05}$ &         $0.97 \pm 0.04$ &           $1.4 \pm 0.2$ &         $8_{-3}^{+4}$ &           $0.3 \pm 0.1$ \\
2M19352118+4207199 &   6698670 &  6760 &   $5850 \pm 80$ &  $-0.13 \pm 0.01$ &         $4.26 \pm 0.03$ &  $0.96_{-0.05}^{+0.06}$ &         $1.20 \pm 0.03$ &  $0.78_{-0.07}^{+0.08}$ &             $9 \pm 3$ &  $0.13_{-0.07}^{+0.06}$ \\
\enddata
\tablenotetext{a}{These are the calibrated values from the ASPCAP pipeline. All other parameters ($M_\star$, $R_\star$, $\rho_\star$, Age, and $A_V$) are parameters in the SED fit using \texttt{EXOFASTv2} using MIST isochrones.}
}
\end{deluxetable*}

\begin{splitdeluxetable*}{ccccccccBccccccc}
    {\tabletypesize{\tiny}
    \tablecaption{Model parameters for the KOI systems. \label{tab:koiorbital}}
    \tablehead{
    \colhead{APOGEE ID}  &
    \colhead{KIC ID}  &
    \colhead{KOI}  &
    \colhead{$P$}  &
    \colhead{$T_0$}  &
    \colhead{$e$}  &
    \colhead{$\omega_\star$}  &
    \colhead{$K$}  &
    \colhead{$\gamma_{\mathrm{APOGEE}}$}  &
    \colhead{$\gamma_{\mathrm{HPF}}$}  &
    \colhead{$\gamma_{\mathrm{SOPHIE}}$}  &
    \colhead{$i$}  &
    \colhead{$a/R_{\star}$} &
    \colhead{$R_{2}/R_{\star}$}  &
    \colhead{$A_{T}$} 
    \\
    \colhead{} & 
    \colhead{} & 
    \colhead{} & 
    \colhead{(days)} & 
    \colhead{($\mathrm{BJD_{TDB}}$)} & 
    \colhead{} & 
    \colhead{(deg)} & 
    \colhead{($\mathrm{km~s^{-1}}$)} & 
    \colhead{($\mathrm{km~s^{-1}}$)} & 
    \colhead{($\mathrm{km~s^{-1}}$)} & 
    \colhead{($\mathrm{km~s^{-1}}$)} & 
    \colhead{(deg)} & 
    \colhead{} & 
    \colhead{} & 
    \colhead{(ppm)}
    }
    \startdata
2M18523991+4524110 &   9071386 &    23 &           $4.693295 \pm 0.000004$ &            $2455077.80838 \pm 0.00004$ &  $0.0009_{-0.0007}^{+0.0033}$ &  $-60_{-30}^{+151}$ &           $17.7 \pm 0.2$ &           $-55.1 \pm 0.1$ & \nodata                   & \nodata                          &         $86.51 \pm 0.06$ &  $7.11_{-0.02}^{+0.03}$ &           $0.1131 \pm 0.0001$ &    $273 \pm 7$ \\
2M19395458+3840421 &   3558981 &    52 &         $2.9878624 \pm 0.0000003$ &            $2455598.58385 \pm 0.00005$ &      $0.005_{-0.003}^{+0.010}$ &   $-40_{-40}^{+97}$ &           $26.2 \pm 0.4$ &     $-46.8_{-0.2}^{+0.3}$ & \nodata                   & \nodata                          &           $85.7 \pm 0.1$ &  $8.79_{-0.09}^{+0.12}$ &           $0.2018 \pm 0.0009$ &  $1950 \pm 30$ \\
2M19480226+5022203 &  11974540 &   129 &          $24.669193 \pm 0.000003$ &              $2454965.8633 \pm 0.0001$ &               $0.08 \pm 0.01$ &    $160_{-300}^{+20}$ &            $5.9 \pm 0.1$ &            $16.6 \pm 0.1$ & \nodata                   &            $15.6 \pm 0.2$ &         $89.22 \pm 0.07$ &          $27.2 \pm 0.7$ &           $0.0796 \pm 0.0002$ & \nodata               \\
2M19492647+4025473 &   5297298 &   130 &          $34.193602 \pm 0.000003$ &            $2455432.81176 \pm 0.00006$ &             $0.495 \pm 0.008$ &       $136_{-2}^{+1}$ &           $11.0 \pm 0.2$ &          $20.92 \pm 0.09$ & \nodata                   & \nodata                          &         $87.53 \pm 0.08$ &          $27.6 \pm 0.4$ &           $0.1141 \pm 0.0002$ & \nodata               \\
2M19424111+4035566 &   5376836 &   182 &         $3.4794244 \pm 0.0000002$ &            $2455512.58740 \pm 0.00002$ &     $0.001_{-0.001}^{+0.007}$ &  $-40_{-60}^{+130}$ &           $20.2 \pm 0.4$ &           $-59.8 \pm 0.4$ & \nodata                   & \nodata                          &         $84.40 \pm 0.05$ &         $7.00 \pm 0.03$ &           $0.1365 \pm 0.0001$ &   $480 \pm 10$ \\
2M19485138+4139505 &   6305192 &   219 &           $8.025118 \pm 0.000002$ &              $2454965.4689 \pm 0.0003$ &               $0.26 \pm 0.02$ &           $-63 \pm 4$ &    $3.47_{-0.09}^{+0.10}$ &          $-8.07 \pm 0.07$ & \nodata                   &          $-8.41 \pm 0.06$ &     $87.4_{-0.3}^{+0.5}$ &    $11.8_{-0.5}^{+1.0}$ &  $0.0539_{-0.0011}^{+0.0005}$ & \nodata               \\
2M19223275+3842276 &   3642741 &   242 &           $7.258448 \pm 0.000002$ &              $2455951.2350 \pm 0.0001$ &     $0.007_{-0.005}^{+0.008}$ &    $0_{-100}^{+68}$ &   $7.01_{-0.07}^{+0.08}$ &         $-44.56 \pm 0.06$ & \nodata                   & \nodata                          &     $88.2_{-0.6}^{+0.9}$ &           $9.8 \pm 0.3$ &           $0.0560 \pm 0.0005$ & \nodata               \\
2M19073111+3922421 &   4247092 &   403 &            $21.05649 \pm 0.00002$ &                $2456541.250 \pm 0.002$ &             $0.644 \pm 0.004$ &           $165 \pm 1$ &           $16.8 \pm 0.3$ &           $-22.6 \pm 0.1$ & \nodata                   & \nodata                          &           $83.9 \pm 0.6$ &    $23.9_{-0.6}^{+0.7}$ &                 $0.3 \pm 0.1$ & \nodata               \\
2M19331345+4136229 &   6289650 &   415 &             $166.7879 \pm 0.0001$ &              $2455078.1422 \pm 0.0005$ &             $0.701 \pm 0.002$ &        $44.9 \pm 0.3$ &          $3.36 \pm 0.01$ &          $-1.26 \pm 0.03$ & \nodata                   &        $-1.481 \pm 0.006$ &           $89.2 \pm 0.3$ &          $98_{-7}^{+6}$ &             $0.064 \pm 0.001$ & \nodata               \\
2M19043647+4519572 &   9008220 &   466 &           $9.391039 \pm 0.000004$ &              $2455003.5390 \pm 0.0003$ &     $0.027_{-0.007}^{+0.008}$ &          $110 \pm 10$ &          $7.95 \pm 0.08$ &         $-56.50 \pm 0.06$ & \nodata                   &  $-56.75_{-0.08}^{+0.09}$ &           $85.4 \pm 0.2$ &          $12.1 \pm 0.4$ &             $0.073 \pm 0.004$ & \nodata               \\
2M19214782+3951172 &   4742414 &   631 &          $15.458053 \pm 0.000003$ &     $2455006.7820_{-0.0001}^{+0.0002}$ &             $0.193 \pm 0.004$ &           $-37 \pm 2$ &          $6.74 \pm 0.04$ &         $-56.43 \pm 0.05$ &  $-56.45 \pm 0.03$ & \nodata                          &           $89.2 \pm 0.2$ &          $15.5 \pm 0.2$ &           $0.0599 \pm 0.0002$ & \nodata               \\
2M19371604+5004488 &  11818800 &   777 &  $40.41940_{-0.00004}^{+0.00005}$ &                $2455006.566 \pm 0.001$ &        $0.70_{-0.03}^{+0.05}$ &           $107 \pm 4$ &     $15.0_{-0.9}^{+1.8}$ &      $-3.8_{-0.2}^{+0.3}$ & \nodata                   & \nodata                          &               $83 \pm 1$ &          $24_{-1}^{+2}$ &     $0.088_{-0.002}^{+0.005}$ & \nodata               \\
2M19473316+4123459 &   6061119 &   846 &          $27.807565 \pm 0.000004$ &              $2455659.2861 \pm 0.0001$ &               $0.37 \pm 0.02$ &       $147_{-6}^{+5}$ &      $9.1_{-0.3}^{+0.4}$ &     $-16.8_{-0.2}^{+0.3}$ & \nodata                   & \nodata                          &     $88.1_{-0.2}^{+0.1}$ &              $33 \pm 1$ &           $0.1619 \pm 0.0004$ & \nodata               \\
2M19270249+4156386 &   6522242 &   855 &          $41.408310 \pm 0.000006$ &              $2455028.7868 \pm 0.0001$ &             $0.148 \pm 0.003$ &           $118 \pm 1$ &          $6.06 \pm 0.02$ &         $-94.87 \pm 0.07$ & \nodata                   &         $-94.89 \pm 0.02$ &  $89.78_{-0.07}^{+0.09}$ &    $61.7_{-0.8}^{+0.7}$ &             $0.136 \pm 0.001$ & \nodata               \\
2M19001520+4410043 &   8218274 &  1064 &       $1.18735246 \pm 0.00000003$ &  $2455754.86785_{-0.00006}^{+0.00007}$ &               $0.04 \pm 0.02$ &     $120_{-20}^{+30}$ &     $23.7_{-0.5}^{+0.4}$ &             $4.4 \pm 0.4$ & \nodata                   & \nodata                          &           $80.6 \pm 0.4$ &  $3.19_{-0.06}^{+0.05}$ &  $0.1144_{-0.0003}^{+0.0002}$ & \nodata               \\
2M18535277+4503088 &   8801343 &  1247 &           $2.739877 \pm 0.000001$ &            $2455808.69158 \pm 0.00001$ &     $0.006_{-0.005}^{+0.007}$ &      $-83_{-4}^{+60}$ &           $25.6 \pm 0.4$ &             $3.2 \pm 0.2$ & \nodata                   & \nodata                          &  $81.93_{-0.09}^{+0.11}$ &         $4.86 \pm 0.03$ &           $0.1371 \pm 0.0001$ &    $668 \pm 4$ \\
2M19160484+4807113 &  10790387 &  1288 &           $117.93111 \pm 0.00007$ &              $2455052.7037 \pm 0.0005$ &               $0.69 \pm 0.01$ &             $4 \pm 2$ &            $4.1 \pm 0.1$ &           $6.84 \pm 0.08$ & \nodata                   &             $6.5 \pm 0.1$ &           $89.6 \pm 0.2$ &             $109 \pm 7$ &     $0.085_{-0.001}^{+0.002}$ & \nodata               \\
2M19320489+4230318 &   7037540 &  1347 &          $14.405857 \pm 0.000001$ &            $2455582.34229 \pm 0.00004$ &  $0.0055_{-0.0006}^{+0.0022}$ &   $-110_{-50}^{+300}$ &           $12.4 \pm 0.2$ &           $-32.4 \pm 0.1$ & \nodata                   & \nodata                          &   $89.89_{-0.1}^{+0.07}$ &          $23.2 \pm 0.1$ &           $0.1587 \pm 0.0002$ &   $530 \pm 10$ \\
2M19282877+4255540 &   7363829 &  1356 &               $787.432 \pm 0.002$ &                $2455168.816 \pm 0.002$ &               $0.66 \pm 0.02$ &       $107_{-7}^{+8}$ &            $2.4 \pm 0.1$ &          $13.83 \pm 0.07$ & \nodata                   & \nodata                          &  $89.72_{-0.09}^{+0.08}$ &            $220 \pm 10$ &   $0.0733_{-0.0009}^{+0.0010}$ & \nodata               \\
2M19460177+4927262 &  11517719 &  1416 &         $2.4957813 \pm 0.0000005$ &            $2456073.96726 \pm 0.00002$ &  $0.0007_{-0.0004}^{+0.0014}$ &  $-40_{-40}^{+110}$ &           $24.5 \pm 0.4$ &           $-43.5 \pm 0.3$ & \nodata                   & \nodata                          &           $88.3 \pm 0.2$ &         $5.24 \pm 0.01$ &           $0.1459 \pm 0.0001$ &    $914 \pm 9$ \\
2M19191325+4629301 &   9705459 &  1448 &         $2.4865864 \pm 0.0000004$ &            $2456212.88989 \pm 0.00002$ &     $0.002_{-0.002}^{+0.007}$ &  $-70_{-10}^{+132}$ &           $23.3 \pm 0.3$ &           $-54.6 \pm 0.2$ & \nodata                   & \nodata                          &  $86.89_{-0.05}^{+0.06}$ &  $7.74_{-0.03}^{+0.05}$ &           $0.1886 \pm 0.0003$ &   $890 \pm 20$ \\
2M19344052+4622453 &   9653622 &  2513 &            $19.00547 \pm 0.00005$ &                $2454977.152 \pm 0.003$ &               $0.36 \pm 0.03$ &      $-154_{-8}^{+7}$ &            $2.0 \pm 0.1$ &   $11.12_{-0.06}^{+0.07}$ & \nodata                   & \nodata                          &     $89.7_{-2.4}^{+0.2}$ &       $100_{-80}^{+30}$ &     $0.022_{-0.002}^{+0.204}$ & \nodata               \\
2M19254244+4209507 &   6690171 &  3320 &            $85.06240 \pm 0.00003$ &              $2455832.8434 \pm 0.0001$ &        $0.27_{-0.03}^{+0.04}$ &           $116 \pm 2$ &      $7.6_{-0.1}^{+0.2}$ &    $12.27_{-0.10}^{+0.09}$ & \nodata                   & \nodata                          &  $89.34_{-0.06}^{+0.04}$ &              $97 \pm 3$ &     $0.225_{-0.005}^{+0.007}$ & \nodata               \\
2M19273337+3921423 &   4263529 &  3358 &          $10.104042 \pm 0.000004$ &              $2455652.9675 \pm 0.0002$ &             $0.009 \pm 0.006$ &     $-70_{-10}^{+30}$ &   $16.66_{-0.09}^{+0.1}$ &           $8.87 \pm 0.06$ & \nodata                   & \nodata                          &     $86.9_{-0.4}^{+0.3}$ &          $23.1 \pm 0.5$ &                 $0.3 \pm 0.1$ & \nodata               \\
2M19520793+3952594 &   4773392 &  4367 &               $170.996 \pm 0.002$ &                  $2454996.97 \pm 0.01$ &        $0.85_{-0.01}^{+0.02}$ &           $-57 \pm 2$ &     $11.5_{-0.4}^{+0.9}$ &     $-20.4_{-0.3}^{+0.2}$ & \nodata                   & \nodata                          &           $89.2 \pm 0.1$ &             $102 \pm 5$ &                 $0.3 \pm 0.1$ & \nodata               \\
2M19543478+4217089 &   6805414 &  5329 &             $200.2348 \pm 0.0005$ &                $2455138.446 \pm 0.002$ &        $0.66_{-0.05}^{+0.08}$ &           $-59 \pm 5$ &                $6 \pm 1$ &     $-24.5_{-0.7}^{+0.6}$ & \nodata                   & \nodata                          &  $89.71_{-0.04}^{+0.03}$ &         $100_{-6}^{+7}$ &  $0.1012_{-0.0009}^{+0.0008}$ & \nodata               \\
2M19480000+4117241 &   5979863 &  6018 &        $16.6218543 \pm 0.0000009$ &            $2455697.91363 \pm 0.00004$ &     $0.119_{-0.011}^{+0.009}$ &        $75_{-2}^{+1}$ &           $14.4 \pm 0.2$ &            $10.0 \pm 0.1$ & \nodata                   & \nodata                          &         $88.80 \pm 0.03$ &    $31.1_{-0.3}^{+0.4}$ &           $0.2242 \pm 0.0005$ &  $1290 \pm 30$ \\
2M19352118+4207199 &   6698670 &  6760 &            $10.81593 \pm 0.00003$ &            $2455351.34940 \pm 0.00003$ &  $0.1529_{-0.0005}^{+0.0006}$ &             $9 \pm 1$ &  $16.67_{-0.07}^{+0.06}$ &  $-38.48_{-0.04}^{+0.05}$ & \nodata                   & \nodata                          &         $89.28 \pm 0.03$ &        $19.74 \pm 0.08$ &           $0.1971 \pm 0.0003$ &   $830 \pm 20$ \\
    \enddata
    }
\end{splitdeluxetable*}

\begin{deluxetable*}{ccccccccc}
{\tabletypesize{\footnotesize}
\rotate
\tablecaption{Derived physical parameters for the KOI systems. \label{tab:koiphysical}}
\tablehead{
\colhead{APOGEE ID}  &
\colhead{KIC ID}  &
\colhead{KOI}  &
\colhead{$R_{2}$} &
\colhead{$M_{2}$} &
\colhead{$q$} &
\colhead{$a$} &
\colhead{$\log g_{2}$} &
\colhead{$\rho_{2}$}
\\
\colhead{} & 
\colhead{} & 
\colhead{} & 
\colhead{($\mathrm{R_{J}}$)} & 
\colhead{($\mathrm{M_{J}}$)} & 
\colhead{} & 
\colhead{(au)} & 
\colhead{(dex)} & 
\colhead{($\mathrm{g~cm^{-3}}$)}
}
\startdata
2M18523991+4524110 &   9071386 &    23 &         $2.22 \pm 0.06$ &   $204_{-8}^{+6}$ &   $0.13_{-0.010}^{+0.007}$ &         $0.067 \pm 0.002$ &         $5.036 \pm 0.005$ &              $23 \pm 2$ \\
2M19395458+3840421 &   3558981 &    52 &  $1.91_{-0.05}^{+0.06}$ &       $195 \pm 5$ & $0.211_{-0.007}^{+0.008}$ &         $0.040 \pm 0.001$ &           $5.09 \pm 0.01$ &              $35 \pm 3$ \\
2M19480226+5022203 &  11974540 &   129 &         $1.31 \pm 0.04$ &   $103_{-7}^{+4}$ & $0.078_{-0.009}^{+0.006}$ &         $0.214 \pm 0.008$ &           $5.30 \pm 0.03$ &          $57_{-6}^{+5}$ \\
2M19492647+4025473 &   5297298 &   130 &         $2.03 \pm 0.06$ &  $187_{-8}^{+10}$ &    $0.15_{-0.01}^{+0.02}$ &         $0.235 \pm 0.008$ &           $5.08 \pm 0.01$ &              $28 \pm 3$ \\
2M19424111+4035566 &   5376836 &   182 &         $2.05 \pm 0.06$ &   $174_{-7}^{+8}$ &           $0.16 \pm 0.01$ &         $0.050 \pm 0.002$ &           $5.05 \pm 0.01$ &          $25_{-2}^{+3}$ \\
2M19485138+4139505 &   6305192 &   219 &     $0.8_{-0.2}^{+0.3}$ &    $37_{-2}^{+4}$ & $0.031_{-0.004}^{+0.006}$ &    $0.08_{-0.02}^{+0.03}$ &    $5.16_{-0.05}^{+0.08}$ &      $100_{-60}^{+100}$ \\
2M19223275+3842276 &   3642741 &   242 &         $0.97 \pm 0.04$ &    $77_{-3}^{+5}$ & $0.063_{-0.004}^{+0.008}$ &         $0.081 \pm 0.004$ &           $5.33 \pm 0.04$ &            $110 \pm 10$ \\
2M19073111+3922421 &   4247092 &   403 &               $4 \pm 2$ &  $208_{-9}^{+10}$ &    $0.18_{-0.01}^{+0.02}$ &         $0.166 \pm 0.007$ &       $4.5_{-0.3}^{+0.4}$ &               $3 \pm 4$ \\
2M19331345+4136229 &   6289650 &   415 &  $0.83_{-0.03}^{+0.04}$ &    $65_{-1}^{+2}$ & $0.066_{-0.003}^{+0.004}$ &    $0.61_{-0.05}^{+0.04}$ &    $5.39_{-0.09}^{+0.06}$ &            $140 \pm 20$ \\
2M19043647+4519572 &   9008220 &   466 &         $1.19 \pm 0.08$ &    $92_{-3}^{+4}$ & $0.081_{-0.005}^{+0.006}$ &         $0.095 \pm 0.005$ &    $5.23_{-0.07}^{+0.06}$ &             $70 \pm 10$ \\
2M19214782+3951172 &   4742414 &   631 &         $1.11 \pm 0.03$ &        $92 \pm 3$ & $0.078_{-0.004}^{+0.005}$ &         $0.137 \pm 0.005$ &    $5.32_{-0.02}^{+0.01}$ &              $83 \pm 8$ \\
2M19371604+5004488 &  11818800 &   777 &     $1.9_{-0.1}^{+0.2}$ & $190_{-20}^{+30}$ &    $0.21_{-0.02}^{+0.03}$ &           $0.25 \pm 0.02$ &    $5.17_{-0.06}^{+0.07}$ &          $32_{-6}^{+9}$ \\
2M19473316+4123459 &   6061119 &   846 &           $2.3 \pm 0.1$ &  $150_{-9}^{+10}$ &           $0.12 \pm 0.01$ &           $0.23 \pm 0.02$ &           $4.98 \pm 0.04$ &          $15_{-2}^{+3}$ \\
2M19270249+4156386 &   6522242 &   855 &  $1.20_{-0.04}^{+0.05}$ &    $93_{-1}^{+2}$ & $0.114_{-0.003}^{+0.004}$ &           $0.26 \pm 0.01$ &           $5.34 \pm 0.02$ &              $67 \pm 8$ \\
2M19001520+4410043 &   8218274 &  1064 &         $1.83 \pm 0.06$ &   $167_{-7}^{+6}$ & $0.115_{-0.008}^{+0.007}$ &       $0.0244 \pm 0.0009$ &           $5.06 \pm 0.02$ &              $34 \pm 4$ \\
2M18535277+4503088 &   8801343 &  1247 &         $2.98 \pm 0.09$ &  $235_{-8}^{+20}$ &    $0.17_{-0.01}^{+0.02}$ & $0.050_{-0.001}^{+0.002}$ & $4.935_{-0.008}^{+0.009}$ &              $11 \pm 1$ \\
2M19160484+4807113 &  10790387 &  1288 &  $0.98_{-0.04}^{+0.05}$ &        $78 \pm 4$ &         $0.071 \pm 0.006$ &    $0.60_{-0.04}^{+0.05}$ &    $5.49_{-0.08}^{+0.07}$ &       $100_{-10}^{+20}$ \\
2M19320489+4230318 &   7037540 &  1347 &         $1.77 \pm 0.07$ &   $152_{-4}^{+5}$ &  $0.162_{-0.007}^{+0.010}$ &         $0.124 \pm 0.005$ & $5.126_{-0.007}^{+0.006}$ &              $34 \pm 4$ \\
2M19282877+4255540 &   7363829 &  1356 &         $1.19 \pm 0.07$ &    $93_{-6}^{+7}$ & $0.082_{-0.007}^{+0.008}$ &             $1.7 \pm 0.1$ &    $5.17_{-0.06}^{+0.05}$ &             $70 \pm 10$ \\
2M19460177+4927262 &  11517719 &  1416 &           $2.6 \pm 0.2$ &   $175_{-7}^{+8}$ &           $0.18 \pm 0.01$ &         $0.045 \pm 0.004$ &         $4.965 \pm 0.008$ &              $12 \pm 3$ \\
2M19191325+4629301 &   9705459 &  1448 &  $1.91_{-0.07}^{+0.08}$ &   $163_{-4}^{+5}$ & $0.169_{-0.006}^{+0.008}$ & $0.038_{-0.001}^{+0.002}$ & $5.062_{-0.007}^{+0.008}$ &          $29_{-3}^{+4}$ \\
2M19344052+4622453 &   9653622 &  2513 &     $0.25_{-0.05}^{+2.00}$ &        $23 \pm 2$ & $0.025_{-0.003}^{+0.004}$ &       $0.5_{-0.4}^{+0.2}$ &           $7_{-3}^{+0.2}$ & $2000_{-1000}^{+50000}$ \\
2M19254244+4209507 &   6690171 &  3320 &  $1.86_{-0.07}^{+0.08}$ &   $160_{-5}^{+7}$ &  $0.178_{-0.008}^{+0.010}$ &           $0.38 \pm 0.02$ &    $5.06_{-0.04}^{+0.03}$ &              $31 \pm 4$ \\
2M19273337+3921423 &   4263529 &  3358 &               $2 \pm 1$ &   $177_{-3}^{+4}$ & $0.204_{-0.007}^{+0.009}$ &         $0.093 \pm 0.003$ &       $4.9_{-0.4}^{+0.5}$ &             $20 \pm 20$ \\
2M19520793+3952594 &   4773392 &  4367 &               $4 \pm 2$ & $200_{-10}^{+30}$ &           $0.16 \pm 0.02$ &           $0.69 \pm 0.04$ &       $4.5_{-0.3}^{+0.5}$ &           $4_{-5}^{+6}$ \\
2M19543478+4217089 &   6805414 &  5329 &         $1.64 \pm 0.09$ & $170_{-30}^{+40}$ &    $0.12_{-0.02}^{+0.03}$ &    $0.77_{-0.06}^{+0.07}$ &             $5.2 \pm 0.1$ &             $50 \pm 10$ \\
2M19480000+4117241 &   5979863 &  6018 &         $2.13 \pm 0.08$ &   $194_{-6}^{+7}$ &           $0.20 \pm 0.01$ & $0.141_{-0.005}^{+0.006}$ &           $5.08 \pm 0.01$ &              $25 \pm 3$ \\
2M19352118+4207199 &   6698670 &  6760 &         $2.31 \pm 0.06$ &   $197_{-6}^{+8}$ &           $0.20 \pm 0.01$ &         $0.111 \pm 0.003$ &         $5.046 \pm 0.004$ &              $20 \pm 2$ \\
\enddata
}
\end{deluxetable*}

\appendix
\section{Photometric detrending}\label{app:a}
We detrended all \textit{Kepler} photometry before fitting with \texttt{EXOFASTv2} using a Gaussian process noise model with the approximate quasi-periodic kernel presented in \cite{Foreman-Mackey2017} of the form:
\small
\begin{equation}
 k(\tau) = \frac{B}{2 + C} e^{-\tau/L} \left[ \cos \left( \frac{2 \pi \tau}{P_\mathrm{GP}} \right) + (1 + C) \right],
 \label{eq:kernelperiodic}
\end{equation}
\normalsize
where $\tau$ is the time of observation while $B$, $C$, $L$, and $P_{\mathrm{GP}}$ are the hyperparameters of the covariance function. $B$ and $C$ represent the weight of the exponential term with a decay constant of $L$ (in days). $P_{\mathrm{GP}}$ determines the periodicity of the quasi-periodic oscillations, which is interpreted as the stellar rotation period. This kernel is able to reproduce the behavior of a more traditional quasi-periodic covariance function and has allowed for computationally efficient inference of stellar rotation periods even for large datasets that are not uniformly sampled \citep[e.g.,][]{Angus2018}. For the purposes of detrending, we make no distinction between activity induced variability (e.g., star spots) or phase modulations due to the binary nature of a system (e.g., Doppler boosting, stellar ellipsoidal distortion, reflection). 

\begin{figure}[!ht]
    \figurenum{6}
    \epsscale{1.15}
    \plotone{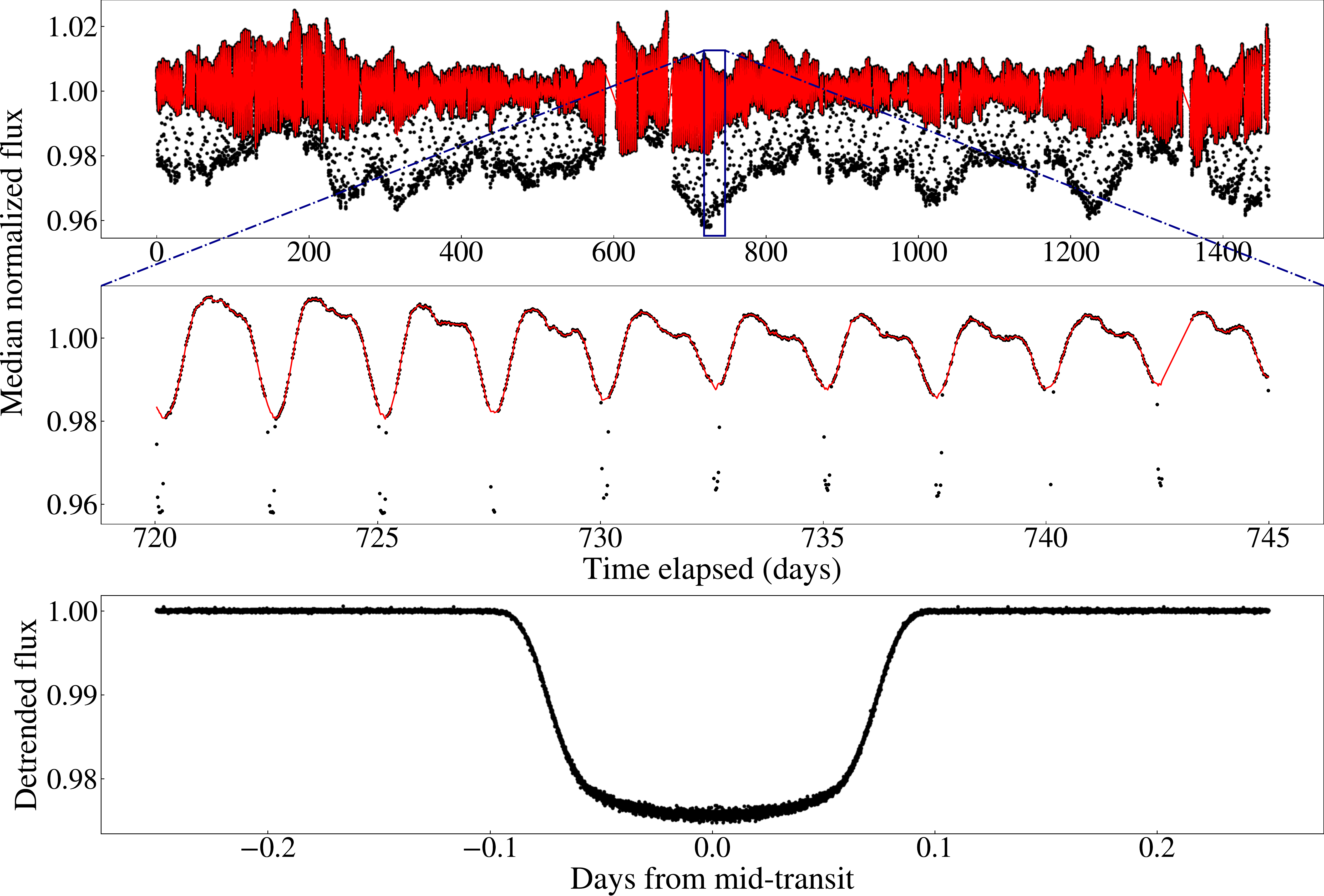}
    \caption{\textbf{Gaussian process detrending}. The example above shows the detrending result for one quarter of long-cadence data for KOI-1416. The out-of-transit photometric variability in the \textit{Kepler} light curve can be modeled using a quasi-periodic kernel. The top panel shows the flux (black points) normalized to the median value while the middle panel is a magnification to the region highlighted by the rectangle. The red line is the maximum \textit{a posteriori} model. The bottom panel shows the phase-folded photometry, once the baseline found by the Gaussian process is removed. No additional processing (e.g., sigma clipping) is performed. \label{fig:exampledetrend}}
\end{figure}

For each system, we use the parameters from \textit{Kepler} DR25 candidate list \citep{Thompson2018} to excise a window of $\pm1.5$ times the transit duration around each transit before fitting the Gaussian process model. We estimate the maximum \textit{a posteriori} parameters for the GP model using the \texttt{L-BFGS-B} non-linear optimization routine implemented in \texttt{scipy} \citep{Virtanen2020}. Each quarter of \textit{Kepler} data is processed separately. An example of detrending with a GP is shown in Figure \ref{fig:exampledetrend}.

\section{Photometric rotation period}\label{app:rotation}
We use the available \textit{Kepler} photometry to derive the stellar rotation period. The \textit{Kepler} team notes \citep[see Section 5.15 in][]{VanCleve2016} that long period signals may be attenuated in the PDCSAP flux and suggest searching for long-period signals in the SAP flux after detrending with the co-trending basis vectors \citep[CBVs;][]{Aigrain2017,Cui2019}. We employ the \texttt{ARC2}\footnote{\url{https://github.com/OxES/OxKeplerSC}} pipeline \citep{Aigrain2017} to correct for systematics in the \textit{Kepler} SAP light curve. The \texttt{ARC2} pipeline detects and removes isolated discontinuities and the instrument systematic trends from the photometry by using the CBVs \citep[see][]{Kinemuchi2012}. All light curves derived from \texttt{ARC2} used seven CBVs for detrending.

\begin{figure}[!ht]
    \figurenum{7a}
    \epsscale{1.15}
    \plotone{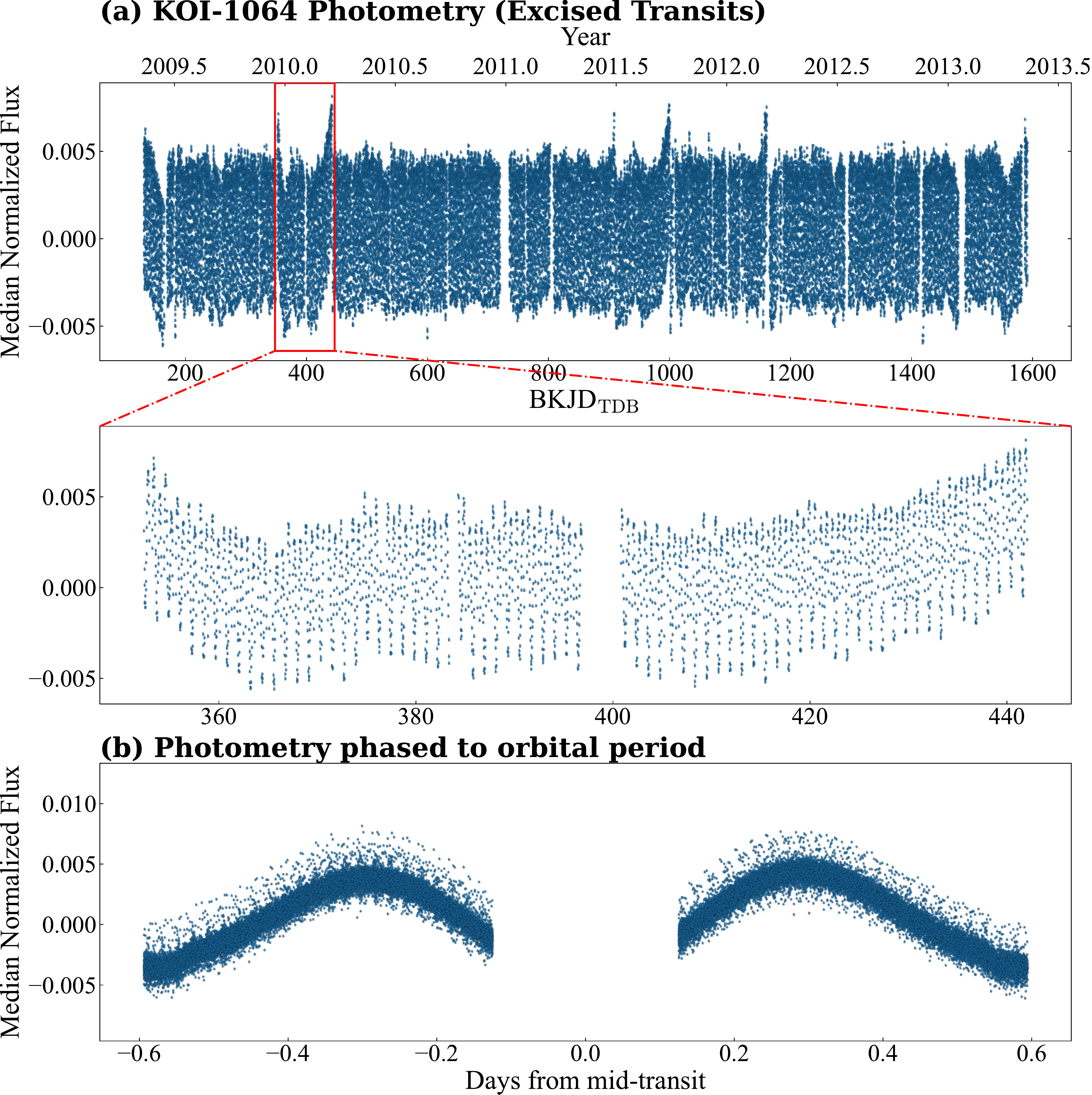}
    \caption{\textbf{(a)} The \texttt{ARC2} corrected light curve for KOI-1064, after excising the transits, with an inset displaying the quarter 2 data. \textbf{(b)} The out-of-transit photometric variability after phasing to the orbital ephemeris. This is an example system where the variability is consistent with ellipsoidal variations. The complete figure set for the KOIs (28 images) is available in the online journal. \label{fig:koirot1}}
\end{figure}

\begin{figure}[!ht]
    \figurenum{7b}
    \epsscale{1.15}
    \plotone{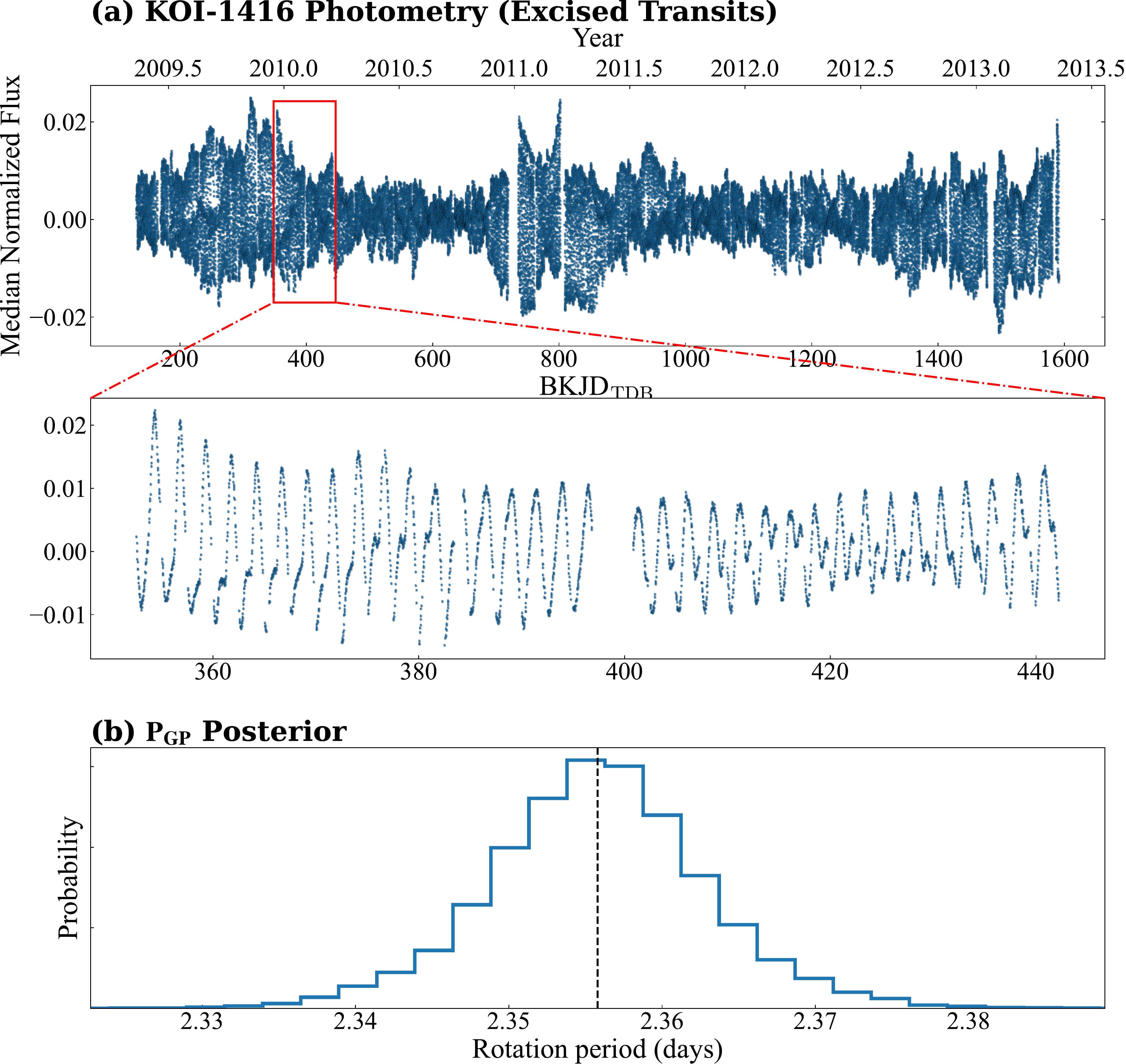}
    \caption{Figure 7 (cont.). \textbf{(a)} The \texttt{ARC2} corrected light curve for KOI-1416, after excising the transits, with an inset displaying a subset of the \textit{Kepler} data. \textbf{(b)} The posterior distribution for the Gaussian process period, which we interpret as a measurement of the stellar rotation period. This is an example system where the variability is different from the measured orbital period. \label{fig:koirot2}}
\end{figure}

To derive the rotation period and an estimate of its uncertainty, we modeled the \textit{Kepler} photometry using the \texttt{juliet} analysis package \citep{Espinoza2019}. We adopt the photometric model from Equation \ref{eq:kernelperiodic} and perform the parameter estimation using the dynamic nested-sampling algorithm \texttt{dynesty} \citep{Speagle2020}. We placed a broad uniform prior on the rotation period of $1-1500$ days. Table \ref{tab:rotation} lists the rotation periods for the KOIs. We do not report a rotation period for systems where the period of the Gaussian process coincides with the orbital period (KOI-23,KOI-855, KOI-1064, KOI-1247). A few systems (KOI-130, KOI-415, KOI-1288, KOI-1347, KOI-1356, KOI-2513) had no significantly detected rotation period ($P_{GP}/\sigma_{P_{GP}} < 5$) but are included in Table \ref{tab:rotation} for reference. Figure \ref{fig:koirot1} is an example of a system where the observed photometric modulation is due to ellipsoidal variations. Figure \ref{fig:koirot2} displays a system where the measured period of the Gaussian process model differs from the orbital period and we interpret the period of the Gaussian process to reflect the rotation period of such stars. 

\begin{deluxetable}{llccc}
{\tabletypesize{\normalsize }
\tablecaption{Stellar rotation period of the KOI systems. \label{tab:rotation}}
\tablehead{
\colhead{APOGEE ID} &
\colhead{KIC ID} &
\colhead{KOI ID}  &
\colhead{Rotation Period$^a$} &
\colhead{Orbital Period$^b$}
\\
\colhead{} & 
\colhead{} & 
\colhead{} &
\colhead{(d)} &
\colhead{(d)} 
}
\startdata
2M18523991+4524110 &   9071386 &    23 &       \nodata                           &   $4.693$ \\
2M19395458+3840421 &   3558981 &    52 &                $3.345 \pm 0.009$ &   $2.988$ \\
2M19480226+5022203 &  11974540 &   129 &                $1.553 \pm 0.005$ &  $24.669$ \\
2M19492647+4025473 &   5297298 &   130 &                  $3.37 \pm 0.04$ &  $34.194$ \\
2M19424111+4035566 &   5376836 &   182 &                $3.475 \pm 0.009$ &   $3.479$ \\
2M19485138+4139505 &   6305192 &   219 &              $7.3_{-0.1}^{+0.2}$ &   $8.025$ \\
2M19223275+3842276 &   3642741 &   242 &                    $8.4 \pm 0.2$ &   $7.258$ \\
2M19073111+3922421 &   4247092 &   403 &              $2.6323 \pm 0.0004$ &  $21.056$ \\
2M19331345+4136229 &   6289650 &   415 &              $16.8_{-0.2}^{+40}$ & $166.788$ \\
2M19043647+4519572 &   9008220 &   466 &                   $12.5 \pm 0.2$ &   $9.391$ \\
2M19214782+3951172 &   4742414 &   631 &                  $8.09 \pm 0.08$ &  $15.458$ \\
2M19371604+5004488 &  11818800 &   777 &                 $11.01 \pm 0.07$ &  $40.420$ \\
2M19473316+4123459 &   6061119 &   846 &                  $2.61 \pm 0.01$ &  $27.808$ \\
2M19270249+4156386 &   6522242 &   855 &   \nodata         &  $41.408$ \\
2M19001520+4410043 &   8218274 &  1064 & \nodata  &   $1.187$ \\
2M18535277+4503088 &   8801343 &  1247 &    \nodata                              &   $2.740$ \\
2M19160484+4807113 &  10790387 &  1288 &            $6.79_{-0.04}^{+0.1}$ & $117.931$ \\
2M19320489+4230318 &   7037540 &  1347 &                 $10_{-10}^{+50}$ &  $14.406$ \\
2M19282877+4255540 &   7363829 &  1356 &                 $50_{-50}^{+30}$ & $384.026$ \\
2M19460177+4927262 &  11517719 &  1416 &                $2.356 \pm 0.007$ &   $2.496$ \\
2M19191325+4629301 &   9705459 &  1448 &                $2.744 \pm 0.006$ &   $2.487$ \\
2M19344052+4622453 &   9653622 &  2513 &           $15.01_{-0.03}^{+0.5}$ &  $19.005$ \\
2M19254244+4209507 &   6690171 &  3320 &             $14.9_{-0.3}^{+0.2}$ &  $85.062$ \\
2M19273337+3921423 &   4263529 &  3358 &          $15.25_{-0.07}^{+0.08}$ &  $10.104$ \\
2M19520793+3952594 &   4773392 &  4367 &                  $3.01 \pm 0.01$ & $170.996$ \\
2M19543478+4217089 &   6805414 &  5329 &                  $2.68 \pm 0.01$ & $200.235$ \\
2M19480000+4117241 &   5979863 &  6018 &                   $10.3 \pm 0.4$ &  $16.622$ \\
2M19352118+4207199 &   6698670 &  6760 &                  $7.0_{-2.0}^{+0.2}$ &  $10.816$ \\
\enddata
\tablenotetext{a}{Empty rows indicate the photometric variability occurs at the orbital period.}
\tablenotetext{b}{The period of the orbit as listed in the DR25 KOI supplemental table.}
}
\end{deluxetable}

\include{fig1set}

\include{fig2set}

\bibliography{combined}
\bibliographystyle{aasjournal}

\end{CJK*}
\end{document}

%% file: fig1set.tex
\figsetstart
\figsetnum{2}
\figsettitle{Photometry and RVs for the sample of 28 KOIs}
The Kepler photometry and RVs for all systems analyzed in this work.

\figsetgrpstart
\figsetgrpnum{2.1}
\figsetgrptitle{KOI--23}
\figsetplot{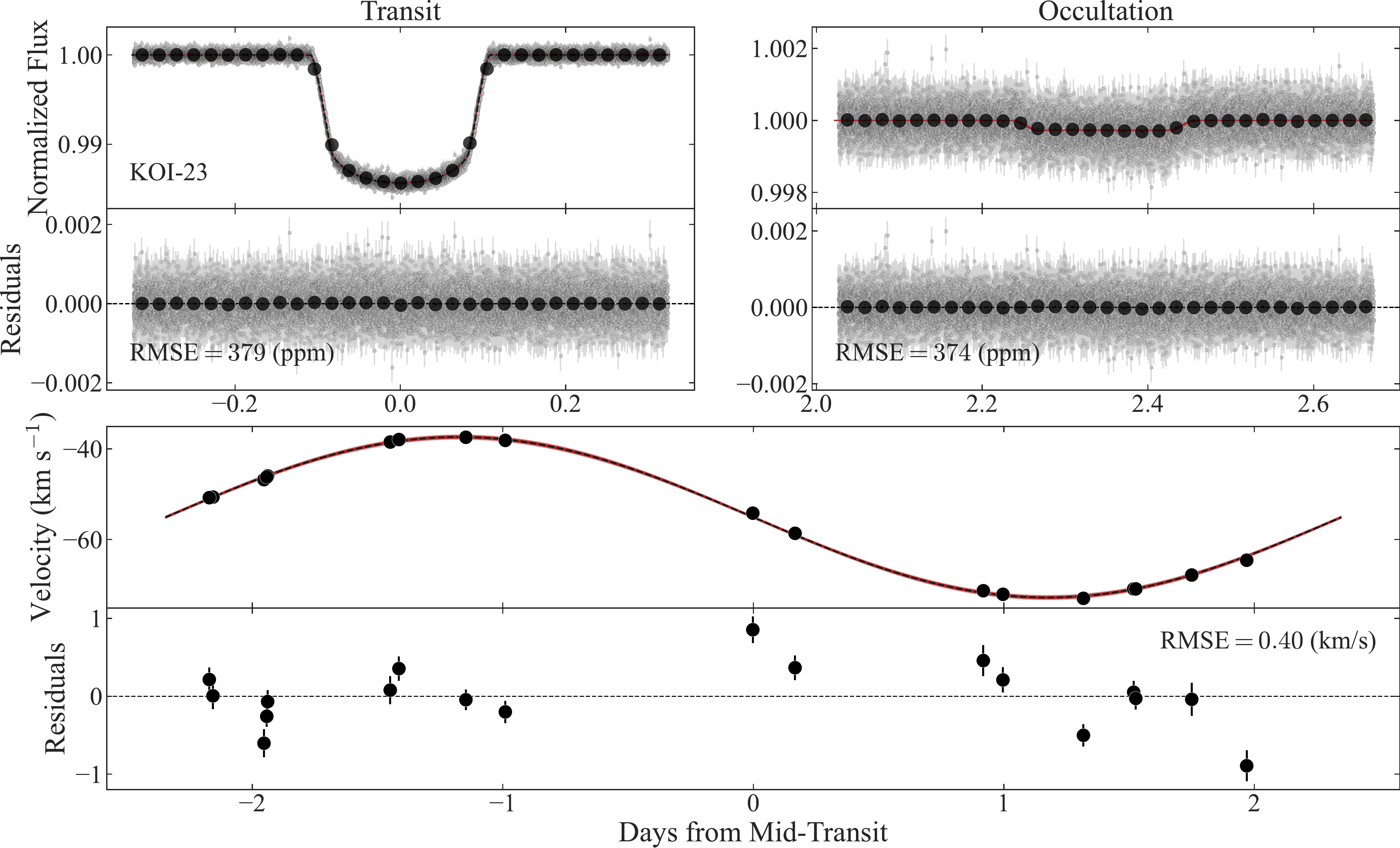}
\figsetgrpnote{\textbf{Top.} The Kepler photometry center on the transit (left) and occultation (right) for KOI-23 after phase-folding to the derived ephemeris. The large circles represent 30 min bins of the raw data. \textbf{Bottom}. The RVs after phase-folding the data to the derived ephemeris. In each panel, the $1\sigma$ (darkest), $2\sigma$, and $3\sigma$ (brightest) extent of the models are shown for reference.}
\figsetgrpend

\figsetgrpstart
\figsetgrpnum{2.2}
\figsetgrptitle{KOI--52}
\figsetplot{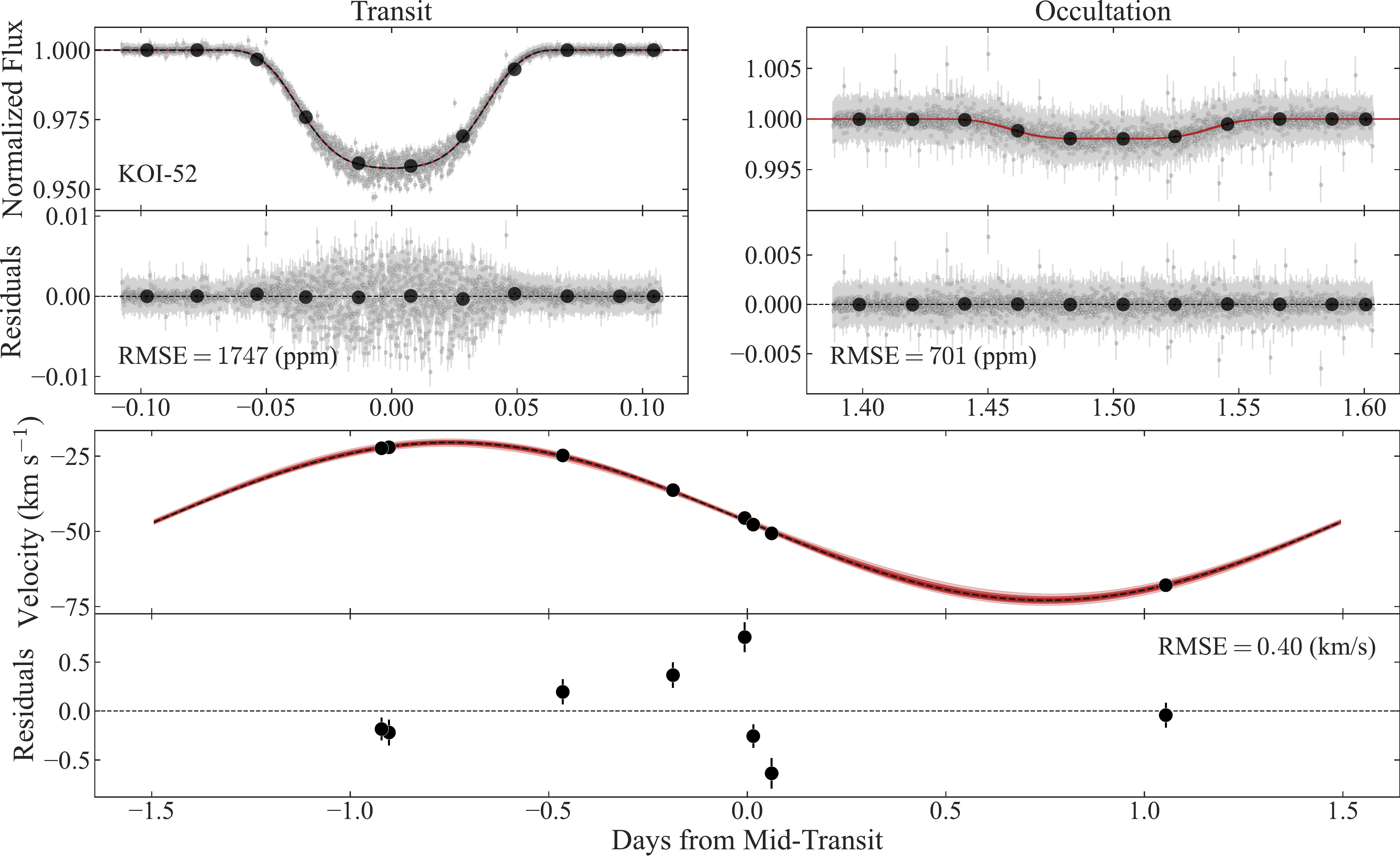}
\figsetgrpnote{\textbf{Top.} The Kepler photometry center on the transit (left) and occultation (right) for KOI-52 after phase-folding to the derived ephemeris. The large circles represent 30 min bins of the raw data. \textbf{Bottom}. The RVs after phase-folding the data to the derived ephemeris. In each panel, the $1\sigma$ (darkest), $2\sigma$, and $3\sigma$ (brightest) extent of the models are shown for reference.}
\figsetgrpend

\figsetgrpstart
\figsetgrpnum{2.3}
\figsetgrptitle{KOI--129}
\figsetplot{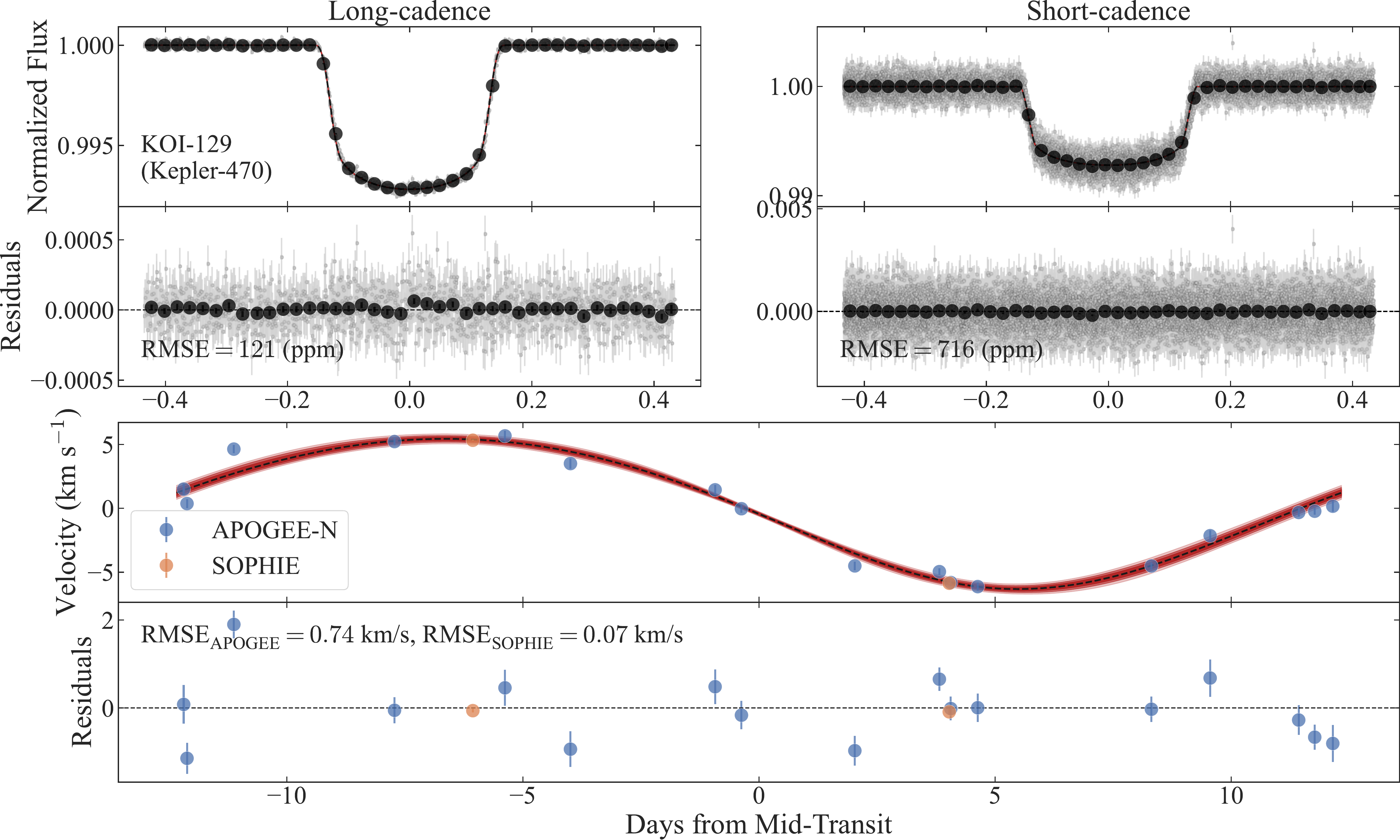}
\figsetgrpnote{\textbf{Top.} The long-cadence (left) and short-cadence (right) Kepler photometry for KOI-129 after phase-folding to the derived ephemeris. The large circles represent 30 min bins of the raw data. \textbf{Bottom}. The RVs after removing instrumental offsets and phase-folding the data to the derived ephemeris. In each panel, the $1\sigma$ (darkest), $2\sigma$, and $3\sigma$ (brightest) extent of the models are shown for reference.}
\figsetgrpend

\figsetgrpstart
\figsetgrpnum{2.4}
\figsetgrptitle{KOI--130}
\figsetplot{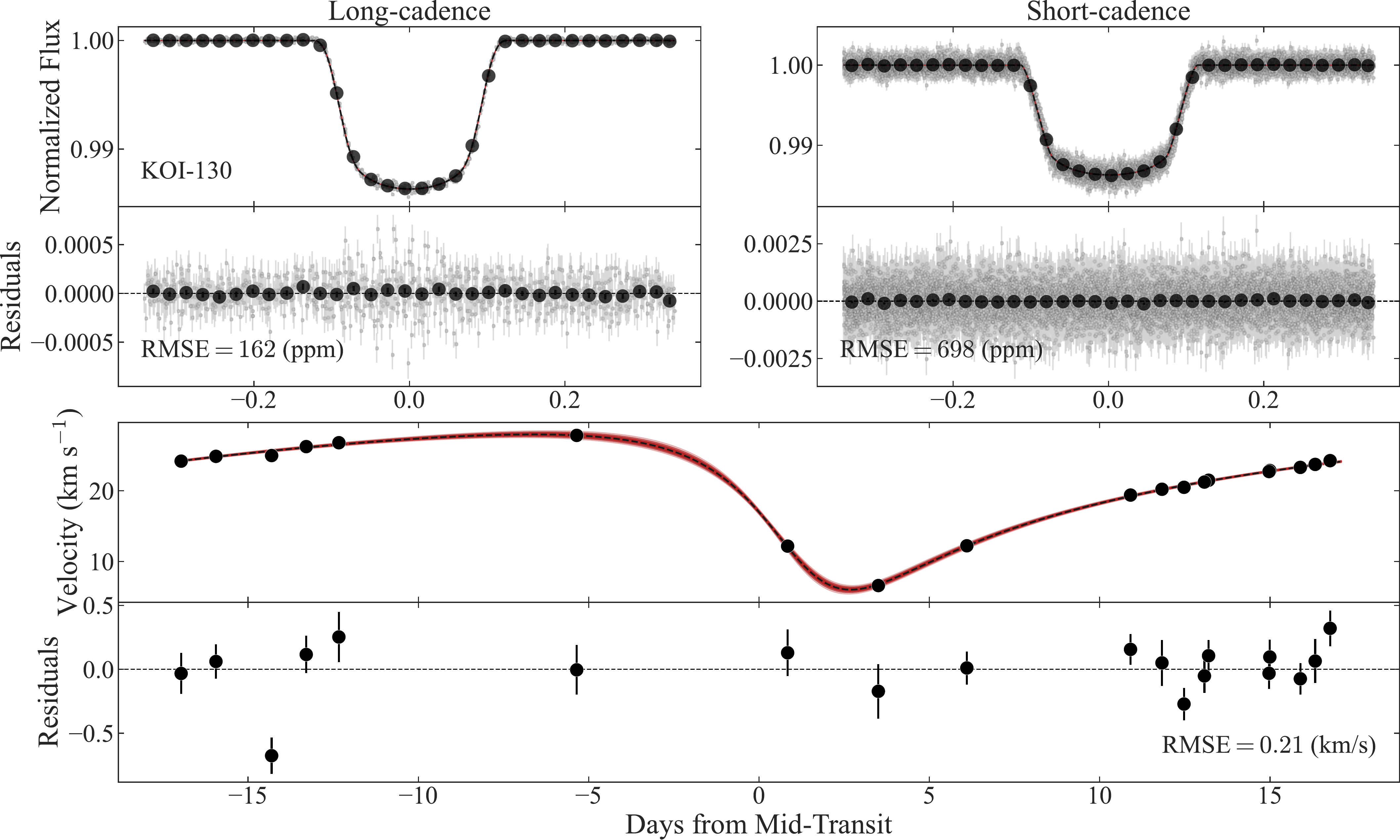}
\figsetgrpnote{\textbf{Top.} The long-cadence (left) and short-cadence (right) Kepler photometry for KOI-130 after phase-folding to the derived ephemeris. The large circles represent 30 min bins of the raw data. \textbf{Bottom}. The RVs after phase-folding the data to the derived ephemeris. In each panel, the $1\sigma$ (darkest), $2\sigma$, and $3\sigma$ (brightest) extent of the models are shown for reference.}
\figsetgrpend

\figsetgrpstart
\figsetgrpnum{2.5}
\figsetgrptitle{KOI--182}
\figsetplot{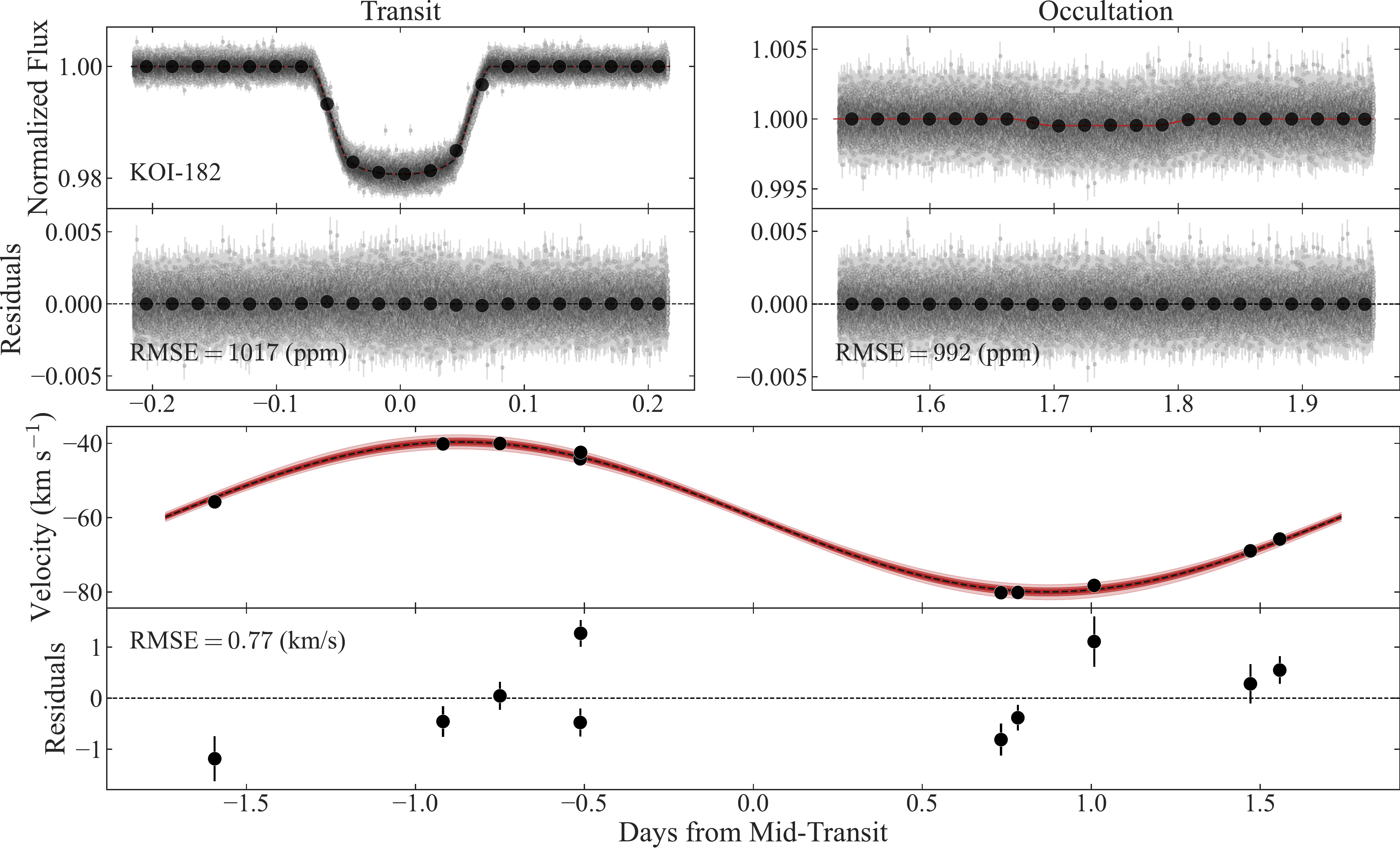}
\figsetgrpnote{\textbf{Top.} The Kepler photometry center on the transit (left) and occultation (right) for KOI-182 after phase-folding to the derived ephemeris. The large circles represent 30 min bins of the raw data. \textbf{Bottom}. The RVs after phase-folding the data to the derived ephemeris. In each panel, the $1\sigma$ (darkest), $2\sigma$, and $3\sigma$ (brightest) extent of the models are shown for reference.}
\figsetgrpend

\figsetgrpstart
\figsetgrpnum{2.6}
\figsetgrptitle{KOI--219}
\figsetplot{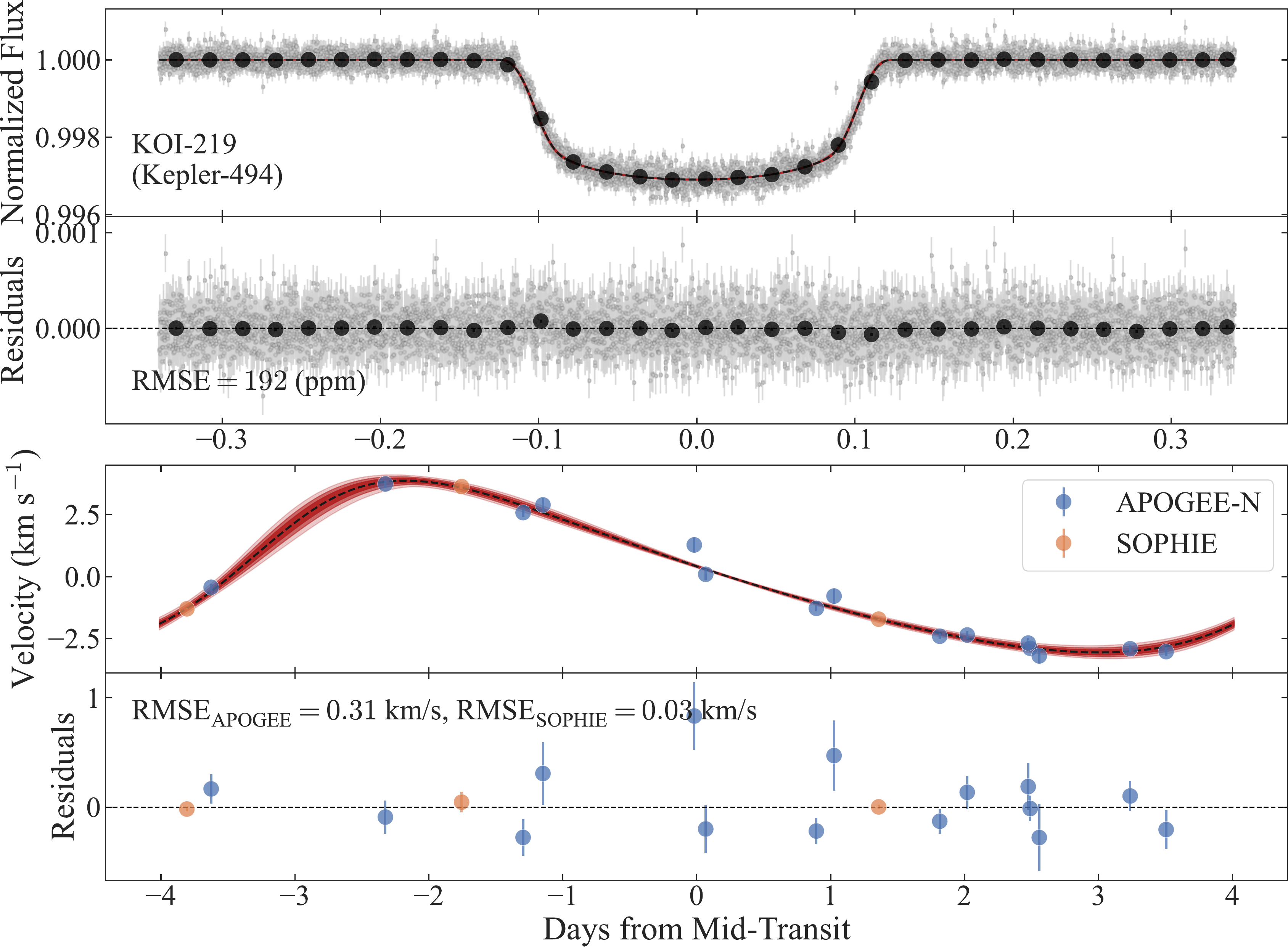}
\figsetgrpnote{\textbf{Top.} The Kepler photometry for KOI-219 after phase-folding to the derived ephemeris. The large circles represent 30 min bins of the raw data. \textbf{Bottom}. The RVs after removing instrumental offsets and phase-folding the data to the derived ephemeris. In each panel, the $1\sigma$ (darkest), $2\sigma$, and $3\sigma$ (brightest) extent of the models are shown for reference.}
\figsetgrpend

\figsetgrpstart
\figsetgrpnum{2.7}
\figsetgrptitle{KOI--242}
\figsetplot{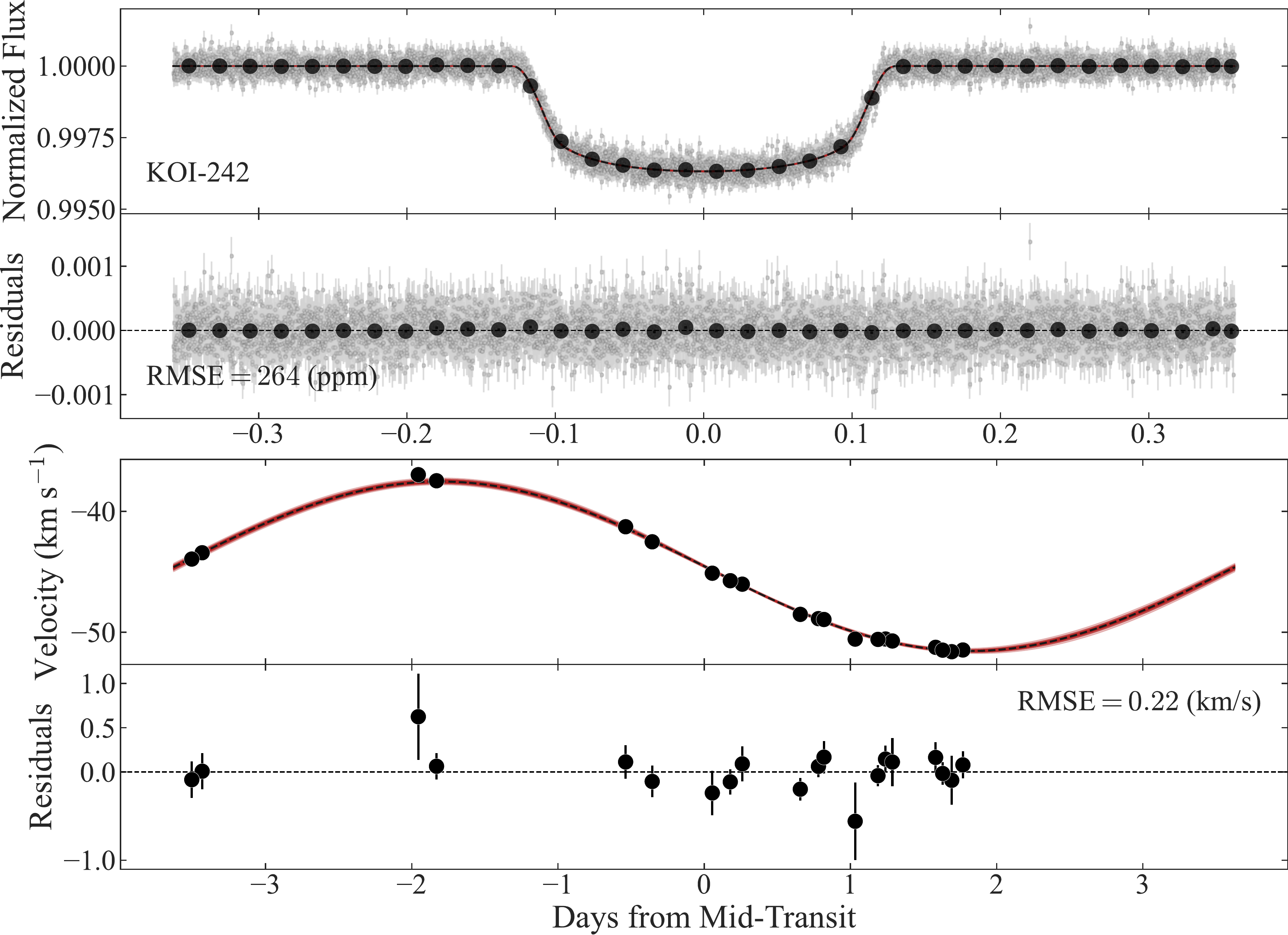}
\figsetgrpnote{\textbf{Top.} The Kepler photometry for KOI-242 after phase-folding to the derived ephemeris. The large circles represent 30 min bins of the raw data. \textbf{Bottom}. The RVs after phase-folding the data to the derived ephemeris. In each panel, the $1\sigma$ (darkest), $2\sigma$, and $3\sigma$ (brightest) extent of the models are shown for reference.}
\figsetgrpend

\figsetgrpstart
\figsetgrpnum{2.8}
\figsetgrptitle{KOI--403}
\figsetplot{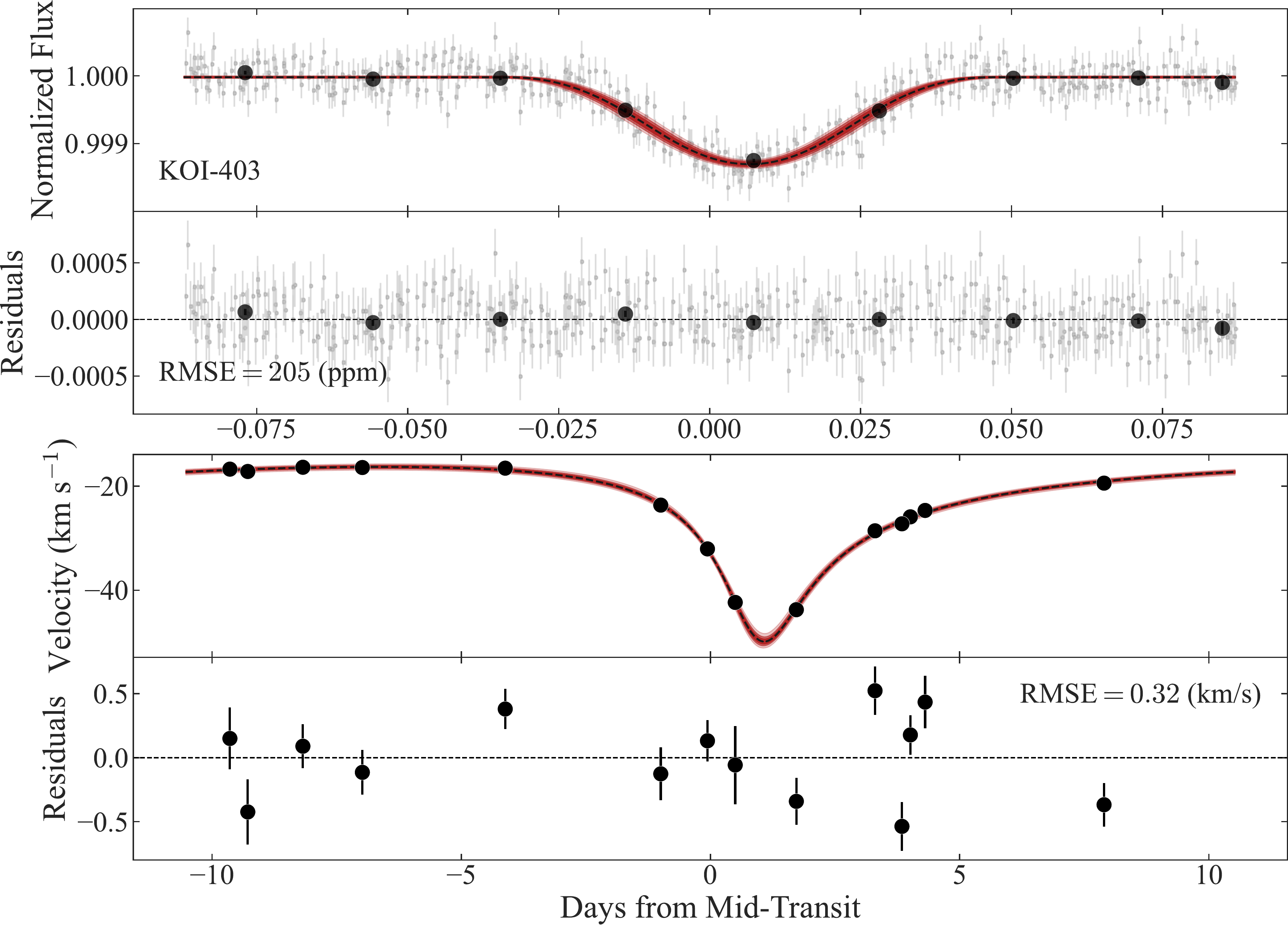}
\figsetgrpnote{\textbf{Top.} The Kepler photometry for KOI-403 after phase-folding to the derived ephemeris. The large circles represent 30 min bins of the raw data. \textbf{Bottom}. The RVs after phase-folding the data to the derived ephemeris. In each panel, the $1\sigma$ (darkest), $2\sigma$, and $3\sigma$ (brightest) extent of the models are shown for reference.}
\figsetgrpend

\figsetgrpstart
\figsetgrpnum{2.9}
\figsetgrptitle{KOI--415}
\figsetplot{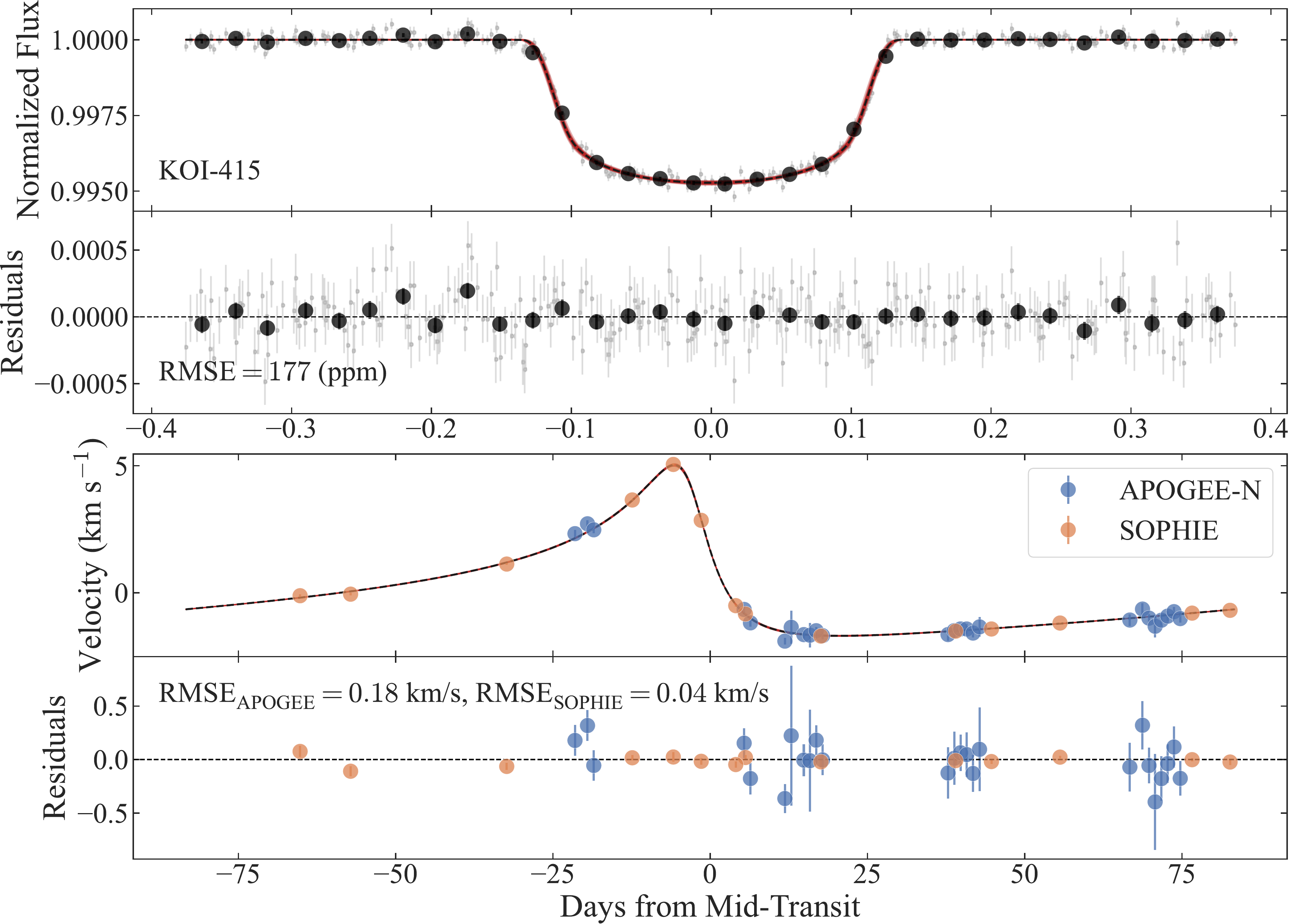}
\figsetgrpnote{\textbf{Top.} The Kepler photometry for KOI-415 after phase-folding to the derived ephemeris. The large circles represent 30 min bins of the raw data. \textbf{Bottom}. The RVs after removing instrumental offsets and phase-folding the data to the derived ephemeris. In each panel, the $1\sigma$ (darkest), $2\sigma$, and $3\sigma$ (brightest) extent of the models are shown for reference.}
\figsetgrpend

\figsetgrpstart
\figsetgrpnum{2.10}
\figsetgrptitle{KOI--466}
\figsetplot{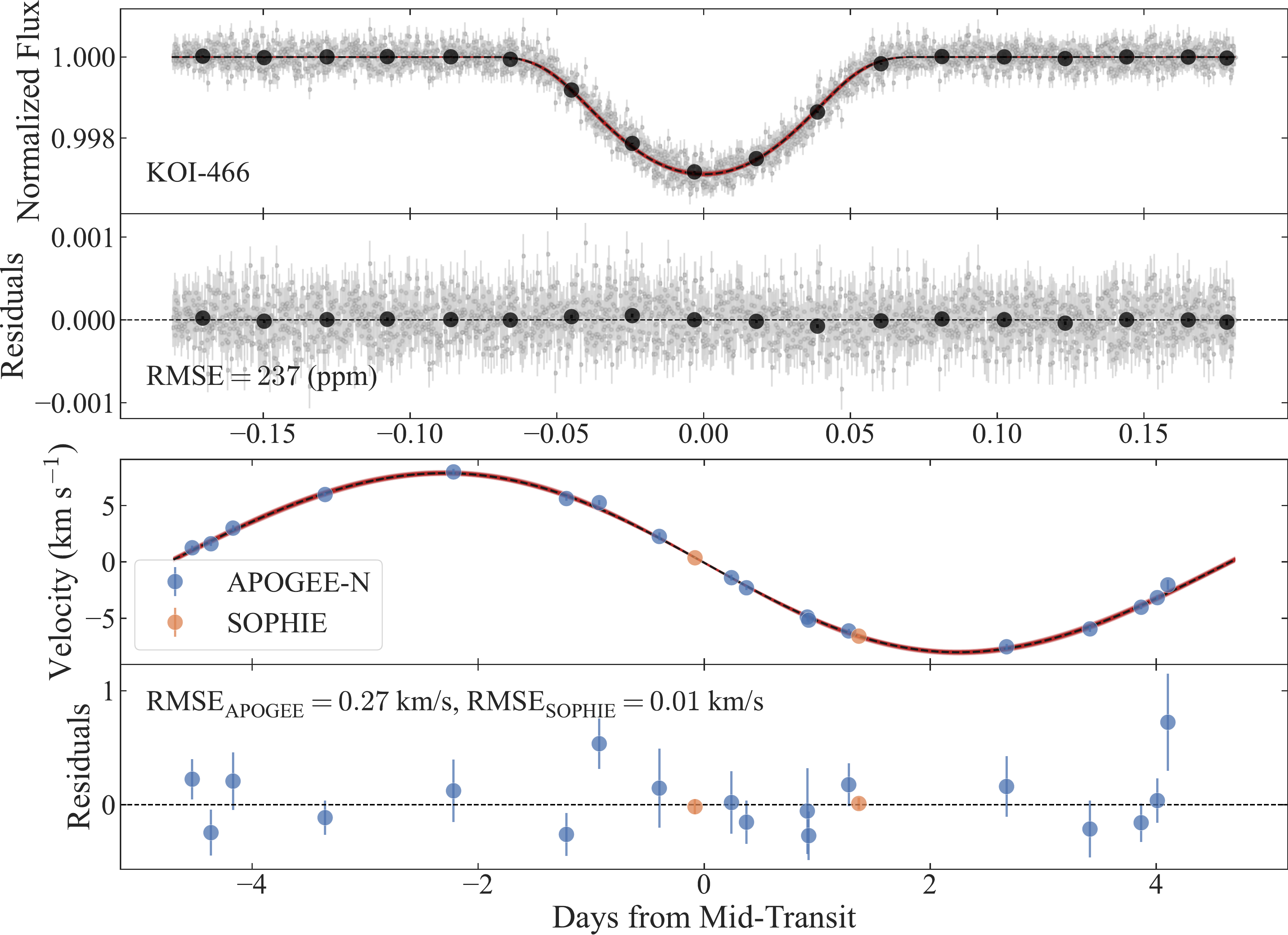}
\figsetgrpnote{\textbf{Top.} The Kepler photometry for KOI-466 after phase-folding to the derived ephemeris. The large circles represent 30 min bins of the raw data. \textbf{Bottom}. The RVs after removing instrumental offsets and phase-folding the data to the derived ephemeris. In each panel, the $1\sigma$ (darkest), $2\sigma$, and $3\sigma$ (brightest) extent of the models are shown for reference.}
\figsetgrpend

\figsetgrpstart
\figsetgrpnum{2.11}
\figsetgrptitle{KOI--631}
\figsetplot{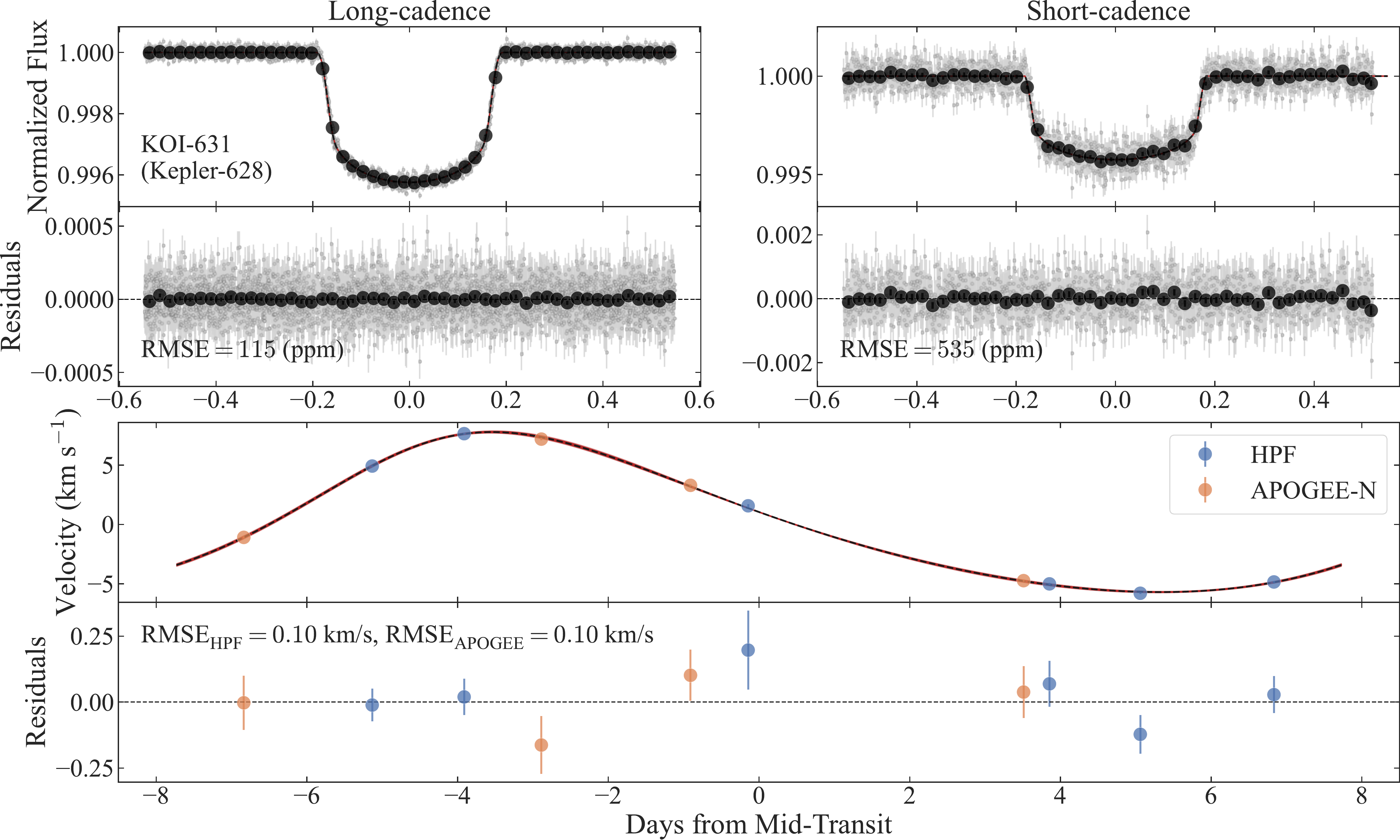}
\figsetgrpnote{\textbf{Top.} The long-cadence (left) and short-cadence (right) Kepler photometry for KOI-631 after phase-folding to the derived ephemeris. The large circles represent 30 min bins of the raw data. \textbf{Bottom}. The RVs after removing instrumental offsets and phase-folding the data to the derived ephemeris. In each panel, the $1\sigma$ (darkest), $2\sigma$, and $3\sigma$ (brightest) extent of the models are shown for reference.}
\figsetgrpend

\figsetgrpstart
\figsetgrpnum{2.12}
\figsetgrptitle{KOI--777}
\figsetplot{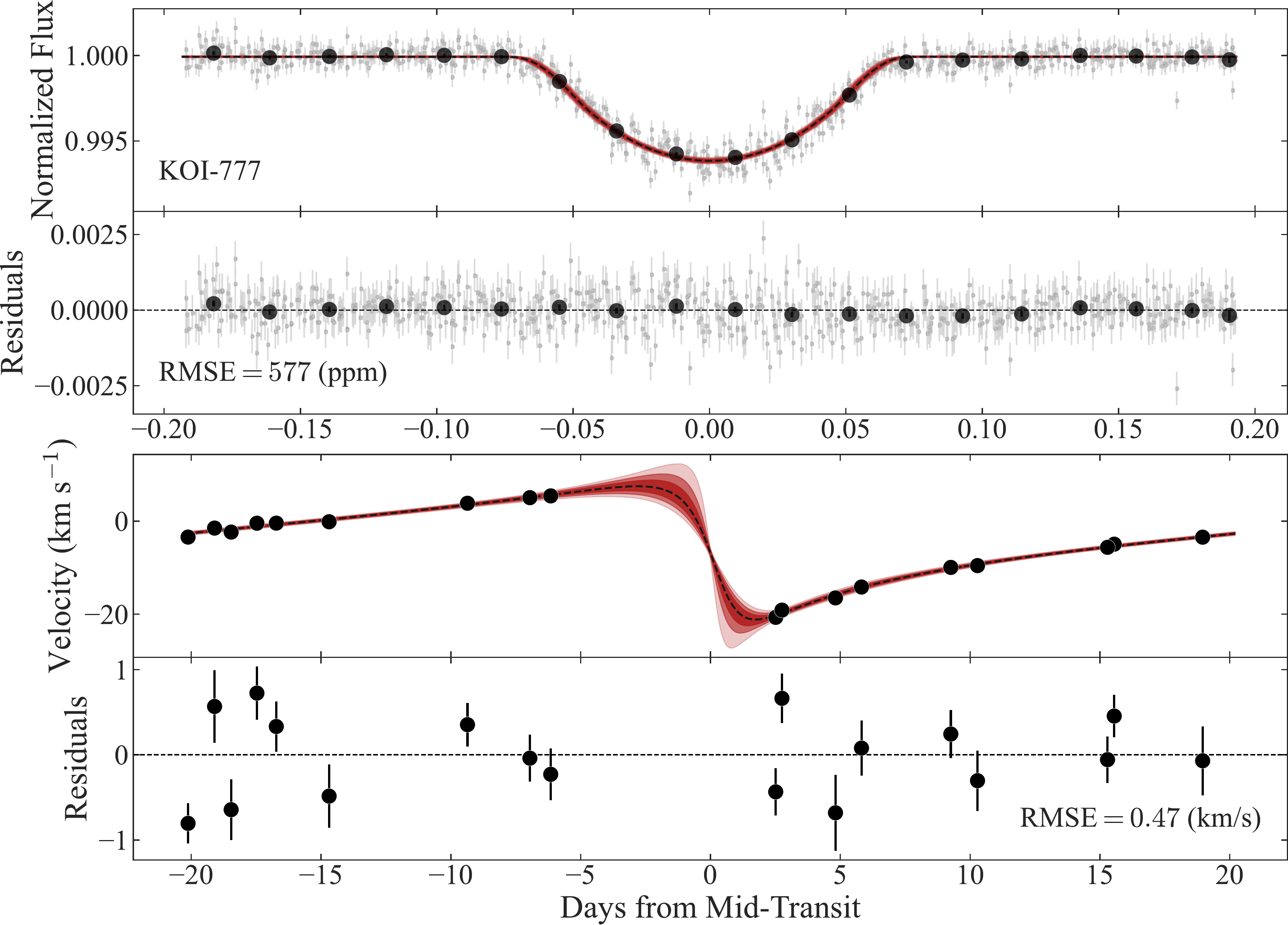}
\figsetgrpnote{\textbf{Top.} The Kepler photometry for KOI-777 after phase-folding to the derived ephemeris. The large circles represent 30 min bins of the raw data. \textbf{Bottom}. The RVs after phase-folding the data to the derived ephemeris. In each panel, the $1\sigma$ (darkest), $2\sigma$, and $3\sigma$ (brightest) extent of the models are shown for reference.}
\figsetgrpend

\figsetgrpstart
\figsetgrpnum{2.13}
\figsetgrptitle{KOI--846}
\figsetplot{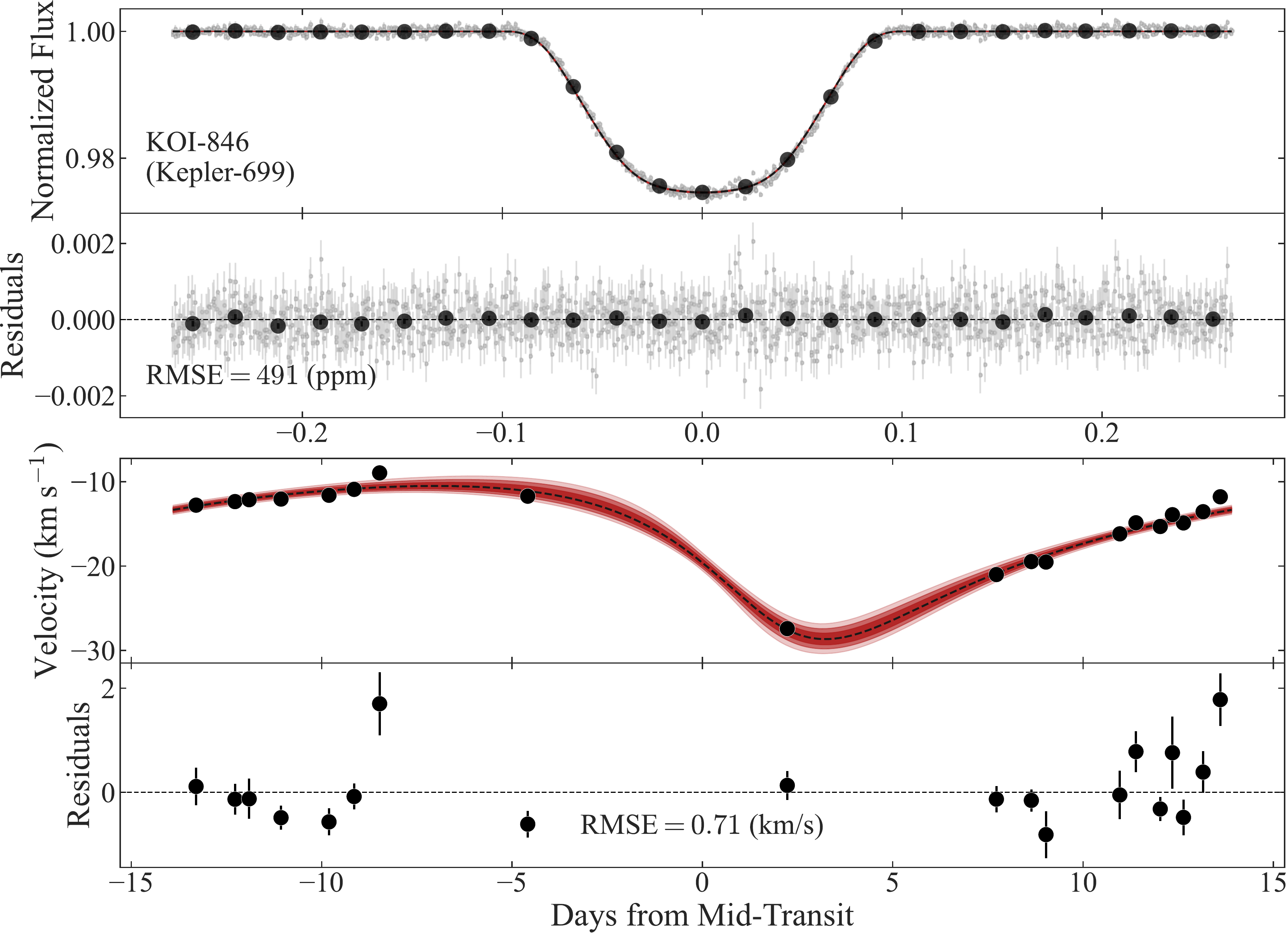}
\figsetgrpnote{\textbf{Top.} The Kepler photometry for KOI-846 after phase-folding to the derived ephemeris. The large circles represent 30 min bins of the raw data. \textbf{Bottom}. The RVs after phase-folding the data to the derived ephemeris. In each panel, the $1\sigma$ (darkest), $2\sigma$, and $3\sigma$ (brightest) extent of the models are shown for reference.}
\figsetgrpend

\figsetgrpstart
\figsetgrpnum{2.14}
\figsetgrptitle{KOI--855}
\figsetplot{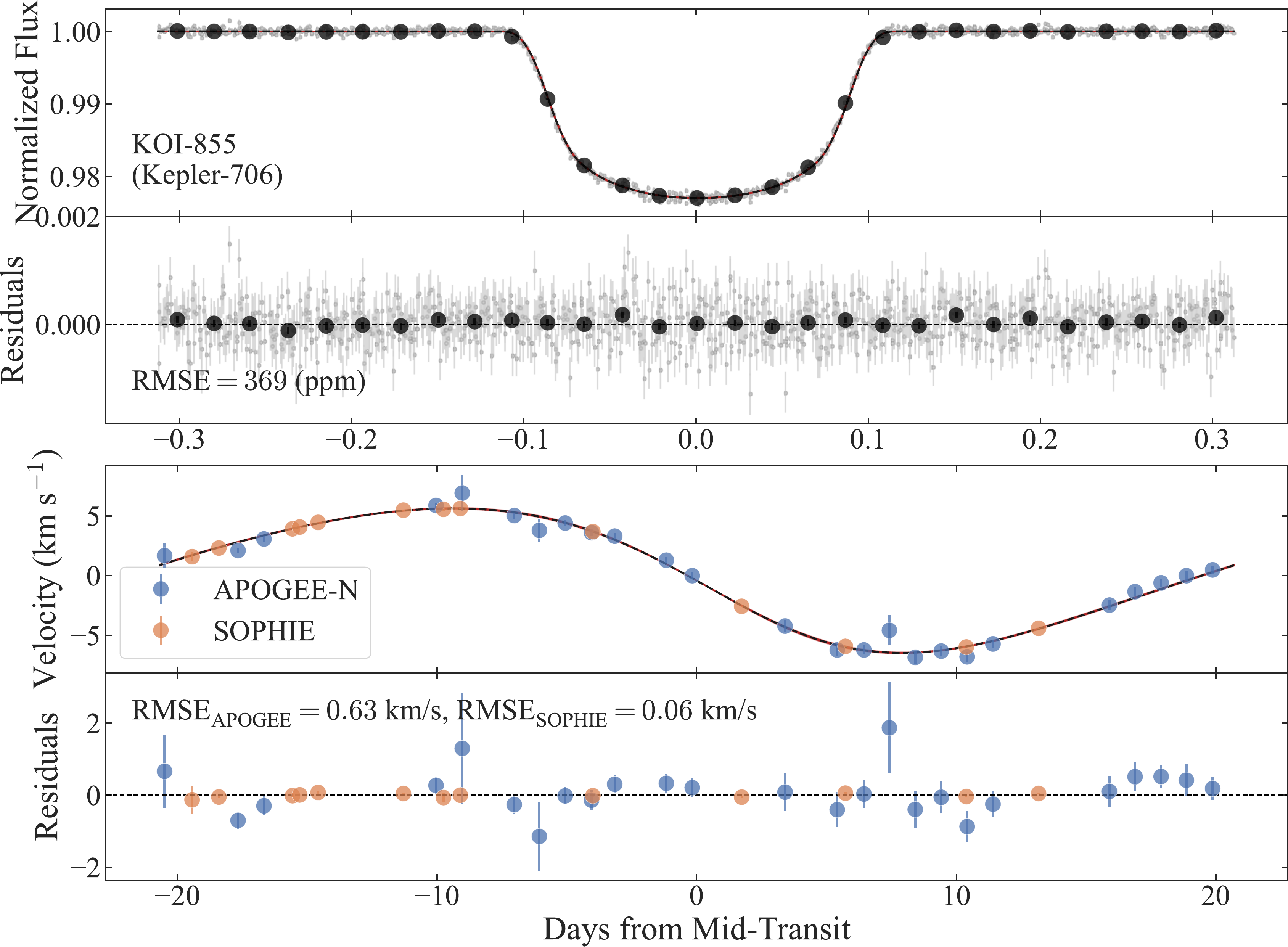}
\figsetgrpnote{\textbf{Top.} The Kepler photometry for KOI-855 after phase-folding to the derived ephemeris. The large circles represent 30 min bins of the raw data. \textbf{Bottom}. The RVs after removing instrumental offsets and phase-folding the data to the derived ephemeris. In each panel, the $1\sigma$ (darkest), $2\sigma$, and $3\sigma$ (brightest) extent of the models are shown for reference.}
\figsetgrpend

\figsetgrpstart
\figsetgrpnum{2.15}
\figsetgrptitle{KOI--1064}
\figsetplot{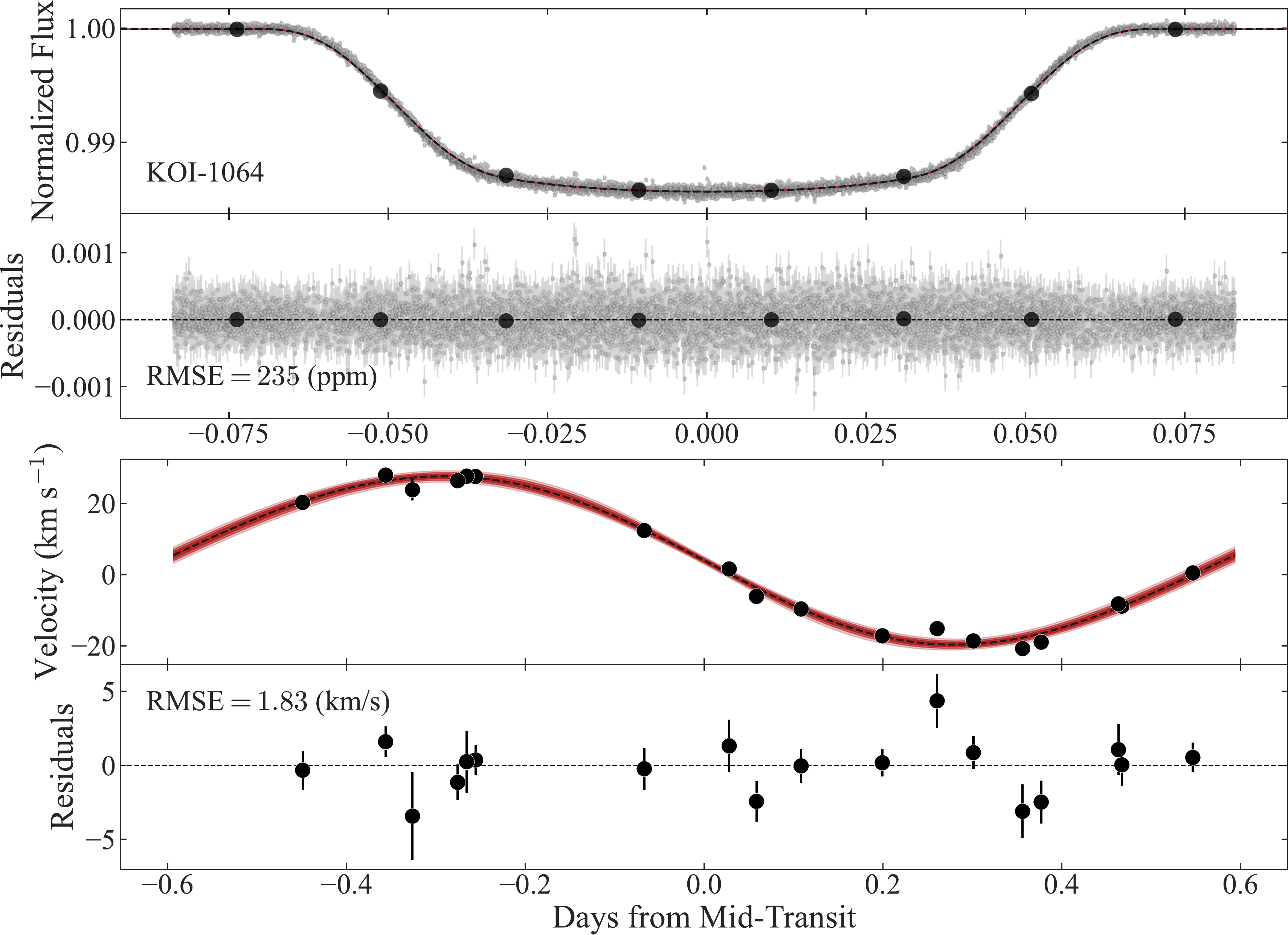}
\figsetgrpnote{\textbf{Top.} The Kepler photometry for KOI-1064 after phase-folding to the derived ephemeris. The large circles represent 30 min bins of the raw data. \textbf{Bottom}. The RVs after phase-folding the data to the derived ephemeris. In each panel, the $1\sigma$ (darkest), $2\sigma$, and $3\sigma$ (brightest) extent of the models are shown for reference.}
\figsetgrpend

\figsetgrpstart
\figsetgrpnum{2.16}
\figsetgrptitle{KOI--1247}
\figsetplot{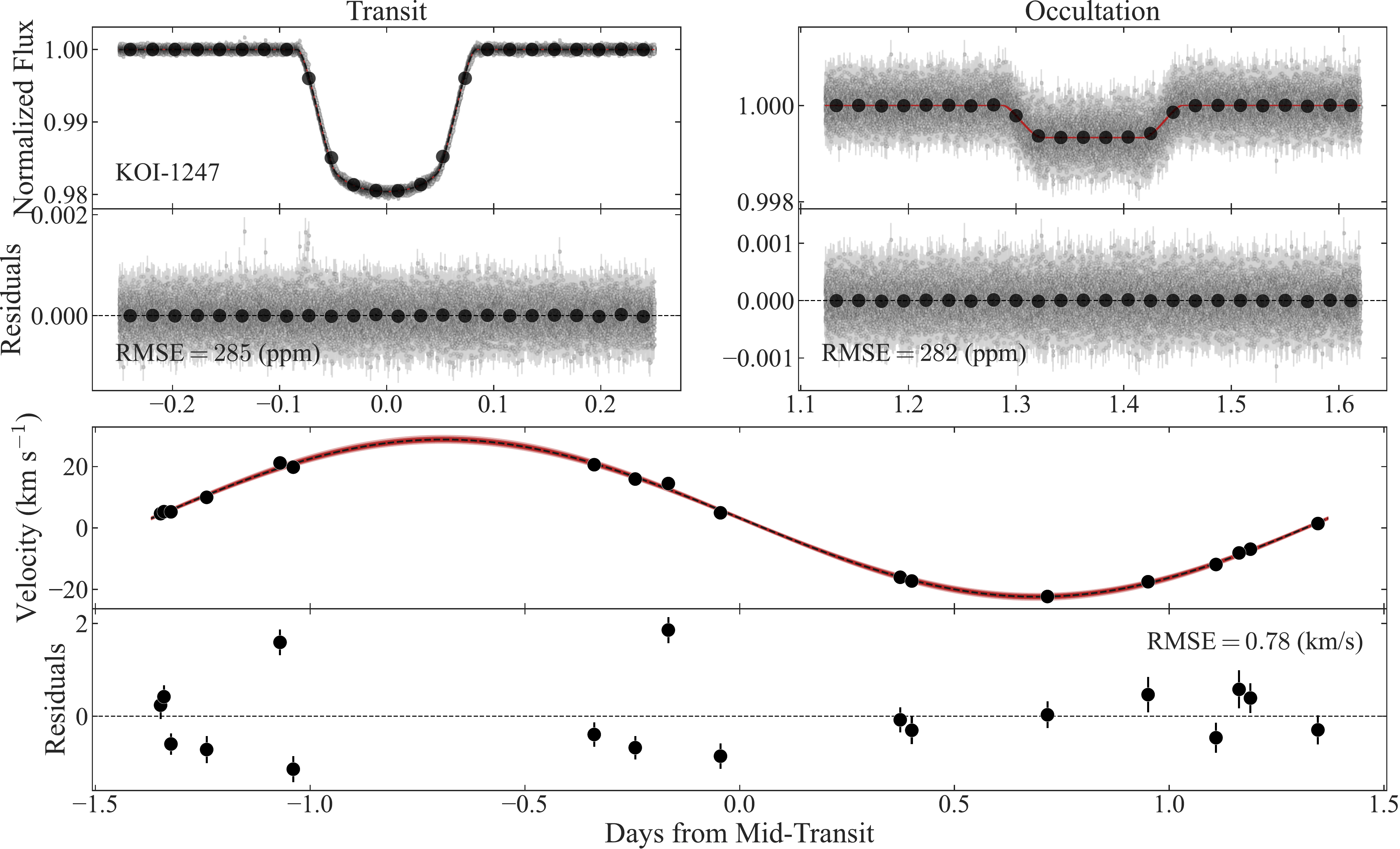}
\figsetgrpnote{\textbf{Top.} The Kepler photometry center on the transit (left) and occultation (right) for KOI-1247 after phase-folding to the derived ephemeris. The large circles represent 30 min bins of the raw data. \textbf{Bottom}. The RVs after phase-folding the data to the derived ephemeris. In each panel, the $1\sigma$ (darkest), $2\sigma$, and $3\sigma$ (brightest) extent of the models are shown for reference.}
\figsetgrpend

\figsetgrpstart
\figsetgrpnum{2.17}
\figsetgrptitle{KOI--1288}
\figsetplot{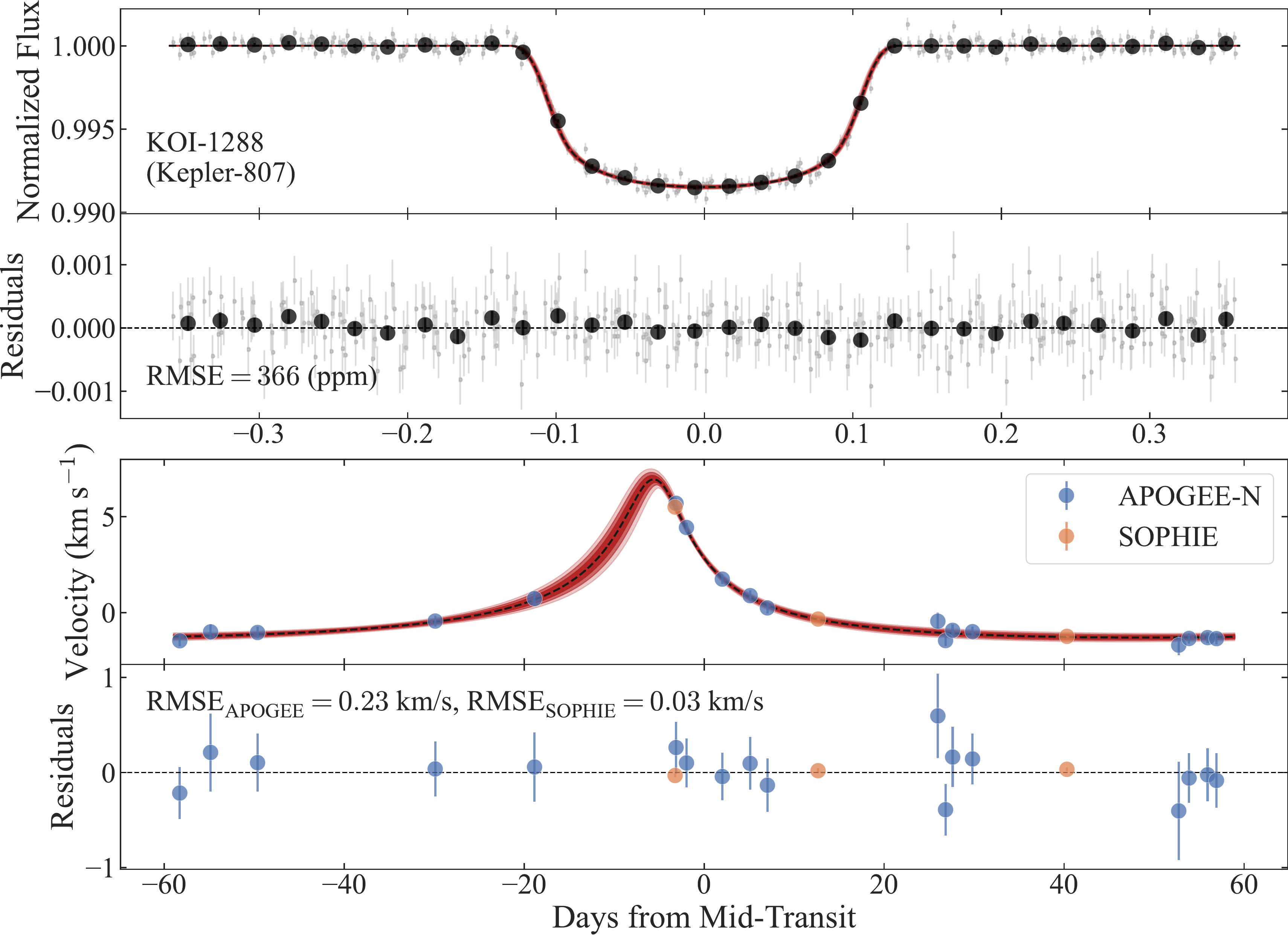}
\figsetgrpnote{\textbf{Top.} The Kepler photometry for KOI-1288 after phase-folding to the derived ephemeris. The large circles represent 30 min bins of the raw data. \textbf{Bottom}. The RVs after removing instrumental offsets and phase-folding the data to the derived ephemeris. In each panel, the $1\sigma$ (darkest), $2\sigma$, and $3\sigma$ (brightest) extent of the models are shown for reference.}
\figsetgrpend

\figsetgrpstart
\figsetgrpnum{2.18}
\figsetgrptitle{KOI--1347}
\figsetplot{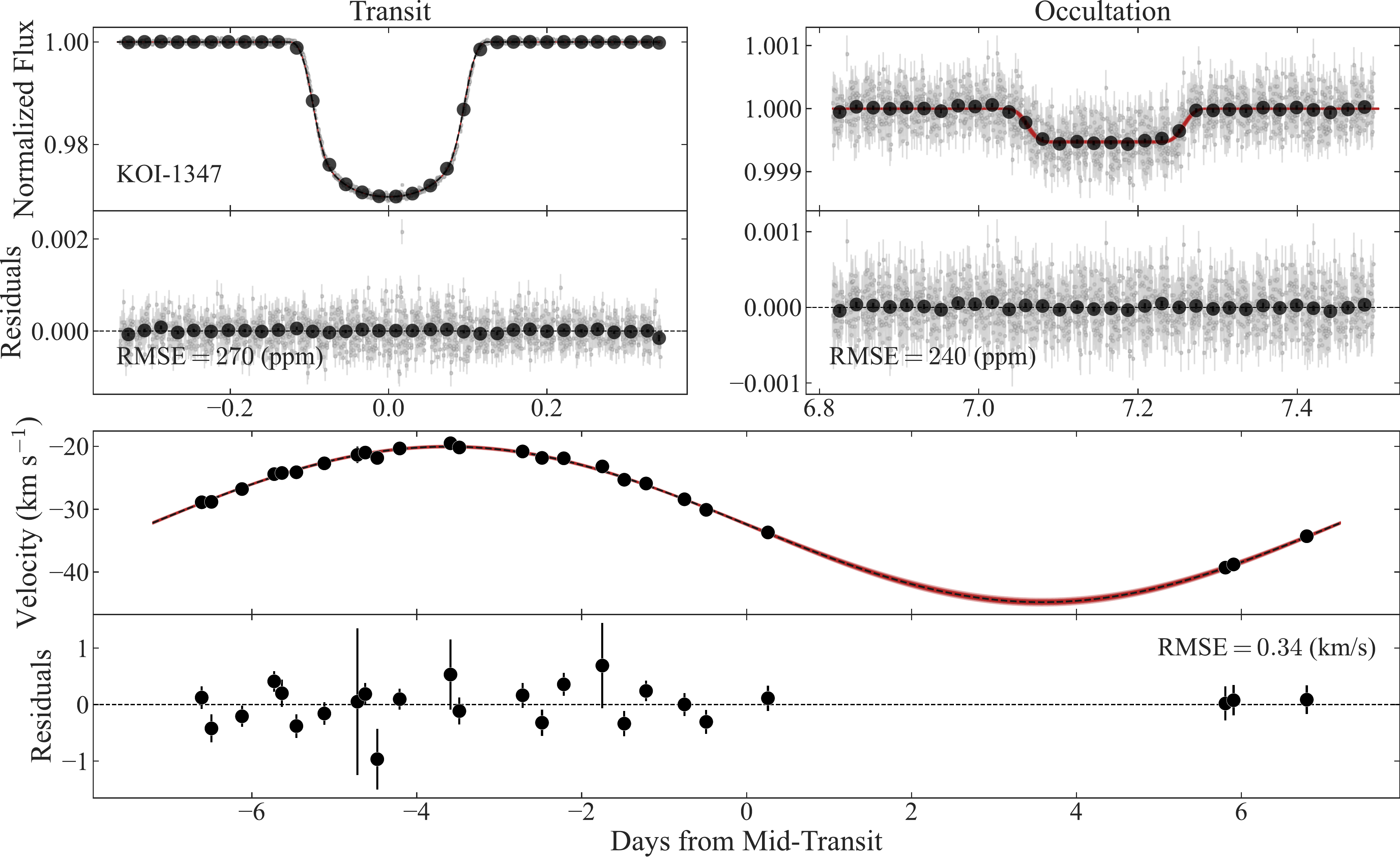}
\figsetgrpnote{\textbf{Top.} The Kepler photometry center on the transit (left) and occultation (right) for KOI-1347 after phase-folding to the derived ephemeris. The large circles represent 30 min bins of the raw data. \textbf{Bottom}. The RVs after phase-folding the data to the derived ephemeris. In each panel, the $1\sigma$ (darkest), $2\sigma$, and $3\sigma$ (brightest) extent of the models are shown for reference.}
\figsetgrpend

\figsetgrpstart
\figsetgrpnum{2.19}
\figsetgrptitle{KOI--1356}
\figsetplot{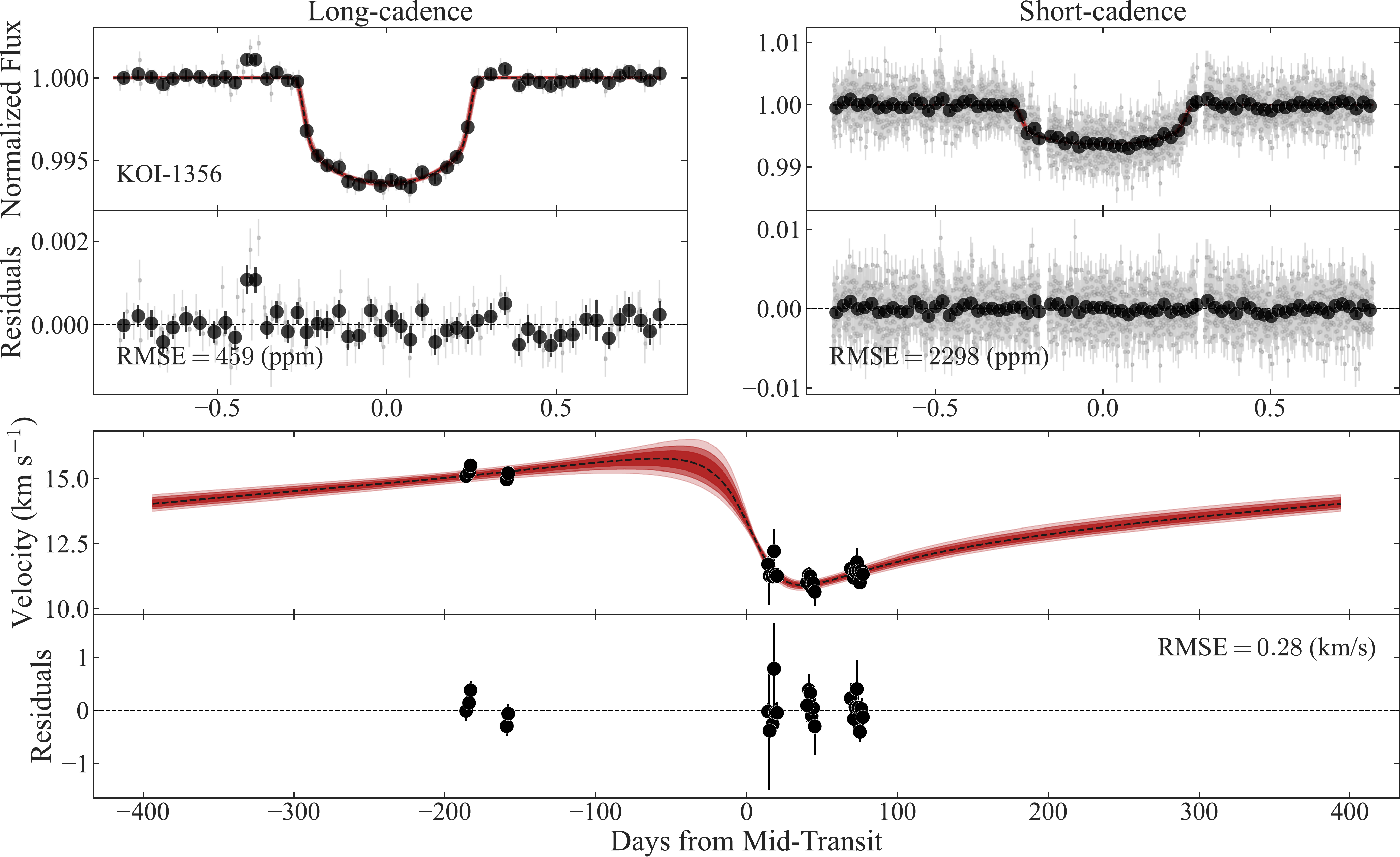}
\figsetgrpnote{\textbf{Top.} The long-cadence (left) and short-cadence (right) Kepler photometry for KOI-1356 after phase-folding to the derived ephemeris. The large circles represent 30 min bins of the raw data. \textbf{Bottom}. The RVs after phase-folding the data to the derived ephemeris. In each panel, the $1\sigma$ (darkest), $2\sigma$, and $3\sigma$ (brightest) extent of the models are shown for reference.}
\figsetgrpend

\figsetgrpstart
\figsetgrpnum{2.20}
\figsetgrptitle{KOI--1416}
\figsetplot{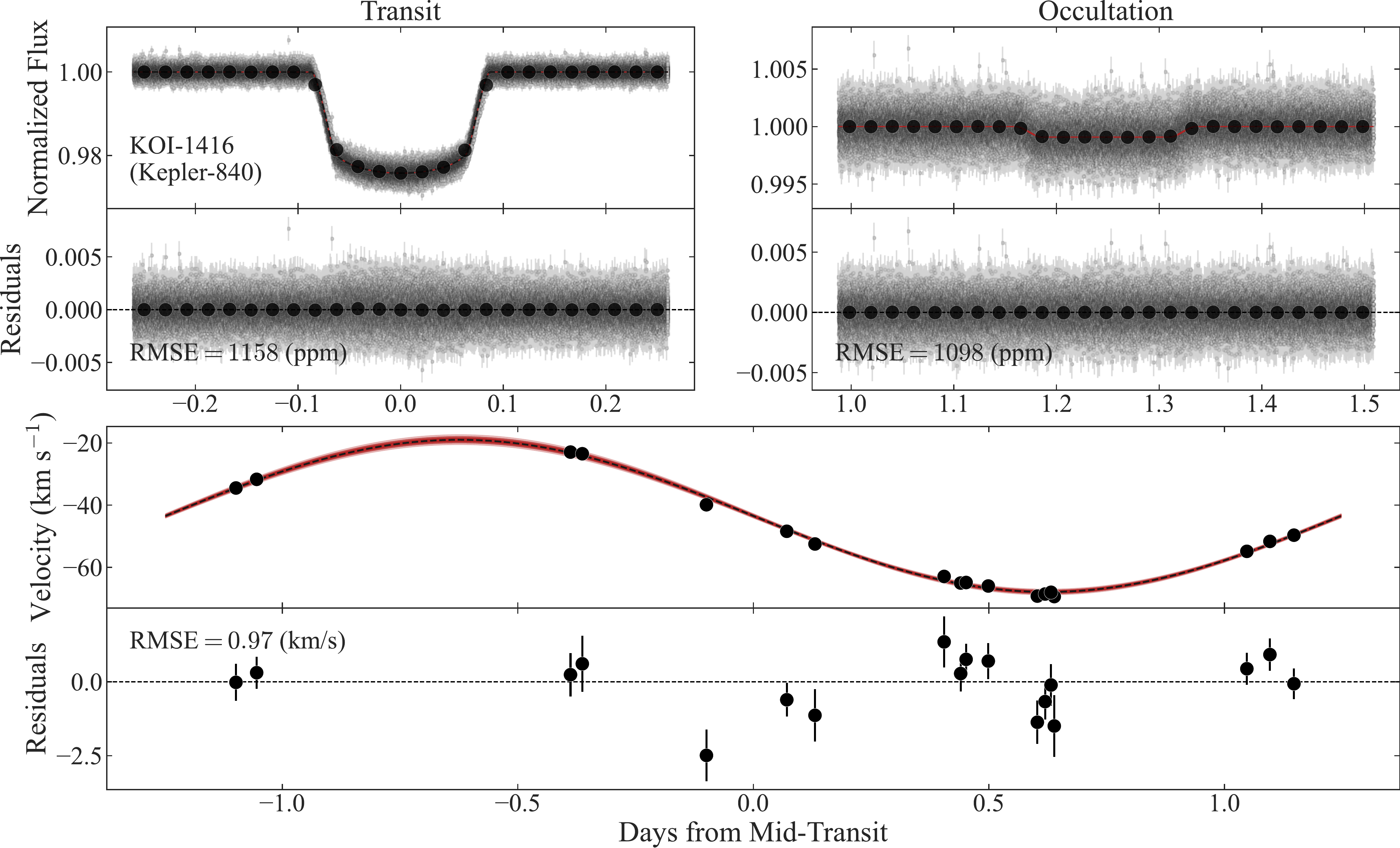}
\figsetgrpnote{\textbf{Top.} The Kepler photometry center on the transit (left) and occultation (right) for KOI-1416 after phase-folding to the derived ephemeris. The large circles represent 30 min bins of the raw data. \textbf{Bottom}. The RVs after phase-folding the data to the derived ephemeris. In each panel, the $1\sigma$ (darkest), $2\sigma$, and $3\sigma$ (brightest) extent of the models are shown for reference.}
\figsetgrpend

\figsetgrpstart
\figsetgrpnum{2.21}
\figsetgrptitle{KOI--1448}
\figsetplot{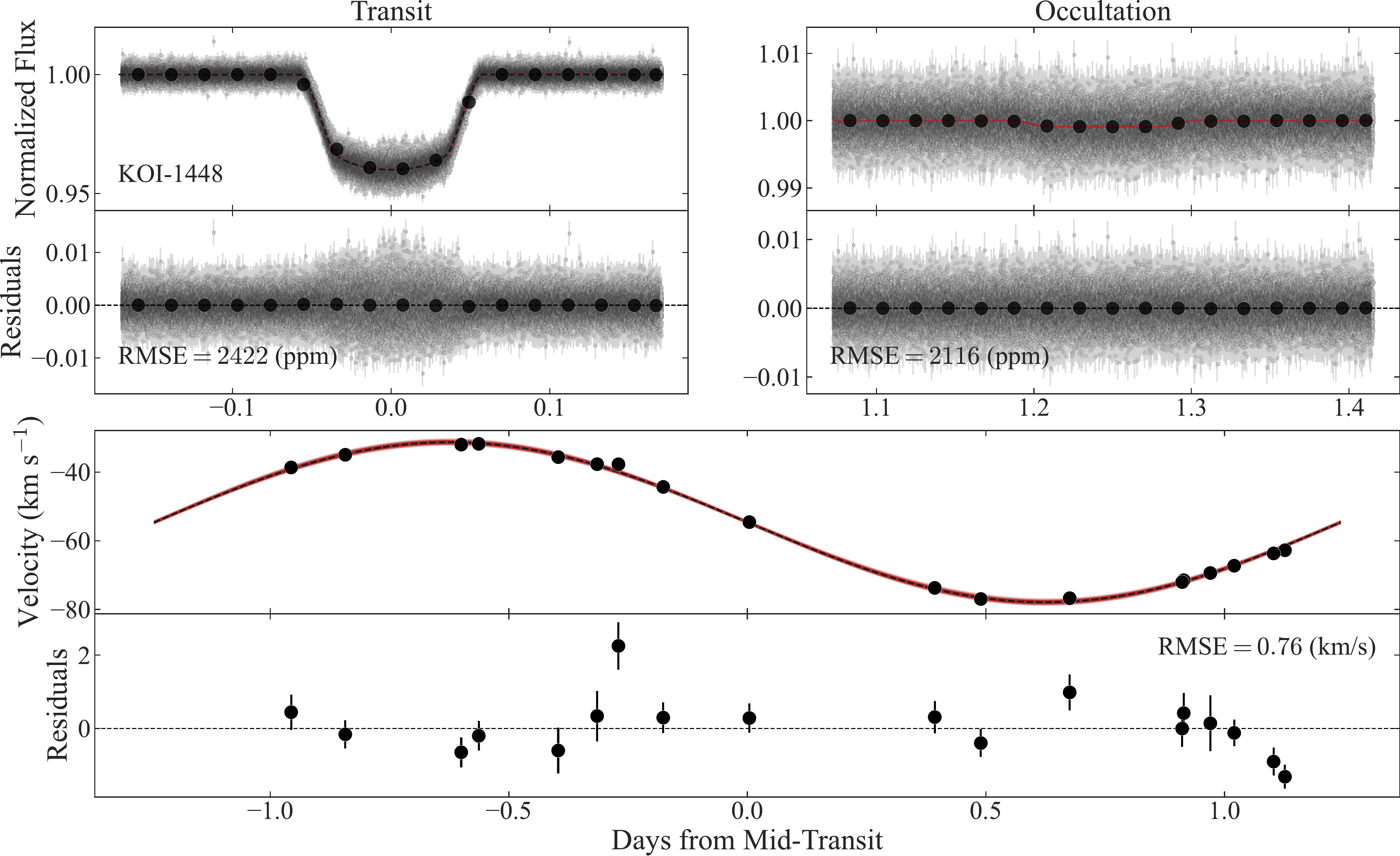}
\figsetgrpnote{\textbf{Top.} The Kepler photometry center on the transit (left) and occultation (right) for KOI-1448 after phase-folding to the derived ephemeris. The large circles represent 30 min bins of the raw data. \textbf{Bottom}. The RVs after phase-folding the data to the derived ephemeris. In each panel, the $1\sigma$ (darkest), $2\sigma$, and $3\sigma$ (brightest) extent of the models are shown for reference.}
\figsetgrpend

\figsetgrpstart
\figsetgrpnum{2.22}
\figsetgrptitle{KOI--2513}
\figsetplot{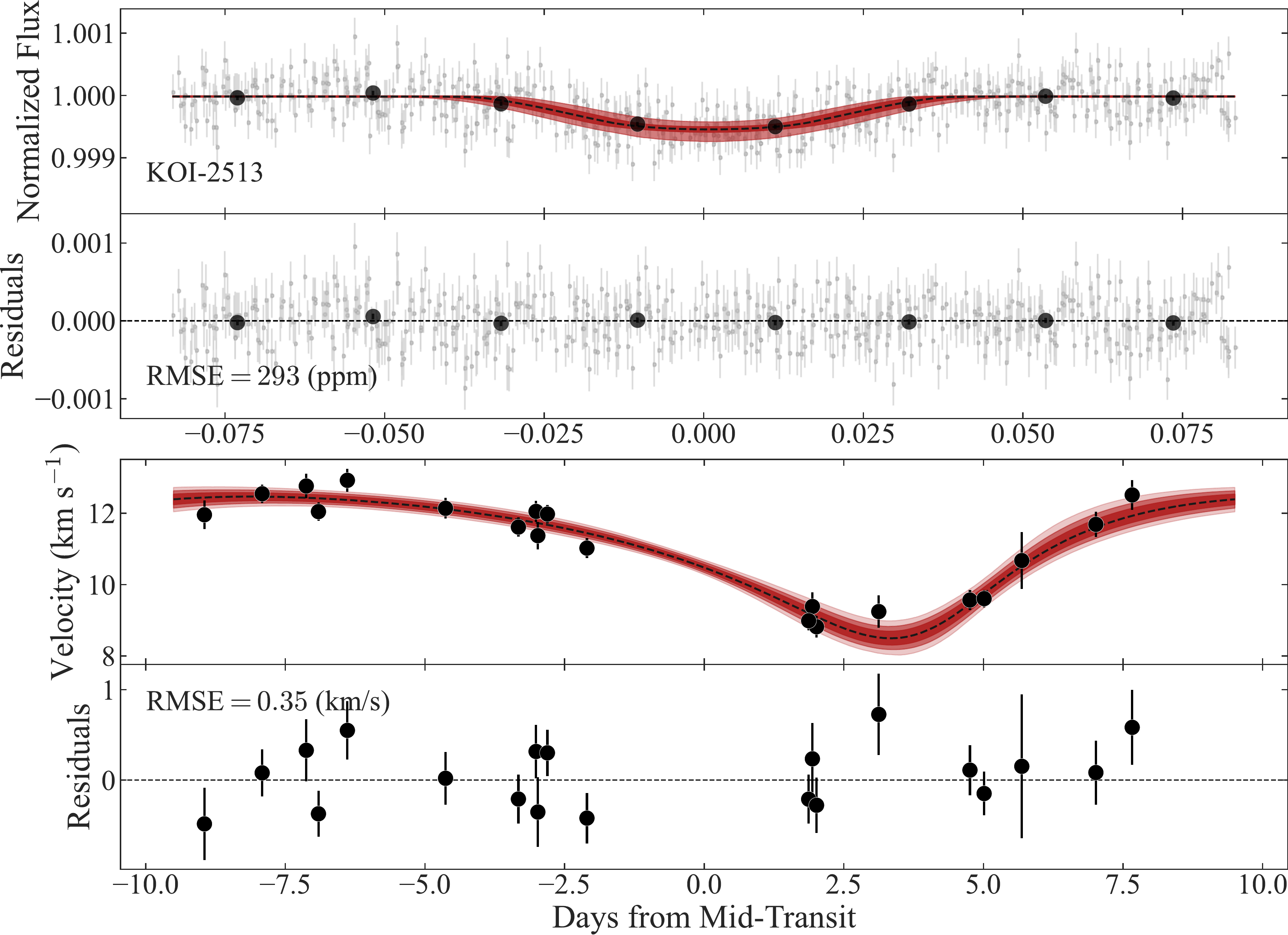}
\figsetgrpnote{\textbf{Top.} The Kepler photometry for KOI-2513 after phase-folding to the derived ephemeris. The large circles represent 30 min bins of the raw data. \textbf{Bottom}. The RVs after phase-folding the data to the derived ephemeris. In each panel, the $1\sigma$ (darkest), $2\sigma$, and $3\sigma$ (brightest) extent of the models are shown for reference.}
\figsetgrpend

\figsetgrpstart
\figsetgrpnum{2.23}
\figsetgrptitle{KOI--3320}
\figsetplot{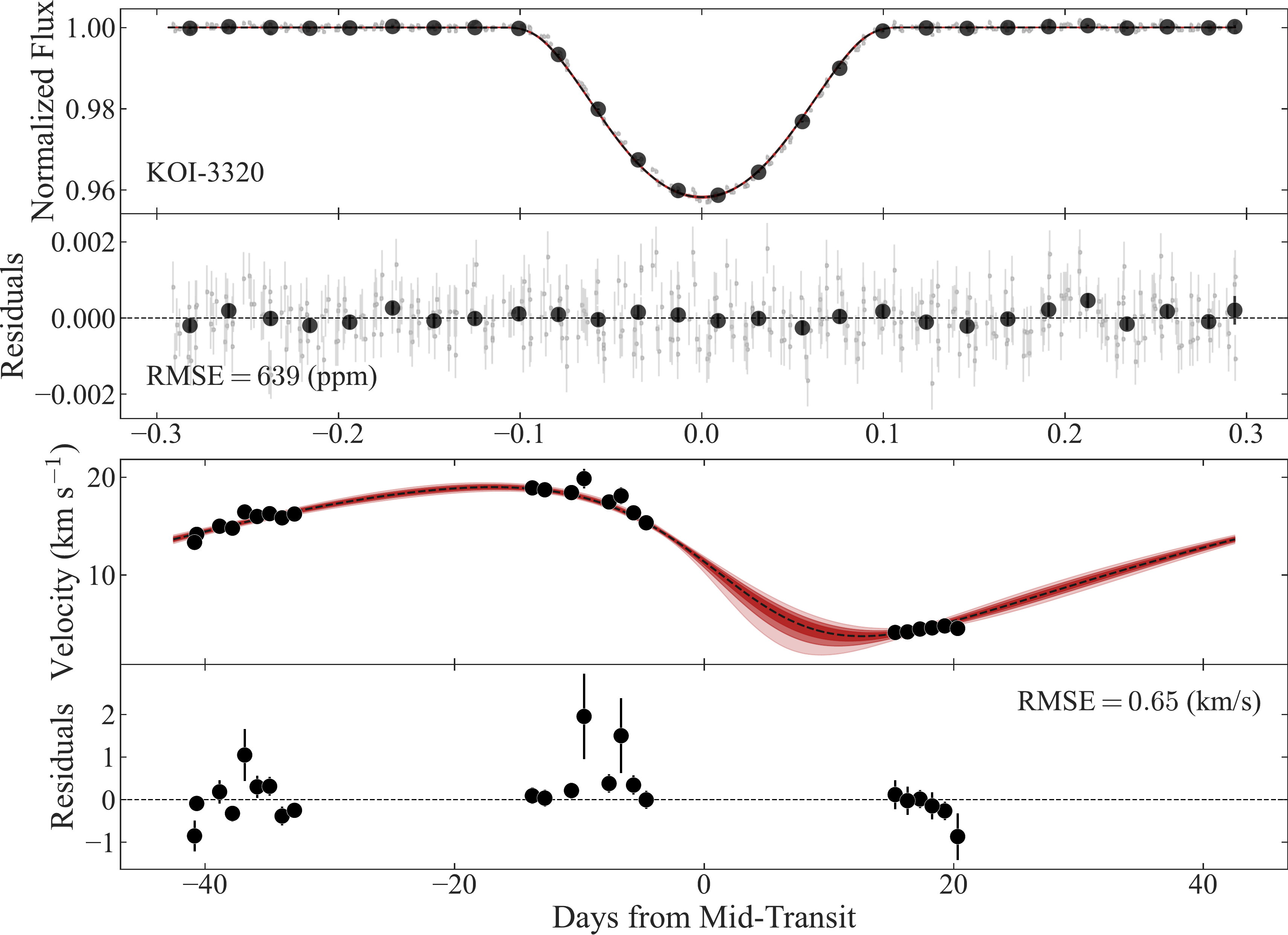}
\figsetgrpnote{\textbf{Top.} The Kepler photometry for KOI-3320 after phase-folding to the derived ephemeris. The large circles represent 30 min bins of the raw data. \textbf{Bottom}. The RVs after phase-folding the data to the derived ephemeris. In each panel, the $1\sigma$ (darkest), $2\sigma$, and $3\sigma$ (brightest) extent of the models are shown for reference.}
\figsetgrpend

\figsetgrpstart
\figsetgrpnum{2.24}
\figsetgrptitle{KOI--3358}
\figsetplot{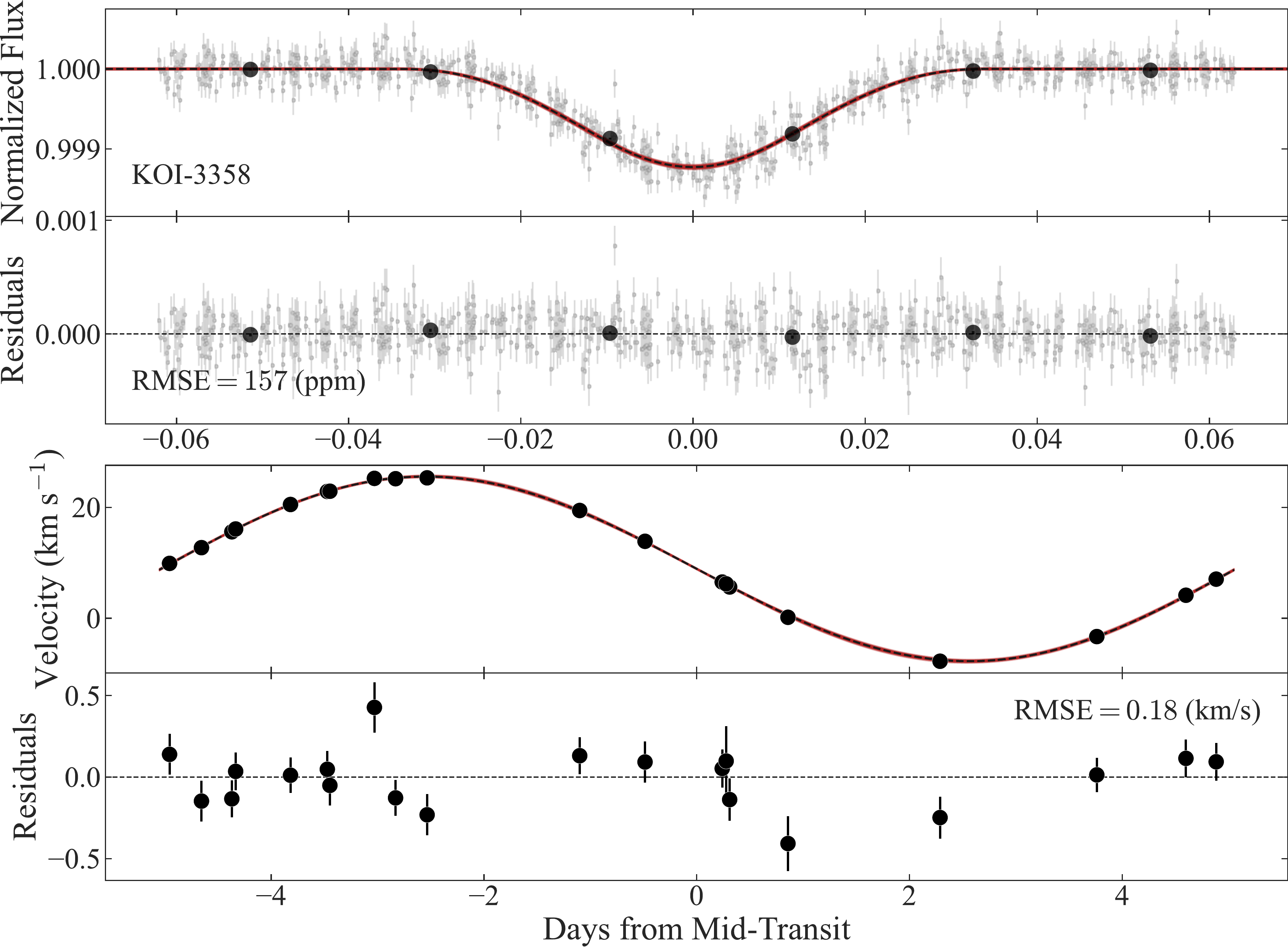}
\figsetgrpnote{\textbf{Top.} The Kepler photometry for KOI-3358 after phase-folding to the derived ephemeris. The large circles represent 30 min bins of the raw data. \textbf{Bottom}. The RVs after phase-folding the data to the derived ephemeris. In each panel, the $1\sigma$ (darkest), $2\sigma$, and $3\sigma$ (brightest) extent of the models are shown for reference.}
\figsetgrpend

\figsetgrpstart
\figsetgrpnum{2.25}
\figsetgrptitle{KOI--4367}
\figsetplot{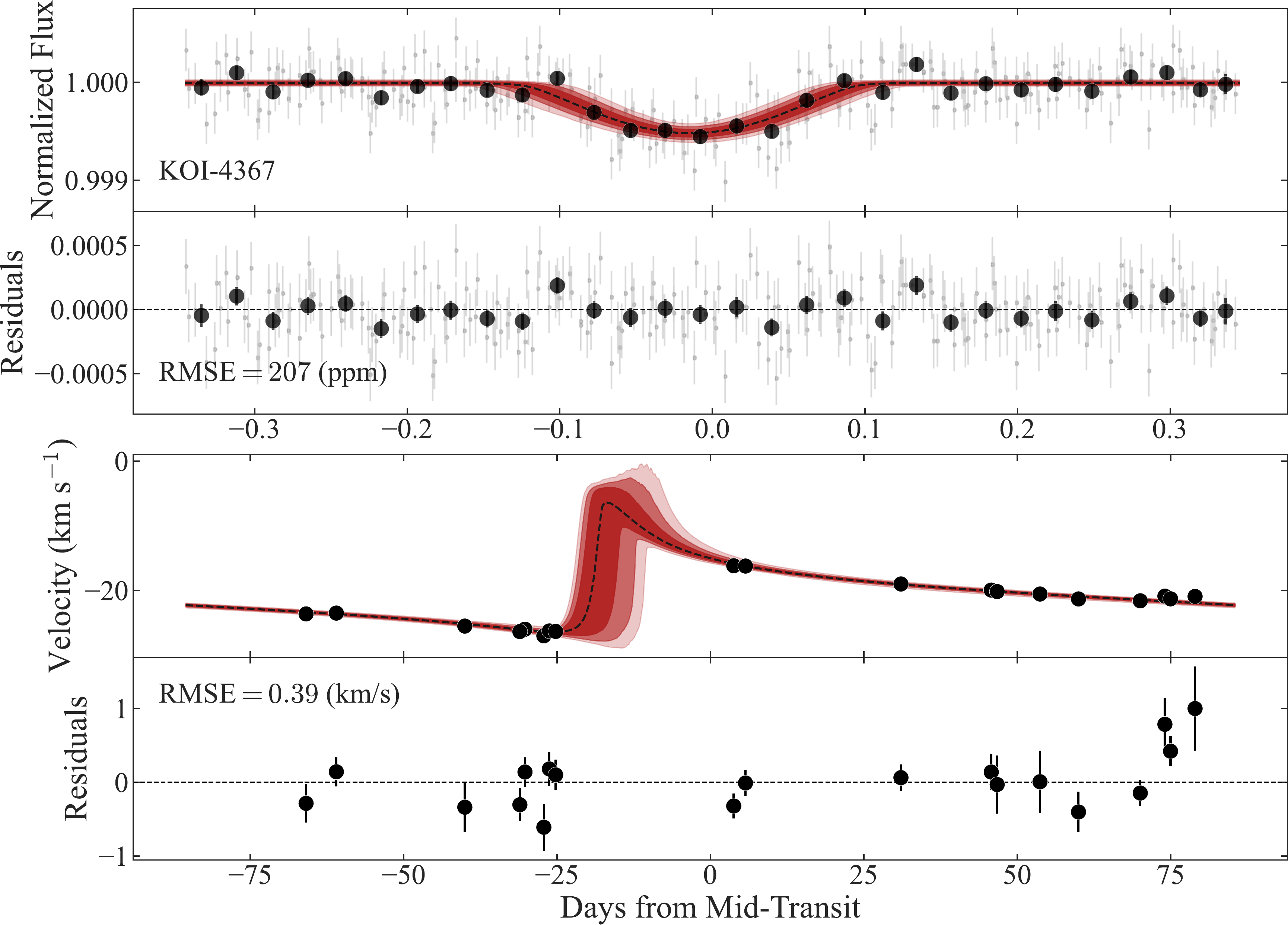}
\figsetgrpnote{\textbf{Top.} The Kepler photometry for KOI-4367 after phase-folding to the derived ephemeris. The large circles represent 30 min bins of the raw data. \textbf{Bottom}. The RVs after phase-folding the data to the derived ephemeris. In each panel, the $1\sigma$ (darkest), $2\sigma$, and $3\sigma$ (brightest) extent of the models are shown for reference.}
\figsetgrpend

\figsetgrpstart
\figsetgrpnum{2.26}
\figsetgrptitle{KOI--5329}
\figsetplot{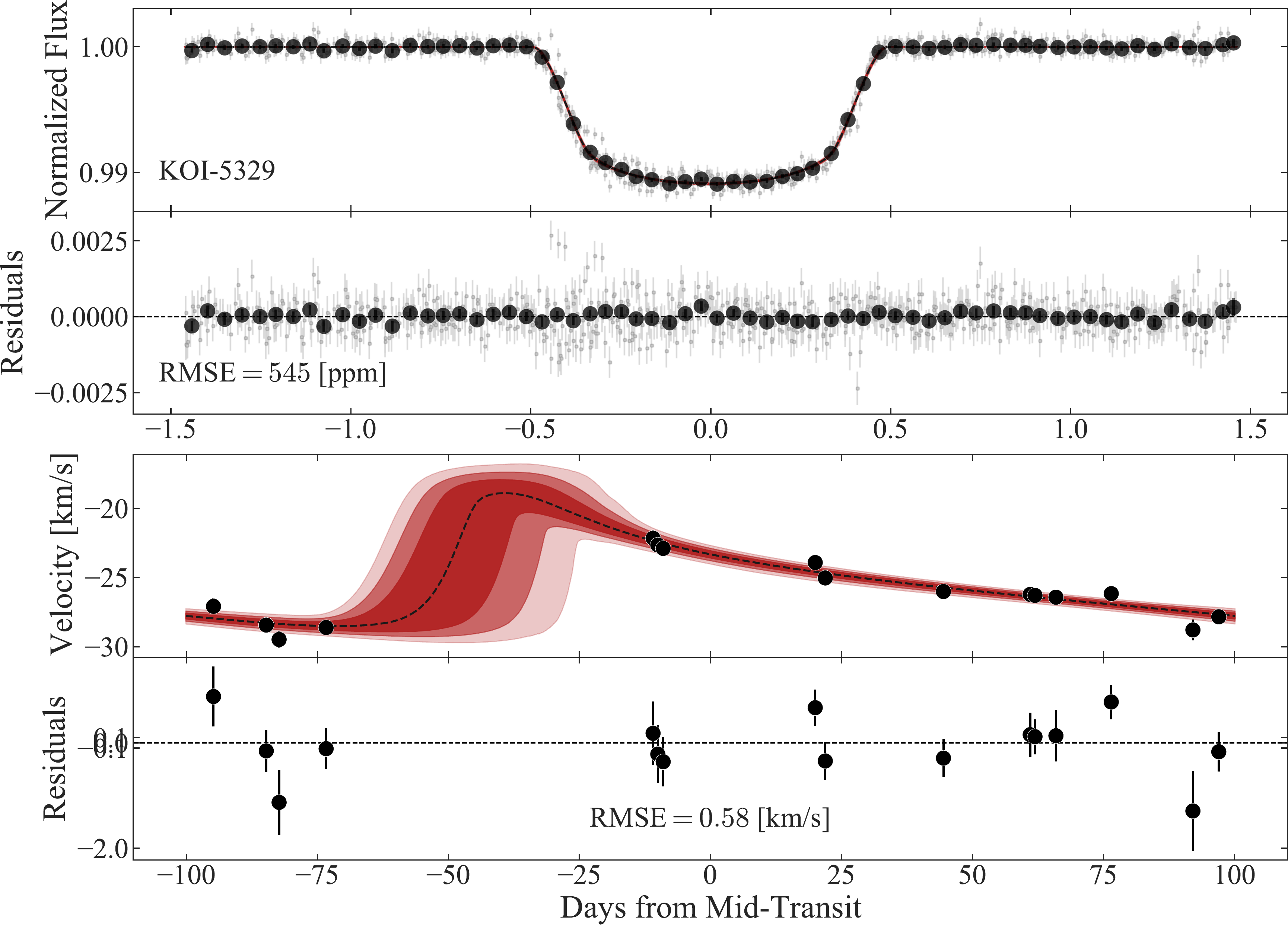}
\figsetgrpnote{\textbf{Top.} The Kepler photometry for KOI-5329 after phase-folding to the derived ephemeris. The large circles represent 30 min bins of the raw data. \textbf{Bottom}. The RVs after phase-folding the data to the derived ephemeris. In each panel, the $1\sigma$ (darkest), $2\sigma$, and $3\sigma$ (brightest) extent of the models are shown for reference.}
\figsetgrpend

\figsetgrpstart
\figsetgrpnum{2.27}
\figsetgrptitle{KOI--6018}
\figsetplot{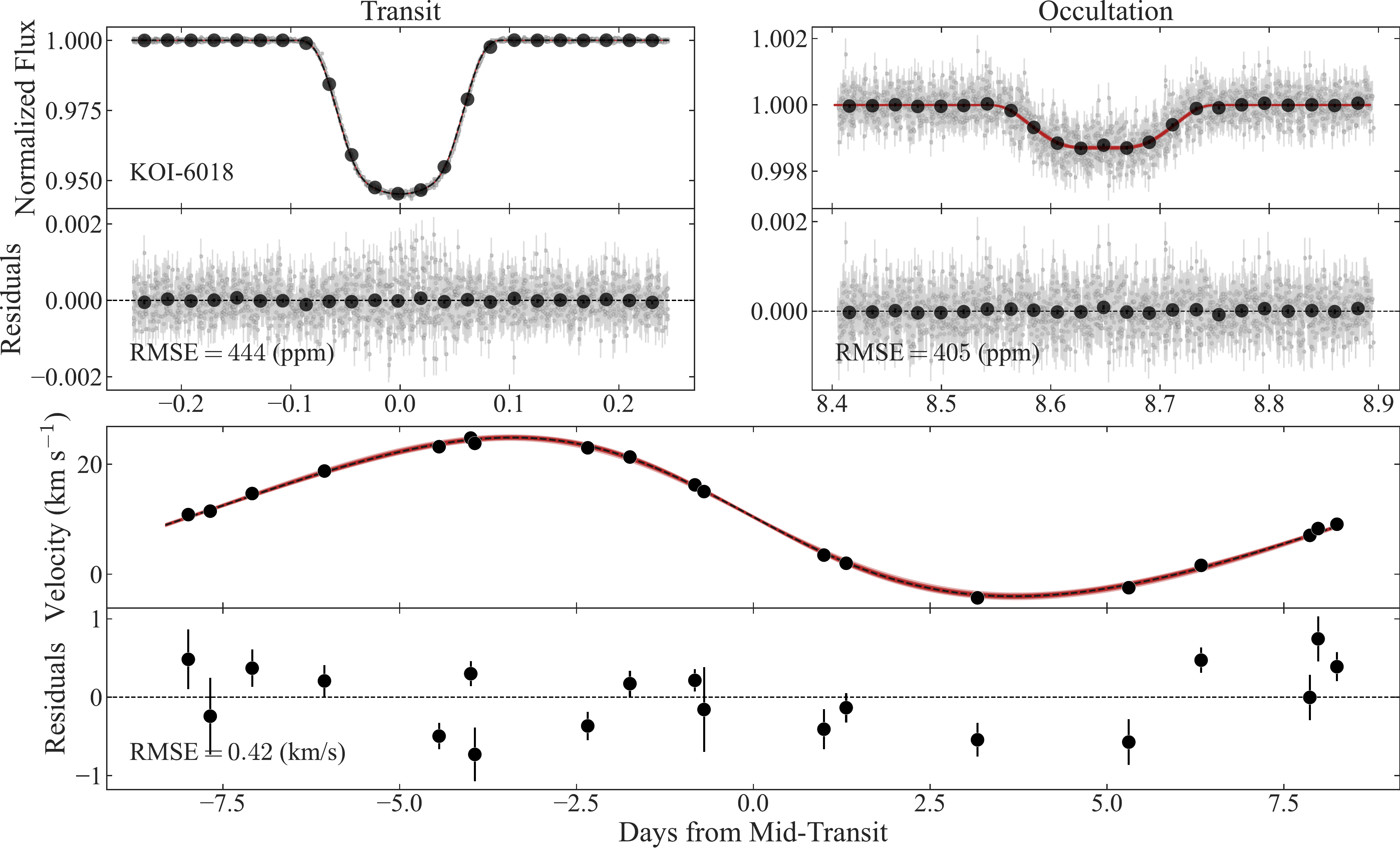}
\figsetgrpnote{\textbf{Top.} The Kepler photometry center on the transit (left) and occultation (right) for KOI-6018 after phase-folding to the derived ephemeris. The large circles represent 30 min bins of the raw data. \textbf{Bottom}. The RVs after phase-folding the data to the derived ephemeris. In each panel, the $1\sigma$ (darkest), $2\sigma$, and $3\sigma$ (brightest) extent of the models are shown for reference.}
\figsetgrpend

\figsetgrpstart
\figsetgrpnum{2.28}
\figsetgrptitle{KOI--6760}
\figsetplot{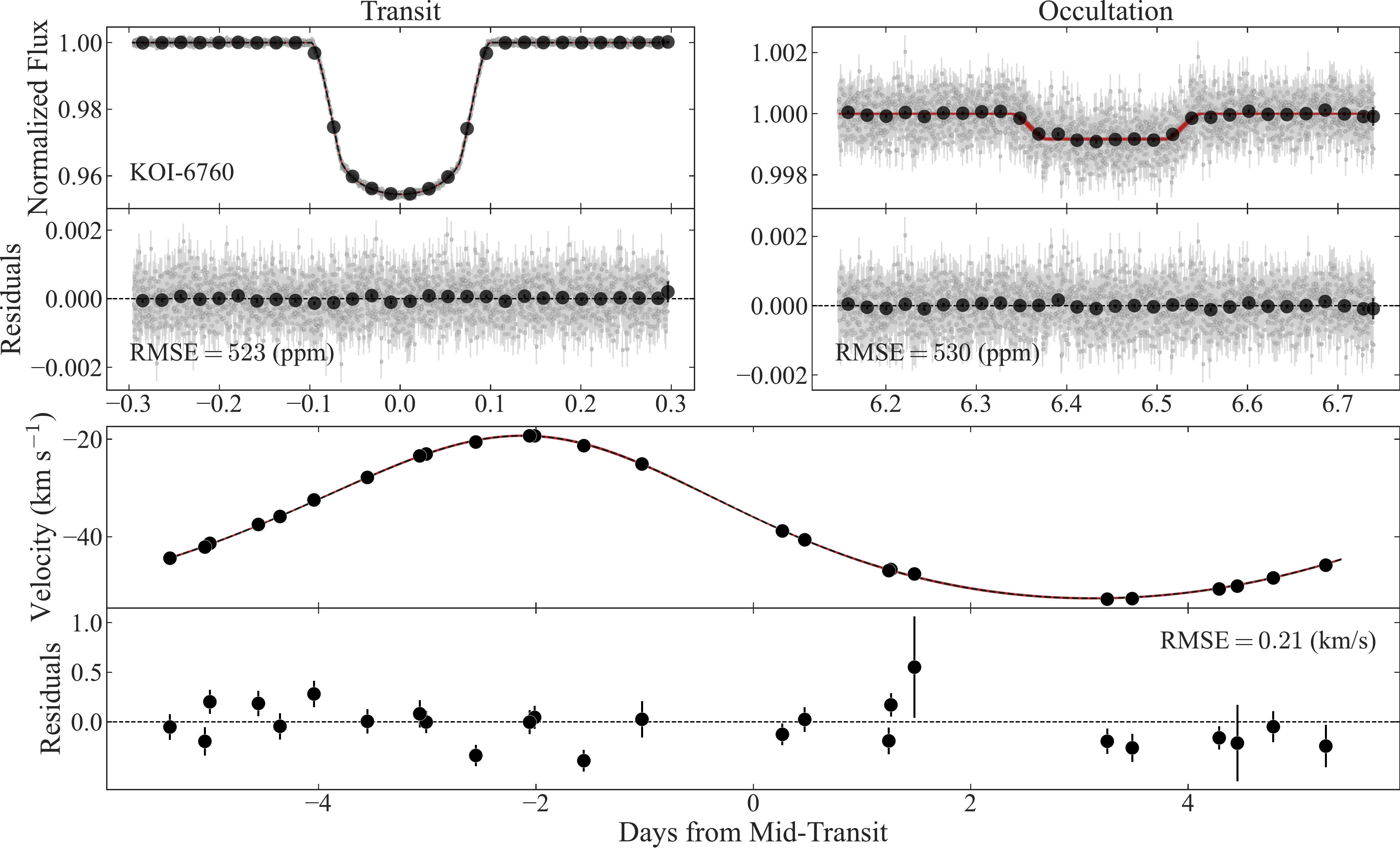}
\figsetgrpnote{\textbf{Top.} The Kepler photometry center on the transit (left) and occultation (right) for KOI-6760 after phase-folding to the derived ephemeris. The large circles represent 30 min bins of the raw data. \textbf{Bottom}. The RVs after phase-folding the data to the derived ephemeris. In each panel, the $1\sigma$ (darkest), $2\sigma$, and $3\sigma$ (brightest) extent of the models are shown for reference.}
\figsetgrpend

\figsetend

%% file: fig2set.tex
\figsetstart
\figsetnum{7}
\figsettitle{Photometric variability for the sample of 28 KOIs}
The \texttt{ARC2} corrected light curve for the systems analyzed in this work. 

\figsetgrpstart
\figsetgrpnum{7.1}
\figsetgrptitle{KOI--23}
\figsetplot{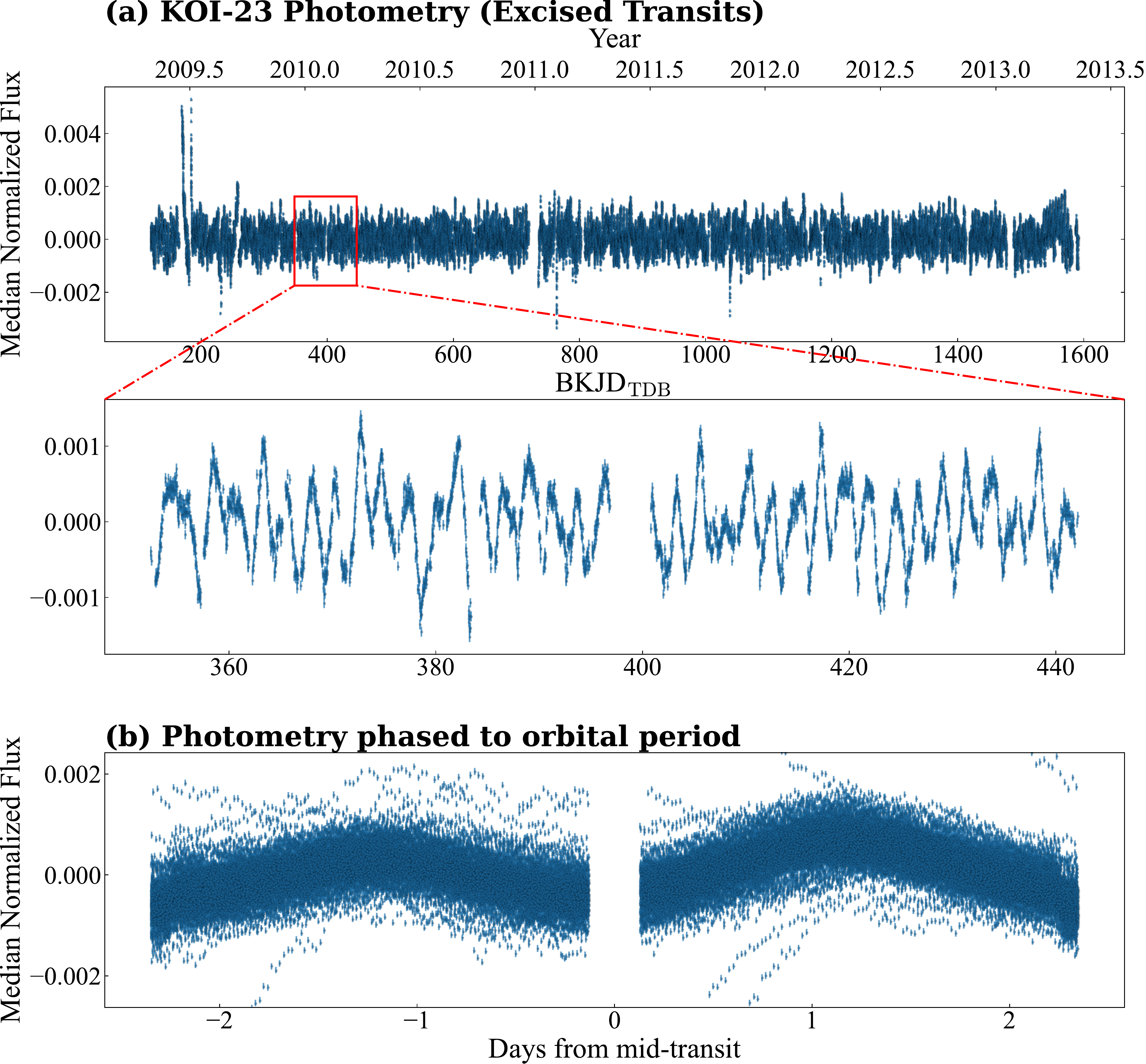}
\figsetgrpnote{\textbf{(a)} The \texttt{ARC2} corrected light curve for KOI-23, after excising the transits, with an inset displaying the quarter 2 data. \textbf{(b)} The out-of-transit photometric variability after phasing to the orbital ephemeris. The derived rotation period is identical (within the $1\sigma$ uncertainty) to the orbital period.}
\figsetgrpend

\figsetgrpstart
\figsetgrpnum{7.2}
\figsetgrptitle{KOI--52}
\figsetplot{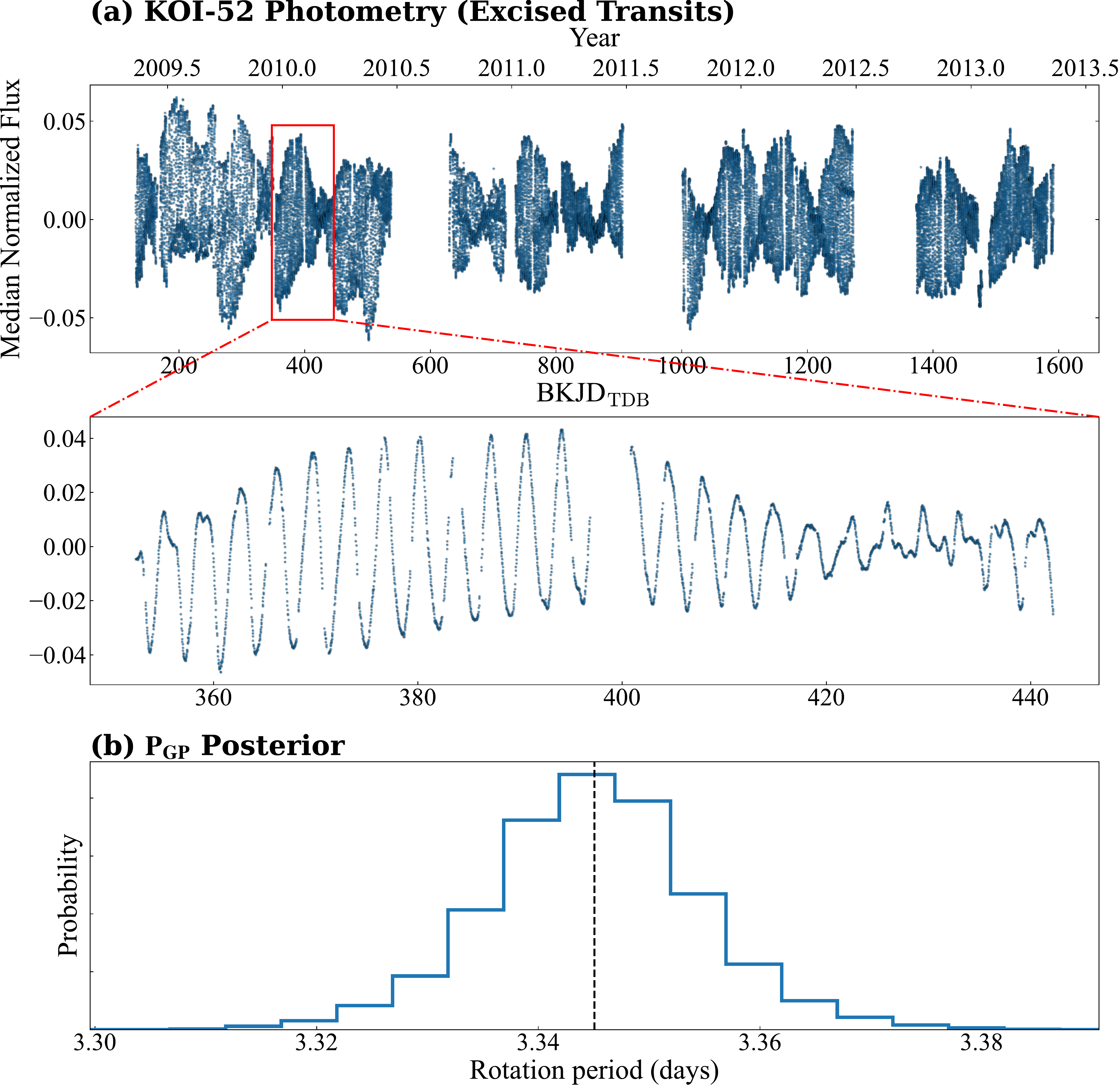}
\figsetgrpnote{\textbf{(a)} The \texttt{ARC2} corrected light curve for KOI-52, after excising the transits, with an inset displaying a subset of the Kepler data. \textbf{(b)} The posterior distribution for the Gaussian process period, which we interpret as a measurement of the stellar rotation period. }
\figsetgrpend

\figsetgrpstart
\figsetgrpnum{7.3}
\figsetgrptitle{KOI--129}
\figsetplot{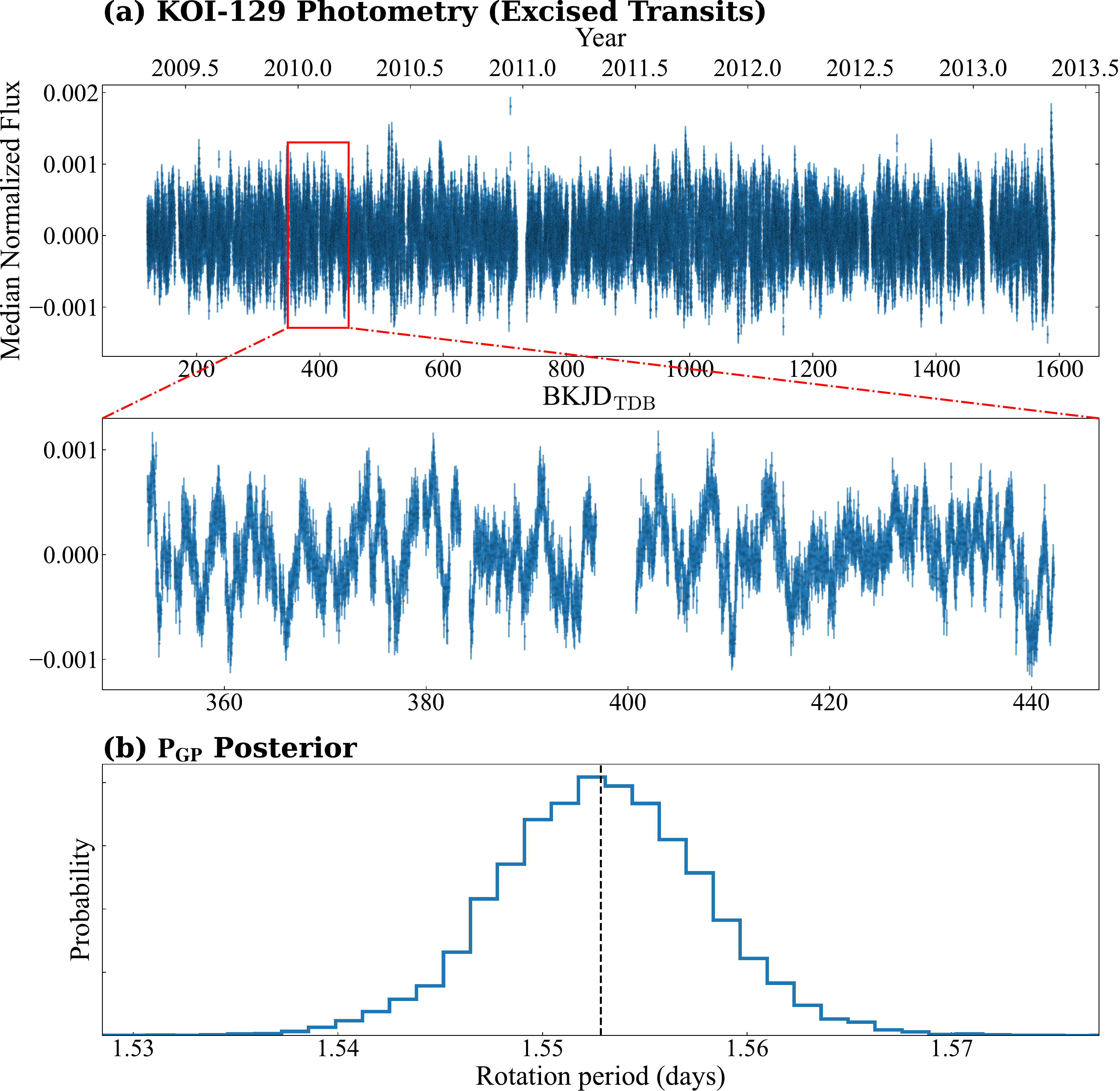}
\figsetgrpnote{\textbf{(a)} The \texttt{ARC2} corrected light curve for KOI-129, after excising the transits, with an inset displaying a subset of the Kepler data. \textbf{(b)} The posterior distribution for the Gaussian process period, which we interpret as a measurement of the stellar rotation period. }
\figsetgrpend

\figsetgrpstart
\figsetgrpnum{7.4}
\figsetgrptitle{KOI--130}
\figsetplot{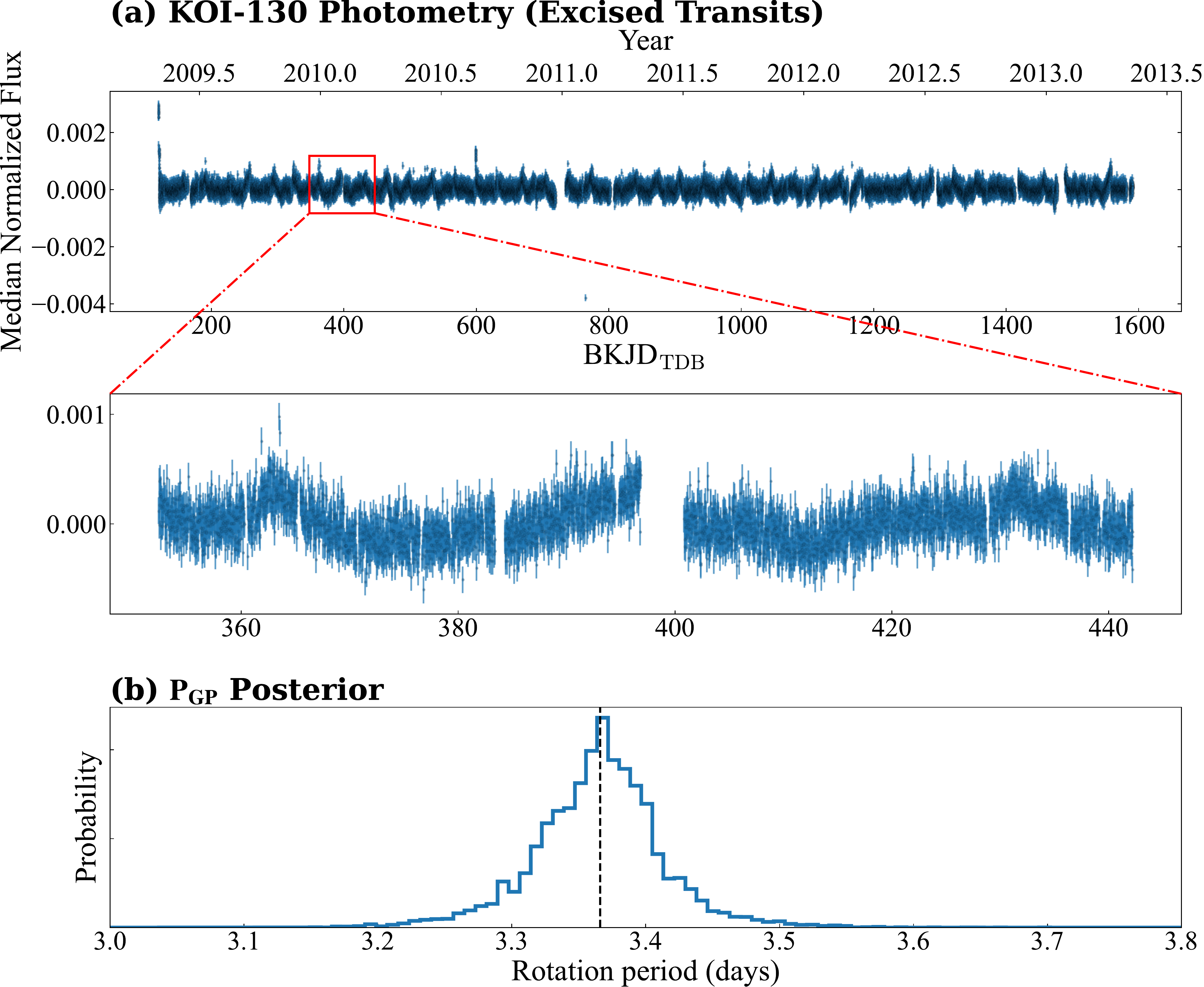}
\figsetgrpnote{\textbf{(a)} The \texttt{ARC2} corrected light curve for KOI-130, after excising the transits, with an inset displaying a subset of the Kepler data. \textbf{(b)} The posterior distribution for the Gaussian process period, which we interpret as a measurement of the stellar rotation period. }
\figsetgrpend

\figsetgrpstart
\figsetgrpnum{7.5}
\figsetgrptitle{KOI--182}
\figsetplot{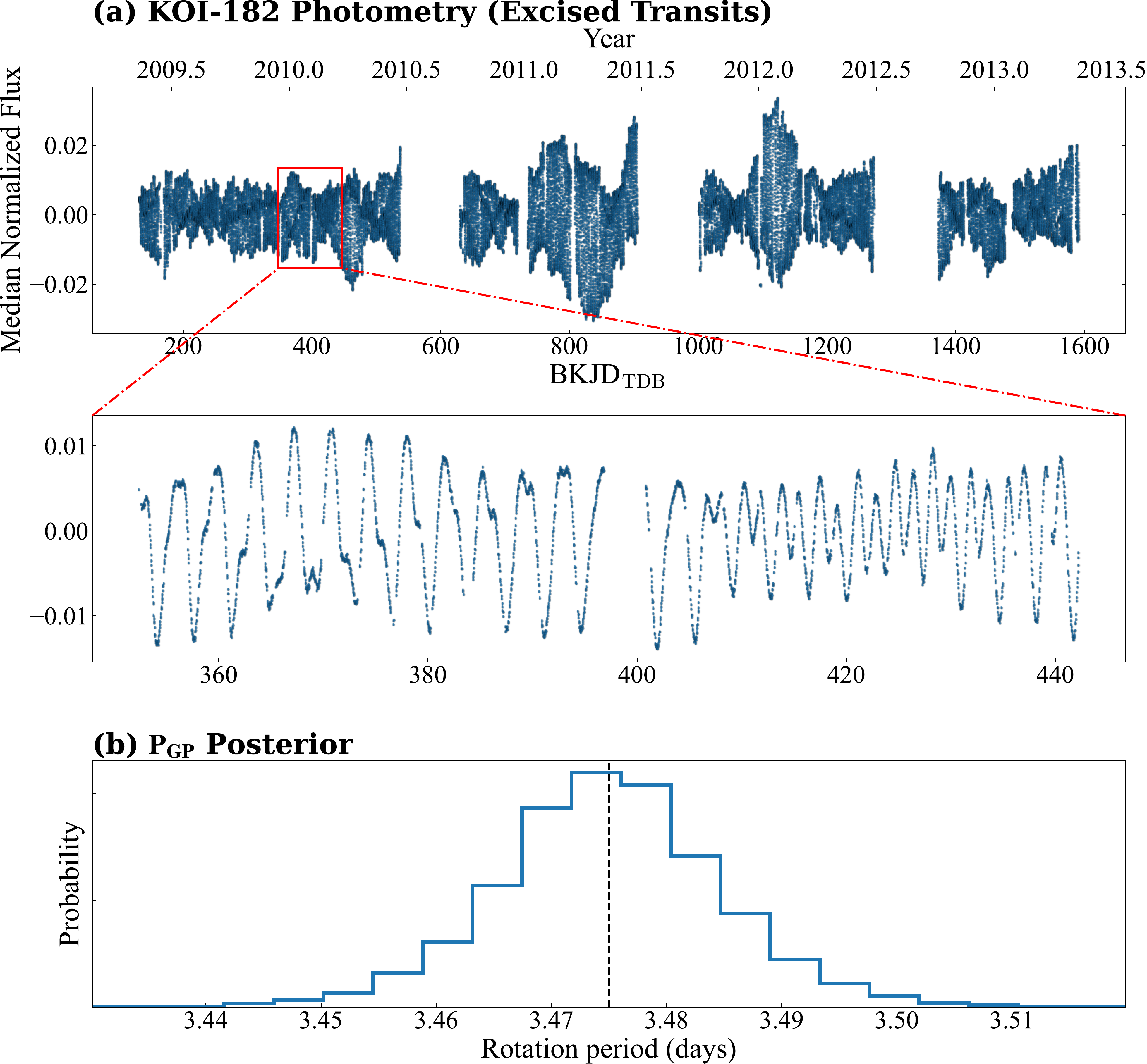}
\figsetgrpnote{\textbf{(a)} The \texttt{ARC2} corrected light curve for KOI-182, after excising the transits, with an inset displaying a subset of the Kepler data. \textbf{(b)} The posterior distribution for the Gaussian process period, which we interpret as a measurement of the stellar rotation period. }
\figsetgrpend

\figsetgrpstart
\figsetgrpnum{7.6}
\figsetgrptitle{KOI--219}
\figsetplot{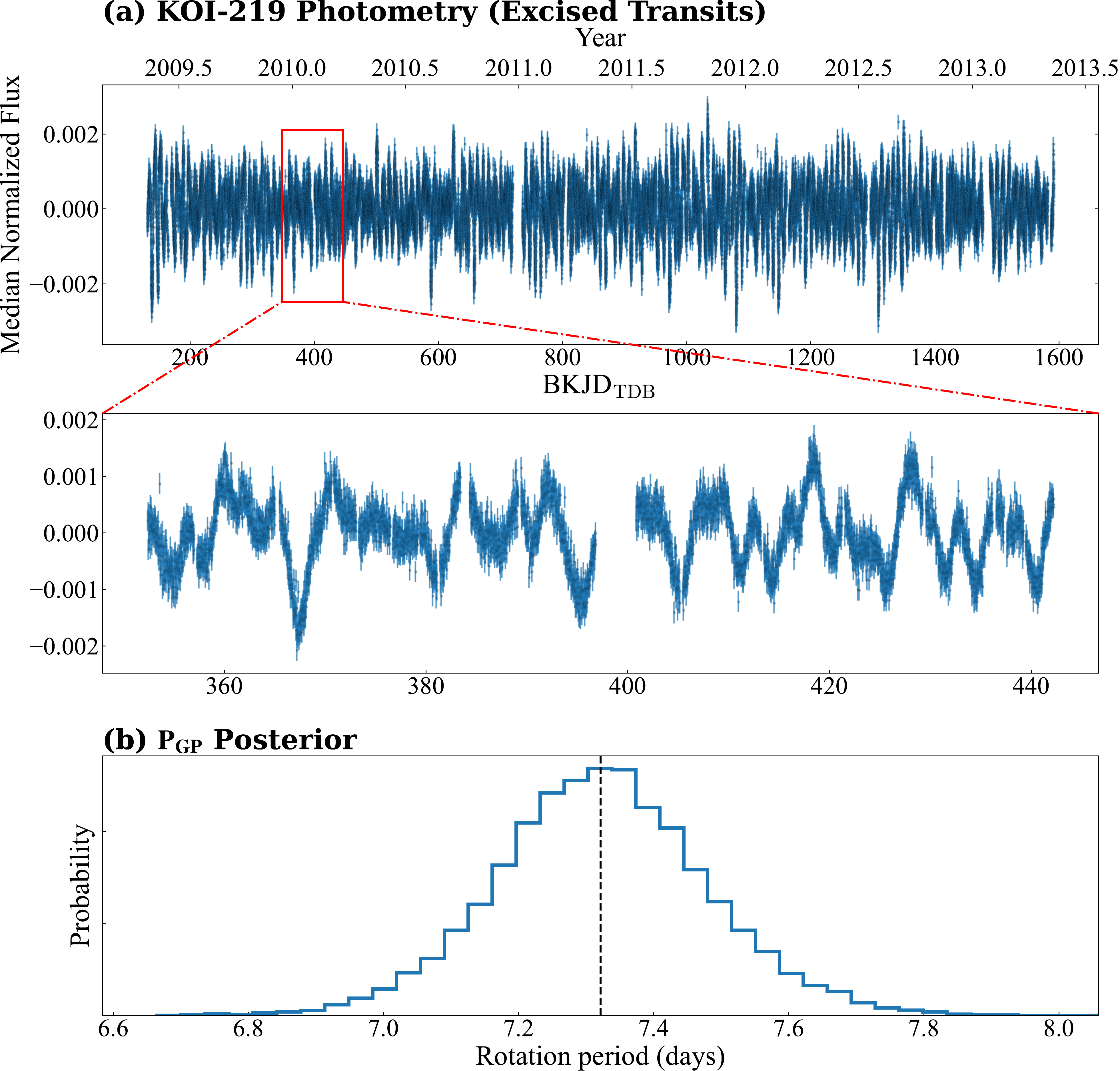}
\figsetgrpnote{\textbf{(a)} The \texttt{ARC2} corrected light curve for KOI-219, after excising the transits, with an inset displaying a subset of the Kepler data. \textbf{(b)} The posterior distribution for the Gaussian process period, which we interpret as a measurement of the stellar rotation period. }
\figsetgrpend

\figsetgrpstart
\figsetgrpnum{7.7}
\figsetgrptitle{KOI--242}
\figsetplot{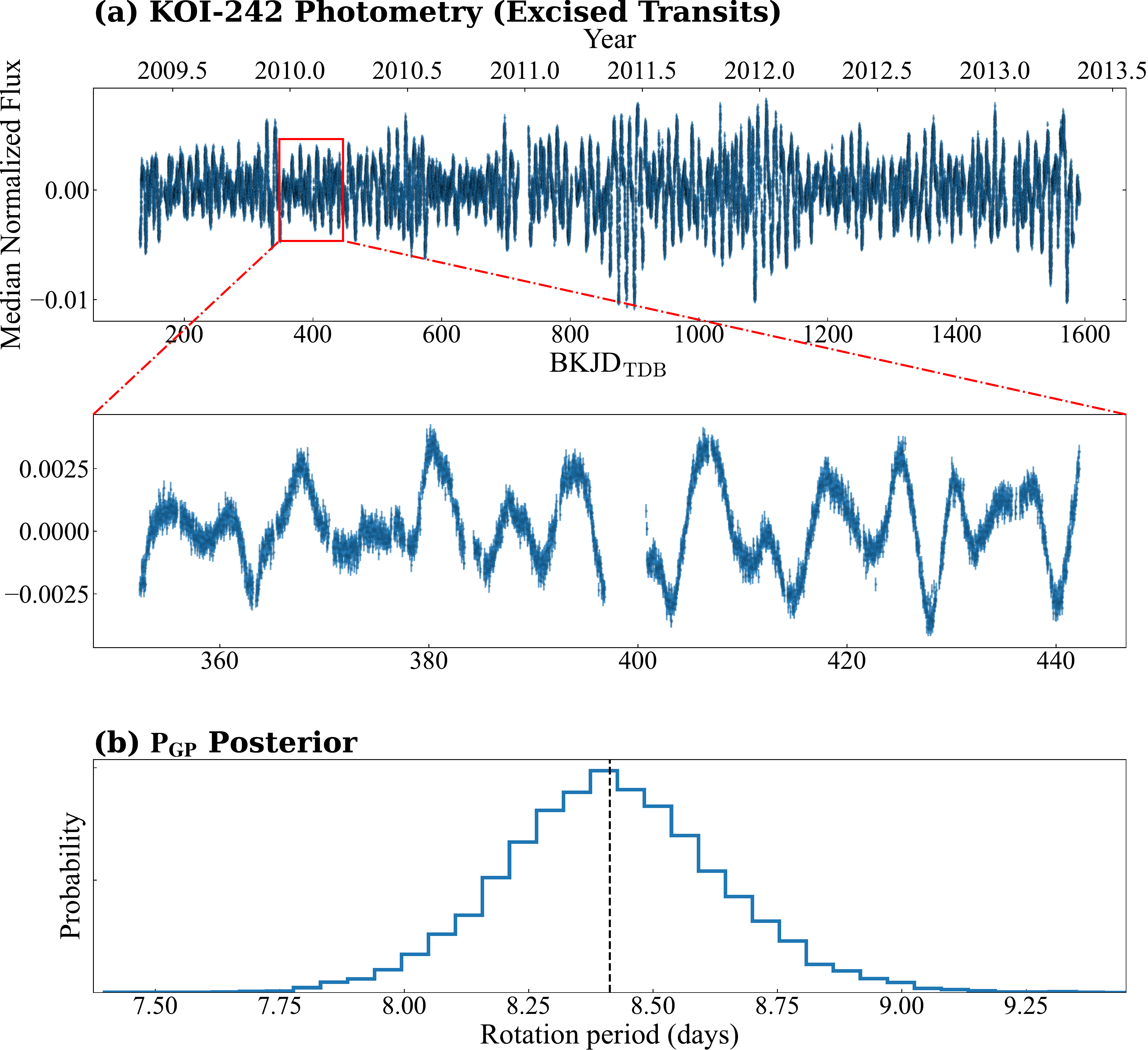}
\figsetgrpnote{\textbf{(a)} The \texttt{ARC2} corrected light curve for KOI-242, after excising the transits, with an inset displaying a subset of the Kepler data. \textbf{(b)} The posterior distribution for the Gaussian process period, which we interpret as a measurement of the stellar rotation period. }
\figsetgrpend

\figsetgrpstart
\figsetgrpnum{7.8}
\figsetgrptitle{KOI--403}
\figsetplot{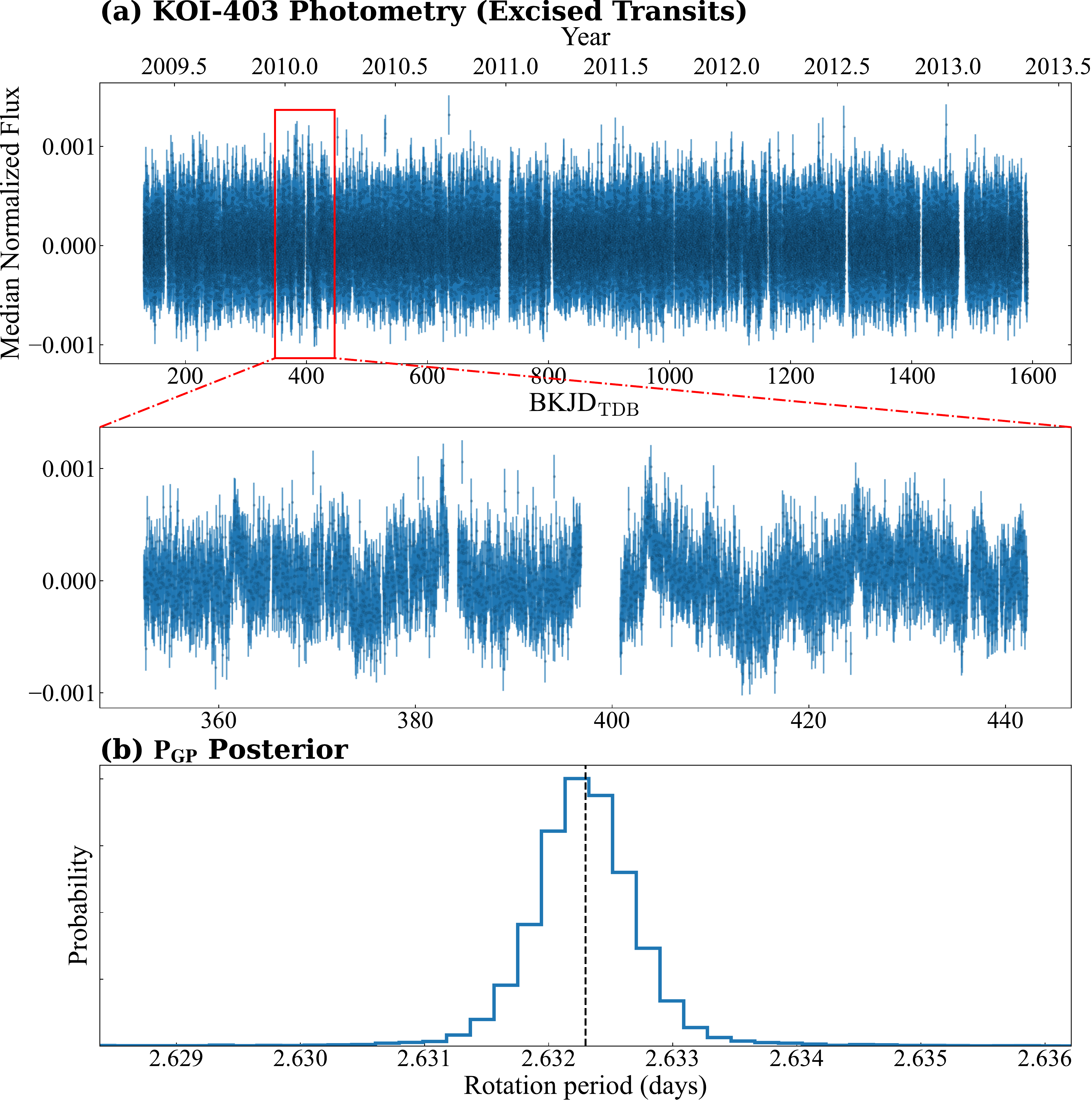}
\figsetgrpnote{\textbf{(a)} The \texttt{ARC2} corrected light curve for KOI-403, after excising the transits, with an inset displaying a subset of the Kepler data. \textbf{(b)} The posterior distribution for the Gaussian process period, which we interpret as a measurement of the stellar rotation period. }
\figsetgrpend

\figsetgrpstart
\figsetgrpnum{7.9}
\figsetgrptitle{KOI--415}
\figsetplot{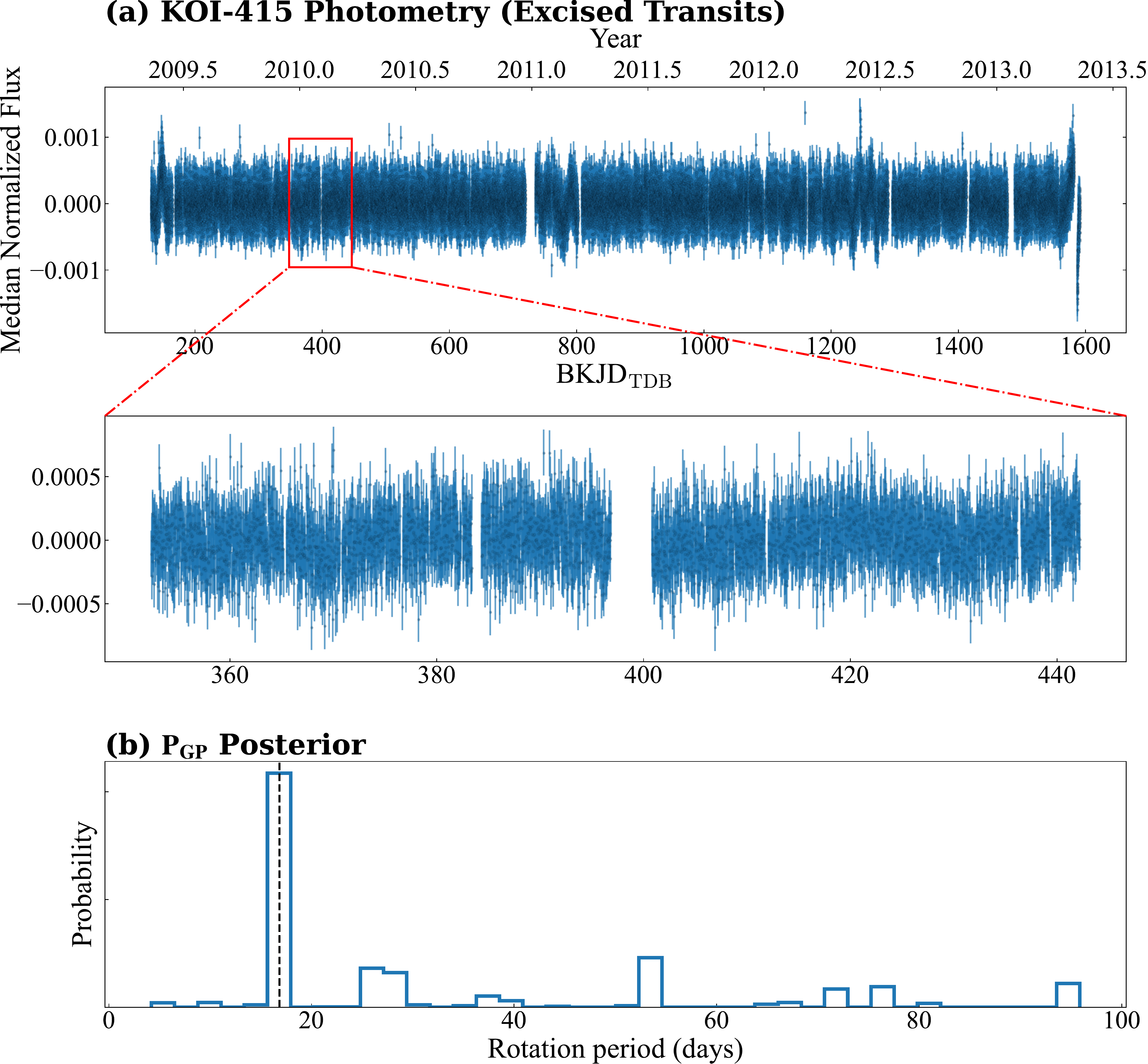}
\figsetgrpnote{\textbf{(a)} The \texttt{ARC2} corrected light curve for KOI-415, after excising the transits, with an inset displaying a subset of the Kepler data. \textbf{(b)} The posterior distribution for the Gaussian process period, which we interpret as a measurement of the stellar rotation period. }
\figsetgrpend

\figsetgrpstart
\figsetgrpnum{7.10}
\figsetgrptitle{KOI--466}
\figsetplot{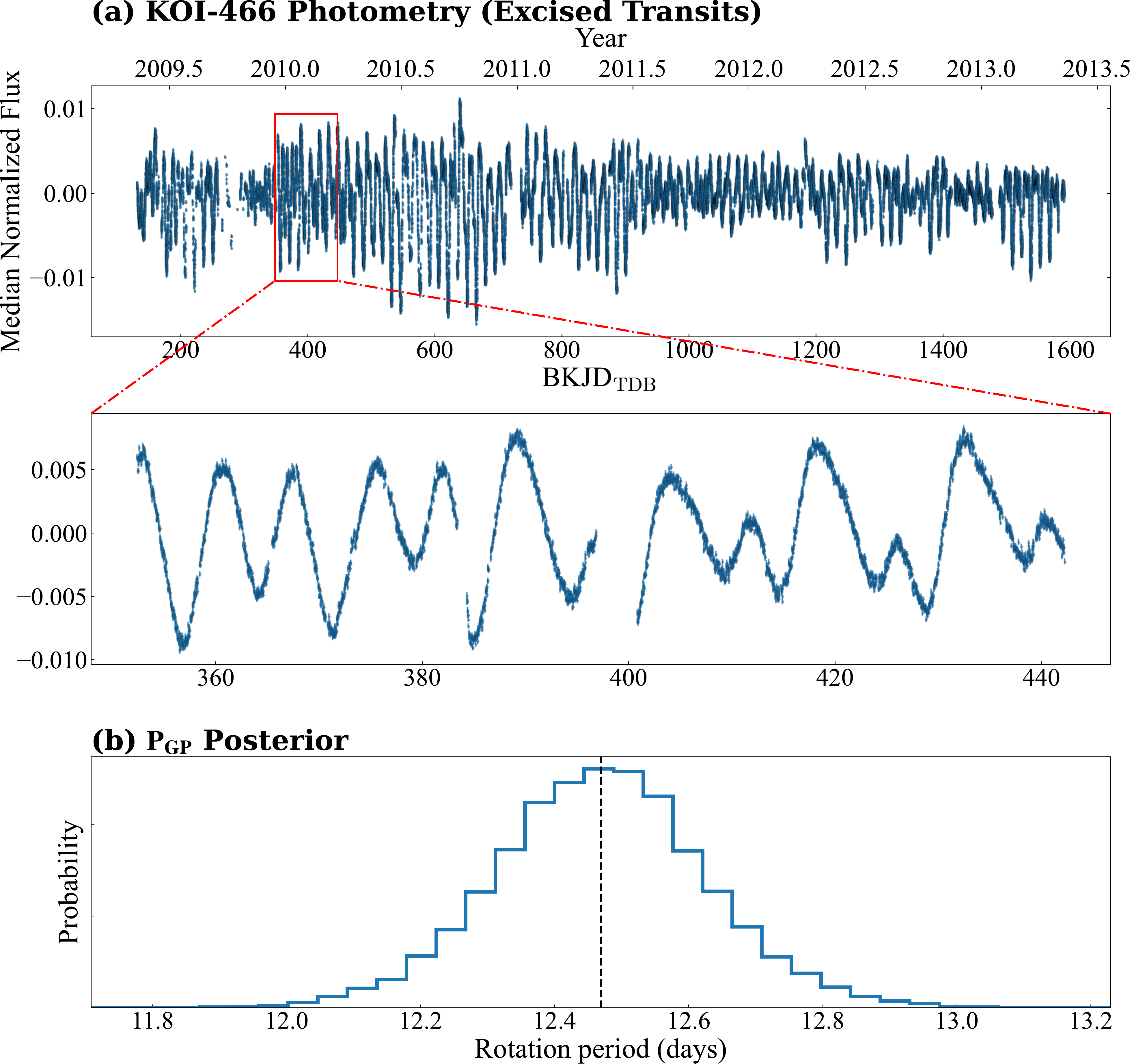}
\figsetgrpnote{\textbf{(a)} The \texttt{ARC2} corrected light curve for KOI-466, after excising the transits, with an inset displaying a subset of the Kepler data. \textbf{(b)} The posterior distribution for the Gaussian process period, which we interpret as a measurement of the stellar rotation period. }
\figsetgrpend

\figsetgrpstart
\figsetgrpnum{7.11}
\figsetgrptitle{KOI--631}
\figsetplot{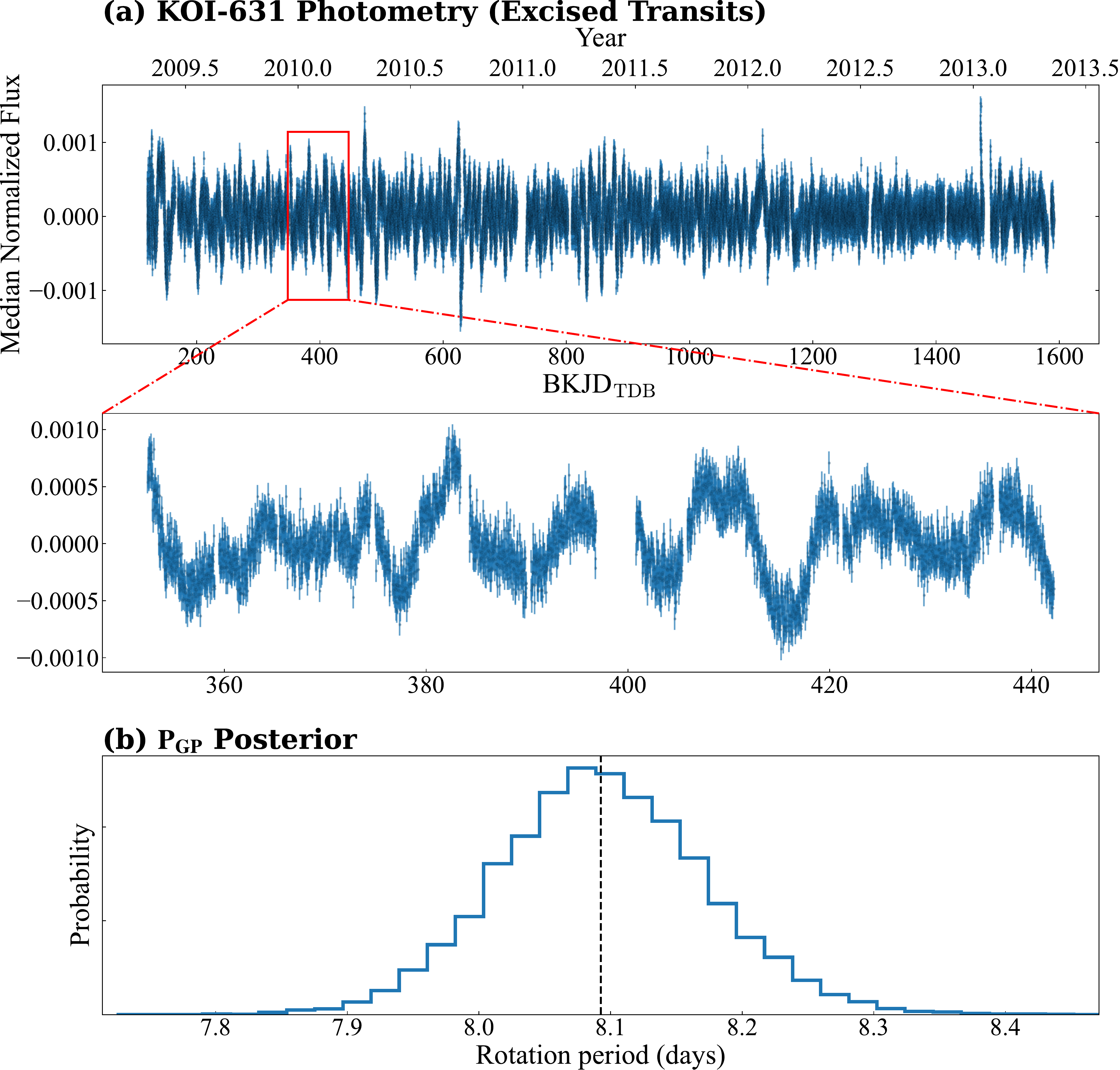}
\figsetgrpnote{\textbf{(a)} The \texttt{ARC2} corrected light curve for KOI-631, after excising the transits, with an inset displaying a subset of the Kepler data. \textbf{(b)} The posterior distribution for the Gaussian process period, which we interpret as a measurement of the stellar rotation period. }
\figsetgrpend

\figsetgrpstart
\figsetgrpnum{7.12}
\figsetgrptitle{KOI--777}
\figsetplot{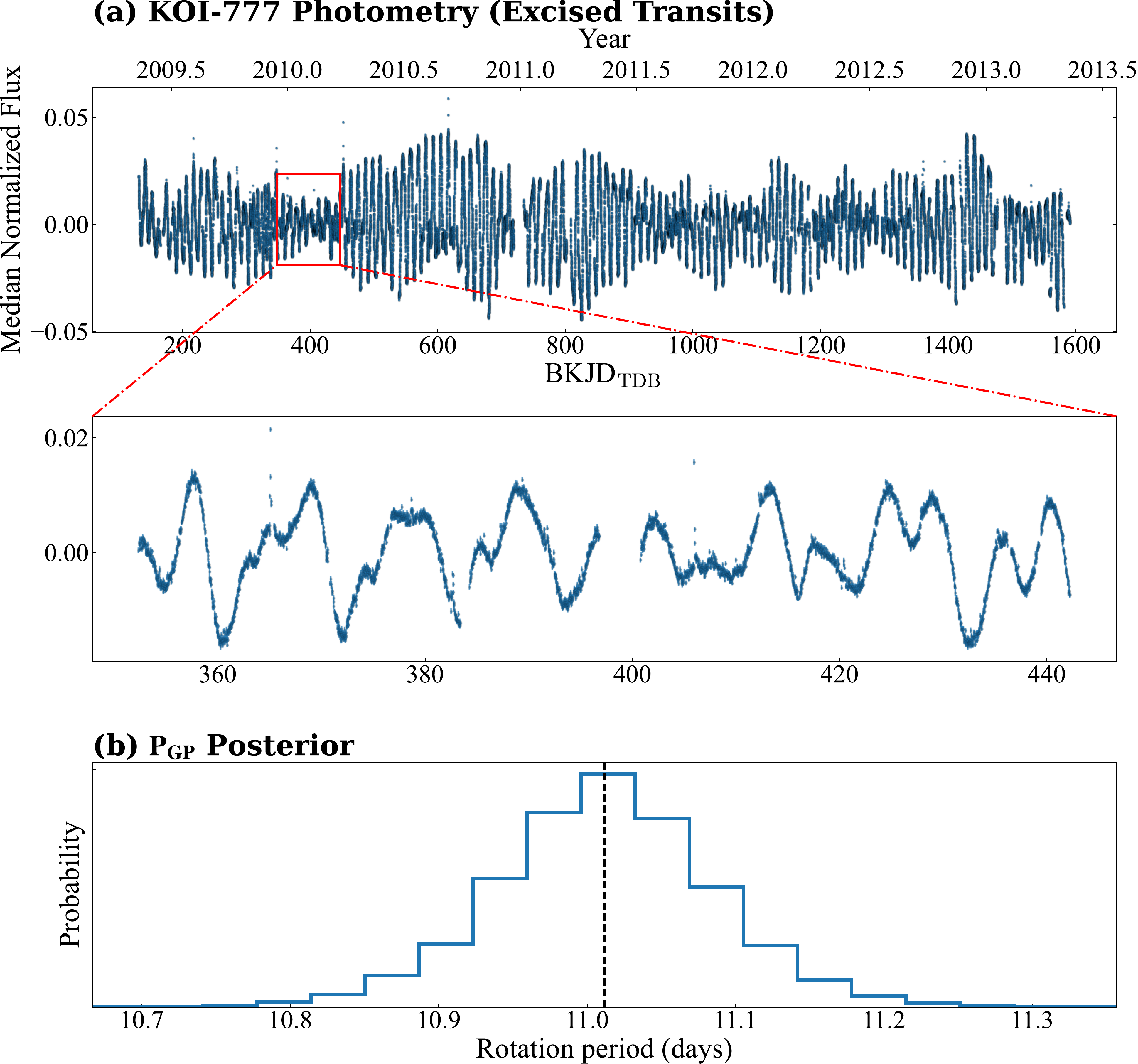}
\figsetgrpnote{\textbf{(a)} The \texttt{ARC2} corrected light curve for KOI-777, after excising the transits, with an inset displaying a subset of the Kepler data. \textbf{(b)} The posterior distribution for the Gaussian process period, which we interpret as a measurement of the stellar rotation period. }
\figsetgrpend

\figsetgrpstart
\figsetgrpnum{7.13}
\figsetgrptitle{KOI--846}
\figsetplot{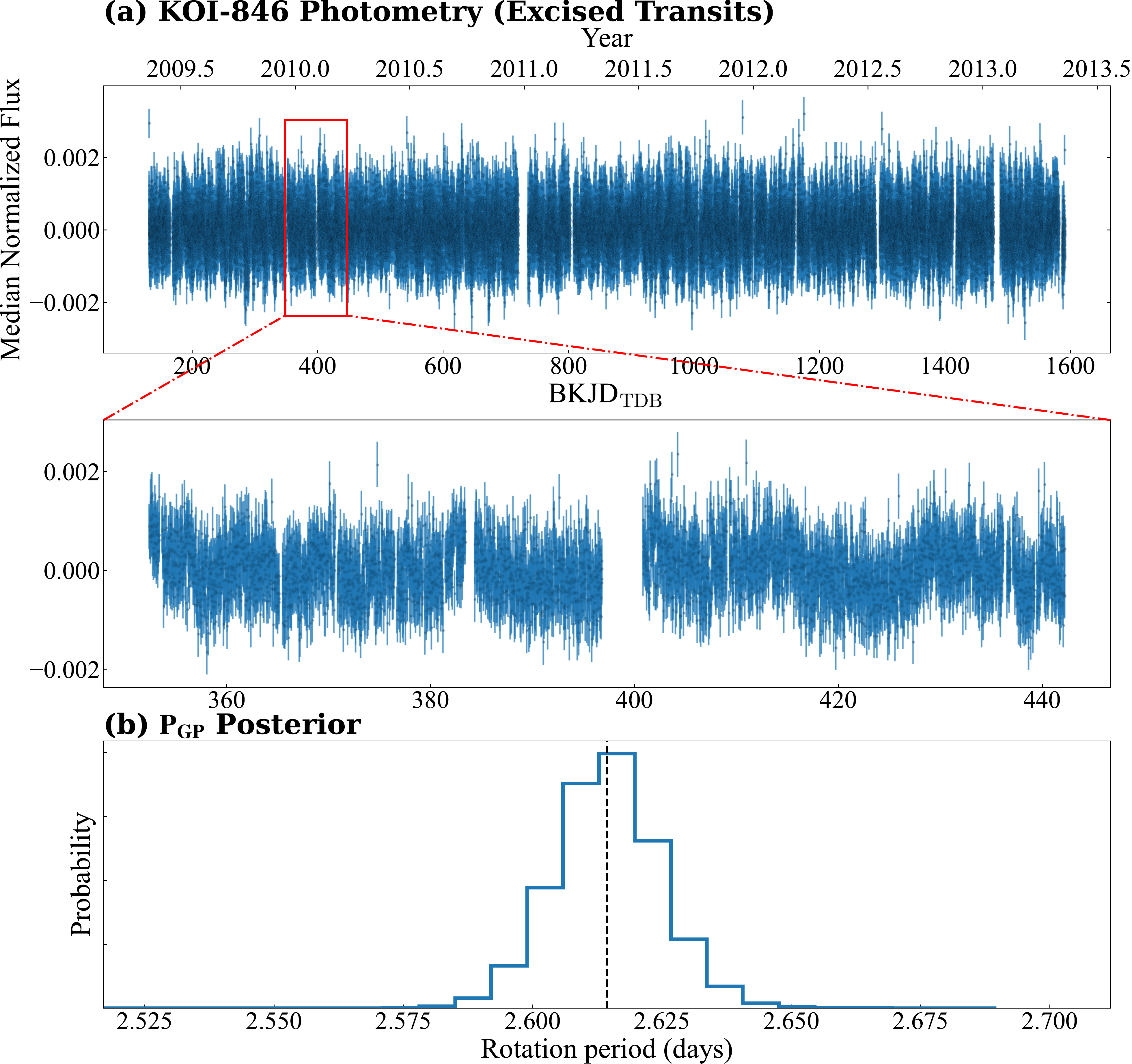}
\figsetgrpnote{\textbf{(a)} The \texttt{ARC2} corrected light curve for KOI-846, after excising the transits, with an inset displaying a subset of the Kepler data. \textbf{(b)} The posterior distribution for the Gaussian process period, which we interpret as a measurement of the stellar rotation period. }
\figsetgrpend

\figsetgrpstart
\figsetgrpnum{7.14}
\figsetgrptitle{KOI--855}
\figsetplot{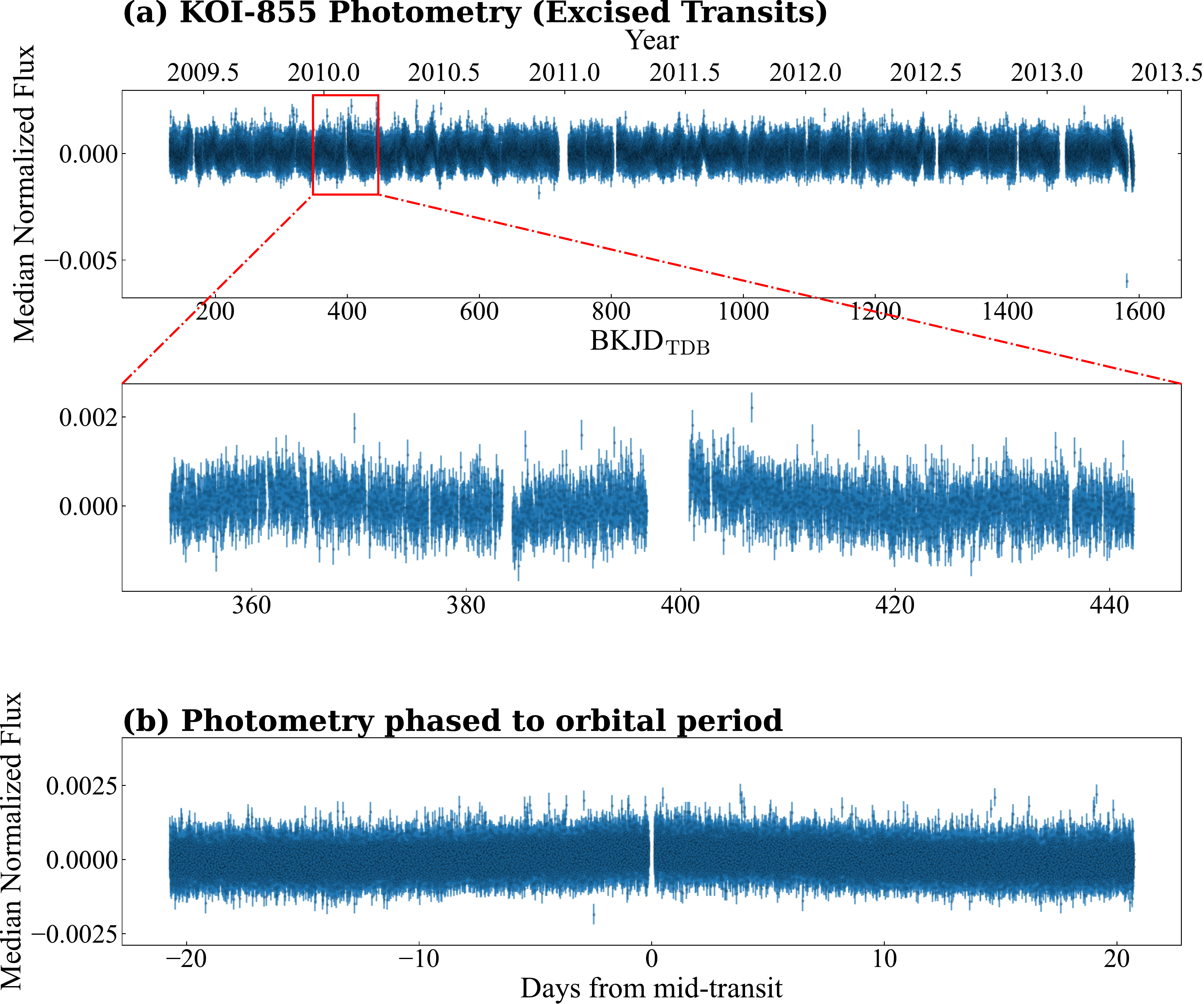}
\figsetgrpnote{\textbf{(a)} The \texttt{ARC2} corrected light curve for KOI-855, after excising the transits, with an inset displaying the quarter 2 data. \textbf{(b)} The out-of-transit photometric variability after phasing to the orbital ephemeris. The derived rotation period is identical (within the $1\sigma$ uncertainty) to the orbital period.}
\figsetgrpend

\figsetgrpstart
\figsetgrpnum{7.15}
\figsetgrptitle{KOI--1064}
\figsetplot{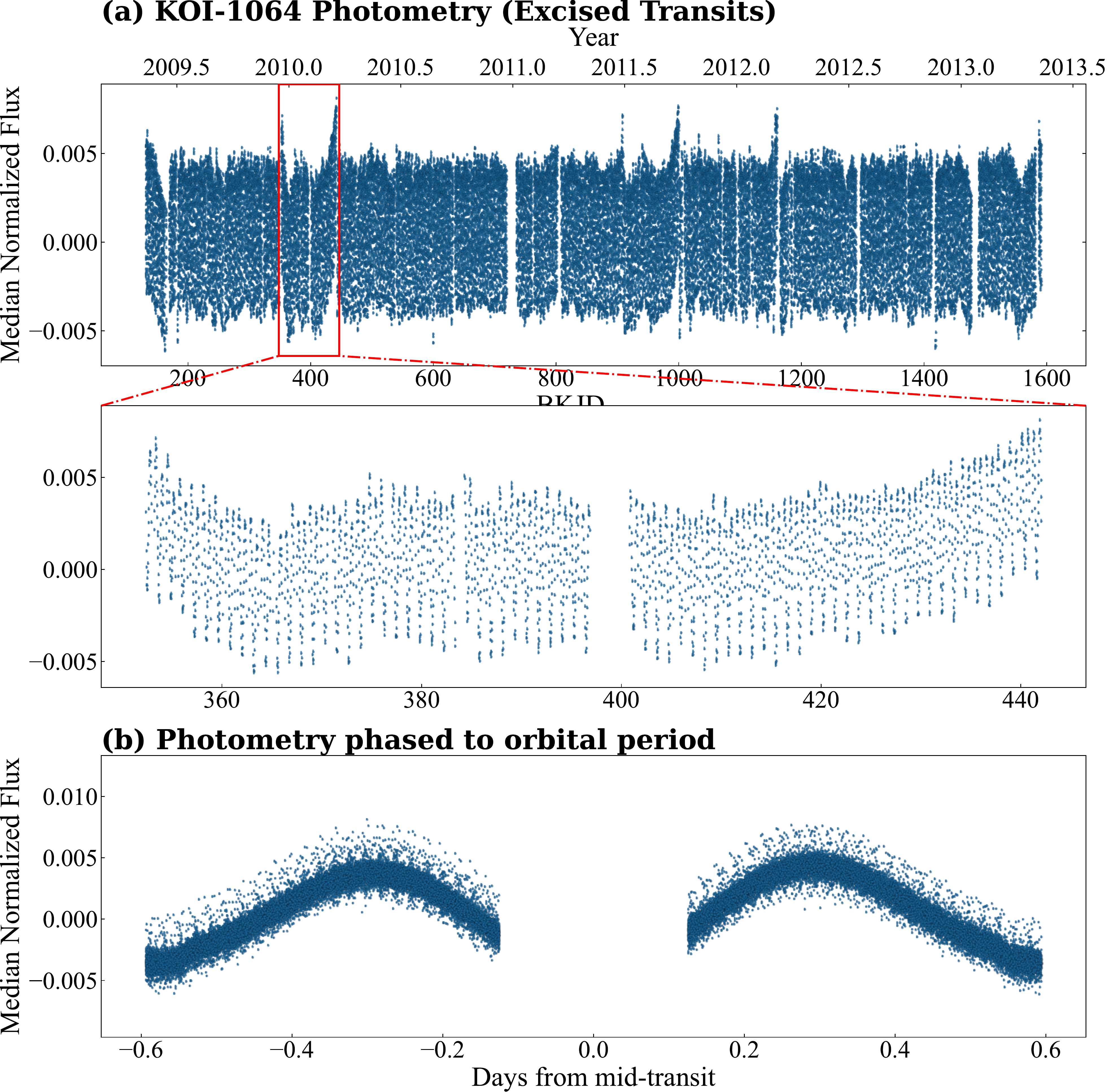}
\figsetgrpnote{\textbf{(a)} The \texttt{ARC2} corrected light curve for KOI-1064, after excising the transits, with an inset displaying the quarter 2 data. \textbf{(b)} The out-of-transit photometric variability after phasing to the orbital ephemeris. The derived rotation period is identical (within the $1\sigma$ uncertainty) to the orbital period.}
\figsetgrpend

\figsetgrpstart
\figsetgrpnum{7.16}
\figsetgrptitle{KOI--1247}
\figsetplot{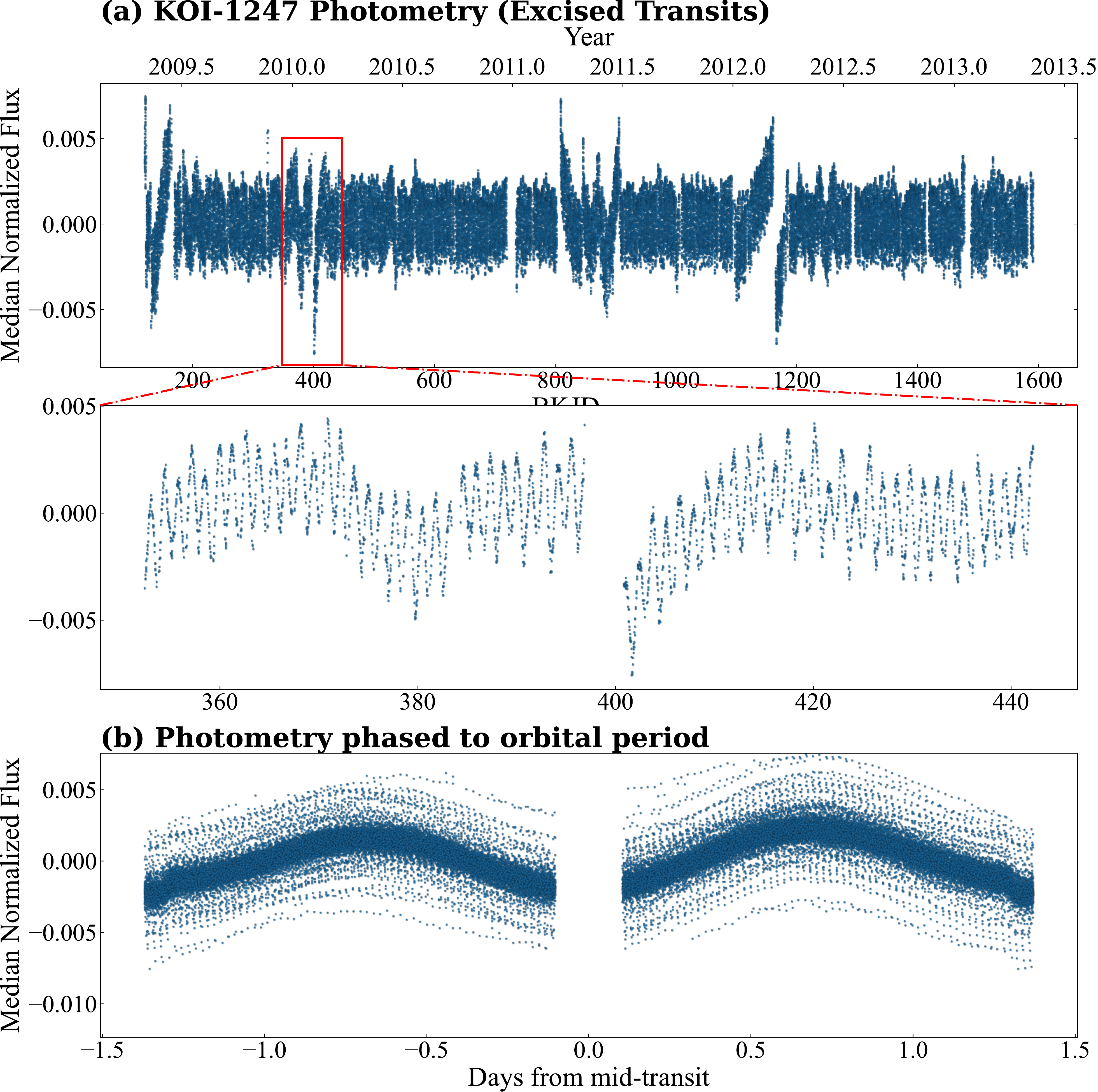}
\figsetgrpnote{\textbf{(a)} The \texttt{ARC2} corrected light curve for KOI-1247, after excising the transits, with an inset displaying the quarter 2 data. \textbf{(b)} The out-of-transit photometric variability after phasing to the orbital ephemeris. The derived rotation period is identical (within the $1\sigma$ uncertainty) to the orbital period.}
\figsetgrpend

\figsetgrpstart
\figsetgrpnum{7.17}
\figsetgrptitle{KOI--1288}
\figsetplot{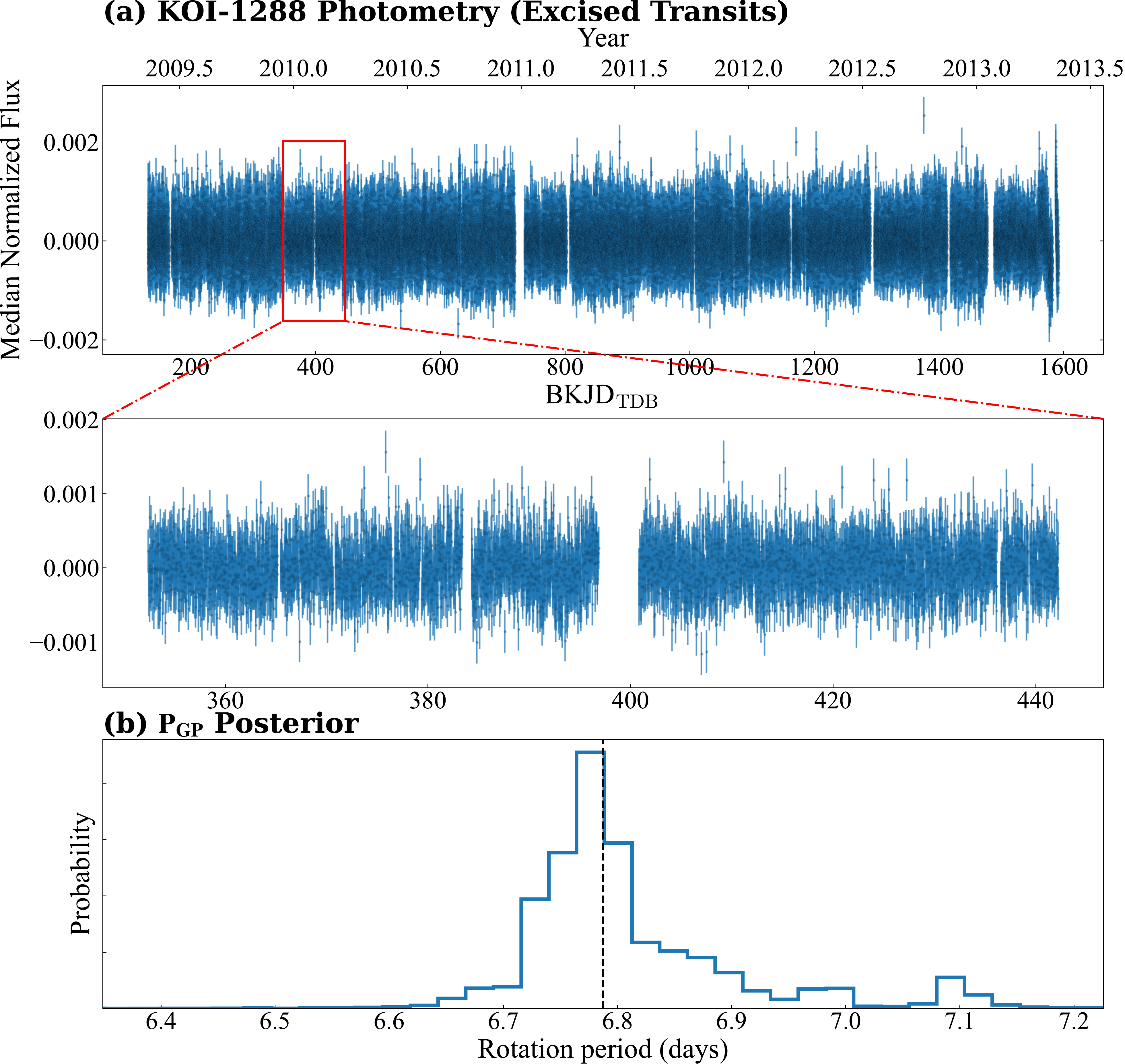}
\figsetgrpnote{\textbf{(a)} The \texttt{ARC2} corrected light curve for KOI-1288, after excising the transits, with an inset displaying a subset of the Kepler data. \textbf{(b)} The posterior distribution for the Gaussian process period, which we interpret as a measurement of the stellar rotation period. }
\figsetgrpend

\figsetgrpstart
\figsetgrpnum{7.18}
\figsetgrptitle{KOI--1347}
\figsetplot{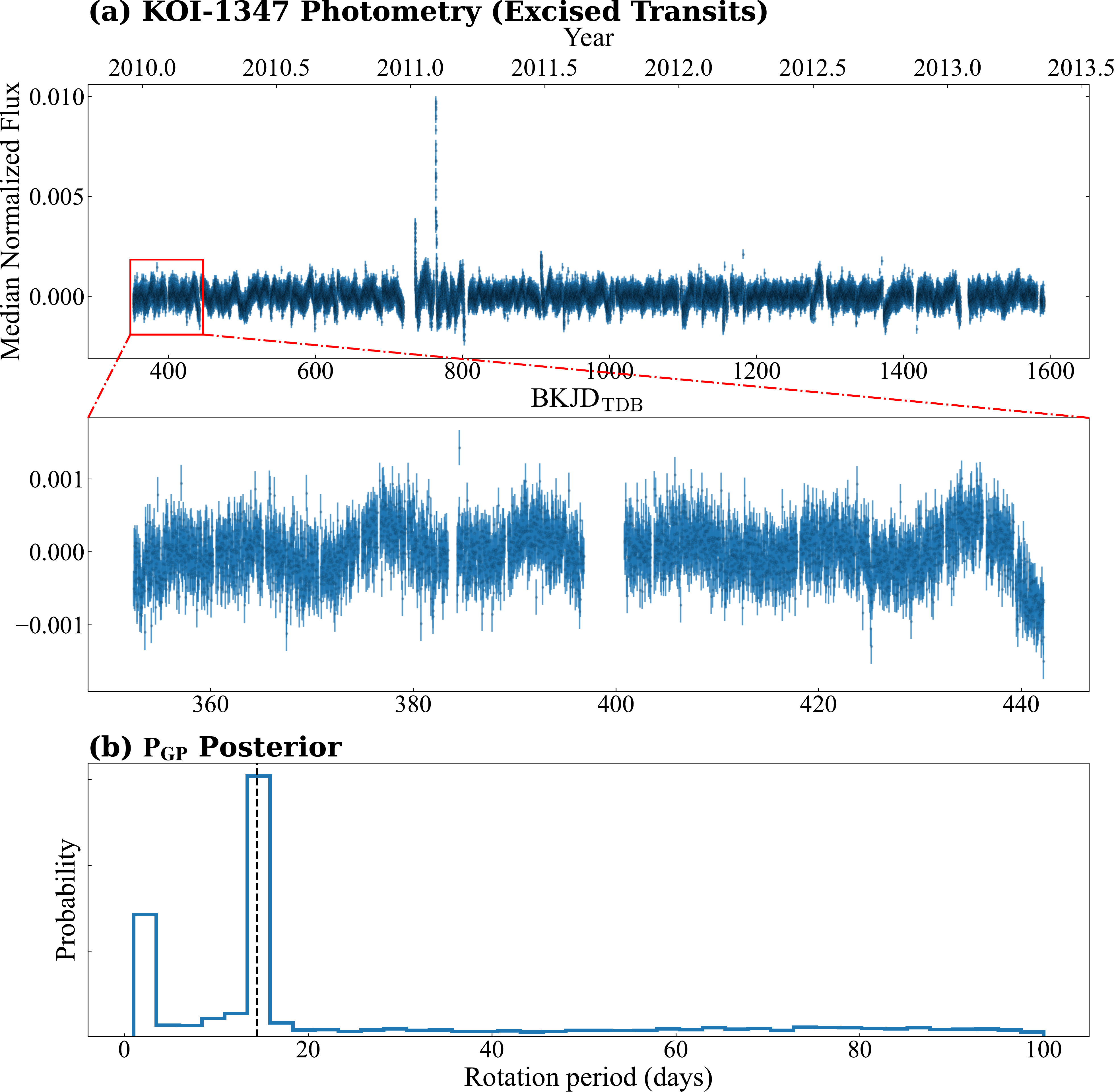}
\figsetgrpnote{\textbf{(a)} The \texttt{ARC2} corrected light curve for KOI-1347, after excising the transits, with an inset displaying a subset of the Kepler data. \textbf{(b)} The posterior distribution for the Gaussian process period, which we interpret as a measurement of the stellar rotation period. }
\figsetgrpend

\figsetgrpstart
\figsetgrpnum{7.19}
\figsetgrptitle{KOI--1356}
\figsetplot{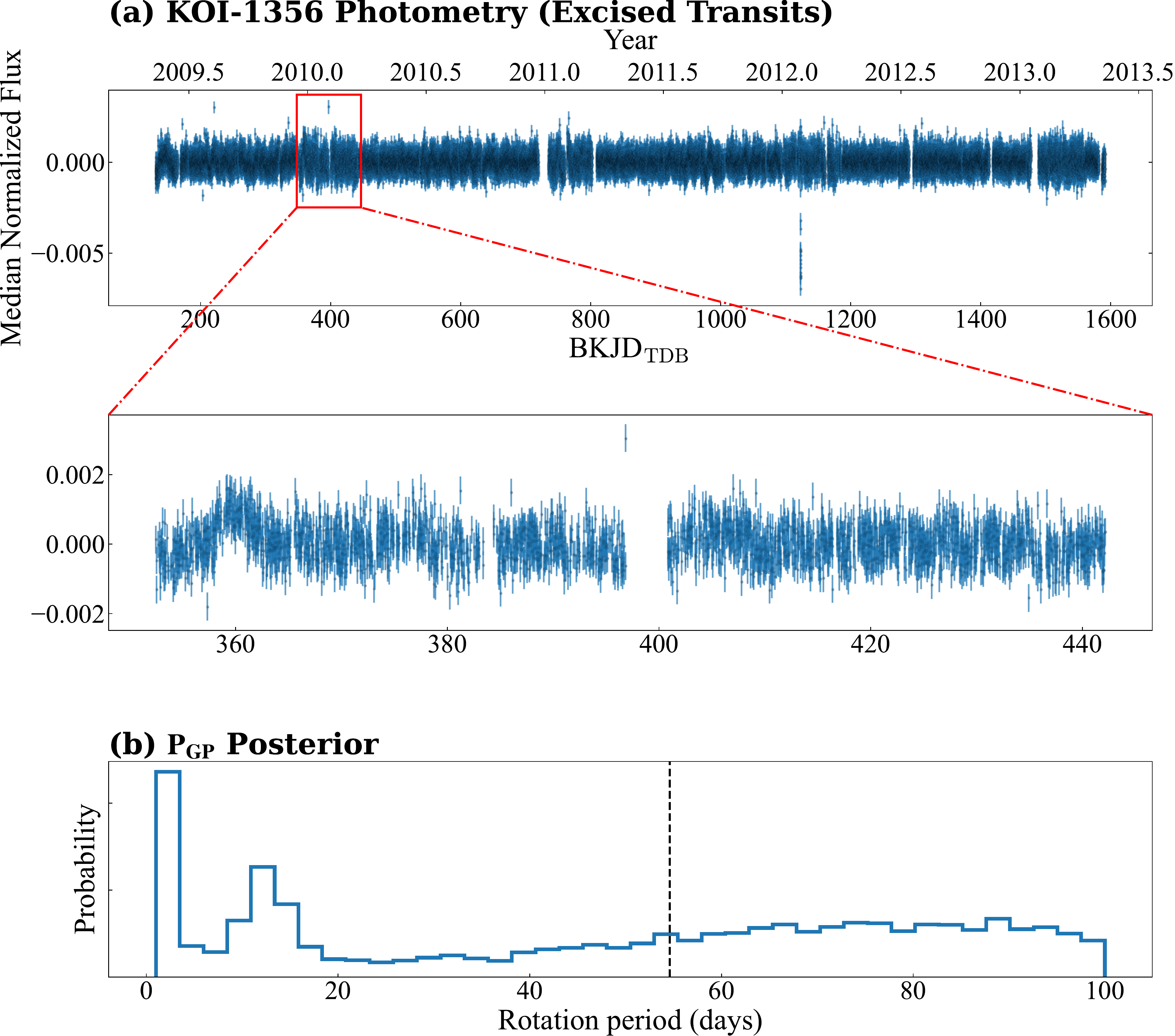}
\figsetgrpnote{\textbf{(a)} The \texttt{ARC2} corrected light curve for KOI-1356, after excising the transits, with an inset displaying a subset of the Kepler data. \textbf{(b)} The posterior distribution for the Gaussian process period, which we interpret as a measurement of the stellar rotation period. }
\figsetgrpend

\figsetgrpstart
\figsetgrpnum{7.20}
\figsetgrptitle{KOI--1416}
\figsetplot{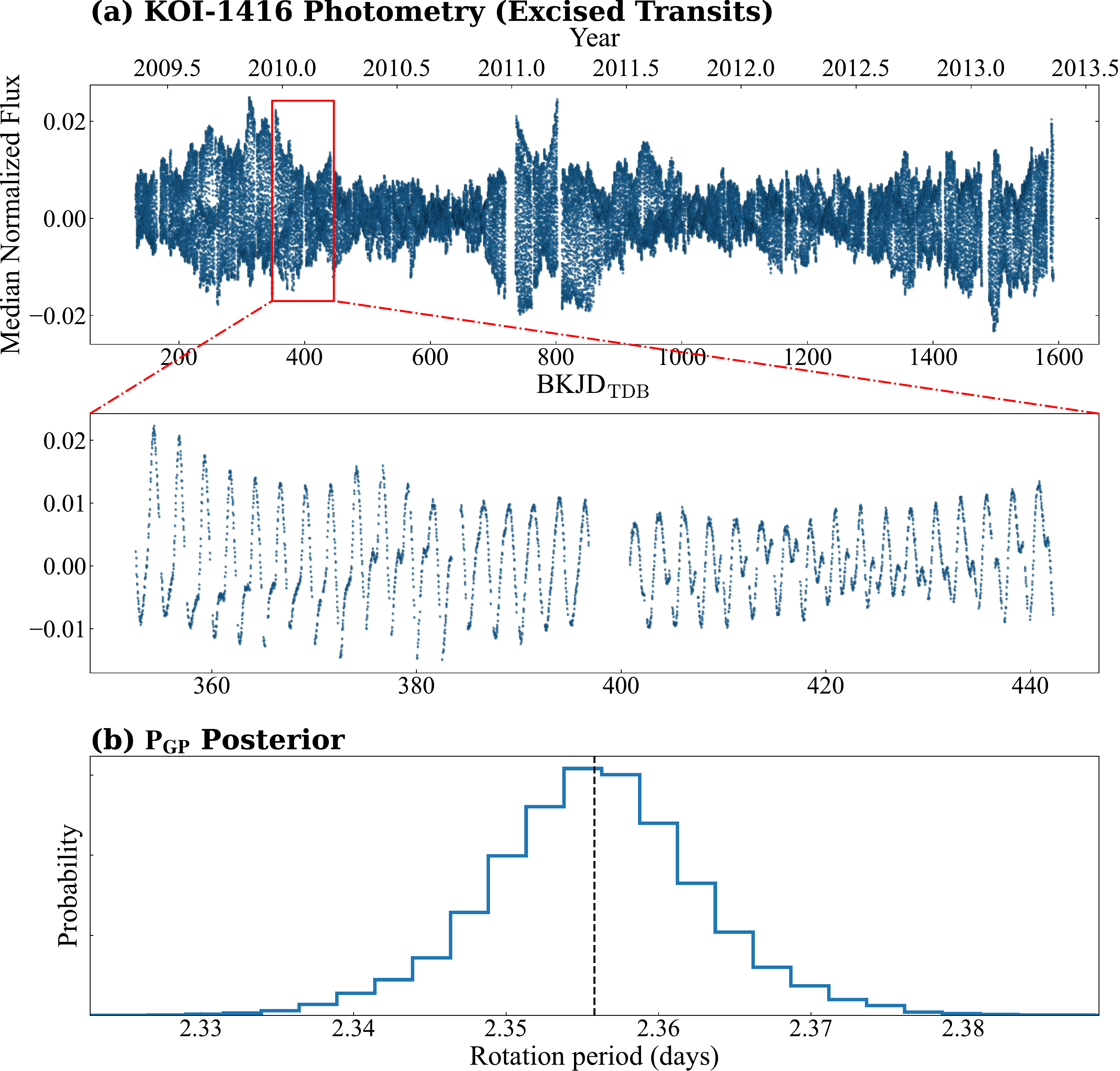}
\figsetgrpnote{\textbf{(a)} The \texttt{ARC2} corrected light curve for KOI-1416, after excising the transits, with an inset displaying a subset of the Kepler data. \textbf{(b)} The posterior distribution for the Gaussian process period, which we interpret as a measurement of the stellar rotation period. }
\figsetgrpend

\figsetgrpstart
\figsetgrpnum{7.21}
\figsetgrptitle{KOI--1448}
\figsetplot{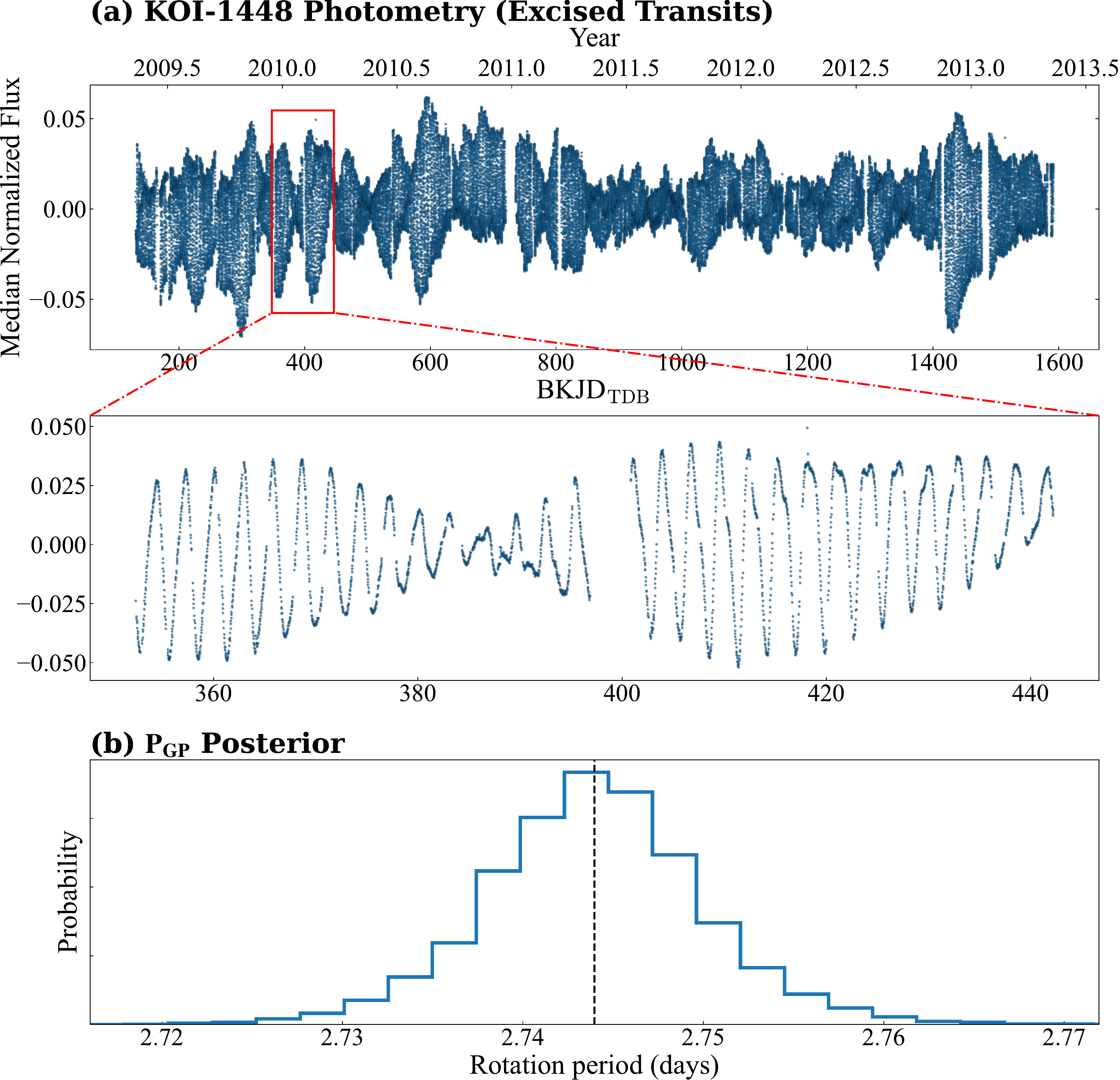}
\figsetgrpnote{\textbf{(a)} The \texttt{ARC2} corrected light curve for KOI-1448, after excising the transits, with an inset displaying a subset of the Kepler data. \textbf{(b)} The posterior distribution for the Gaussian process period, which we interpret as a measurement of the stellar rotation period. }
\figsetgrpend

\figsetgrpstart
\figsetgrpnum{7.22}
\figsetgrptitle{KOI--2513}
\figsetplot{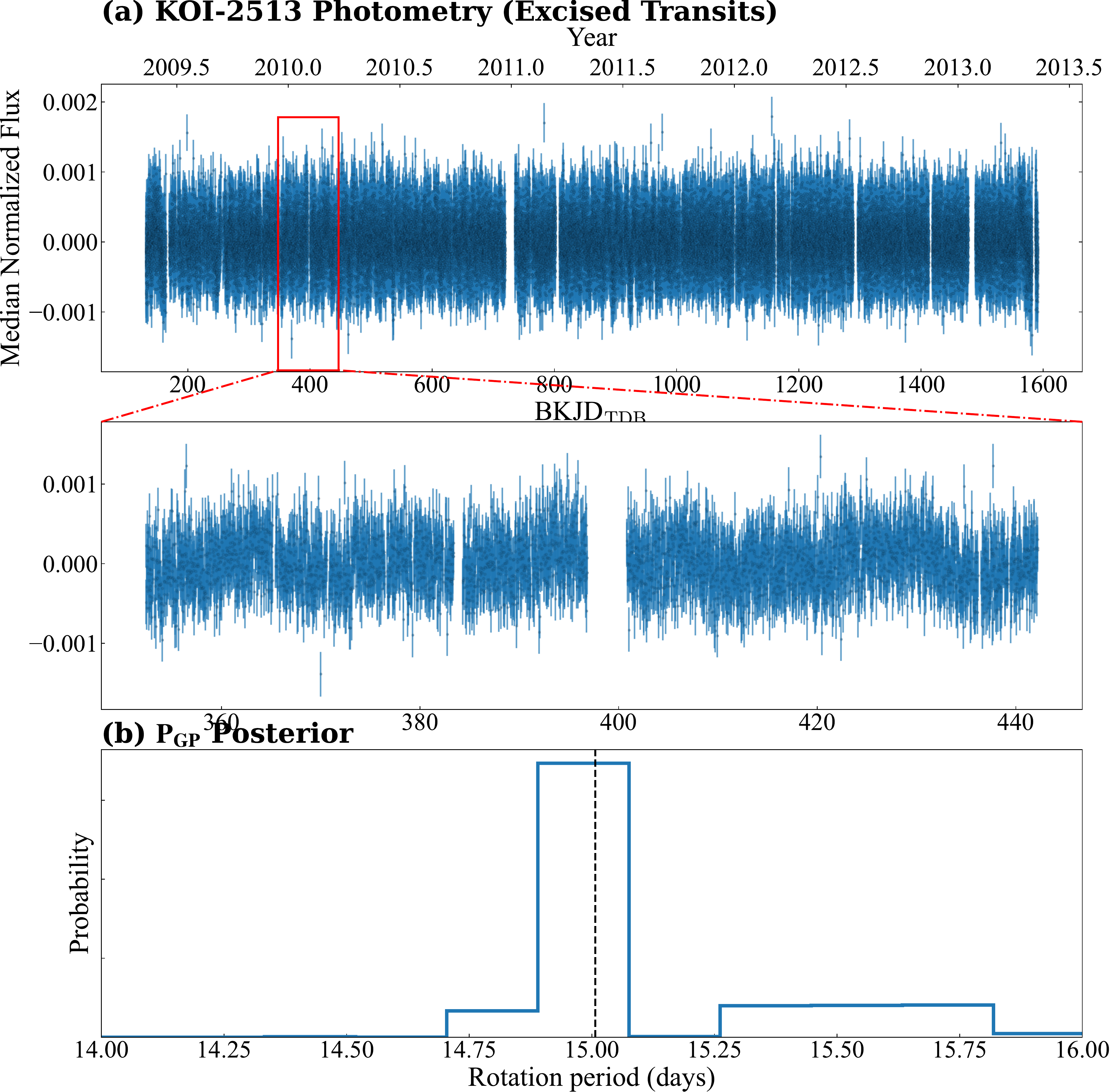}
\figsetgrpnote{\textbf{(a)} The \texttt{ARC2} corrected light curve for KOI-2513, after excising the transits, with an inset displaying a subset of the Kepler data. \textbf{(b)} The posterior distribution for the Gaussian process period, which we interpret as a measurement of the stellar rotation period. }
\figsetgrpend

\figsetgrpstart
\figsetgrpnum{7.23}
\figsetgrptitle{KOI--3320}
\figsetplot{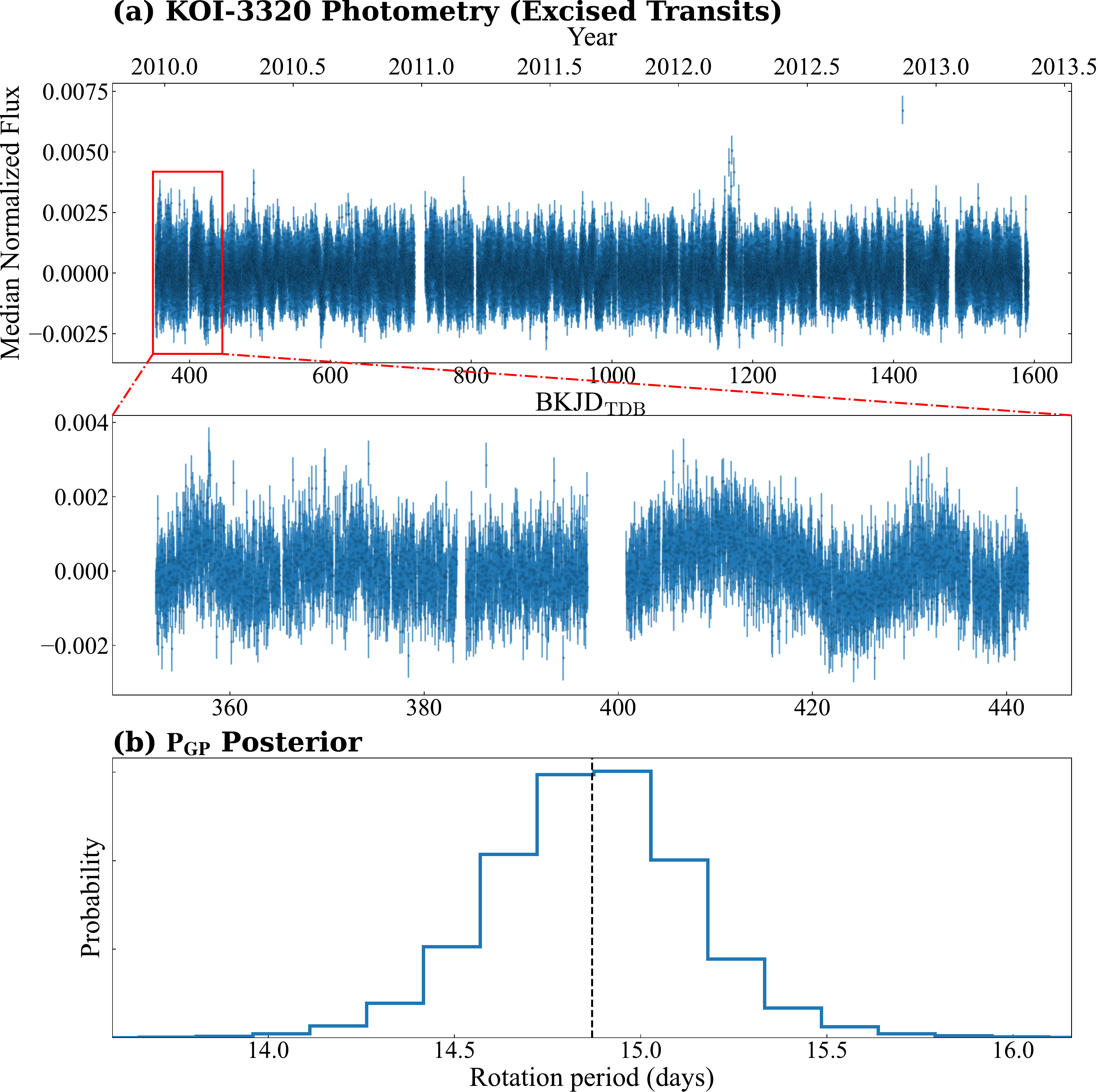}
\figsetgrpnote{\textbf{(a)} The \texttt{ARC2} corrected light curve for KOI-3320, after excising the transits, with an inset displaying a subset of the Kepler data. \textbf{(b)} The posterior distribution for the Gaussian process period, which we interpret as a measurement of the stellar rotation period. }
\figsetgrpend

\figsetgrpstart
\figsetgrpnum{7.24}
\figsetgrptitle{KOI--3358}
\figsetplot{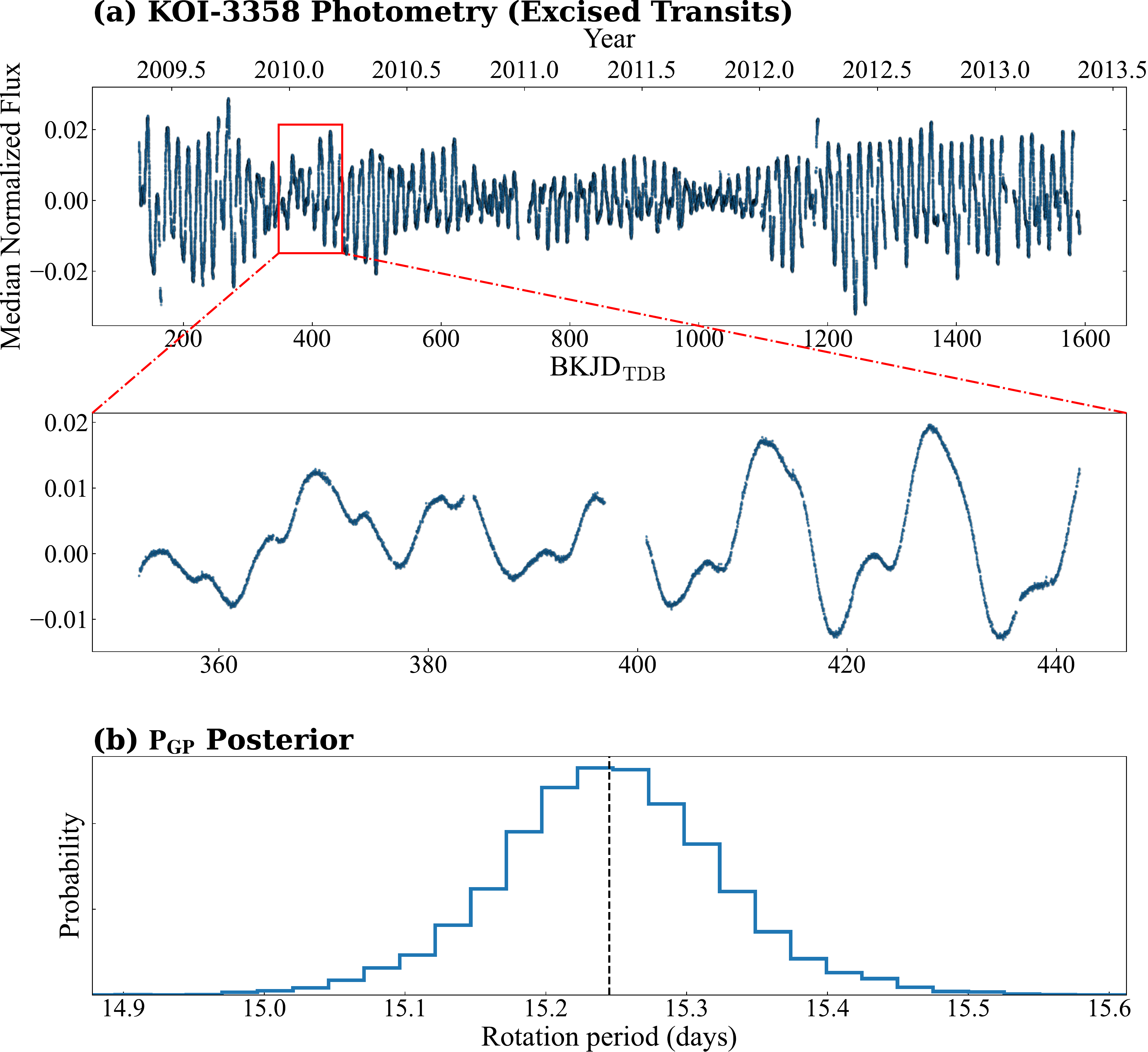}
\figsetgrpnote{\textbf{(a)} The \texttt{ARC2} corrected light curve for KOI-3358, after excising the transits, with an inset displaying a subset of the Kepler data. \textbf{(b)} The posterior distribution for the Gaussian process period, which we interpret as a measurement of the stellar rotation period. }
\figsetgrpend

\figsetgrpstart
\figsetgrpnum{7.25}
\figsetgrptitle{KOI--4367}
\figsetplot{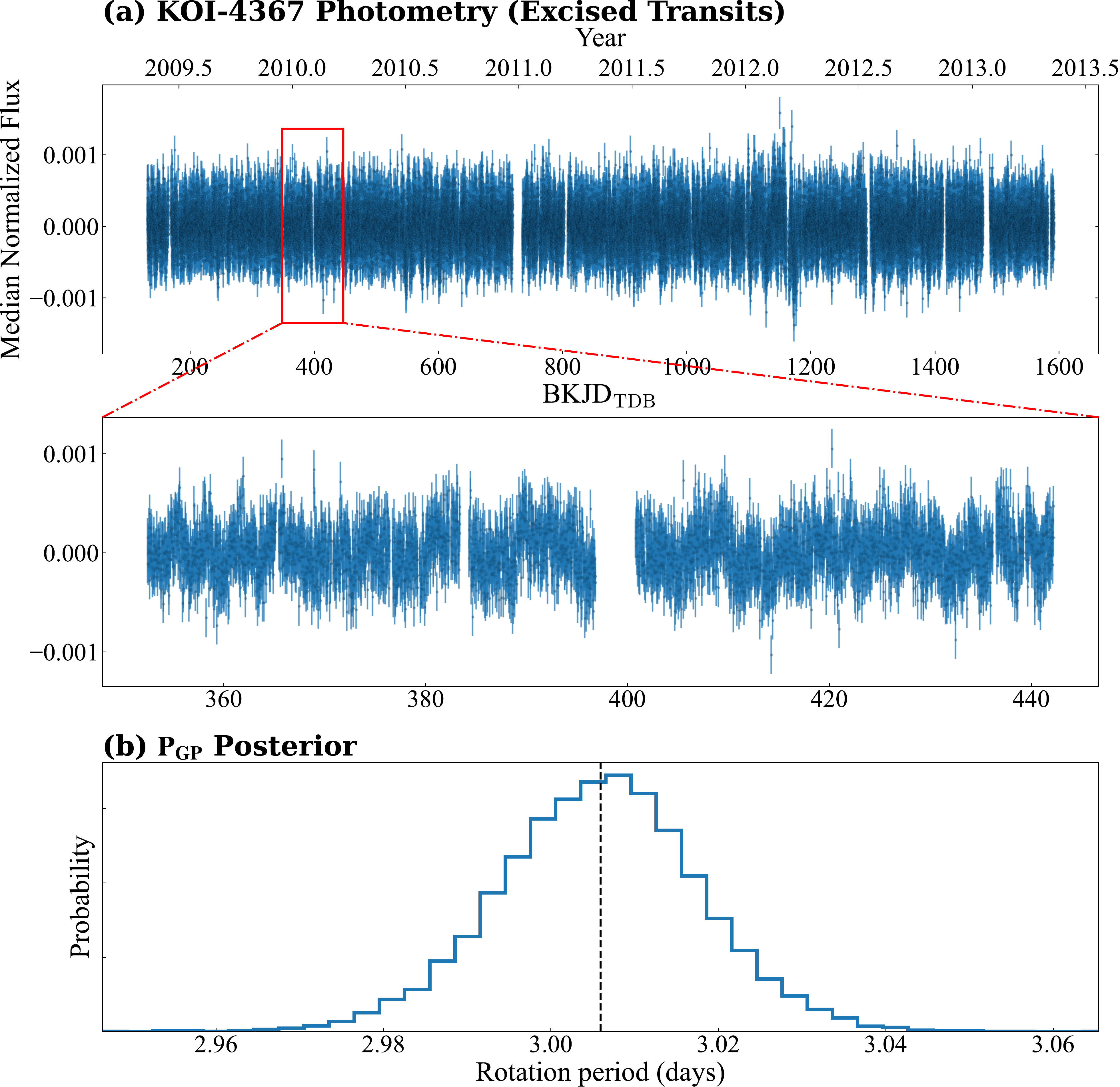}
\figsetgrpnote{\textbf{(a)} The \texttt{ARC2} corrected light curve for KOI-4367, after excising the transits, with an inset displaying a subset of the Kepler data. \textbf{(b)} The posterior distribution for the Gaussian process period, which we interpret as a measurement of the stellar rotation period. }
\figsetgrpend

\figsetgrpstart
\figsetgrpnum{7.26}
\figsetgrptitle{KOI--5329}
\figsetplot{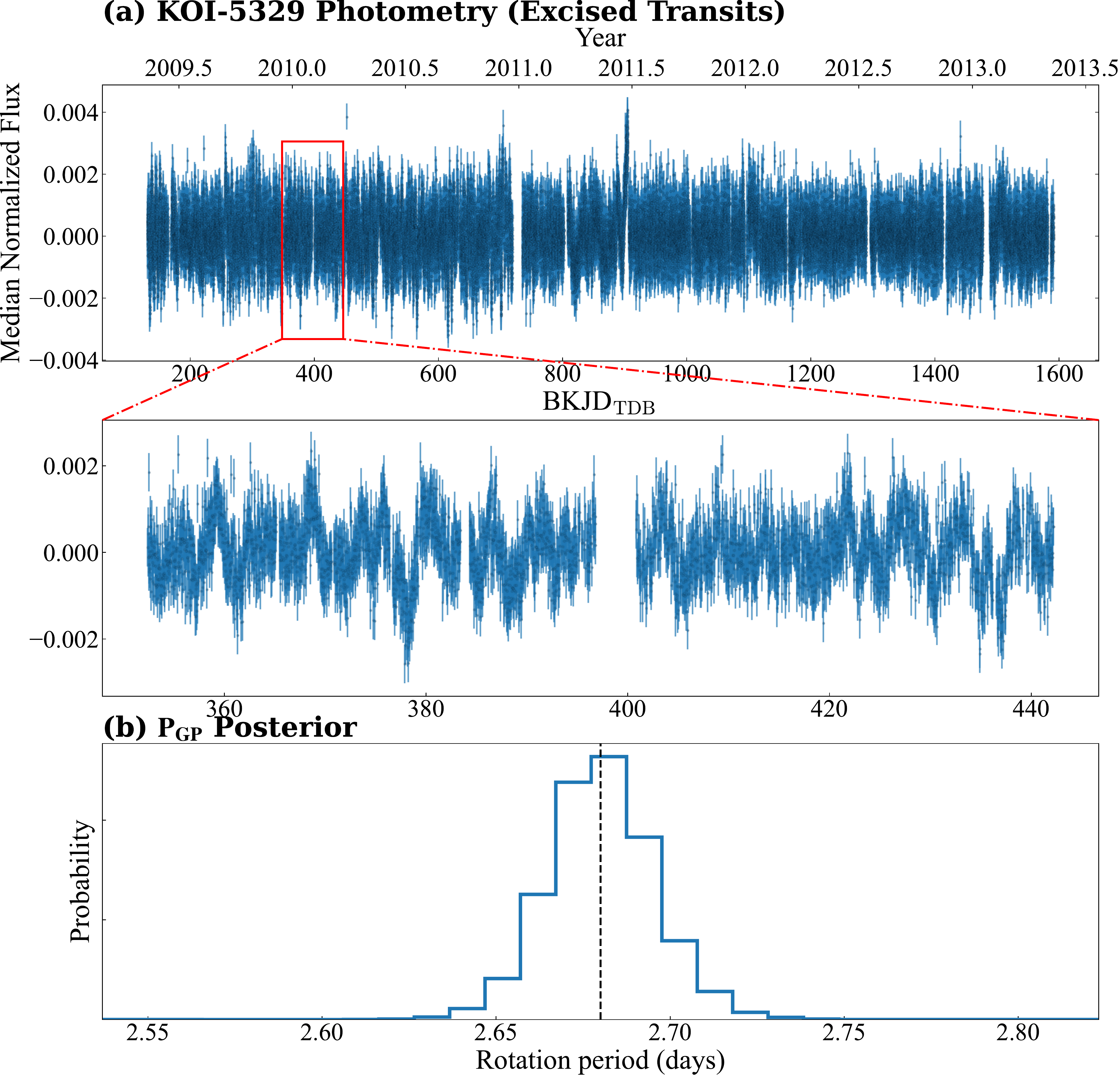}
\figsetgrpnote{\textbf{(a)} The \texttt{ARC2} corrected light curve for KOI-5329, after excising the transits, with an inset displaying a subset of the Kepler data. \textbf{(b)} The posterior distribution for the Gaussian process period, which we interpret as a measurement of the stellar rotation period. }
\figsetgrpend

\figsetgrpstart
\figsetgrpnum{7.27}
\figsetgrptitle{KOI--6018}
\figsetplot{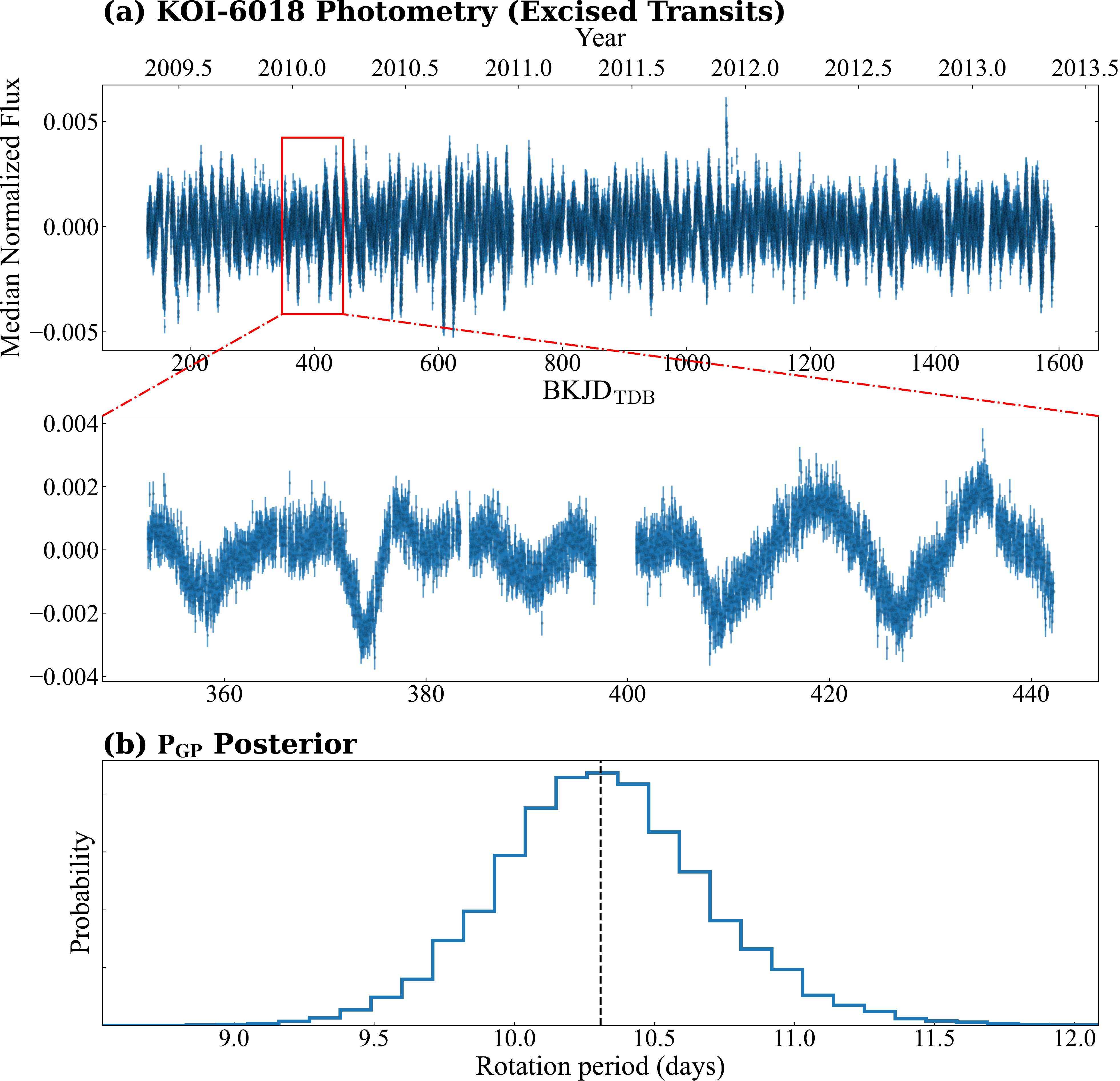}
\figsetgrpnote{\textbf{(a)} The \texttt{ARC2} corrected light curve for KOI-6018, after excising the transits, with an inset displaying a subset of the Kepler data. \textbf{(b)} The posterior distribution for the Gaussian process period, which we interpret as a measurement of the stellar rotation period. }
\figsetgrpend

\figsetgrpstart
\figsetgrpnum{7.28}
\figsetgrptitle{KOI--6760}
\figsetplot{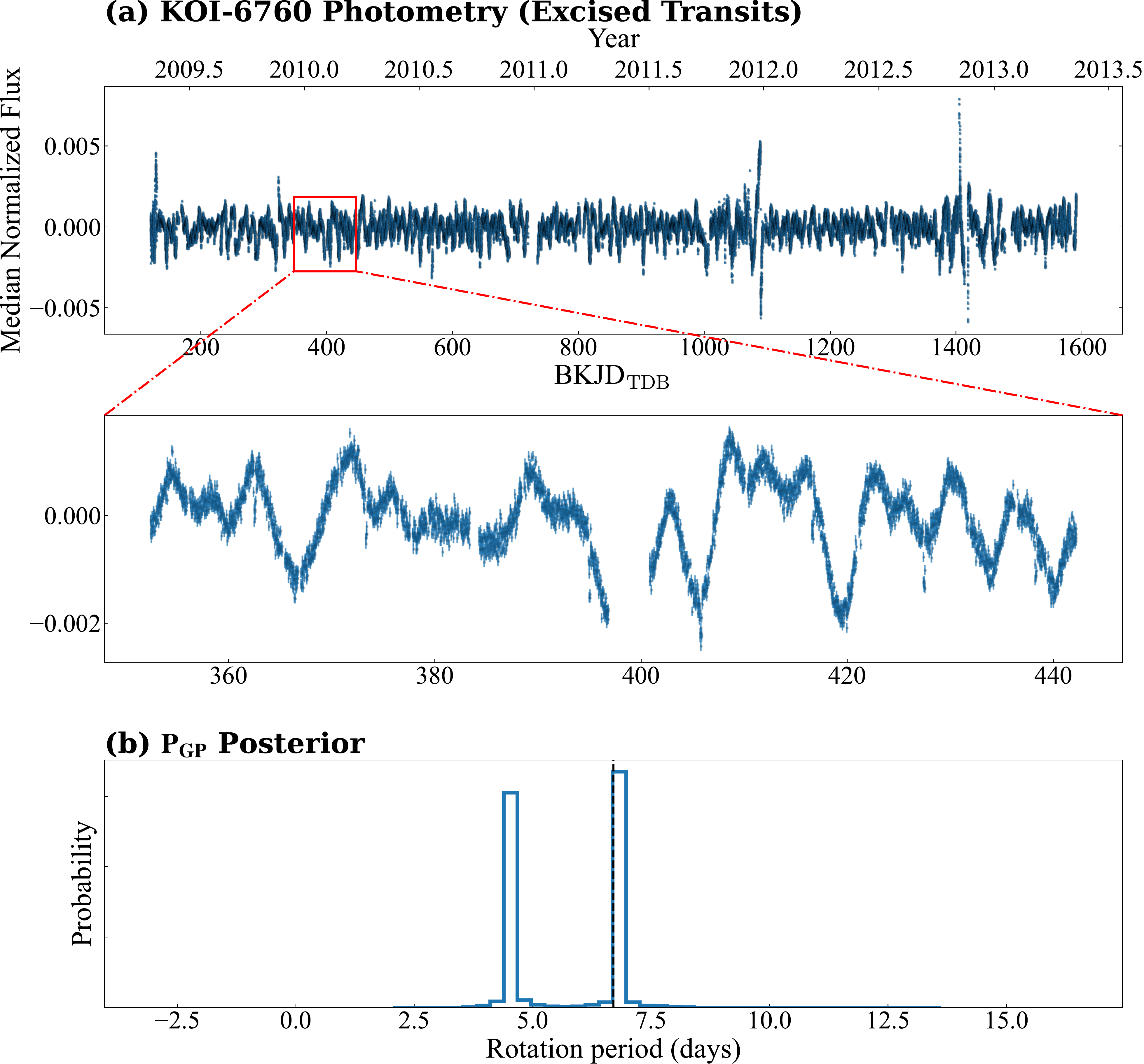}
\figsetgrpnote{\textbf{(a)} The \texttt{ARC2} corrected light curve for KOI-6760, after excising the transits, with an inset displaying a subset of the Kepler data. \textbf{(b)} The posterior distribution for the Gaussian process period, which we interpret as a measurement of the stellar rotation period. }
\figsetgrpend

\figsetend